\documentclass[pdfpagelabels,11pt,a4paper]{book}
\usepackage{nameref}

\usepackage{booktabs}
\usepackage{makeidx}\makeindex
\usepackage[scaled=0.9]{beramono}
\usepackage[T1]{fontenc}
\usepackage[protrusion=true,expansion=true]{microtype}
\usepackage[draft=false,hidelinks]{hyperref}
\usepackage{rotating}
\usepackage{graphicx}
\usepackage{tikz-qtree}
\usepackage{pdflscape}
\usepackage[left=3.5cm, right=2.5cm, top=2.5cm, bottom=2.5cm]{geometry}
\usepackage{amsmath}
\usepackage{cleveref}
\usepackage{amssymb}
\usepackage{ulem}
\usepackage{attrib}
\usepackage{xcolor}
\newcommand\addition[1]{#1}
\newcommand\correction[1]{#1}
\usepackage{emptypage}
\usepackage{etoolbox}
\usepackage{textcomp}
\usepackage{fancyvrb}
\usepackage{multicol}
\usepackage{makeidx}
\usepackage{upquote,textcomp}
\newcommand\upquote[1]{\textquotesingle#1\textquotesingle}
\usepackage{breakcites}
\usepackage{xspace}
\usepackage{enumitem}
\usepackage{ulem}
\setlist{  
  listparindent=\parindent,
  parsep=0pt,
}
\fvset{fontsize=\footnotesize}
\RecustomVerbatimEnvironment{verbatim}{Verbatim}{}
\usepackage{setspace}
\AtBeginEnvironment{quotation}{\itshape}
\usepackage{regexpatch}
\makeatletter
\xpatchcmd{\chaptermark}{\MakeUppercase}{}{}{}%
\xpatchcmd{\sectionmark}{\MakeUppercase}{}{}{}%
\xpatchcmd*{\tableofcontents}{\MakeUppercase}{}{}{}%
\makeatother
\usepackage{fancyhdr}
\renewcommand{\chaptermark}[1]{\markboth{#1}{}}
\renewcommand{\sectionmark}[1]{\markright{#1}}
\newcommand{\pletrec}{\textbf{let rec}\xspace}
\newcommand{\plet}{\textbf{let}\xspace}
\newcommand{\pin}{\textbf{in}\xspace}
\newcommand{\pmatch}{\textbf{match}\xspace}
\newcommand{\pwith}{\textbf{with}\xspace}
\newcommand{\pif}{\textbf{if}\xspace}
\newcommand{\pfun}{\textbf{fun}\xspace}

\newcommand{\pthen}{\textbf{then}\xspace}
\newcommand{\pelse}{\textbf{else}\xspace}
\newcommand{\praise}{\textbf{raise}\xspace}
\newcommand{\ptry}{\textbf{try}\xspace}
\newcommand{\ptype}{\textbf{type}\xspace}
\newcommand{\pof}{\textbf{of}\xspace}

\newcommand{\pexception}{\textbf{exception}\xspace}
\newcommand{\pwhile}{\textbf{while}\xspace}
\newcommand{\pfor}{\textbf{for}\xspace}
\newcommand{\pto}{\textbf{to}\xspace}

\newcommand{\pdo}{\textbf{do}\xspace}
\newcommand{\pdone}{\textbf{done}\xspace}

\newcommand{\pval}{\textbf{val}\xspace}

\makeindex
\begin{document}

\frontmatter

\title{Debugging Functional Programs by Interpretation}
\author{{}\\{}\\{}\\{}\\{}\\{}\\Thesis submitted for the degree of\\{}\\Doctor of Philosophy\\{}\\at the University of Leicester\\{}\\{}\\{}\\{}\\{}\\{}\\John Whitington\\{}\\Department of Informatics\\{}\\University of Leicester\\{}\\{\small July 2020} \\{}\\{}}
\date{}\maketitle

\cleardoublepage
\thispagestyle{empty}
\phantom{foo}
\vspace{2in}
\begin{center}
{\Large Debugging Functional Programs by Interpretation}\\
\vspace{4mm}
John Whitington\\
\vspace{10mm}
{\Large Abstract}
\end{center}
Motivated by experience in programming and in the teaching of programming, we make another assault on the longstanding problem of debugging. Having explored why debuggers are not used as widely as one might expect, especially in functional programming environments, we define the characteristics of a debugger which make it usable and thus likely to be widely used. We present work on a new debugger for the functional programming language OCaml which operates by direct interpretation of the program source, allowing the printing out of individual steps of the program's evaluation, and discuss its technical implementation and practical use.

It has two parts: a stand-alone debugger which can run OCaml programs by interpretation and so allow their behaviour to be inspected; and an OCaml syntax extension, which allows the part of a program under scrutiny to be interpreted in the same fashion as the stand-alone debugger whilst the rest of the program runs natively. We show how this latter mechanism can create a source-level debugging system that has the characteristics of a usable debugger and so may eventually be expected to be suitable for widespread adoption.

\newpage
\cleardoublepage
\thispagestyle{empty}
\phantom{foo}
\vspace{2in}
\begin{center}
{\Large Acknowledgments}
\end{center}
This research could not have been undertaken nor this thesis written without the continual support and good humour of my supervisors Tom Ridge, Neil Walkinshaw and José Miguel Rojas Siles. It was funded by the Engineering and Physical Sciences Research Council, part of UK Research and Innovation.

\newpage
\tableofcontents
\newpage
\chapter*{List of figures}

\begin{tabular}{@{}lp{14.25cm}@{}}
1. & Handwritten diagrams for length and append functions\dotfill \pageref{F1}\\ 
2. & Handwritten diagram for the factorial function\dotfill \pageref{F2}\\ 
3. & Two computer visualizations of the factorial function\dotfill \pageref{F3}\\ 
4. & Handwritten diagram for insertion sort\dotfill \pageref{F4}\\ 
5. & The CodeCenter C interpreter\dotfill \pageref{F5}\\ 
6. & Handwritten diagram for a function using exceptions for control flow\dotfill \pageref{F6}\\ 
7. & Two handwritten diagrams of a pattern-match\dotfill \pageref{F7}\\ 
8. & Pattern matching by pattern matching\dotfill \pageref{F8}\\ 
9. & \addition{Interpreting inside the Standard Library}\dotfill \pageref{F9}\\
10. & \addition{Growth of time and space usage in \textsf{OCamli}}\dotfill \pageref{F10}\\
11. & A shim for a function\dotfill \pageref{F11}\\
12. & Compilation of a program to bytecode\dotfill \pageref{F12}\\
13. & Evaluation of a bytecode program\dotfill \pageref{F13}\\
14. & Decompilation of a bytecode program\dotfill \pageref{F14}\\
15. & Decompilation of a partly-evaluated bytecode program\dotfill \pageref{F15}\\
\end{tabular}

\mainmatter

\newpage
\thispagestyle{empty}
\topskip0pt
\vspace*{\fill}
\begin{quotation}\textit{\Large I am in much dismay at having got into so amazing a quagmire \& botheration with these Numbers, that I cannot possibly get the thing done today. \ldots\ I am now going out on horseback. Tant mieux. \textrm{\begin{flushright}--- Lovelace to Babbage, July 1843\end{flushright}}}\end{quotation}
\vspace*{\fill}

\chapter{Introduction}
\label{chap:introduction}

\begin{quotation}\textit{\large All beginnings are delightful; the threshold is the place to pause.\textup{\begin{flushright}--- Goethe\end{flushright}}}\end{quotation}

\vspace{20pt}

\noindent This thesis addresses the question of debugging. Since the beginning of the computer age, the concept has been known, discussed, and solutions have been engineered. And yet, many programmers never touch a debugger. Is this a fundamental problem -- is the whole notion of debugging tools a mirage -- or is it simply that the right solutions have yet to be found?

And so, we have another go at the problem of debugging. It might be argued that ``having another go at something'' does not constitute research. We disagree wholeheartedly. To re-tread a path well-trodden with failure or limited success is not simply to repeat. To take the experience of the past  with the technologies of today, beat back the weeds, and see if we can put some proper paving down, is worth the effort -- and an essential part of the research process.

We concentrate our efforts on functional programming languages: long a niche area with promise; suddenly in the past ten years a mainstream field. There are certain characteristics of the functional way which are amenable to a different kind of debugging, one which we will show alleviates many previous barriers to practical debuggers.

We have produced some new debugging tools which show promise.  We have learnt a great deal about the possibilities and inherent limitations of debugging. But, before all that, let us pause, and set the work in context.

\section{Motivation from teaching}

\noindent When teaching functional programming\index{functional programming!teaching}, we like to draw diagrams on paper\index{evaluation!handwritten diagrams of} like that shown in figure 1.
\begin{figure}
{\centering\hspace{3mm}\includegraphics[width=0.4\textwidth]{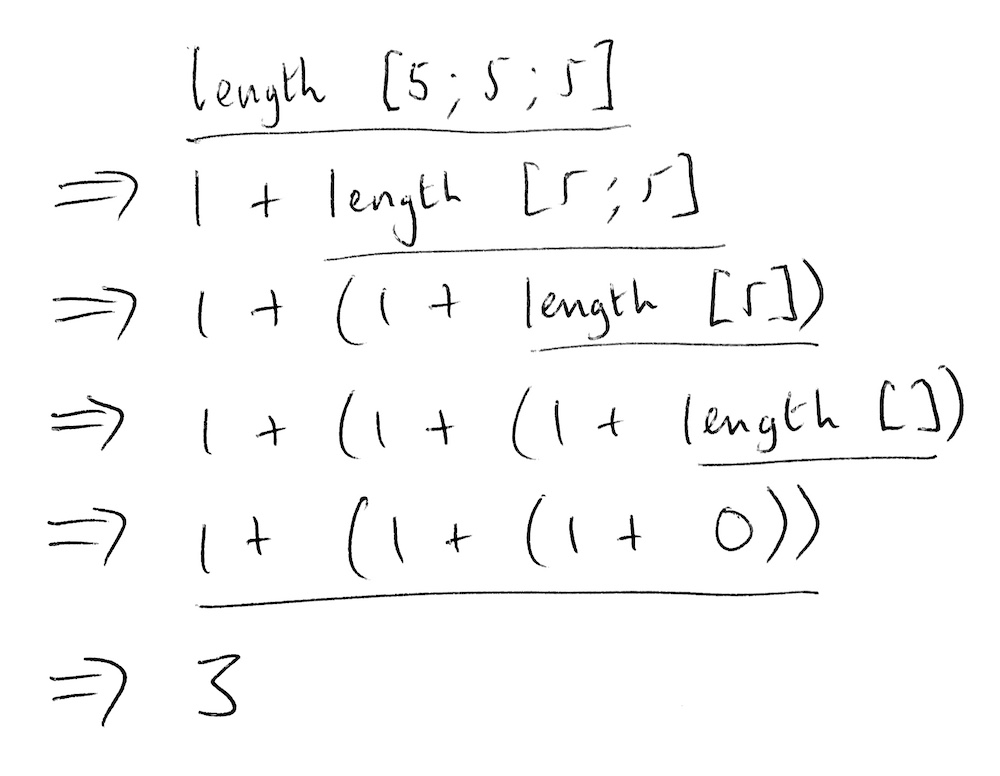}\includegraphics[width=0.55\textwidth]{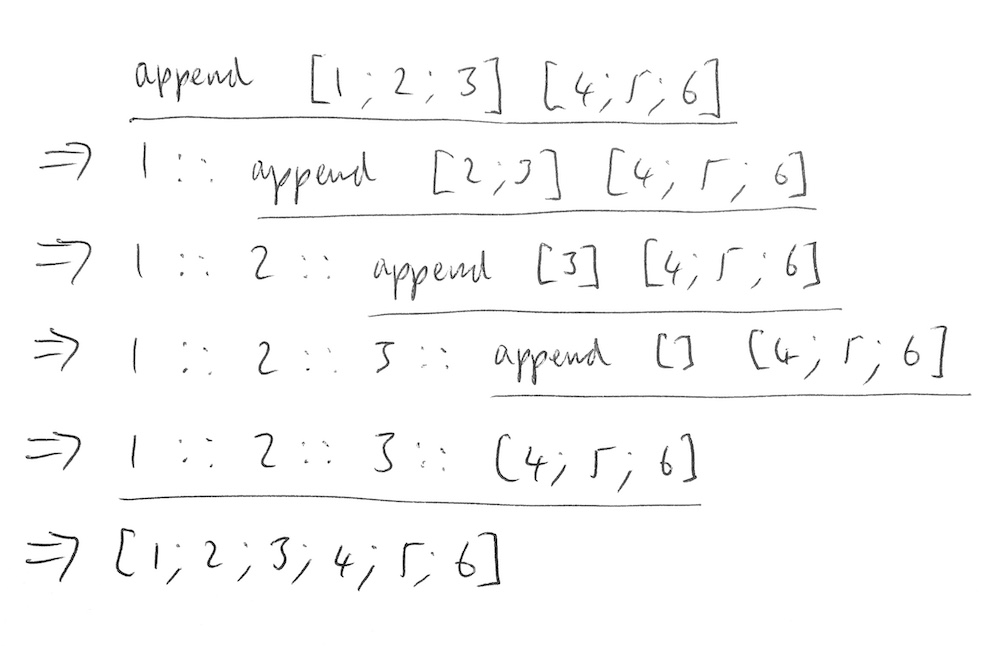}}

{\label{F1}\centering\singlespacing\noindent\small\hspace{-3mm} Figure 1. Handwritten diagrams for finding the length of a list and for appending two lists.\par}
\end{figure}
Such diagrams help to visualize the computation\index{computation!visualization of}\index{visualization} not only in terms of what it computes, but the number of steps taken (roughly the number of lines in the diagram), the stack space\index{stack space} required for a function which is not tail-recursive\index{tail recursion} (roughly the width of the diagram), and the subexpression being evaluated at any step (the underlined portion). They can also be useful to the more experienced programmer as a debugging aid. Stepping through a computation is one of the building blocks of debugging, though this is typically done with a compiled form, where much information is not available. Stepping through in source form means we have all information to hand.

What makes it possible to draw these diagrams on paper is the \index{functional programming!immutable nature of} immutable nature of most functional programs. Stepping through an imperative program\index{program!imperative} on paper involves drawing little boxes as storage locations and updating them, rubbing out and writing in new values as the program runs. However, whilst we need only write the source code of the imperative program once, in our functional diagrams we have had to repeatedly write out almost the same expression line after line. Such repetition suggests automation via computer would be useful. Notice also that our diagrams, when written by hand, implicitly omit information\index{elision!in handwritten diagrams}, such as the definitions of the functions and some of their internal workings. Thus, such automation will entail significant visualization and efficiency challenges, especially with large programs.

\section{Motivation from the working programmer}

The dream of debugging is this: having observed a misbehaviour caused by a bug, we quickly and easily assemble all relevant information, both about the program source and the full trace of the program's operation, and, describing ourselves concisely to the computer, we narrow the circumstances of the failure down, again and again reducing the search space, until we have the bug in our grasp, and understand it fully. The fix is then often easy.

And yet, this dream so often falls apart. Our program has been compiled and only limited debugging information is available, creating a lossy map back to the source locations, and we are in the dark. Our tools might not operate on all of the program for obscure technical reasons. We may lack a way of converting our intuition about what might cause the misbehaviour into a command to give the debugger to aid in the narrowing-down process. We throw the debugger away and insert a few print statements\index{print statement}\index{debugging!by print statement} and spend an  afternoon on what, in our dream world, we could fix in minutes.

We hope to show that, whilst debugging will always remain a somewhat mercurial, individual process, there is plenty of scope for improvement in the tools -- provided we stick steadfastly to some important principles.

\section{Research questions}

\index{research questions}It is useful, when beginning a piece of research, to write down the questions we have in our mind. As we refine the scope during literature review and, of course, due to discoveries during the design and implementation of our tools, we will doubtless drop some of these, add others, and change emphasis. Nonetheless, we list now the research questions, as they were written down early in the project. We shall return to them when evaluating our work.

\vspace{10pt}

\noindent \textbf{It appears that debuggers are not as widely used as one might expect, despite being common for decades. Why?} This appears to be doubly true for functional programming languages. How does debugging practice vary among languages (compiled vs  interpreted, stateful vs stateless)? What can we learn from debugging theory and practice since the dawn of the computer age?

\bigskip
\noindent \textbf{Can we find a good way to visualize functional program execution?} Is the automatic production of such diagrams always going to be inferior to drawing them on paper? How can we deal with scale? How can we show exceptions? What about imperative and mutable features? What are the practicalities of directly interpreting an \index{Abstract Syntax Tree}Abstract Syntax Tree? Can the direct interpretation of the AST of a program ever or always have the same time or space complexity as running the compiled program?

\bigskip
\noindent \textbf{Is such an interpreter useful for debugging?} Taking into account of the current practice of debugging, is an interpretive debugger better than what is already available? If so, why, and in what ways?

\bigskip
\noindent \textbf{Is there an alternative abstract machine\index{abstract machine} which might allow for this kind of visualized debugging?} That is to say, are we condemned to interpret the AST in the simplest way if we want to be able to properly visualize the evaluation in a human-readable manner? We know that this may have greater complexity of running time than well-known abstract machines. Can we design a \index{bytecode}bytecode that retains the ability to produce source code for the running computation upon demand, but which is much faster than brute interpretation and maybe even close to that of a normal bytecode system?

\bigskip
\noindent \textbf{Could we, instead, build an interpreter which can work alongside the native code execution of a program, interpreting only when required?} We could compile a program in a slightly different manner. It would run as usual, but when it comes to a part we wish to debug, it would begin interpretation. After this part, it would return to native code, as if nothing had happened.

\section{Thesis}

It is as well to foreshadow our central claims now, though we have not yet given support for them. We claim:

\begin{itemize}
\item That not many programmers use debuggers, even though they exist.
\item That (almost) everyone could benefit from a debugger.
\item That the reasons for this disparity are frequently incidental\index{incidental complexity}\index{intrinsic complexity}  rather than intrinsic and include:
\begin{itemize}
\item The inability of debuggers working on compiled programs to properly reflect their workings at a source code level.
\item The requirement to adapt build systems\index{build system}, tools, and environments to ensure everything is ready for the debugger to be used at any time.
\end{itemize}
\item That many of these barriers melt away in the presence of an interpreter ranking equally to a native compiler, with the same language and \index{toolchain}toolchain support.
\item That the huge disadvantage of slowness\index{interpretation!speed of} which comes with this approach:
\begin{itemize}
\item Can be ameliorated more that one might expect.
\item Is not a show-stopper for most uses, since the external steps of debugging such as case reduction are still in place.
\item May be obviated by finding a way to produce mixed native/interpreted programs so that, in any case, the need for interpretation is much reduced.
\end{itemize}
\item That such an interpretive approach is particularly suited to functional programs due to the mental model of calculation.
\end{itemize}

\noindent Our thesis is this: an interpretive debugger for a functional language is technically practicable and might be expected to make debugger use more widespread, routine and productive.

\section{Contributions}

The contribution of a practical tool-based piece of research such as this should be twofold: a new program or piece of code; and a new approach to a problem, exemplified by that tool, but which has reusable lessons for future researchers and implementers. Our principal practical, or technical contributions are:

\begin{itemize}
\item A direct interpreter for syntax trees of the statically-typed strict functional language \index{OCaml}OCaml  \cite{ocaml}, demonstrating how such interpretation might work for real, sizeable programs.
\item A mechanism by which the program is run natively, save for the  focus of debugging, which is interpreted, forming a single executable which prints out step-by-step evaluation only of a given part of the program.
\end{itemize}

\noindent The implementation of these tools is unfinished; but enough exploratory work has been done to be reasonably sure that finishing them is a practical, not a research task. In terms of our wider contribution, our research output has also created:

\begin{itemize}
\item A narrative literature review of the history of debugging -- in particular the very early development of debugging -- from many perspectives, putting modern debugging into context.
\item A survey of modern debugging systems for functional programming languages, sketching the similarities and differences between systems for lazy and strict languages, and following progress over time.
\item A collation of the characteristics which make a debugging system usable, and thus likely to be used. These rules, we believe, transcend our focus on functional programming, and should apply more widely.
\end{itemize}

\section{Note}

This thesis contains paragraphs whose content is similar or, in some cases, identical to those in our recent papers \cite{whitington, eptcs}.

\section{Summary}

We have given a little preview and summary of the work in this thesis, without recourse to too much technical detail.\label{decision1} \addition{We have described various concrete decisions: (1) to restrict ourselves to functional programming; (2) to work by analogy to drawn diagrams; and (3) to try to build an interpreter which ranks equally with the compiler. The first two of these decisions were taken right at the beginning of the project and, not being countermanded by the evidence gathered later, still stem from that time. The third, as an explicit aim, emerged only after the literature review. And so the justification for this latter decision must wait for chapter 3.}

Before coming to that technical detail, we pause again, to review the history and present practice of debugging.

\chapter{Related work}
\label{chap:related}

\begin{quotation}\textit{\large We come mentally of age when we discover that the great minds of the past, whom we have patronized, are not less intelligent than we are because they happen to be dead. \textrm{\begin{flushright}--- \textup{Cyril Connolly,} The Unquiet Grave\end{flushright}}}\end{quotation}

\bigskip

\noindent In order to choose how to attack our problem, we shall need to know what has come before and to what extent it has succeeded or failed. We begin with an informal \index{literature survey}\index{survey of literature}survey of the academic literature in debugging from just after the Second World War. Then, we shall look at recent work in functional languages, since that is our particular area of interest. \addition{Finally, we shall look at modern debuggers for imperative languages, seeing if there is anything to learn from them which we might apply to the functional world.}

\section{Seventy years of debugging 1949--2019}

We take a broad look at debugging literature of all kinds over the last seventy years, seeking to tease out the principles which survive and problems which resist the enormous changes associated with the so-called computer revolution. We shall see the same ideas, or rhymes of ideas, again and again. This gives us hope -- the problems are agreed upon, but also trepidation. If no-one has solved them yet, why should we imagine we can?

\subsection{Debugging defined}

Before we discuss debuggers, and in particular our new debugger, we ought to make sure our basic definitions are secure. What is  \index{debugging!definition of}debugging? In ``A Software Debugging Glossary'', Johnson defines it like this: \textit{debugging: 1) n. the process of \index{bug!isolation}isolating and \index{bug!correcting}correcting mistakes in computer programs. 2) adj. having to do with debugging 
\cite{johnson1982software}}. Most important here is the word `process'. Already we can see the separation between isolation and correction, and that there is an order: isolate first, then correct. In his survey ``Testing: principles and practice'', Schach uses one method of isolation, that of testing, to define debugging: \textit{The correction of a fault exposed by testing is termed \textit{debugging}  \cite{schach1996testing}}. If we use a sufficiently broad definition of testing, to include accidentally coming across a fault as well as structured, active testing, this is as useful as the previous definition. Ko contrasts again detection of errors (called testing) and debugging: \textit{Whereas verification and testing detect the \textit{presence} of errors, debugging is the process of finding and removing errors 
\cite{ko2011state}}. So, we begin debugging when and only when we have found a fault. Parnin and Orso break down the process into three subprocesses, and define debugging as the sum total of those:

\begin{quotation}
The first activity, \textit{fault localization}, consists of identifying the program statement(s) responsible for the failure. The second activity, \textit{fault understanding}, involves understanding the root cause of the failure. Finally, \textit{fault correction} is determining how to modify the code to remove such root cause. \index{fault!localization}Fault localization, understanding, and correction are referred to collectively with the term \textit{debugging}.
\cite{Parnin:2011:ADT:2001420.2001445}
\end{quotation}

\noindent Scowen subdivides debugging tools into two kinds, based upon when they do their work in relation to the point at which the existence of a bug is identified by the programmer:

\begin{quotation}
Debugging tools can be classified as \index{debugger!active}active or \index{debugger!passive}passive. An active tool is one which enables the programmer to specify what he wants after he has realized there is an error\ldots\ A passive tool works automatically without any effort from the programmer, e.g.\ failure messages, store post-mortem,  etc.
\cite{scowen1972debugging}
\end{quotation}

\noindent In a nod to debugging as an essentially human process, in which boundaries are blurred, and in which rigour and instinct play equal roles, Hailpern and Santhanam advise us: \textit{
Note that the terms `debugging', `testing', and `verification' are not mutually-exclusive activities, especially in everyday practice \cite{Hailpern:2002:SDT:1660992.1660994}}. In contrast, in line with the wholly rigorous procedures beloved of certification authorities, here is Ko's precis of the relevant \index{IEEE definitions}\index{bug!IEEE definition of}IEEE definitions of the conditions leading to the need for debugging -- what we and our quotations have hitherto referred to as faults or errors:

\begin{quotation}
In this paper, we use definitions of \index{error!IEEE definition of}\textit{error}, \index{fault!IEEE definition of}\textit{fault} and \index{failure!IEEE definition of}\textit{failure} from IEEE standard 610.12-1990. A \textit{failure} occurs when a program's output does not match documented output requirements, or the programmer's mental model of output requirements. Failures are ultimately the result of a \textit{fault}, which is a run-time state of a program that either is or \textit{appears to be} incorrect (as in assuming a lack of output from a debugging print statement to mean the code is not reached). Faults occur as a result of \textit{errors}, which are program fragments that do not comply with documented program specifications or the programmer's mental model of specifications (such as a missing increment or misspelled variable name). Failures are usually the first indication to the programmer that one or more errors exist in the program, although some errors are found before they cause failures, since they may be similar to errors already found or may be apparent when inspecting code. While a failure guarantees one or more faults exist and a fault guarantees one or more errors exist, errors do not always cause faults, and faults do not always cause failures. 
\cite{ko2003development} summarising \cite{IEEE}
\end{quotation}

\noindent It is worth remembering that debugging may (and frequently does) exist, of course, entirely in the absence of debuggers. From simply staring at code until one finds the fault, to experimenting with multiple inputs or inserting print statements\index{debugging!by print statement}\index{print statement} to localise the problem. In fact, one of our claims is that not many people use debuggers (or at least not regularly), and one of our questions is simply ``Why Not?''. We must answer these questions first if we are to devote time to constructing yet another debugger. Here is Taylor, on the practical uses of print statements:

\begin{quotation}
The use of print statements\index{debugging!by print statement}\index{print statement} is one of the most common techniques adopted by programmers using an imperative language, when faced with a program whose behaviour they do not understand. By using print statements programmers can achieve a number of goals: they can confirm that a piece of code is being executed; they can trace the state of a number of variables; they can trace when a procedure is being called and with what arguments, and so on. \cite{taylor-thesis}
\end{quotation}

\noindent Whilst considering what debugging is, it is worth wondering whether it could be eliminated. So far, this has proven largely to be a false dawn -- even today, very few programs are formally proved correct (in any event, one may need to debug the proofs).

\begin{quotation}
Considerable interest has been shown in recent years in the development of methods for proving that a given computer program has certain properties. If this avenue of research proves successful, we may one day see the virtual elimination or at least diminution in importance of the program debugging process.
\cite{evans1966line}
\end{quotation}

\begin{quotation}
Debugging would be unnecessary if programs could be \index{program!proof of correctness}\index{proof}proved to be correct \ldots\ and debugging is likely to be necessary for some time yet.
\cite{scowen1972debugging}
\end{quotation}

\noindent Dijkstra gives a similar argument:

\begin{quotation}
If you want more effective programmers, you will discover that they should not waste their time debugging, they should not introduce the bugs to start with.
\cite{dijkstra1}
\end{quotation}

\begin{quotation}
Already now, debugging strikes me as putting the cart before the horse: instead of looking for more elaborate debugging aids, I would rather try to identify and remove the more productive bug-generators!
\cite{dijkstra2}
\end{quotation}

\noindent Now that we have defined debugging, let us continue on by defining debuggers (and in terms better than ``something which helps one debug''). Again, let us begin with Johnson: \textit{debugger: n. a collection of software tools to aid in debugging\index{debugger!definition of a}  \cite{johnson1982software}}. It seems very common that the debugger is considered a separate tool, and not part of the programming language or its compiler. Yes, a compiler\index{compiler} may add special information to an executable to help the debugger, and yes, a compiler may ship with a debugger as part of an Integrated Development Environment, but programmers tend to think of the debugger, however convenient to use, as a separate entity. Satterthwaite explains the role in a little more detail: \textit{\ldots a \textit{debugging tool} is a service provided by the computing system to reduce the number of aspects of a program's behaviour which are misunderstood or poorly understood by the user 
\cite{satterthwaite1972debugging}}. However, he is careful to point out an essential tension: we only need debuggers because we are inadequate programmers: \textit{Debugging tools are not intended to relieve the programmer of any obligations to analyse his problem carefully or to construct his program in a disciplined manner  \cite{satterthwaite1972debugging}}. Gaines agrees but, even writing in 1969, it is clear to him that, whilst \index{programming!by experimentation}experimental programming is usually undesirable, it happens anyway: \textit{Occasionally it is difficult to distinguish between debugging and writing of programs. This arises because the programmer is writing a program `experimentally' \cite{Gaines:1969:DCP:905460}}

We must try as hard as we can, using talent, discipline and all the tools of modern programming languages such as compilers and type systems, but in the end, there will likely always be the need for debugging as a process distinct from, and happening after, programming itself.

\subsection{Early thoughts}

Before continuing on to a detailed discussion of debugging theory and practice, it will be instructive to look at some of the very earliest published descriptions of what we now call debugging from those working near the beginning of the computer age, just after the Second World War. These early papers, being at the very beginning of the field, are written in a straightforward, practical way. They are often the first thoughts about topics and definitions and words we now take for granted. Looking back at how these thoughts were first formed can give an interesting perspective to our work on modern-day debugging. Our principal sources will be Maurice Wilkes and Stan Gill's descriptions of their work on the \index{EDSAC computer}EDSAC computer \cite{wilkes1949programme,Gill538}, and Ira Diehm's 1952 ACM National Meeting article ``Computer aids to code checking'' \cite{diehm1952computer}.

Writing in 1949, Wilkes has to define even the word programming for us, bringing into focus that, at that time, building the physical computer was the primary difficulty, and providing for it the correct sequence of instructions to make it perform the correct calculations for solving a given problem a secondary concern.

\begin{quotation}
A good deal has been written about the design and construction of high-speed automatic calculating machines, but little has been said about the detailed steps which are necessary to prepare a problem for the machine and to obtain a solution -- a process which is usually referred to as `programming'.
\cite{wilkes1949programme}
\end{quotation}

\noindent We are reminded also of the division of labour. The job of the computer to silently and quickly obey a set of instructions; that of the programmer to provide the correct list of instructions. Again, this seems obvious now, but was perhaps less so at the very beginning.

\begin{quotation}
The EDSAC is designed on the supposition that the programmer is responsible for making sure that his programme is correct and that the operations called for are correctly carried out by the machine.
\cite{wilkes1949edsac}
\end{quotation}

\noindent This division of labour is what we would now call an abstraction. Writing decades later, in his 1985 book ``Memoirs of a Computer Pioneer'', Wilkes explains that naivety about the relative difficulty of building a computer and programming it, and its implication: that debugging was important.

\begin{quotation}
As soon as we started programming, we found out to our surprise that it wasn't as easy to get programs right as we had thought. Debugging had to be discovered. I can remember the exact instant when I realized that a large part of my life from then on was going to be spent in finding mistakes in my own programs. \cite{wilkesmemoirs}
\end{quotation}

\noindent Writing in 1951, after only two years working with the machine they had built, the experiences of programming and debugging were already well-known, and recognisable in modern terms.

\begin{quotation}
 Experience has shown that such mistakes are much more difficult to avoid than might be expected. It is, in fact, rare for a program to work correctly the first time it is tried, and often several attempts must be made before all errors are eliminated.
\cite{wilkes1951preparation}
\end{quotation}

\noindent The notion that most programs will require debugging was also discussed by Deihm and so we can imagine that, by the early 1950s, it was widely understood. \textit{When a complex routine is tried on a computer for the first time, it is seldom found to be free from error  \cite{diehm1952computer}}. Wilkes was well aware that programming, even then, involved two tasks with regard to correctness: trying to get the program right the first time and, knowing that it probably would not be, finding and fixing the remaining mistakes after the program had been fed to the computer and a misbehaviour noticed.

\begin{quotation}
 Since much machine time can be lost in this way a major preoccupation of the EDSAC group at the present time is the development of techniques for avoiding errors, detecting them before the tape is put on the machine, and locating any which remain undetected with a minimum expenditure of machine time.
\cite{wilkes1951preparation}
\end{quotation}

\noindent Even then, it was clear that the diagnosis of the problem was much harder than simply knowing that there was a problem. This is another example of an observation which seems obvious in retrospect, but must not have been so at the time.

\begin{quotation}
The difficulty lies not in detecting the presence of a mistake, but in diagnosing it. In practice its presence is nearly always obvious, for the character of most programmes is such that even a slight error will usually have an extensive effect. \cite{Gill538}
\end{quotation}

\noindent Gill goes on to explain that, even in the days of early machine languages, many of the mistakes could be found beforehand with simple checks. Of course, these days we expect these ``simple checks'' to be done automatically by the use of compilers, type-checkers and so forth.

\begin{quotation}
\ldots experience with the EDSAC has shown that although a high proportion of mistakes can be removed by preliminary checking, there frequently remain mistakes which could only have been detected in the early stages by prolonged and laborious study. Some attention, therefore, has been given to the problem of dealing with mistakes after the programme has been tried and found to fail. \cite{Gill538}
\end{quotation}

\noindent Having looked at early echoes of our modern debugging experience in these old texts, let us consider some of the differences. For example, Gill has to draw a distinction between program runs whose result is incorrect due to a faulty program being supplied and those which are incorrect due to the \index{computer!physical malfunctioning of}computer itself malfunctioning silently:

\begin{quotation}
Two kinds of mistakes, or blunders, arise in the use of an automatic digital computing machine: (i) those resulting from faults in the machine itself, and (ii) those arising because the orders or data presented to the machine are not those required to obtain the results sought. \cite{Gill538}
\end{quotation}

\noindent The same was true even for an output device. Wilkes explains how a number sent to an output device to be printed could be read back in again to check it survived the transport (this was before the invention of error correcting codes):

\begin{quotation}
Included in the order code \textup{[instruction set]} is a special order which causes the number set up on the teleprinter during the last output cycle to be read and placed in the store. The programmer can then take such steps as are necessary to verify that this number is the same as the one he intended to print.
\cite{wilkes1949edsac}
\end{quotation}

\noindent Let us turn to the first mentions of specific mechanisms for debugging in these texts. So-called \index{debugging!by single order operation}\index{single order operation}single order operation is the most primitive method of debugging once the program was in the computer (as opposed to checking the program manually beforehand).\! \textit{All high-speed computers are equipped with a  means of causing orders to be obeyed singly, at the press of a button, to enable the progress of a calculation to be followed by eye \cite{Gill538}}. Of course, such single order operation is rather like doing a dry run of the program on paper before feeding it to the machine. Gill cautions that even for the small programs they were writing then, such a method of debugging was hardly natural or productive: \textit{Single order operation is a useful facility for the maintenance engineer, but the programmer can only regard it as a last resort \cite{Gill538}.}

It is clear from Diehm's paper that, even in these early moments, the thought that the computer itself may be able to help with debugging in more sophisticated ways than single order mode was present.

\begin{quotation}
Careful proofreading and clerical checking are obvious but important methods of eliminating errors before going to the computer. On the other hand, our machines are intended to help to eliminate such drudgeries, so that we are interested in how the machines themselves can be used to analyze coding errors.
\cite{diehm1952computer}
\end{quotation}

\noindent Indeed, a precursor to \index{debugging!by tracing}\index{tracing}tracing or \index{logging}logging was present in the early use of the EDSAC, where the primitive output device could be used to write a record of certain register or data values during the run of the program.

\begin{quotation}
It might be thought that a good way of finding errors in a program would be to make the machine proceed order by order under the control of the ``Single E.P.'' button \ldots\ and to study the numbers in the machine by watching the monitors attached to the arithmetical unit and store. This, however, usually turns out to be a very slow and inefficient process, especially as the numbers are displayed in binary form. Methods have therefore been developed which permit the machine to proceed unhindered by the operator, whilst printing on the teleprinter a permanent record that can be studied at leisure, and that will assist in understanding the nature of the mistake.
\cite{wilkes1951preparation}
\end{quotation}

\noindent The utility of abstraction (through subroutines), and the intuition that testing,  debugging and abstraction are interlinked but separate human activities are both clear, even in early texts.

\begin{quotation}
Library subroutines are all checked on the machine before being put into the library and are presumably free from error. This in itself would be a sufficient reason for having a library, quite apart from any other considerations. When subroutines are specially made for a particular program it is good practice to test them beforehand by means of short programs constructed for the purpose.
\cite{wilkes1951preparation}
\end{quotation}

\noindent Some of these early feelings turn out to be less relevant given the passage of time and the great increase in computing power and abstraction mechanisms which have resulted. For example, the following paragraph feels rather foreign, perhaps because of the fact that machine errors are rarer, and that machine time is no longer at a premium.

\begin{quotation}
The important principles to be followed, I believe, are that the procedures to be used at the computer should be planned in advance and mechanized as much as possible. There should be no attempt at human analysis of errors while at the computer. It is my feeling that the computer should not be operated by the programmer, for he will not follow a predetermined plan, but will make on-the-spot improvisations which are usually regretted later, and will leave the machine idle while he speculates on possible causes of errors. The operation is best done by another person, preferably one who devotes most of his time to computer operation and is, therefore, completely familiar with the controls and has an intuitive feeling, gained from experience, of possible machine errors.
\cite{diehm1952computer}
\end{quotation}

\noindent Already we recognise terminology which is familiar today: \index{breakpoint}\index{debugging!by breakpoints}\textit{The term `break-point' is used to mean a special programmed halt which may be overridden by a manual switch  \cite{diehm1952computer}}. Here are some of Diehm's early facilities, provided by auxiliary routines loaded into the computer at known addresses before the program is loaded:

{\itshape\noindent\begin{itemize}
\item An interpretive routine which provides a complete history of a specified cell \textup{[storage location]}.
\item An interpretive routine which determines the path which the control has taken through another routine.
\item A routine which determines the total effect, on the high speed memory, of each of several chosen sections of the routine being tested. \cite{diehm1952computer}
\end{itemize}}

\noindent Already the problems of scale are evident. It is no use having reams of debugging output if the programmer is overwhelmed by it:

\begin{quotation}
What one tries to achieve in designing such auxiliary routines is to program the machine to select the pertinent information rather than to read out large quantities of data which must be searched through by the programmer.
\cite{diehm1952computer}
\end{quotation}

\noindent The early debugging process of the EDSAC, as described in ``The Preparation of Programs for an Electronic Digital Computer'' (arguably the world's first textbook on computer programming) include:

\begin{itemize}
\item A list of \textit{``points to be checked''} for the manual proofreading of programs.
\item The \textit{`post-mortem'} method of debugging where a program is stopped and then, without clearing the memory, a little debugging program tape is loaded which can print out parts of the store.
\item \textit{``Method using extra output orders''} which is rather like inserting print statements today. They suggest using these extra output orders in the original program, rather than waiting until a bug is found.
\item Subroutines for checking programs, where the original program is effectively interpreted or emulated by a debugging routine wrapped around it.
\end{itemize}

\noindent For another perspective of early debugging experience, here is Brooks, looking back on a long career, recalling:

\begin{quotation}
Early machines had relatively poor input-output equipment and long input-output delays. Typically, the machine read and wrote paper tape or magnetic tape and off-line facilities were used for tape preparation and printing. This made tape input-output intolerably awkward for debugging, so the console was used instead. Thus debugging was designed to allow as many trials as possible per machine session.

The programmer carefully designed his debugging procedure -- planning where to stop, what memory locations to examine, what to find there, and what to do if he didn't. This meticulous programming of himself as a debugging machine might well take half as long as writing the computer program to be debugged.
\cite{Brooks:1995:MM:207583}
\end{quotation}

\noindent The advent of \index{debugging!online}on-line debugging (where the programmer can interact with the debugger during the debugging process in a conversational style) came with larger and more powerful computers in the 1960s. Evans and Darley describe the similarities and differences:

\begin{quotation}
What's so new about on-line debugging? Nothing really; current on-line debugging techniques are the result of a gradual development from the days when \index{debugging!at the console}debugging at the computer console was the norm, as it has remained for small computers over the years. Debugging methods based on single-stepping through parts of a program and on examination and modification of memory registers by means of console lights and switches were the natural precursors of today's more sophisticated techniques, and there is no sharp dividing line at any stage of the progression. Perhaps the critical step was the replacement of console lights and switches by some typewriter-like device as the principal means of communication between user and machine.
\cite{evans1966line}
\end{quotation}

\noindent The description of the debugging system FLIT (FLexowriter Interrogation Tape) for the early 1960s TX-O system at MIT provides an insight into early online debugging:

\begin{quotation}
With FLIT, for the first time, it was possible for the user to examine and modify his program in terms of the symbols used in his source program and, in fact, to examine and change the contents of registers in a form almost identical to that used in the corresponding assembly language.
\cite{evans1966line}
\end{quotation}

\noindent The typical experience of offline and online debugging with such a system is described in the Programming Manual for the DEC PDP-8 `DDT' debugger:

\begin{quotation}
Users of most computers, especially large-scale ones, are familiar with the procedure of submitting a new program for a computer run, waiting for it to be processed (which may take anywhere from a few hours to several days), and finally receiving the compilation and/or assembly listings, a list or dump of the contents of each core memory cell at the time the run was terminated, and perhaps a storage map giving the addresses of the symbols used in the programs. The user may get a few remarks from the computer operator regarding the failure of the program to run properly. If the user is present in the machine room when his program is processed, he may get additional information from the console lights, motion of tapes, etc., but his correcting must be done away from the computer. Getting a program to work under these conditions takes considerable time.

DDT (DEC Debugging Tape) helps shorten this debugging time by allowing the user to work on his program at the computer, to control its execution, and to make corrections to the program or its data. For example, tracking down a subtle error in a complex section of coding is a laborious and frustrating job by hand; but with the breakpoint facility of DDT-8, the user can interrupt the operation of his program at any point and examine the state of the machine. In this way, sources of trouble can be located quickly.
\cite{ddt}
\end{quotation}

\noindent Note the disadvantages of \index{debugging!offline}\index{debugging!online}offline debugging and the advantages of online debugging given here. Balzer, writing in 1969 after a decade of compiled, high-level languages, explains the complications created by this shift -- the need to map back to the source from the executable.

\begin{quotation}
The debugging\index{debugging!in high- and low-level languages} systems for higher-level languages are much more complex than those for assembly code. They must locate the symbol table, find the beginning and end of source-level statements, and determine some way to extract the dynamic information -- needed for debugging -- about the program's behavior, which is now hidden in a sequence of machine instructions rather than being the obvious result of one machine instruction.
\cite{balzer1969exdams}
\end{quotation}

\noindent Some voices, by the 1960s, saw that debugging would become ever more important, as systems became larger and more complicated:

\begin{quotation}
Oedipus is based on the premise that debugging is a central problem, perhaps \textup{the} central problem, in the implementation and in the operation of any large programming system. Furthermore, conventional debugging aids (octal snaps and post mortems) are hopelessly inadequate in the face of dynamic storage allocation and recursion. \cite{brown1965operating}\end{quotation}

\noindent Dynamic storage allocation and recursion are two important aspects of typical functional programming systems, incidentally. Let us move on now, to think about why so much of this difficulty in debugging persists. 

\subsection{The difficulty of debugging}\index{debugging!difficulty of}

Is debugging difficult, or is it just that programming is difficult, and debugging is really just programming? The consensus in the literature is that debugging can, for the most part, be thought of as a separate activity, especially when it occurs after the program is thought to be complete. As early as the 1960s this observation is common: \textit{Programmers seem to spend more time trying to find out why the programs they have written do not work than all other activities put together \cite{Gaines:1969:DCP:905460}}. Or, more pithily, from ``The Elements of Programming Style'': \textit{Debugging is twice as hard as writing the program in the first place \cite{kern-debug}}. These sentiments are still widespread in recent times, showing that all the improvements in programming languages do not seem to have translated into a fundamental shift in the difficulty of debugging.

\begin{quotation}
Debugging is notoriously difficult and extremely time consuming. \cite{Parnin:2011:ADT:2001420.2001445}
\end{quotation}

\begin{quotation}
Nobody would claim that debugging software is easy: all too often it proceeds by trial-and-error experiments in which programmers examine the behaviour if the system and form hypotheses that could explain what they see.
\cite{harris2002dependable}
\end{quotation}

\noindent In his book on debugging ``Why programs fail: a guide to systematic debugging'' Zeller explains that whilst many things have changed, the difficulty of debugging has been ever present: 

\begin{quotation}
Improved programming languages and tools can supplant, but not eliminate debugging, by statically identifying errors and by dynamically detecting invariant violations.
\cite{zeller2009programs}
\end{quotation}

\noindent Harris points out another set of modern challenges, that of debugging communicating, distributed systems:

\begin{quotation}\index{debugging!distributed systems}\index{distributed system}
These problems \textup{[of debugging]} are exacerbated when developing distributed, peer-to-peer or multi-processor applications, or when unreliable network links form part of the system under test. Environments for pervasive computing take this to an extreme, allowing user-supplied code to run or migrate within and around the network.
\cite{harris2002dependable}
\end{quotation}

\noindent There has been an enormous increase in the scale of computer programs we now routinely write using modern, structured languages, so to have expected otherwise may have been too much. Plainly the improvements to programming language toolchains such as \index{programming!structured}\index{structured programming}structured programming and \index{type safety}type safety have lead to better programs and better programming, but improvements to debugging seem at best to keep pace -- the war is one of attrition. So why are programming and debugging intrinsically hard? In his landmark software management retrospective ``The Mythical Man-month'', Brooks explains:

\begin{quotation}
First, one must perform perfectly. The computer resembles the magic of legend in this respect, too. If one character, one pause of the incantation is not strictly in proper form, the magic doesn't work. Human beings are not accustomed to being perfect, and few areas of human activity demand it. Adjusting to the requirement for perfection is, I think, the most difficult part of learning to program.
\cite{Brooks:1995:MM:207583}
\end{quotation}

\noindent With regard to debugging in particular, it is a task which is unpleasant to most people as well as difficult. When a bug is found, the programmer is apt to be annoyed rather than intrigued, since their failure has been exposed. If someone else has written the program one has to debug, understanding it is another unpleasant task. Katz, in his analysis of bug-location strategies, writes: \textit{Programmers often report that instead of debugging someone else's program, they would rather write their own version  \cite{katz1987debugging}}. In his 1979 book ``The Art of Software Testing'' -- which is mostly about testing but a lot about debugging too -- Myers writes: \textit{Debugging appears to be the single part of the software-production process that programmers seem to abhor the most  \cite{myers1979art}}. Brooks explains why:

\begin{quotation}
The next woe is that designing grand concepts is fun; finding nitty little bugs is just work. With any creative activity come dreary hours of tedious, painstaking labor, and programming is no exception.
\cite{Brooks:1995:MM:207583}
\end{quotation}

\noindent But why specifically is debugging difficult? A common theme amongst explanations in the literature is that the abstractions\index{abstraction} we associate with good programming practice -- and so modularity and what Dijkstra called ``separation of concerns'' --  break down under the requirements of debugging: \textit{Often, the code relevant to the task is scattered across many modules, increasing the difficulty of the task \cite{robillard2004effective}}. This includes, of course, not just the present state of the code, but the project's history. From the LaToza et al.\ study of the work habits of programmers:

\begin{quotation}
Developers must know or obtain a variety of information to successfully understand and edit code -- what code to change, how design decisions are scattered across code, the rationale or history behind decisions, the slice affecting a variable's value, the owner responsible for editing the code, other developers currently editing it, which changes will break code elsewhere, and which changes elsewhere affect it. \cite{latoza2006maintaining}
\end{quotation}

\begin{quotation}
When debugging, programmers view programs in ways that need not conform to the programs' textual or modular structures. In particular, the statements in a slice may be scattered throughout the code of the larger program and yet experienced programmers routinely extract the slices from a program.
\cite{Weiser:1982:PUS:358557.358577}
\end{quotation}

\noindent The term \index{slicing}`slicing' refers to calculating the part of the program which can lead to the alteration of a value at a given storage location:

\begin{quotation}
Computer programmers break apart large programs into smaller coherent pieces. Each of these pieces: functions, subroutines, modules, or abstract data types, is usually a contiguous piece of program text. The experiment reported here shows that programmers also routinely break programs into one kind of coherent piece which is not contiguous. When debugging unfamiliar programs programmers use program pieces called \textit{slices} which are sets of statements related by their flow of data. The statements in a slice are not necessarily textually contiguous, but may be scattered through a program. \cite{Weiser:1982:PUS:358557.358577}\end{quotation}

\noindent There are many kinds of slices, categorized by Penney:

\begin{quotation}
The \textit{forward slice} consists of all program statements affected by the slicing criterion, whereas the \textit{backward slice} represents those statements upon which the criterion depends. \textit{Static} slices are based on the static analysis of source code. \textit{Dynamic} slices take account of program input and particular test cases. \cite{penney2000augmenting}\end{quotation}

\noindent In his paper ``Some psychological evidence on how people debug computer programs'', in which he develops what he calls ``a gross theory of debugging'', Gould writes:

\begin{quotation}
Debugging computer programs is difficult for several reasons. The programmer must simultaneously keep track of several aspects of the program's detailed procedure specification, but the ability to do this is severely restricted (e.g.\ Yntema 1963 \textup{[A study on human memory recall]}). Second, the variety within all examples of many other systems that are regularly diagnosed (e.g.\ automotive, human anatomical, or plumbing systems). Third, debugging (as well as writing) a program requires a degree of precision because computer systems are unrelenting in their demands for accuracy.
\cite{gould1975some}
\end{quotation}

\noindent Looking more closely at the debugging process, Ko et al.\ discuss debugging as a concretion, beginning with the abstract notion of the bug and finding concrete debugging actions to take to clarify and debug it.

\begin{quotation}
What makes debugging difficult in general is that programmers typically begin the process with a `why' question about their program's behavior, but must translate this question into a series of actions and queries using low-level tools such as breakpoints and print statements.
\cite{ko2011state}
\end{quotation}

\noindent Gaines again, on the spectrum of difficulty, bug-to-bug: is it perhaps this \index{bugs!heterogeneity of}heterogeneity which makes debugging a task where repeatability is limited?

\begin{quotation}
The degree of difficulty the programmer experiences in isolating a bug once he has noticed an error depends on the nature of the bug and the ease with which he can obtain additional information about intermediate states in the computation.
\cite{Gaines:1969:DCP:905460} 
\end{quotation}

\noindent The gulf we have already mentioned, between the amount of effort needed to find a bug and the amount of effort needed to fix it, is explained by Metzger in a way which also validates our instinct that, whilst programming and debugging are often intertwined, they are separate in nature:

\begin{quotation}
Coding, designing, analyzing, and testing are all constructive activities. They each produce a tangible result. Coding produces source code. Designing produces design documents. Analysis produces a variety of documents, depending on the methodology used. Testing produces test cases and reports on the success or failure of tests.

In contrast, debugging is primarily a cognitive activity. The end result is knowledge of why there is a problem and what must be done to correct it. There will be a source change, but it may only involve adding or deleting a single character or word. The constructive output of debugging is often disproportionate to the effort expended.
\cite{metzger2004debugging}
\end{quotation}

\noindent There are fundamental problems to these kinds of analyses and conversations, in terms of how we talk to each other about our processes.

\begin{quotation}
An attempt to find out what people do when they are debugging by asking a number of programmers has only served to demonstrate the complexity of the problem and the general inability, well-known to psychologists, of people to describe what they do when they are involved in complex mental activity.
\cite{Gaines:1969:DCP:905460} 
\end{quotation}

\noindent But this should not prevent us attempting such analyses. We often know it when we see it. Myers suggests some reasons for the unpleasantness of debugging\index{debugging!unpleasantness of} (and such unpleasantness and difficulty are closely linked, for most people):

\begin{description}
\item[Psychologically difficult] \textit{``...because it is an indication that they \textup{[the programmer]} are less than perfect''}.
\item[Mentally taxing] Both intrinsically, and because of external pressures (getting a release to a customer, \textit{``self-induced pressure''} or other stressors).
\item[Location unclear] \index{bug!location of}\textit{``The location of the error is potentially any statement in the program''}. Myers contrasts this with fault-location in another area -- physical systems such as vehicle maintenance -- where it is often easy to know which subsystem is likely at fault given the symptom observed. 
\item[Lack of resources] \textit{``\ldots comparatively little research, literature, and formal instruction exists on the process of debugging.''} \cite{myers1979art}
\end{description}

\noindent So debugging is difficult, but we have been working on tools to aid debugging for decades, so how do we explain the persistence of the debugging problem?

\subsection{Our lack of progress}

As early as 1965, surprise was being expressed that the shift from machine code to assembly language to compiled languages to block-structured compiled languages, and the simultaneous vast improvements in computing power and cost had not led to as great a reduction in the frequency or severity of bugs. Halpern, in ``Computer programming: the debugging epoch opens'', writes:

\begin{quotation}
That tendency to err that programmers have been noticed to share with other human beings has often been treated as if it were an awkwardness attendant upon programming's adolescence, which like acne would disappear with the craft's coming of age. It has proved otherwise \ldots\ Many of us expected compiler languages to eliminate all bugs except those so glaring as to leap to the first fresh eye cast on the program. \ldots\ 
An unfriendly behaviorist studying programmers might conclude that we deliberately elaborate our tasks so as to keep the bug rate constant. \cite{halpern1965computer}
\end{quotation}

\noindent Balzer explains that this surprise was widespread in the industry, as became clear when more and more large programs were being written in modern compiled languages:

\begin{quotation}
With the advent of the higher-level algebraic languages, the computer industry expected to be relieved of the detailed programming required at the assembly-language level. This expectation has largely been realized. Many systems are now being built in higher-level languages (most notably MULTICS).

However, the ability to debug programs has advanced but little with the increased use of these high-level languages.
\cite{balzer1969exdams}
\end{quotation}

\noindent Hamlet suggests a possible reason -- the lack of equivalent progress in debugging para\-digms, suggesting that high-level languages might require different kinds of debugging tools rather than mere analogs of low-level ones.

\begin{quotation}
Debugging techniques originated with low-level programming languages, where the memory dump and interactive word-by-word examination of memory were the primary tools. `High-level' debugging is often no more than low-level techniques adapted to high-level languages.
\cite{hamlet1983}
\end{quotation}

\noindent It is fascinating to see Halpern writing on the same topic again, forty years later:

\begin{quotation}
The most remarkable thing about debugging today is how little it differs from debugging at the dawn of modern computing age, half a century ago. \ldots\ We've made little progress in debugging methods in half a century, with the result that projects everywhere are bogged down because of buggy software.
\cite{halpern2005assertive}
\end{quotation}

\noindent Thirty years later, \index{debugger!lack of use}debuggers were still not widely used, even in difficult domains. In a paper on debugging practices for complex legacy systems, Regelson and Anderson write: \textit{The major item noted by survey respondents was that few people really have learned to use the capabilities of their debuggers \cite{regelson1994debugging}}. Still later, in recent years, debugging in industry is sporadic. Parnin and Orso, writing specifically about automated debugging techniques, say:

\begin{quotation}
Although potentially useful, most of these \textup{[debugging]} techniques have yet to demonstrate their practical effectiveness. One common limitation of existing approaches, for instance, is their reliance on a set of strong assumptions on how developers behave when debugging.
\cite{Parnin:2011:ADT:2001420.2001445}
\end{quotation}

\noindent This ``reliance on a strong set of assumptions'' as the key to understanding why people do not use debuggers is a theme we shall return to again and again in this thesis. Hailpern and Santhanam tie this into the wider issue of immaturity with regard to software development practices:

\begin{quotation}
\ldots we observe that due to the informal nature of software development as a whole, the prevalent practices in the industry are still immature, even in areas where improved technology exists.
\cite{Hailpern:2002:SDT:1660992.1660994}
\end{quotation}

\begin{quotation}
Designing a debugger which will deal with the whole debugging knowledge is still a challenge.
\cite{ducasse1988review}
\end{quotation}

\noindent Most working programmers today, reading the five ``maturity levels'' from IBM's Capability Maturity Model \cite{ibmcmm} would recognise their own workplaces as being only at level one or two -- the more haphazard and less mature:

\begin{quotation}
\noindent 1. Initial

\noindent The software process is characterized as ad hoc, and occasionally even chaotic. Few processes are defined, and success depends on individual effort.   

\medskip

\noindent 2. Repeatable

\noindent Basic project management processes are established to track cost, schedule, and functionality. The necessary process discipline is in place to repeat earlier successes on projects with similar applications.

\medskip

\noindent 3. Defined

\noindent The software process for both management and engineering activities is documented, standardized, and integrated into a standard software process for the organization. All projects use an approved, tailored version of the organization's standard software process for developing and maintaining software.

\medskip

\noindent 4. Managed

\noindent Detailed measures of the software process and product quality are collected. Both the software process and products are quantitatively understood and controlled.

\medskip

\noindent 5. Optimizing

\noindent Continuous process improvement is enabled by quantitative feedback from the process and from piloting innovative ideas and technologies. \cite{ibmcmm}
\end{quotation}

\noindent In academia, too, the intangibility of the debugging problem is recognised:

\begin{quotation}
Even today, debugging remains very much an art. Much of the computer science community has largely ignored the debugging problem. Eisenstadt studied 59 anecdotal debugging experiences and his conclusions were as follows: Just over 50 per cent of the problems resulted from the time and space chasm between symptom and root cause or inadequate debugging tools.
\cite{Hailpern:2002:SDT:1660992.1660994}
\end{quotation}

\noindent Although, the proliferation of debugging papers does not necessarily address these practical issues:\textit{ \dots\ only 3 out of 111 papers on the technique of program slicing based techniques \cite{Weiser:1982:PUS:358557.358577} have considered issues with the use of the techniques in practice \cite{Parnin:2011:ADT:2001420.2001445}}. We shall confront this in our work too, having to evaluate our own new tool, even though it is unlikely to be in widespread use.

We shall address debugging in teaching in detail a little later, but to illustrate that the same challenges persist there, we can look at two recent pieces of work. McCauley et al., in a review of debugging literature from an education perspective, write about the lack of progress in that area too:

\begin{quotation}
Debugging is an important skill that continues to be both difficult for novice programmers to learn and challenging for computer science educators to teach. These challenges persist despite a wealth of important research on the subject dating back as far as the mid 1970s. Although the tools and languages novices use for writing programs today are notably different from those employed decades earlier, the basic problem-solving and pragmatic skills necessary to debug them effectively are largely similar.
\cite{mccauley2008debugging}
\end{quotation}

\noindent And, in many cases, such education is simply omitted. It is not clear, in fact, that debugging is even really possible to teach (or, if it is, that, just like learning to drive, most of the education happens after the formal learning process is over). Siegmund et al.\ describe the debugging knowledge of subjects in their study:

\begin{quotation}
Only half of our participants mentioned receiving debugging education, indicating that educators still assume that debugging is a minor part of software development or that students will learn it by themselves. However, there also was no significant difference between participants with and without education, indicating that the existing courses and trainings \textup{[sic]} are indeed not more effective in teaching important debugging skills than self-learning.
\cite{6983851}
\end{quotation}

\noindent Finally, we must quote at length the 1997 rallying cry of Henry Lieberman in his introduction to a special issue of the Communications of the ACM ``The Debugging Scandal and What to Do About It'':  

\begin{quotation}
Debugging is the dirty little secret of computer science. Despite all the progress we have made in the last thirty years: faster computers, networking, easy-to-use graphical interfaces, and everything else, we still face some embarrassing facts. First, all too often, computer programs don't work as they should. This makes software development costly. Too much buggy software reaches end users, leading to needless expense and frustration. That's unfortunate, but what is surprising is the fact that when something does go wrong, the people who write these programs still have no good ways of figuring out exactly what went wrong. Debugging is still, as it was thirty years ago, largely a matter of trial and error. 
What borders on scandal is the fact that the computer science community as a whole has largely ignored the debugging problem. This is inexcusable, considering the vast economic cost of debugging and emotional toll buggy software takes on users and programmers. Today's commercial programming environments provide debugging tools that are little better than the tools that came with programming environments thirty years ago. It is a sad commentary on the state of the art that many programmers name ``inserting print statements'' as their debugging technique of choice.
\cite{lieberman}
\end{quotation}

\noindent Does this lack of progress in debugging over the years reflect a genuine problem which has not yet been attacked in the right way, or simply that debugging is fundamentally not very tractable? We hope that this thesis will add some weight to one of those piles of evidence.

\subsection{Teaching debugging}

Debugging is a common difficulty among those \index{debugging!learning}\index{programming!learning}learning to program \cite{badiozamanydebugging}. It is not often taught as a separate skill in university Computer Science courses, nor touched upon more than tangentially in most programming textbooks. Again, this is a longstanding situation: \textit{It is interesting to note than \textup{[sic]} an examination of 17 introductory programming texts revealed only 4 in which the subject of debugging received more than a few sentences \cite{Gaines:1969:DCP:905460}}.

In Ahmadzadeh et al's ``Analysis of patterns in debugging among novice computer science students'', the authors find that it is possible (indeed common) to learn a decent amount about programming without one's ability at debugging keeping up. This is perhaps another facet of the programming/debugging dichotomy to which we have already alluded.

\begin{quotation}
We discovered that many students with a good understanding of programming do not acquire the skills to debug programs effectively, and this is a major impediment to their producing working code of any complexity.
\cite{ahmadzadeh2005analysis}
\end{quotation}

\noindent Students seem to find debugging to be a job of roughly the same character and difficulty as industrial programmers, even though they are typically working on much smaller codebases. For example, in their study ``Debugging from the student perspective'', Fitzgerald et al.\ find:

\begin{quotation}
Students in this multi-institutional study report that finding bugs is harder than fixing them. \ldots\ Hypothesizing about the cause of bugs is an underdeveloped skill.
\cite{fitzgerald2010debugging}
\end{quotation}

\noindent In particular, the study asks students to break down the debugging task in their minds into stages, and tabulates the difficulties encountered:

\bigskip
\begin{tabular}{ll}
Troubleshooting stage & Subjects who found this stage most difficult\\\hline
Understanding the code & 4 (19\%)\\
Testing & 3 (14\%)\\
Finding the problem & 12 (57\%)\\
Fixing the problem & 2 (10\%)\\
\end{tabular}
\bigskip

\noindent Murphy et al., in a qualitative study of novice debugging strategies, say debugging is fundamentally a hard skill in the initial stages of learning: \textit{Similar to new drivers who must learn to steer, accelerate, brake, etc. all at once, novice debuggers must apply many new skills simultaneously \cite{Murphy:2008:DGB:1352322.1352191}}. Pea, in ``Language-independent conceptual `bugs' in novice programming'' agrees. \textit{
The novice programmer works \textit{intuitively} and pursues many blind alleys in learning the formal skill of programming \cite{pea1986language}}. Vessey breaks down this difference between experts and novices\index{debugging!novices vs experts}, observing:

{\itshape\begin{enumerate}
\item \begin{enumerate}\item Experts use breadth-first approaches to debugging and, at the same time, adopt a system view of the problem area.
\item Experts are proficient at chunking programs and hence display smooth-flowing approaches to debugging.\end{enumerate}
\item \begin{enumerate}\item Novices use breadth-first approaches to debugging but are deficient in their ability to think in system terms.
\item Novices use depth-first approaches to debugging.
\item Novices are less proficient at chunking programs and hence display erratic approaches to debugging.
\end{enumerate}\hspace{-13pt}\cite{vessey1985expertise}
\end{enumerate}}

\noindent Writing about end-user programming (for example, programming a set-top box to record a TV program), which is another form of novice programming, Ko et al.\ look at the informality of debugging strategies employed by such users:

\begin{quotation}
...because end users often prioritize their external goals over software reliability, debugging strategies often involve ``quick and dirty'' solutions, such as modifying their code until it appears to work. In the process of remedying existing errors, such strategies often lead to additional errors.
\cite{ko2011state}
\end{quotation}

\noindent Regelson and Anderson give another reason, which may be more related to the difficulty of teaching rather than learning, but this is much the same problem in practice: \textit{Different people exhibit very different levels of debugging skill \cite{regelson1994debugging}}. Pea makes an interesting argument for the idea that  misconceptions about how a computer operates need `un-learning' before one can program and debug effectively:

\begin{quotation}
This article argues for the existence of persistent conceptual `bugs' in how novices program and understand programs. These bugs are not specific to how a given programming language, but appear to be language-independent. \ldots\ It is suggested that these classes of conceptual bugs are rooted in a `superbug', the default strategy that there is a hidden mind somewhere in the programming language that has intelligent interpretive powers.
\cite{pea1986language}
\end{quotation}

\noindent There are clearly many similarities between the difficulties industrial programmers of all abilities have in performing the task of debugging and the problems students only beginning their programming journey encounter. We might expect such similarities if the problem of debugging is fundamental or intrinsic, rather than incidental.

\subsection{What makes a good debugger?}

\noindent Many of the extracts from the literature we have already shown allude to what makes a good debugger, if only by implication. But it is useful to look to see if there are any more explicit commentary on the matter. Satterthwaite, in developing ``A Philosophy of Design for Debugging'' says:
\textit{Since debugging, as usually understood, is more a practical than a theoretical problem, proposed solutions must be evaluated within a framework of practical constraints
\cite{satterthwaite1972debugging}}. Brady, in a paper about a debugging tool for experienced users, explains that prior systems tried too hard to be approachable for novices, and the loquacity of their commands alienated the experienced. He writes: \textit{In a debugging program it is of prime importance that the program be simple, flexible, and highly efficient to use
\cite{brady1968writing}}. Evans and Darley, in their 1966 survey of online debugging systems, concur, explaining that when designing the interface to a breakpoint-based debugger: \textit{Here, as in other aspects of on-line work, convenience is critical
\cite{evans1966line}}. Eisenstadt lists three principles:

{\itshape\begin{itemize}
\item Allow full functionality at all times. Debugging environments that prevent access to certain facilities make matters worse.
\item Viewers should be provided for ``key players'' (any evaluable expression) rather than just for `variables'.
\item Provide a variety of navigation tools at different levels of granularity. \cite{eisenstadt1997my}\end{itemize}}

\noindent Grishman echoes the first principle, suggesting that a debugger\index{debugger!applicability} is at its best when it is at its most widely applicable, situationally:

\begin{quotation}
\ldots to facilitate maintenance, the same program was to be useable in both batch and interactive modes. Second, to facilitate distribution, the system had to be useable without any modification to the operating system.
\cite{grishman1970debugging}
\end{quotation}

\noindent Eisenstadt's last theme of granularity is touched upon by several other authors. Hamlet notes the chasm between the separated and abstracted structured program written by the programmer, and the typical view of an executable-based debugger: \textit{What the designer has divided and conquered, the debugger sees as an overwhelming monolith
\cite{hamlet1983}}. Scowen sees the same lack of granularity in early flowchart-based debuggers:

\begin{quotation}
Flowcharters have the disadvantage that they present all parts of the program with the same degree of emphasis, rather like a map that shows footpaths and motorways in the same way.
\cite{scowen1972debugging}
\end{quotation}

\noindent The structure of the program has been lost. Since we expect the programmer to think about debugging using the same basic thought-structures with which they think about programming, this disconnect is troubling. Evans and Darley argue that this fineness of control is possible only with online rather than batch debugging:

\begin{quotation}
\ldots a very selective and close control over the execution of portions of one's program and for the examination of intermediate results, together with the possibility of making on-the-spot changes based on them, as desired.
\cite{evans1966line}
\end{quotation}

\noindent Halpern draws the distinction between high-level debuggers which use the source language and those which operate only on executables or dumps:

\begin{quotation}
But the general principles to be observed in implementing the XRAY (as I have dubbed it) are clear: show the programmer the system, not the machine, and do so in his language, not octal or hexadecimal.
\cite{halpern1965computer}
\end{quotation}

\noindent Grishman explains the fundamental choice between \index{source-level debugger}\index{debugger!source-level}source debuggers (such as a source-level interpreter) and \index{object-level debugger}\index{debugger!object-level}object debuggers:

\begin{quotation}
The most important decision in designing a debugging system is whether to process the source language directly (by adding debugging statements to a compiler, or interpreting the source text) or to work from the object code.
\cite{grishman1970debugging}
\end{quotation}

\noindent Grishman also gives an argument for object debuggers operating by \index{debugging!by simulation}simulation rather than by executing the object code directly: 

\begin{quotation}
Simulation provides a far richer set of traces and checks than could a system which executes the object code; in particular, it provides a simple solution to what appears to be the most common plight of the desperate user, ``What part of my program stored \textit{that}?''. \cite{grishman1970debugging}
\end{quotation}

\noindent Zeller imagines the attributes of the ideal debugger, including a conception of the debugging process on a particular bug as something which can be packaged up and passed around:

\begin{quotation}
I think this \textit{debugging process} is the most important factor in debugging. Such a process should be systematic; that it should encourage people to explicitly state their hypotheses, predictions, observations, and outcomes; that it should allow interrupting and resuming sessions; and that it should allow moving a debugging session over to some co-worker at any point. 
\cite{zeller-debugging}
\end{quotation}

\noindent Of course, not everyone agrees on the usefulness of having a debugger at all, preferring the scattering of print statements and ad hoc test harnesses. In some ways, this is a useful standard against which a debugger writer may measure themselves. A debugger, if it is any good, should be so obviously useful as to render such dissent idiosyncratic. Knuth, in ``The Art of Computer Programming'', says:

\begin{quotation}
The most effective debugging techniques seem to be those which are designed and built into the program itself -- many of today's best programmers will devote nearly half of their programs to facilitating the debugging process on the other half; the first half, which usually consists of fairly straight-forward routines, will eventually be thrown away, but the net result is a surprising gain in productivity.
\cite{knuth-taocp}
\end{quotation}

\noindent Some go further and claim that the use of debuggers as a sort of crutch is counterproductive and should be discouraged:

\begin{quotation}
I happen to believe that not having a kernel debugger forces people to
think about their problem on a different level than with a debugger. I
think that without a debugger, you don't get into that mindset where you
know how it behaves, and then you fix it from there. Without a debugger,
you tend to think about problems another way. You want to understand
things on a different level.

It's partly ``source vs binary'', but it's more than that. It's not that you
have to look at the sources (of course you have to -- and any good debugger
will make that easy). It's that you have to look at the level above
sources. At the meaning of things. Without a debugger, you basically have
to go the next step: understand what the program does. Not just that
particular line.
\cite{torvalds}
\end{quotation}

\subsection{Human factors}

If we want to understand why debugging is hard or to look for reasons why debuggers are not as widely used as we might expect, it is important to address not only the technical factors, but the human factors which give rise to this situation. Debugging is as personal, or more so, than programming itself. Gaines (in his thesis ``The Debugging of Computer Programs'', which we have quoted extensively, and which is well worth reading in full) summarises: \textit{
Computer programming is a creative art, best done by individuals, and debugging is the most highly individual aspect of programming \cite{Gaines:1969:DCP:905460}}. Zeller captures the dual nature of debugging as a human experience of highs and lows:

\begin{quotation}
\ldots debugging can be an enjoyable activity that shares the thrill of the hunt and chase found in a good detective novel or video game. On the other hand a protracted, unsuccessful search for a bug in your code quickly loses its charm, particularly when your boss is asking repeatedly about your (lack of) progress.
\cite{zeller2009programs}
\end{quotation}

\noindent The debugging process begins with an understanding that something is wrong (some wrong result or behaviour), and as the debugging process continues, we hold in our minds an ever-narrowing list of possibilities for the source or location of the bug. \textit{The programmer will at any time have some notion of the probability that any given part of the formulation or implementation of the program is in error 
\cite{Gaines:1969:DCP:905460}}. Sometimes this happens quickly, because of positive information narrowing down the scope, sometimes slowly because we are just ticking one item off the list of things which might have caused the bug (but which did  not). Ko et al.\ believe this natural testing of the programmer's theories about the source of the bug is almost universal:

\begin{quotation}
Although the process of debugging can involve a variety of strategies, studies have shown across a range of populations that debugging is fundamentally a hypothesis-driven diagnostic activity.
\cite{ko2011state}
\end{quotation}

\noindent Inexperienced debuggers still use this approach, even if they do not know it, implying that perhaps there is something fundamental or innate to it.

\begin{quotation}
Although all followed a standard approach that can be seen as a simplified scientific method, none of them was aware of this or able to explain his approach without resorting to demonstration.
\cite{6983851}
\end{quotation}

\noindent It appears that there is a wide range of talent levels for debugging, even among equally \index{debugging!talent in} talented or experienced programmers, another piece of evidence to suggest that debugging and programming have fundamental differences as activities.

\begin{quotation}
There is a small number of people who are unusually good at debugging. These people can walk into almost any setting and quickly offer very helpful insights, suggestions, and solutions. \ldots\ 
Years of experience seem to have some impact on people's debugging skill level. However, there are examples of good debuggers who are relatively new to the field and very experienced engineers who are not adept debuggers \cite{regelson1994debugging}
\end{quotation}

\noindent Some of this is undoubtedly not just technical skill but emotional skill -- debugging can be a profoundly humbling experience. Several authors note that the impedance mismatch between the modular, abstracted nature of a program and the unstructured, unknown conception of the bug at the beginning of the \index{bug!isolation}bug-hunting process might be one cause of some of the difficulties of debugging. For example, here are Perera et al.\ on this topic:
\begin{quotation}
\ldots in debugging we aim to understand why some erroneous result was computed by a program. This goal of understanding and explaining computations and their results often runs against our desire to treat a computation as a black box that maps inputs to outputs \ldots\ opening the box exposes a great deal of useful information to the programmer but also presents several implementation and user-interface challenges.
\cite{perera2012functional}
\end{quotation}

\noindent Ducass\'e sees another part of this contrast, this haziness, the question whether a behaviour is a misbehaviour at all:

\begin{quotation}
People sometimes talk about code errors as if they were absolute and well defined concepts whereas they are, in general, only relative and ill-defined \ldots\ For example, depending on the point of view, a particular behavior of a program can be considered an error or a feature. Another example is a piece of code which breaks some company's convention, it may then be considered an error in that company while being considered correct in another place.
\cite{ducasse1993pragmatic}
\end{quotation}

\noindent It is clear that the difficulty of debugging stems not only from the intrinsic difficulty of the technical problem, but also the many wider human factors which cannot help but affect the working programmer.

\subsection{Classifications}

\index{debugger!classification of a}\index{bug!classification}So much for debugging in the abstract. We cannot go much further without considering at least a little, a survey of the concrete kinds of debugging ideas, approaches, and tools which have been pervasive in the literature over the last few decades. Our principal sources will be a  Ducass\'e's debugging survey \cite{ducasse1988review}, Eisenstadt's article on programmers' debugging ``war stories'' \cite{eisenstadt1997my}, and the debugging (not testing) parts of Glenford J. Meyers classic book ``The Art of Software Testing'' \cite{myers1979art}.

One crude but effective method of extracting classifications from a text in the literature is simply to list the section headings, trusting the author's innate need to classify in order to understand. Consider, for example, some of Gaines' suggestions on how we might classify bugs \cite{Gaines:1969:DCP:905460}:

\begin{description}
\item[Point of origin in the programming process] Did the bug originate when the programmer was formulating the program in their mind, or whilst the task of converting these thoughts to a concrete program took place?
\item[Control and computation bugs] Distinguishes bugs which affect only the values of variables without affecting further control flow, from those which affect both.
\item[Bugs resulting from lack of knowledge \ldots\ of operating environment] \textit{``The lack of knowledge may be the result of the programmer's forgetting something he once knew, or getting an incorrect idea about some aspect of his operating environment, as well as being caused by something he was never informed of.''}
\item[Fatal and non-fatal]\index{bugs!fatal and non-fatal}This distinguishes buggy programs which terminate abnormally, due to an exception or assertion or segmentation fault from those which terminate normally but with incorrect results.
\item[The point at which a bug may be detected] During automatically during compilation or loading or execution (say an assertion) of the program, or detected manually at execution by seeing a bad output.
\end{description}

\noindent For an overview of practical kinds of debugging, we turn to the chapter concerning debugging in Myers' book on Testing ``The art of software testing'' \cite{myers1979art}. He lists:

\begin{description}
\item[Debugging by brute force]\index{debugging!by brute force} The most common technique, Myers says, and split into three: a) debugging with a storage dump; b) debugging by \textit{``scattering print statements throughout your program''}; and c) using the debugging tools of the programming environment, such as breakpoints. \textit{``The general problem with these brute-force methods is that they ignore the process of \textup{thinking}.''}

\item[Debugging by induction]\index{debugging!by induction} This is a hypothesis-testing method, often conducted without the use of the computer, consisting of four steps: \textit{``Locate the Pertinent data''}, \textit{``Organize the data''} (to allow the observing of patterns), \textit{``Devise a hypothesis''} and \textit{``Prove the hypothesis''} (by further testing, before jumping in to fix the problem).

\item[Debugging by deduction]\index{debugging!by deduction} By contrast to induction, this method corrals a list of possible causes of the bug, then narrows them down with reasoning. The steps given are \textit{``Enumerate the possible causes or hypotheses''} (the hypotheses \textit{``need not be complete explanations''}), \textit{``Use the data to eliminate possible causes''}, \textit{``Refine the remaining hypothesis''}, and \textit{``Prove the remaining hypothesis''} (same as the last step in Induction above).

\item[Debugging by backtracking]\index{debugging!by backtracking}\index{backtracking}Effective only for small programs, or very self-contained parts of larger programs, this method works backward through the program logic from the first evidence of failure, to work out what antecedent circumstances (states of variables, for example) may or must have caused that failure.

\item[Debugging by testing]\index{debugging!by testing}\index{testing} Despite the distinction between testing and debugging being drawn in much of the literature (and indeed in Myers), in this section it is explained that there are, in fact, uses for testing procedures in the wider context of the debugging process. These are `slim' tests, \textit{``...attempts to cover only a single condition of a few conditions in each test case''}, rather than the `fat' tests \textit{``...attempts to cover many conditions in a few test cases''} used for testing more generally.
\end{description}

\noindent In their analysis of debugging in novice computer science students, Ahmadzadeh et al.\ list four major categories of knowledge used by their subjects when trying to debug faulty programs, written by others, which were presented to them in the study. These are knowledge of:

\begin{enumerate}
\item \textit{The intended program
\item The actual program
\item The use of debugging methods
\item The error itself \cite{ahmadzadeh2005analysis}}
\end{enumerate}

\noindent Ducass\'e et al., in their review of automated debugging systems \cite{ducasse1988review}, give a  classification of debugging knowledge, which echoes in part the classification of Ahmadzadeh above:

\begin{enumerate}
\item \textit{Knowledge of the intended program (program I/O, behaviour and implementation)
\item Knowledge of the actual program (program I/O, behaviour and implementation)
\item Understanding of the programming language
\item General programming expertise
\item Knowledge of the application domain
\item Knowledge of bugs
\item Knowledge of debugging methods}
\end{enumerate}

\noindent Eisenstadt conducted a survey of programmers, in an attempt to harness the \textit{``generally overlooked \ldots\ potential benefit of self-reports by programmers that reflect the phenomenology of debugging''}. He then uses this data to classify sources of difficulty in bugs, methods by which they were round, and their root causes. This classification encompasses much of what is alluded to by others we have already quoted. However, the material is all together in one single study, so we look at selected parts of the classification in some detail (names of headings Eisenstadt's from \cite{eisenstadt1997my}, descriptions ours in precis, except when in quotation marks): 

\begin{itemize}
 \item Dimension 1 (why difficult)
  \begin{description}
   \item[Cause/effect chasm] This describes what happens when the observed misbehaviour is distant in running time or source location from the root cause of the bug.
   \item[Tools inapplicable or hampered] Bugs which refuse to show themselves when debugged, in which the bug itself erases its own cause, or when \textit{``some configuration or memory constraints make it impractical or impossible to use the debugging tool''}.
   \item[Faulty assumption/model] Also known as \textit{``misdirected blame''}. Some basic misunderstanding sends the programmer off on the wrong course, when in fact the bug is elsewhere.
   \item[Spaghetti (unstructured) code] \textit{``Informants sometimes complain about `ugly' code invariably written by `someone else'.''} This is broader than the usual use of the phrase to mean simply ``code with GOTOs'' -- one may create incomprehensible code in any language, no matter how structured.
  \end{description}
 \item Dimension 2 (how found)
  \begin{description}
   \item[Gather data] When a programmer decides \textit{``to find out more''}, in order to locate the bug.
    \begin{description}
     \item[Step and study]Single-stepping through the execution of the program, looking at the changes in variable values.
     \item[Wrap and profile]Wrapping a suspect function in a custom piece of code to print out some data before and after each invocation of the function.
     \item[Print and peruse]Scatter print statements at various suspect points in the code to print out function arguments, data structures, or both.
     \item[Dump and diff]Core dumps or extensive logging off a successful and failing run, and then compare them, say with the Unix tool {\small\texttt{diff}}.
     \item[Conditional break] Insert breakpoints conditional on some behaviour or value, then inspect what information the debugger provides.
     \item[Specialist profiling tool] Tools such as {\small\texttt{valgrind}} which detect illegal memory accesses.
    \end{description}
   \item[Inspeculation]\textit{``A hybrid of `inspection', `simulation', and `speculation'\! .''} \textit{``In other words, either they go away and think about something else for a while, or they spend a lot of time reading through code and thinking about it.''} This is recognisable as the catch-all approach when nothing else is working.  
   \item[Expert recognised clich\'es] When a programmer asks for help from a colleague, and the colleague is able to offer an immediate diagnosis, based on some sense of similar bugs seen in the past. This sense may be conscious or unconscious.
   \item[Controlled experiments] When a clear idea about the root cause of a bug has been obtained, such tests or experiments are used to confirm. Compare with Myers' remarks on testing as a part of debugging above.
  \end{description}
 \item Dimension 3 (root cause categories)
  \begin{description}
   \item[Memory] In low-level software in languages without type safety, or in the case of compiler errors, overwritten memory can cause bugs which are not immediately apparent. In earlier times, before operating system support for separate text and data segments, these were even more common. Even in modern programming languages, memory used up (in particular stack space used up) can lead to such hard-to-diagnose errors.
   \item[Vendor] Compilers or toolchains or operating systems with bugs. Or, bugs in libraries or APIs provided by others.
   \item[Design logic] The algorithm worked correctly, but was misconceived. A common cause is some missing case which was not considered.
   \item[Initialization] Incorrect starting state for a program, data structure or function, rather than incorrect logic per se.
   \item[Variable] Wrong variable used, either a design logic error (see above) or a simple typing blunder (see Lexical below). But self-reports made it hard to distinguish the two, so Eisenstadt used a separate category.
   \item[Lexical] Simple typos, or wrong understanding of operator precedence and other similar faults.
   \item[Unsolved] \textit{``Some informants never solved their problems.''}
   \item[Language] Semantic ambiguity or misunderstanding. \textit{``In one case, an informant reported he thought 256K meant 256,000.''}
   \item[Behaviour] \textit{``The end user's or programmer's subtle behaviour \ldots\ In one case, the bug was caused by an end user's mysteriously pressing several keys at once; in another case, the cause was mischievous code.''}  
  \end{description}
\end{itemize}

\noindent Another interesting classification is that of Knuth, drawn from a diary of ten years of bugs found and fixed in the \TeX\ typesetter \cite{knuth1989errors}. Some 867 bugs are enumerated and classified according to an ad hoc categorisation of 15 kinds. For example:

\begin{quotation}354 Avoid infinite loop when recovering from {\small\texttt{\$\$}} in restricted horizontal mode. \textsection 1138 R
\cite{knuth1989errors}\end{quotation}

\noindent This is bug number 354, at section 1138 in the code, categorised as type R ``Robustness''.

When evaluating our debugger, in chapter \ref{chap:evaluation}, we shall return to some of these lists and classifications, as well as to the commentary in this chapter, to see if our solution has the potential to improve on the state of the art with regard to the fundamental characteristics and difficulties of debugging that we have illuminated through this look at the historical literature.

\section{Debugging in functional programming today}

We plan to build a new debugger for the functional language OCaml \cite{ocaml}, trying to learn lessons from the history of debugging. Some of these lessons will no doubt be language-agnostic, but we expect functional languages to have special requirements. And so, we take a tour of existing debuggers for functional languages and examine to what extent they are usable and used. \label{decision2-1}\addition{In the next section, we will take a brief look at modern debuggers for imperative languages, and debuggers which use visualization (we will look at the literature in the field of software visualization later in this thesis in section 3.6.) There is, regrettably, not enough time or space to look in detail neither at debuggers for imperative languages nor at visualization, so we must specialise early.}

Debuggers for functional languages have often followed the pattern of those for imperative languages, even though the mental model of evaluation as expression-rewriting is different. Concepts such as breakpoints often appear, for example. Such debuggers come in several flavours. Some work by extending low-level executable debuggers such as \index{GDB}GDB \cite{gdb} or \index{LLDB}LLDB \cite{lldb} with extra routines to allow reconstruction of expressions, some modify the program as it is being compiled, inserting information which can be used by a specialised debugging program, and some work simply by providing macros or extra routines for debugging or logging.

Almost all the literature we have thus far reviewed is concerned with the debugging of imperative programs. How does debugging for functional programs differ? One of the great claims of \index{debugging!functional programs}functional programming is that type systems -- and, in particular, \index{type inference}type inference -- remove whole classes of bugs. However, it is important to separate somewhat this claim from the question whether type errors coming from the type inference engine are a form of debugging as such. Often we think of debugging as something which happens once a program can be run -- a program which fails the typechecker cannot be run. Nevertheless, we can consider type inference errors a sort of debugging support. Since the executable is not produced, however, many standard debugging techniques cannot be applied.

Most functional programming languages have a \index{REPL}Read-Eval-Print Loop, which is used not only for learning and programming-by-experimentation but also for light testing and debugging. Debuggers for functional languages aim to provide facilities over and above the REPL. The limitation of the REPL-as-debugger approach is that debugging often occurs due to an unexpected failure in production, rather than something the programmer provokes deliberately (which we would probably call testing). It is worth noting a practical point: in many functional languages, it is possible to build a REPL automatically with all libraries and modules used in a project linked in, for example by typing {\small\texttt{make\negthinspace\ repl}} instead of just {\small\texttt{make}} -- a boon for usability. We shall see that these sorts of practical concerns have an outsize influence upon the utility of a debugging tool. Penney does not consider REPLs an adequate replacement for a real debugger, however:

\begin{quotation}
Unfortunately, this approach can be clumsy and inadequate. Suppose we are debugging a compiler, and we find that a function call \texttt{compile\!\! program} fails. Moreover, we suspect the problem lies in a sub-expression in the definition of \texttt{compile}:

\texttt{transform\! table\! expression}

\noindent We want to supply different test cases to \texttt{transform}. However, it could be extremely difficult to type suitable arguments in by hand: the corresponding expressions may be long and complicated and have consistency requirements that make it difficult to write correct examples.

A tracing tool can supply the information needed much more easily. By placing a breakpoint on the definition of the \texttt{transform} function, we can pick out every call to this directly from the trace of a failing execution. \cite{penney2000augmenting}
\end{quotation}

\noindent The problem of not being able to produce (or edit) complex data for test cases is a scenario which appears again and again in debugging and testing.

Let us now examine contemporary debugging tools in popular functional programming languages.

\subsection{Standard ML}

\index{Standard ML}Refreshingly, debugging was considered, at least in passing, during the early stages of the design of Standard ML, as Hall and O'Donnell quote Milner recalling:

\begin{quotation}
\textit{ML does \textup{not} use lazy evaluation; it calls by value. This was decided for no other reason than inability to see the consequences of lazy evaluation for debugging (remember that we wanted a language which we could use rather than research into), and the interaction with the assignment statement, which we kept in the language for reasons already mentioned.} \cite{hall} \cite{milner}
\end{quotation}

\noindent In fact, there was a primitive multi-typed value printer in early versions of ML, though it was not robust or portable, and does not appear in Standard ML:

\begin{quotation}
Two multi-typed functions are included as quick debugging aids. The function \texttt{print\!\! :\!\!\!\!\! ty\!\! ->\!\! ty} is an identity function, which as a side-effect prints its argument exactly as it would be printed at top-level. The printing caused by \texttt{print(exp)} will depend upon the type ascribed to this \textit{particular} occurrence of \textit{exp}; thus print is not a normal polymorphic function. The function \texttt{makestring\!\! :\!\!\!\!\! ty\!\! ->\!\! string} is similar, but instead of printing it returns as a string what print would produce on the screen. Since top-level printing is not fully specified, programs using these two functions should not be ported between implementations. \cite{harper1986standard}\end{quotation}

\noindent The history of debugging tools for Standard ML has not always followed this pattern. Wadler \cite{Wadler} records the story of Tolmach and Appel's debugger \cite{tolmach}, which was deeply intertwined with the compiler and runtime of \index{SML/NJ}SML/NJ Standard ML. As the compiler implementation evolved, the debugger fell out of step, and is no longer available. Standard ML developers \textit{``must return to older, more manual debugging methods''} \cite{Wadler}. This is a reminder that keeping a tool which is not part of the standard language toolchain up to date requires either frequent modification, or a design which is fundamentally distanced from the language. Of course the spectrum of effort required to update a debugger for a new version of the language is broad. It is likely any tool other than a GDB-style one (operating solely on executables) will always require some updating with each new toolchain release. One practical way to ensure this, if only socially, is to make it part of the toolchain.

The Poly/ML\index{Poly/ML} implementation of Standard ML contains an interactive debugger which operates not in a separate environment, but within the usual REPL. For example, here we set a breakpoint on a list reversal function, and ask for the values associated with some names: 

\medskip
\begin{verbatim}[commandchars=\\\{\}]
Poly/ML 5.7.1 Release                   
> PolyML.Compiler.debug := true;         \textrm{\textit{initialise the debugger}}
val it = (): unit
> fun rev [] = []
#   | rev (h::t) = rev t @ [h];
val rev = fn: \textquotesingle{}a list -> \textquotesingle{}a list
> PolyML.Debug.breakIn "rev";            \textrm{\textit{enable the debugger on our function}}
val it = (): unit
> rev [1, 2, 3, 4];                      
function:rev
debug > h;                               \textrm{\textit{ask for values at the debugger prompt}}
val it = ?: \textquotesingle{}a
debug > t;
val it = [?, ?, ?]: \textquotesingle{}a list
debug > ^CCompilation interrupted        \textrm{\textit{exit the debugger}}

Exception- Interrupt raised

> clearIn "rev";                         \textrm{\textit{disable the debugger for our function}}
val it = (): unit
\end{verbatim}
\medskip

\noindent Notice, though, that even our simple polymorphic list reversal prevents the Poly/ML debugger from printing out the full details of the values we would like to see. One can give the type manually, but giving the wrong type can crash Poly/ML, the documentation advises. Poly/ML also includes a tracer:
\medskip
\begin{verbatim}[commandchars=\\\{\}]
> trace true;                           \textrm{\textit{set tracing on}}
val it = (): unit
> rev [1, 2, 3, 4];                     \textrm{\textit{run our function}}
 rev [?, ?, ?, ?]
  rev [?, ?, ?]
   rev [?, ?]
    rev [?]
     rev []
     rev () = []
    rev () = [?]
   rev () = [?, ?]
  rev () = [?, ?, ?]
 rev () = [?, ?, ?, ?]
val it = [4, 3, 2, 1]: int list
\end{verbatim}
\medskip

\noindent Polymorphism again defeats it, even in our toy scenario.

\subsection{F\#}\index{F\#}

Microsoft's F\# \cite{fsharp} is an example of a functional language tightly integrated into (and shipped by default with) a platform of frameworks, libraries and so on, based on the Common Language Runtime \cite{clr}. Thus, we would expect F\# to be an interesting case when examining debugging functional programs in a broadly imperative scenario. The official guidance on debugging F\# \cite{fsharpdebugging} is, however, a little disheartening:

\begin{quotation}

{\itshape \noindent Debugging F\# is similar to debugging any managed language, with a few exceptions:}

\begin{itemize} 
\item \textit{The Autos window does not display F\# variables.} 

\item \textit{Edit and Continue is not supported for F\#. Editing F\# code during a debugging session is possible but should be avoided. Because code changes are not applied during the debugging session, editing F\# code during debugging will cause a mismatch between the source code and the code being debugged.}

\item \textit{The debugger does not recognise F\# expressions. To enter an expression in a debugger window or a dialog box during F\# debugging, you must translate the expression into C\# syntax. When you translate an F\# expression into C\#, make sure to remember that C\# uses == as the comparison operator for equality and that F\# uses a single =.} \cite{fsharpdebugging}
\end{itemize}
\end{quotation}

\noindent So the advantage of having a full IDE and a widely-used platform for the functional language to sit within is tempered by inadequate support for debugging, at least in this case.

\subsection{OCaml}\index{OCaml}
We surveyed OCaml users informally to ask whether they routinely use debuggers, and if not, why not. The overwhelming result was that debuggers are not widely used. The Haskell community has found the same \cite{marlow}. There was plenty of general assent:

\begin{quotation}
\textit{For sure, a simpler and more robust way to visualise/follow the execution of a program would be a great help to debug OCaml programs.}
\end{quotation}

\noindent More interestingly, several respondents whittled this down to a theme: 

\begin{quotation}
\textit{I use tools that I am familiar with when debugging because I don't want to focus on two things (learning a new tool and tracking down/fixing a bug).}
\end{quotation}

\noindent One coined this the \textit{lack-of-use vicious circle}:

\begin{quotation}
\textit{When you really need a debugger, you're not willing to learn a new tool. When you're willing to learn a new tool, you don't really want to learn a debugger.}
\end{quotation}

\noindent We shall now look at typical methods used for debugging in the OCaml community, in addition to the use of the REPL for debugging-like tasks which we have already highlighted.

\paragraph{Exception backtraces}\index{exception!backtraces}An OCaml program, when appropriately compiled with debugging information, can print out a useful trace of the stack of function calls and exception raises which led to an uncaught exception reaching the top-level and thus leading to the termination of the program:

\begin{verbatim}
$ OCAMLRUNPARAM=b ./a.out 
Fatal error: exception Failure("tl")
Raised at file "pervasives.ml", line 30, characters 22-33
Called from file "example.ml", line 2, characters 24-33
Called from file "example.ml", line 5, characters 2-6
\end{verbatim}

\noindent The system is very much a best-effort service, which can be fooled by OCaml's optimizer, and results vary from architecture to architecture. In addition, it is not possible to build a custom REPL which continues on after the error and stack trace to allow further interactive interrogation of the program's state.

\paragraph{Debugging with print statements}\index{debugging!by print statement}\index{print statement} Inserting print statements is a popular method of informal debugging and logging across multiple languages and platforms. However, OCaml (unlike, for example, Haskell or Java), has no generic mechanism for printing user-defined data types. So one is limited to printing only parts of the data -- such as strings or numbers, or forced to use custom printers, or limited to a library whose purpose is to provide custom printers. Such restrictions can be painful. Nonetheless, inserting print statements is an example of a debugging mechanism which, whilst it may not always be effective, is at least available in almost all circumstances. Perhaps it is this aspect of its usability which explains its enduring popularity. The OCaml community still recommends it (perhaps an indication of the paucity of debugging tools):

\begin{quotation}
``In fact, for complex programs, it is likely the case that the programmer will use explicit printing to find the bugs, since this methodology allows the reduction of the trace material: only useful data are printed and special purpose formats are more suited to get the relevant information, than what can be output automatically by the generic prettyprinter used by the trace mechanism''
\cite{ocamlorgdebug}
\end{quotation}

\paragraph{OCaml tracing}\index{tracing!in OCaml}The OCaml REPL has a very basic tracing mechanism. For example, here we define a simple function and the tracer displays inputs to and outputs from the function as it runs:

\medskip
\begin{verbatim}[commandchars=\\\{\}]
# let rec f x = function 0 -> x | n -> f (succ x) (pred n);;
val f : int -> int -> int = <fun>
# #trace f;;                         \textrm{\textit{enable tracing for our function}}
f is now traced.
# f 0 2;;                            \textrm{\textit{invoke the function}}
f <-- 0
f --> <fun>
f* <-- 2
f <-- 1
f --> <fun>
f* <-- 1
f <-- 2
f --> <fun>
f* <-- 0
f* --> 2
f* --> 2
f* --> 2
- : int = 2
\end{verbatim}
\medskip

\noindent Too much has been lost in the compilation process to provide more information about the evaluation process of the expressions. In particular, currying is not preserved. Values having polymorphic types cannot be printed but appear as {\small\texttt{<poly>}}, a significant obstacle to usability.

\paragraph{OCamldebug} 

The OCamldebug\index{OCamldebug} program is supplied with OCaml. It operates only on compiled and linked bytecode executables, not on native code executables nor on source code. The program must have been compiled with debug information. In addition, one's build process must make a bytecode executable by default, or in addition to a native code one. The stand-out feature of OCamldebug is its ability to `time-travel' -- that is to jump backwards in a program's execution as well as forwards. This is achieved by the use of the Unix fork mechanism. The intention is to make it easier to ``catch the bug in the act''.

The program is run in a sequence of numbered steps. A step is something like a function application or a conditional branch. One may:

\begin{itemize}
\item jump to a numbered step, forward or backward;
\item print out the source code at the current step;
\item inspect a value from the source code;
\item set breakpoints based on source code positions.
\end{itemize}

\noindent As we shall see, there are some limitations. Let us take an example run. We start the debugger with the program {\small\texttt{ocaml\negthinspace\ {-}{-}version}}:

\medskip
\begin{verbatim}[commandchars=\\\{\}]
$ ocamldebug ocaml --version
	OCaml Debugger

(ocd) run
Loading program... done.
The OCaml toplevel, version 4.06.1
Time: 49260                                         
Program exit.                        
\end{verbatim}
\medskip

\noindent We go to time zero, the beginning of the program. We have `time-travelled'. 

\medskip
\begin{verbatim}[commandchars=\\\{\}]
(ocd) go 0
Time: 0
Beginning of program.
\end{verbatim}
\medskip

\noindent We step forward one step at a time. What we see is just module initialisation from OCaml's built-in Standard Library {\small\texttt{Pervasives}}.

\medskip
\begin{verbatim}[commandchars=\\\{\}]
(ocd) step
Time: 1 - pc: 7384 - module Pervasives
26     (Invalid_argument "index out of bounds")<|a|>
(ocd) step
Time: 2 - pc: 7552 - module Pervasives
164   float_of_bits 0x7F_F0_00_00_00_00_00_00L<|a|>
\end{verbatim}
\medskip

\noindent We move into code from the actual program (rather than module initialization) but we are still stuck in Standard Library code, there being no way to ask OCamldebug to show only steps involving the user's main program only.

\medskip
\begin{verbatim}[commandchars=\\\{\}]
(ocd) go 20000
Time: 20000 - pc: 136812 - module Arg
277     else <|b|>if s.[n] = \textquotesingle{} \textquotesingle{} then loop (n+1)
(ocd) step
Time: 20001 - pc: 136828 - module Arg
277     else if s.[n]<|a|> = \textquotesingle{} \textquotesingle{} then loop (n+1)
(ocd) step
Time: 20002 - pc: 136864 - module Arg
278     else <|b|>n
\end{verbatim}
\medskip

\noindent We print some values by giving their names: 

\medskip
\begin{verbatim}[commandchars=\\\{\}]
(ocd) print n                                      \textrm{\textit{ask for value of {\texttt{n}}}}
n: int = 1
(ocd) go 20001
Time: 20001 - pc: 136828 - module Arg
277     else if s.[n]<|a|> =  \textquotesingle{} \textquotesingle{} then loop (n+1)
(ocd) print s                                      \textrm{\textit{ask for value of {\texttt{s}}}}
s: string = " Display this list of options"
(ocd) print loop                                   \textrm{\textit{ask for value of {\texttt{loop}}}}
Unbound identifier loop                           
\end{verbatim}
\medskip

\noindent Some values cannot be found, or are opaque. We cannot alter the values within the debugging environment and re-run the code.

OCamldebug can be used in conjunction with the Emacs text editor \cite{stallman1981emacs} to provide for a smoother debugging experience via shortcuts for debugger commands, and the ability to jump to the source code position of a breakpoint. Again, though, this places a restriction on the programmer's environment if they want the best from the tool. It is also possible to install printers for user-defined data types, although the manual cautions \textit{``For technical reasons, the debugger cannot call printing functions that reside in the program being debugged.''}

\paragraph{OCaml and GDB} It is possible to use a debugger which works on executables (such as GDB) with OCaml, of course, but facilities are limited. The semantic gap between the source text and the executable in the functional model of computation is much wider than when debugging a language such as C. There is, however, an extension  to GDB \cite{libmonda} in development, which allows for limited printing out of OCaml values using the type annotation files left behind during compilation. A similar system \cite{lefessant} is available for the LLVM debugger LLDB. But they operate very much a best-effort service. The advantage, of course, is that they work  on native code executables, and can be attached in situ to processes as and when required.

\subsection{Haskell}

\index{Haskell}In 2005, in a survey of the users of GHC, the most prominent Haskell compiler, \textit{``By far the most common request was for a debugger.''} \cite{marlow}. We shall describe three such debuggers briefly, and then discuss the results of a similar survey undertaken ten years later, in 2015. The paper describing the current debugger (shipped with GHC), says \textit{``The most prominent working debuggers for Haskell are Hat and Hood.''} \cite{marlow}, so we choose those to look at before examining the GHC debugger.

\paragraph{Hat}

The Hat debugger \cite{chitil2002transforming} operates by recompiling programs in such a way that they dump a trace of the whole execution to file as the program runs, whether it ends normally or with an error. After the program has finished, the user runs tools which use the dumped data to explore the execution of the program. A transformed program runs about a hundred times more slowly than the original. However, Hat allows some modules (say the standard library) to be `trusted' and therefore untraced. This also enables Hat users to debug programs which use third-party libraries which Hat has not, or cannot, recompile.

The tools provided include {\small\texttt{hat-observe}} to show the arguments with which each function is called, {\small\texttt{hat-trail}} to explore computations backwards (to answer the question ``Where did my bug come from?''), and {\small\texttt{hat-explore}} to step through computations.

However, there are problems. The trace can be enormous, even for modest program runs. This, together with the tracing slowdown, may restrict the debugging of programs which do not fail (or otherwise end) quickly. Since Hat relies on \index{debugging!by transforming the source}transforming Haskell programs into ones which are semantically equivalent but which also output trace data, it cannot be used with programs which make use of language extensions Hat does not know about. Thus, one use of a recent Haskell extension in a codebase can rule out Hat as a debugger. To debug inside libraries, Hat also requires one to recompile all the libraries in tracing variants, for use with Hat.

\paragraph{Hood} The Hood debugger \cite{Gill00debugginghaskell} works by printing out data structures at various ``observation points'' in the program, rather than using the stepping model of the typical imperative debugger. As with Hat, part of the motivation for its design choices revolves around the extra complication of laziness -- with Haskell's built-in  {\small\texttt{Debug.trace}}, for example, the act of printing something out might change the evaluation order of the program, and therefore suppress a bug, or at least complicate reasoning. Hood allows the user to insert points at which observations about data structures are collected without altering the observable behaviour of the program. The authors show how  this method fits particularly well with the point-free style of functional programming, the observation point acting as a sort of identity function in the middle of a chain of functions. For example, {\small\texttt{consumer\negthinspace\ .\negthinspace\ observe\negthinspace\ "intermediate"\negthinspace\ .\negthinspace\ producer}} as the equivalent to {\small\texttt{consumer\negthinspace\ .\negthinspace\ producer}} but storing the debug information for this observation point under the label {\small\texttt{intermediate}}, from where it may be retrieved later. The Hood tool itself can be used for viewing such information. 

\paragraph{The GHC debugger} The 2007 GHC debugger \cite{marlow} was designed by looking at the flaws of Hat and Hood and trying to avoid them. In particular, the authors list ways in which Hat and Hood are not always available -- for example, not suitable for use on all programs, or being limited to one compiler, or requiring re-compilation of libraries, or not being able to work interactively, or not being able to print polymorphic values. They go so far as to say \textit{``The debugger should work with everything and always be available, even if this means sacrificing functionality.''} We have seen similar observations about what we call \index{accessibility}`accessibility' as the cornerstone of usability in our review of the literature in \cref{chap:related}.

The debugger is used by loading the program into the REPL in the normal way, and using the extended REPL commands provided by the debugger (for example {\small\texttt{:break}}) to control debugging. Values of in-scope names may be inspected, and the program single-stepped.

\paragraph{2015 Survey}

An email survey \cite{fpcomplete} commissioned by a commercial Haskell contractor, targeting 16000 Haskell users (with 1240 replies), asked respondents to finish the following sentence, choosing from a list of words: \textit{``Debugging and profiling: improvements in this would be\ldots''}. Here are the results:

\bigskip
\begin{center}
\begin{tabular}{ll}
crucial & 29\%\\
important & 30\%\\
helpful & 23\%\\
slight help & 9\%\\
no impact & 4\%\\
\end{tabular}
\end{center}
\bigskip

\noindent (Total crucial or important 59\%). A free response field was also provided. Here are a selection of responses:

\begin{quotation}
\noindent\textit{I see this as one of the major blockers to Haskell development. Even with understanding of the language, it is sometimes very difficult to discover why programs behave in certain ways.}
\end{quotation}

\begin{quotation}
\noindent\textit{Debugging Haskell code is like groping in the dark with a hand tied behind your back.}
\end{quotation}

\begin{quotation}
\noindent\textit{Debugging Haskell is still a pain for beginners and hampers adoption.}
\end{quotation}

\begin{quotation}
\noindent\textit{This is of key importance. When \textup{[the]} compiler eliminates \ldots\ most of the basic problems, the most troublesome and complicated issues with logical structure are still to be debugged away.}
\end{quotation}

\begin{quotation}
\noindent\textit{Students complain about the difficulty of debugging Haskell programs (laziness, no printf).}
\end{quotation}

\begin{quotation}
\noindent\textit{To be honest I'm a bit `afraid' of this part of Haskell.}
\end{quotation}

\begin{quotation}
\noindent\textit{The backbone of development lies in debugging. This shouldn't even be a question.}
\end{quotation}

\begin{quotation}
\noindent\textit{I would never be able to convince my coworkers \textup{[to adopt Haskell]} without decent debugging support.}
\end{quotation}

\noindent This last answer alludes to a source of past disillusionment about the apparent lack of progress of the art of programming despite vast improvements in computing power, language design, and compiler tools. As the field advances, old problems are solved only to be replaced with ones which could not have been conceived of unless we had already solved the old ones. Debugging is likely always to be needed, and unlikely to be eliminated in the way envisaged by the pioneers of computing.

\subsection{Lisp}

Common Lisp \cite{commonlisp}\index{Common Lisp}\index{Lisp} has a tracing function similar to the OCaml one we looked at earlier, although there are more sophisticated facilities: the user can ask for certain values to be printed at each step, or for tracing to begin or end only when a certain predicate related to the code holds. The tracer is itself implemented as a Lisp macro.

\medskip
\begin{verbatim}[commandchars=\\\{\}]
[1]> (defun rev (l)
           (cond
             ((null l) \textquotesingle())
             (T (append (rev (cdr l)) (list (car l)))))) 
REV
[2]> (TRACE rev)                                            \textrm{\textit{trace our function}}
;; Tracing function REV.
(REV)x
[3]> (rev \textquotesingle(1 2 3 4))                                       \textrm{\textit{invoke it}}
1. Trace: (REV \textquotesingle(1 2 3 4))
2. Trace: (REV \textquotesingle(2 3 4))
3. Trace: (REV \textquotesingle(3 4))
4. Trace: (REV \textquotesingle(4))
5. Trace: (REV \textquotesingle{}NIL)
5. Trace: REV ==> NIL
4. Trace: REV ==> (4)
3. Trace: REV ==> (4 3)
2. Trace: REV ==> (4 3 2)
1. Trace: REV ==> (4 3 2 1)
(4 3 2 1)
\end{verbatim}
\medskip

\noindent In a similar fashion to the OCaml tracer, only the inputs and outputs are shown, rather than a diagram of the evaluation of the body  of the function.

The Froglet debugger \cite{watt1994froglet} is a modern source-level debugger for Lisp. Its author explains that, at the time it was written, Lisp had been overtaken in terms of debugging facilities by the advent of source-level debuggers for traditional imperative languages. Previously, Lisp had relatively good facilities, due to the pervasive use of S-expressions at a time when debuggers for other languages worked at the machine level. However, this very abstraction of S-expressions was in turn inferior to more modern debuggers which operated at source level.

Racket\index{Racket} \cite{racket}, a modern Scheme\index{Scheme} implementation, contains two debugging tools. The first is  a breakpoint-based debugger with an optional graphical interface. Panes show the stack and the values of local names. When execution is paused at the start of an expression, an alternative value may be substituted for an expression, for experimentation purposes. Similarly, when execution is paused at the end of an expression's evaluation, an alternative return value may be substituted. The second is an ``algebraic stepper'', which can show each step of the evaluation of the program in the source language:

\bigskip
\medskip

{\centering\includegraphics[width=0.75\textwidth]{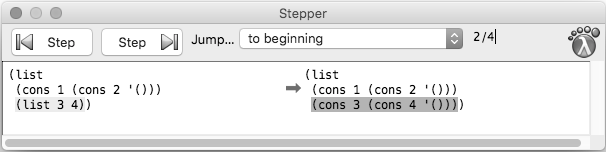}\par}

\bigskip

\noindent Both the debugger and the algebraic stepper use Racket's \index{continuation marks}``continuation marks'' scheme \cite{Clements2001}, which elaborates the program source such that, when it is compiled, enough information remains on the stack to point to, or even reconstruct the expression at the marked points. The debugger works for all Racket programs, but does not show the actual state of the expression being evaluated. The algebraic stepper does show the actual expression, but it only works for the small ``Beginning student'' and ``Intermediate Student'' languages, not the full Racket language.  One cannot debug into other libraries unless the libraries themselves have been compiled with such an elaboration. This may be mitigated somewhat by shipping an optional, elaborated version of the language's standard libraries. Such a  stepping approach, nevertheless, appears to offer a compelling foundation for debugging functional programs, if it could be freed from its limitations.

Some other dynamic languages, for example Smalltalk, have interesting debuggers, but these ideas do not usually translate to the kind of statically-typed batch-compiled environment which is dominant among today's general-purpose languages.

\section{\addition{Modern debuggers for imperative programming}}

\addition{There are two areas of the debugging literature and practice which we have not thus far covered in any detail: visual interfaces to general debugging (that is to say, debugging without the command line); and specialised visual debugging for particular domains (for example, visual debugging of concurrency issues). We address these briefly now, and explain why they do not form a main plank of our thesis.} 

\addition{Modern debuggers which operate in a GUI follow the pattern of DDD \cite{zeller2000debugging,zeller1996ddd}, a graphical interface to the ubiquitous GDB debugger. As the authors explain:}

\begin{quotation}
\addition{Besides ``usual'' features such as viewing source texts and breakpoints, DDD provides a graphical data display, where data structures are displayed as graphs. A simple mouse click dereferences pointers or reveals structure contents. Complex data structures can be explored incrementally and interactively, using automatic layout if preferred. Each time the program stops, the data display reflects the current variable values. \cite{zeller1996ddd}}
\end{quotation}

\noindent\addition{Here is a modern example, the Microsoft Visual Code editor \cite{viscode}, debugging another imperative language, Python:}\par

\bigskip
\begin{centering}
\includegraphics[width=\textwidth]{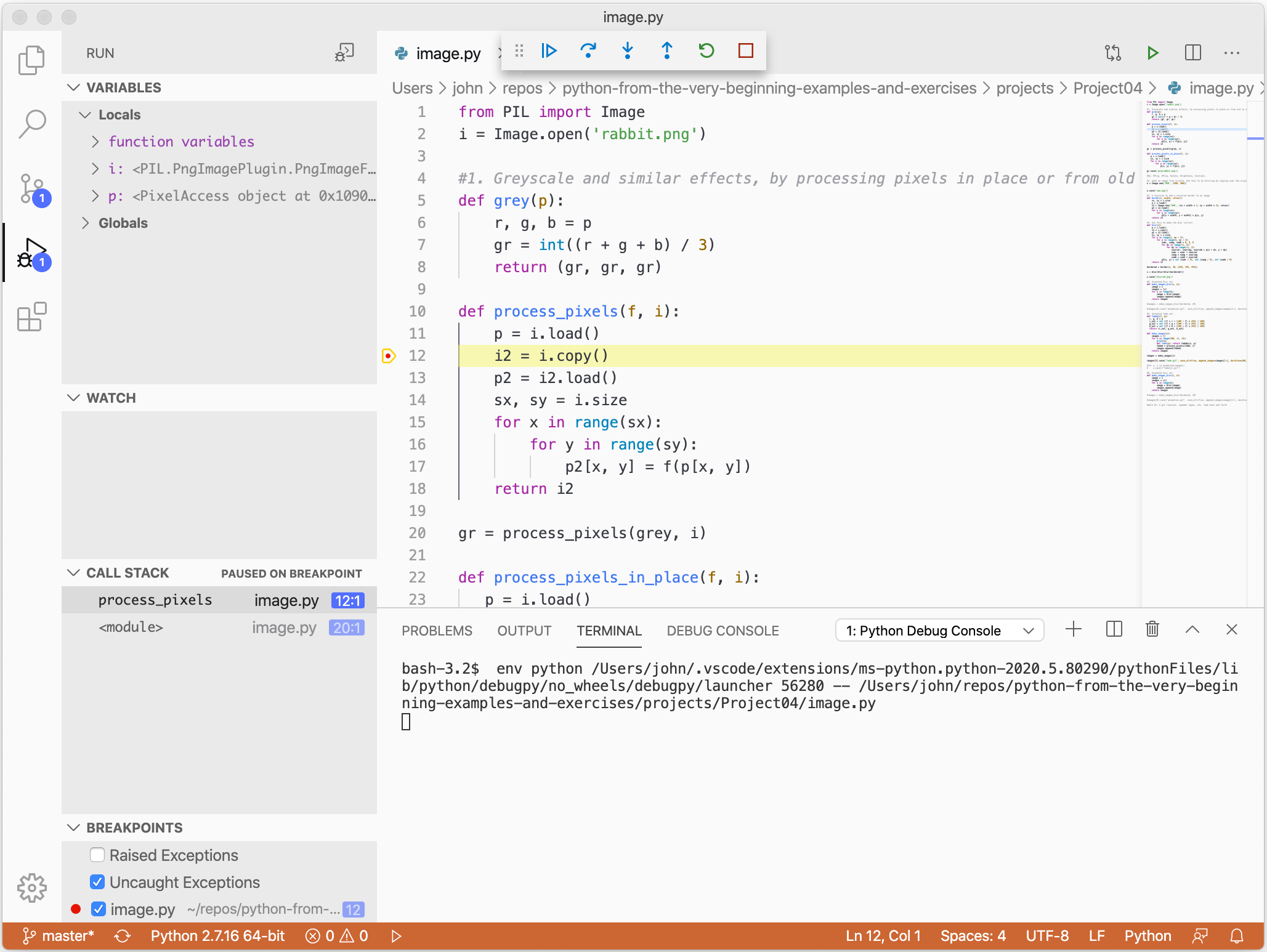}
\end{centering}

\noindent\addition{Debuggers for object-oriented languages, for example the HotWire debugger for C++ \cite{laffra1994hotwire}, often have visualization methods specifically for that idiom. As the authors say:}

\begin{quotation}
\addition{What is required, are filtering techniques that enable us to define and regulate our focus while we are trying to understand why a system behaves as it apparently does. Additionally, the enormous amount of information encountered while inspecting running applications should be represented in a fashion that allows programmers to actually interpret and use it. \cite{laffra1994hotwire}}
\end{quotation}

\noindent\addition{Visual debugging in the sense of drawing charts and graphs and flow diagrams is exemplified by the JIVE debugger \cite{gestwicki2004jive,gestwicki2002interactive} and Alsallakh's tracing debugger \cite{alsallakh2012visual}, both for Java. The authors of JIVE summarise:}

\begin{quotation}
\addition{JIVE provides visualizations of object structures and execution histories. The notation is generally applicable to object-oriented programs but has been customized for Java. Our methodology highlights the fact that objects are environments of program execution, with method activations nested within their proper object contexts. We provide intuitive visualizations of such Java features as objects, static contexts, inner classes, threads, and exceptions. \cite{alsallakh2012visual}}
\end{quotation}

\noindent\addition{We can see an interesting example of how quickly these more tightly-integrated approaches can age:}

\begin{quotation}
\addition{Programs with Swing or AWT interfaces are fully supported in JIVE; the GUIs coexist with the visualization of the program’s execution. \cite{alsallakh2012visual}}
\end{quotation}

\noindent\addition{Neither of these GUI toolkits is in significant use today. The requirement for a debugger to ``move with the times'' in this fashion is avoided by DDD in its dependence only on the well established and well-supported GDB debugger. We should like to avoid it too.}

\addition{Another kind of visualization, for debugging and educational purposes is touched upon in a separate literature review on interpretation in section \ref{interplit}.}

\section{Summary}

\correction{We have taken lessons from the history of debugging since the beginning of the computer era, and surveyed recent systems in our area of interest, functional programming. Now, having done this, it is time to tease out the lessons from this reading, and decide how to build a debugger which might advance the state of the art.} \label{summary2}\addition{Being restricted by time and space, we have been unable to give full attention to the range of modern imperative debuggers, but reading the literature on debugging for functional languages makes a convincing case that they are essentially different and, in any event, our historical review leads us to understand that the issues at stake 
are fundamental enough that they should be well represented there, despite the passage of time.}  

\addition{We shall next decide upon the lessons to take from this literature review -- both positive and negative, and delineate our scope so as to focus the project appropriately.}

\chapter{Approach}
\label{chap:approach}

\begin{quotation}\textit{\large I want everything explained to me or nothing. \textrm{\begin{flushright}--- \textup{Camus,} The Myth of Sisyphus\end{flushright}}}\end{quotation}

\vspace{10pt}

\noindent What can we learn from our review of the theory and practice of debugging? Having identified the attributes of a usable debugger, can we see how to apply that to the functional world, building a new debugger which represents a genuine step forward, and avoiding the mistakes of the past?

In this chapter, we develop the concept for our new debugger, justifying its overall design by reference to the literature review. We include examples of the use of our debugger as it was imagined at the time this approach was developed. We shall see later that the favoured interface to our final debugger turned out rather differently, but still in the spirit of the approach we advocate in this chapter.

\section{Concept}

We intend to tackle the problem of debugging by \index{interpretation}\index{direct interpretation}directly interpreting the program, showing the intermediate steps of evaluation. When we say ``directly interpreting'', we mean just that -- a completely naive \index{step-by-step evaluation}\index{evaluation!step-by-step}step-by-step evaluation of the source code or AST without recourse to any kind of transformation or compilation whether involving  bytecode or not. We make this clarification because the REPL is often referred to as an interpreter though it does not work by interpretation but by compilation of each entered phrase to bytecode. 

Why interpretation? Because it allows the program to be run without the \index{compiler!loss of information in}loss of information inherent in the compilation process. There is no reconstruction of information required, no lossy mapping back and forth between source and executable. It fits the model of functional programming as \index{evaluation!by reduction}evaluation by the reduction of an expression to a value, rather than making the programmer think imperatively. This choice will, we hope, allow for a design which sweeps away many of the disadvantages of existing solutions, replacing them with one big disadvantage -- that interpretation is extremely slow. We shall then work to mitigate that disadvantage to arrive at a usable debugger.

Thus, we are writing a kind of \index{tracing!debugger}\index{debugger!tracing}tracing debugger, but because of the interpretation method we shall have all information available at all times. So the traces will be unusually complete. Here is Penney describing tracing debuggers:

\begin{quotation}Traditionally, tracing the execution of a program means displaying an outline of the sequence of evaluation steps taking an initial program state to the final result. It is more sophisticated than just revealing what point in a program has been reached, usually highlighting the expression under evaluation and giving the user a degree of control with facilities for single-stepping and breakpoints. \cite{penney2000augmenting} \end{quotation}

\noindent Before getting too deeply into this line of thought, let us begin with an example to illustrate the concept before returning to our argument.

\section{Example}

\begin{figure}
{\centering\noindent\includegraphics[width=0.75\textwidth]{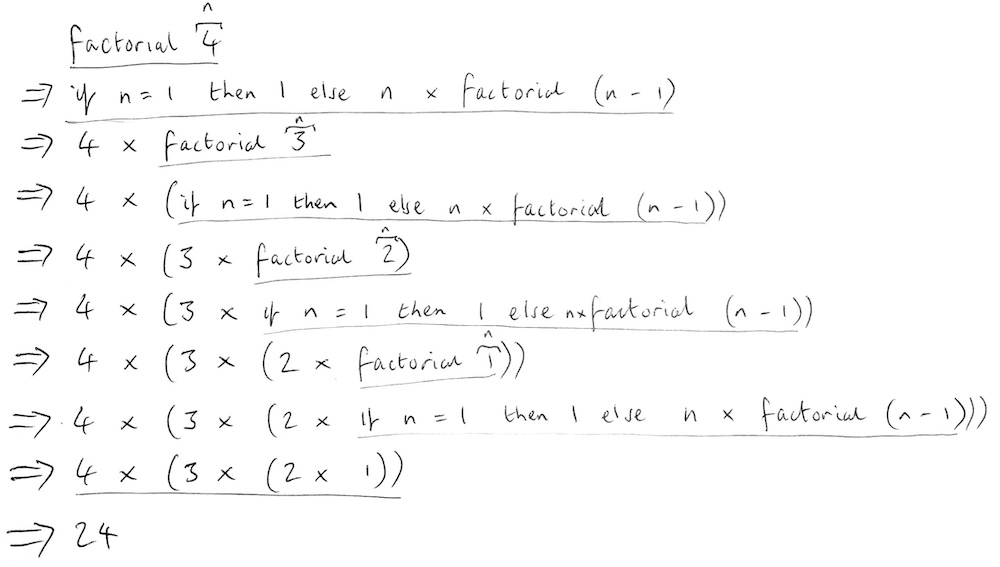}\par}
{\centering\singlespacing\noindent\small\label{F2} Figure 2. Handwritten diagram for the factorial function.\par}\label{factorial}
\end{figure}

\begin{sidewaysfigure}
\begin{landscape}
\scalebox{0.65}{
\begin{minipage}{3\textwidth}
{\ttfamily
~~~~let rec factorial n = if n = 1 then 1 else n * factorial (n - 1) in factorial 4\\
=>  ~let rec factorial n = if n = 1 then 1 else n * factorial (n - 1) in let n = 4 in if n = 1 then 1 else n * factorial (n - 1)\\
=>  ~let rec factorial n = if n = 1 then 1 else n * factorial (n - 1) in let n = 4 in if false then 1 else n * factorial (n - 1)\\
=>  ~let rec factorial n = if n = 1 then 1 else n * factorial (n - 1) in let n = 4 in n * factorial (n - 1)\\
=>  ~let rec factorial n = if n = 1 then 1 else n * factorial (n - 1) in let n = 4 in 4 * factorial (n - 1)\\
=>  ~let rec factorial n = if n = 1 then 1 else n * factorial (n - 1) in 4 * factorial (4 - 1)\\
=>  ~let rec factorial n = if n = 1 then 1 else n * factorial (n - 1) in 4 * factorial 3\\
=>  ~let rec factorial n = if n = 1 then 1 else n * factorial (n - 1) in 4 * (let n = 3 in if n = 1 then 1 else n * factorial (n - 1))\\
=>  ~let rec factorial n = if n = 1 then 1 else n * factorial (n - 1) in 4 * (let n = 3 in if false then 1 else n * factorial (n - 1))\\
=>  ~let rec factorial n = if n = 1 then 1 else n * factorial (n - 1) in 4 * (let n = 3 in n * factorial (n - 1))\\
=>  ~let rec factorial n = if n = 1 then 1 else n * factorial (n - 1) in 4 * (let n = 3 in 3 * factorial (n - 1))\\
=>  ~let rec factorial n = if n = 1 then 1 else n * factorial (n - 1) in 4 * (3 * factorial (3 - 1))\\
=>  ~let rec factorial n = if n = 1 then 1 else n * factorial (n - 1) in 4 * (3 * factorial 2)\\
=>  ~let rec factorial n = if n = 1 then 1 else n * factorial (n - 1) in 4 * (3 * (let n = 2 in if n = 1 then 1 else n * factorial (n - 1)))\\
=>  ~let rec factorial n = if n = 1 then 1 else n * factorial (n - 1) in 4 * (3 * (let n = 2 in if false then 1 else n * factorial (n - 1)))\\
=>  ~let rec factorial n = if n = 1 then 1 else n * factorial (n - 1) in 4 * (3 * (let n = 2 in n * factorial (n - 1)))\\
=>  ~let rec factorial n = if n = 1 then 1 else n * factorial (n - 1) in 4 * (3 * (let n = 2 in 2 * factorial (n - 1)))\\
=>  ~let rec factorial n = if n = 1 then 1 else n * factorial (n - 1) in 4 * (3 * (2 * factorial (2 - 1)))\\
=>  ~let rec factorial n = if n = 1 then 1 else n * factorial (n - 1) in 4 * (3 * (2 * factorial 1))\\
=>  ~let rec factorial n = if n = 1 then 1 else n * factorial (n - 1) in 4 * (3 * (2 * (let n = 1 in if n = 1 then 1 else n * factorial (n - 1))))\\
=>  ~let rec factorial n = if n = 1 then 1 else n * factorial (n - 1) in 4 * (3 * (2 * (let n = 1 in if true then 1 else n * factorial (n - 1))))\\
=>  ~let rec factorial n = if n = 1 then 1 else n * factorial (n - 1) in 4 * (3 * (2 * 1))\\
=>  ~let rec factorial n = if n = 1 then 1 else n * factorial (n - 1) in 4 * (3 * 2)\\
=>  ~let rec factorial n = if n = 1 then 1 else n * factorial (n - 1) in 4 * 6\\
=>  ~24\par}
\bigskip
\bigskip
\bigskip

{\ttfamily~~~~\underline{factorial 4}\\
n = 4 ~=> ~\underline{\textbf{if} n = 1 \textbf{then} 1 \textbf{else} n * factorial (n - 1)}\\
n = 4 ~=> ~n * factorial (\underline{n - 1})\\
\-~~~~~~~=> ~4 * \underline{factorial 3}\\
n = 3 ~=> ~4 * (\underline{\textbf{if} n = 1 \textbf{then} 1 \textbf{else} n * factorial (n - 1)})\\
n = 3 ~=> ~4 * (n * factorial (\underline{n - 1}))\\
\-~~~~~~~=> ~4 * (3 * \underline{factorial 2})\\
n = 2 ~=> ~4 * (3 * (\underline{\textbf{if} n = 1 \textbf{then} 1 \textbf{else} n * factorial (n - 1)}))\\
n = 2 ~=> ~4 * (3 * (n * factorial (\underline{n - 1})))\\
\-~~~~~~~=> ~4 * (3 * (2 * \underline{factorial 1}))\\
n = 1 ~=> ~4 * (3 * (2 * (\underline{\textbf{if} n = 1 \textbf{then} 1 \textbf{else} n * factorial (n - 1)})))\\
\-~~~~~~~=> ~4 * (3 * (\underline{2 * 1}))\\
\-~~~~~~~=>* 24\par}
\end{minipage}}
\bigskip

\label{sideways}
\singlespacing\noindent \label{F3}Figure 3. A naive rendering of the evaluation of {\small\texttt{factorial\negthinspace\ 4}} showing each step of the evaluation, followed by an automatically abridged one, eliding a) parts of the evaluation of the {\small\texttt{\textbf{if}}} construct; b) the definition of a recursive function mentioned in the expression; c) the final portion of arithmetic; and d) trivial operations such as {\small\texttt{3\negthinspace\ -\negthinspace\  1}}. In addition, {\small\texttt{\textbf{let}}} expressions unique in the whole expression are moved to the left, and basic syntax highlighting has been used. The expression to be reduced in each step has been underlined, even if the next line is elided.\end{landscape}\end{sidewaysfigure}

\onehalfspacing

\noindent As is traditional, we consider a program for calculating the factorial of a positive number:

\medskip
{\small\ttfamily
\noindent\texttt{\textbf{let rec} factorial n =}\\
\texttt{\-~~\textbf{if} n = 1 \textbf{then} 1 \textbf{else} n * factorial (n - 1)}\\
\texttt{\textbf{in}}\\
\texttt{\-~~factorial 4}
}
\medskip

\noindent In figure 2 we show how we might write the evaluation of this program on paper. Note that the definition of the function is not shown, and that we have used underlining and annotations for clarity. Now let us consider the same visualization produced by machine. The upper portion of figure 3 shows a naive computer-generated visualization of the evaluation of this program. This is certainly  not how we would write such an evaluation on paper. Although the evaluation shown is self-contained in the sense that each line of it is a valid program, it is hard to see what is going on. It is large, both in width -- how long the expression becomes, and length -- how many lines are needed. Writing each evaluation step over multiple lines as we did with the source program would not only increase the length, but make it difficult to visually compare adjacent lines. We must reduce the amount of information shown, even in this simple case.

Look now at the lower part of figure 3, showing the output of one of our prototype systems. The following changes have been made:\clearpage

\begin{itemize}
\item We removed the definition of the {\small\texttt{factorial}} function itself. As it is recursive, its name will appear in the expression anyway.

\item We avoided printing any reduction step which leads to an expression such as {\small\texttt{\textbf{if}\negthinspace\ false}} or {\small\texttt{\textbf{if}\negthinspace\ true}}.

\item We have not shown the intermediate steps of simple arithmetic which reduce {\small\texttt{4\negthinspace\ *\negthinspace\ (3\negthinspace\ *\negthinspace\ (2\negthinspace\ *\negthinspace\ 1))}} to {\small\texttt{24}}.

\item We have removed trivial arithmetic (e.g.\ subtracting one), even when it involves variable names, such as reducing {\small\texttt{n\negthinspace\ -\negthinspace\ 1}} to {\small\texttt{3}} directly rather than via {\small\texttt{4\negthinspace\ -\negthinspace\ 1}}.

\item We have removed {\small\texttt{\textbf{let}}} bindings which apply to the whole expression to the left-hand side of the {\small\texttt{=>}} arrow to avoid too many {\small\texttt{\textbf{let}\negthinspace\ n\negthinspace\ = \ldots}} instances making the output too wide.

\item We have used simple syntax highlighting in the form of bold for keywords.

\item We have underlined the expression to be reduced at each step.
\end{itemize}

\noindent All changes have been made automatically. Each step is no longer a valid OCaml program, but the increase in readability is significant. Clearly, for larger programs, such elision will be even more important, since the focus needs to be on the currently-evaluating subexpression of a potentially huge expression representing the whole program. Note that all the intervening steps of the computation are performed, but certain lines are not printed. This means that the finer details of the computation may be inspected upon demand.

In the program trace we have already exhibited, it is clear that for realistic programs, the program trace (both its width and its length) may be significant. This issue is discussed in some detail by Taylor \cite{taylor-thesis} and Pajera-Flores \cite{winhipe}. A practical solution must involve providing ways of a) eliding information within a single step, reducing the width; b) eliding whole steps, reducing the length; c) searching the resultant trace, if it is still too large to spot the bug; and d) moving backward and forward through the trace to connect cause and effect in the computation.

Figure 4 shows another example of handwritten evaluation, this time with more radical elisions. This is insertion sort, showing just the top-level \texttt{insert} and \texttt{sort} functions, not their internal workings. This kind of abstraction is perfectly natural when writing such diagrams by hand, but we shall have to carefully instruct the computer how to do it. It is not clear yet, of course, what the best interface to such a visualization system might be, in particular whether such elision may be interactively variable.

\begin{figure}
{\centering\noindent\includegraphics[width=0.8\textwidth]{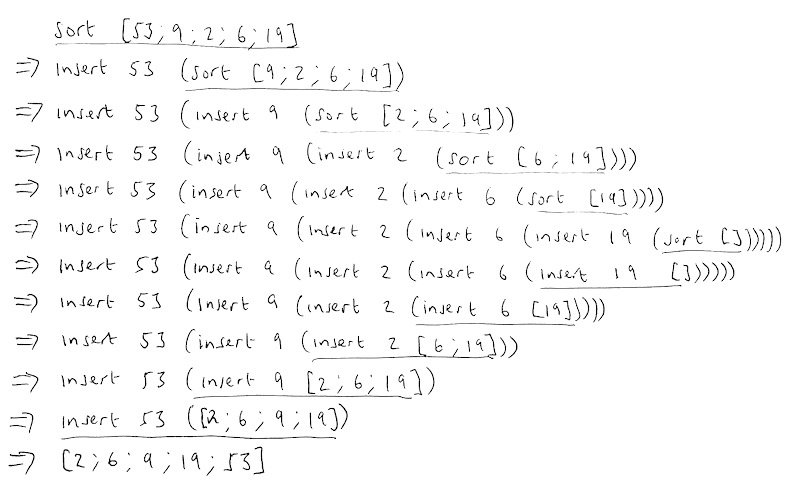}\par}
{\centering\singlespacing\noindent\small\label{F4} Figure 4. Handwritten diagram for insertion sort.\par}
\end{figure}

\section{Rationale}

Existing debuggers suffer, to a lesser or greater
extent, from a lack of what Marlow  calls  
\index{accessibility}\textit{accessibility} \cite{marlow}. They provide only for a subset of the language, or
require changes to be made to build environments, or do not scale well. So there is often one or more fundamental impediments to their use -- they are not \textit{accessible}. A debugger must be as accessible as a compiler. Marlow claims that the most complete Haskell trace debugger, Hat \cite{chitil2002transforming}, remains largely unused due to a lack of such accessibility -- for example, it must be modified to support new third-party libraries. We
intend, then, to bake in the correct design decisions to support widespread
applicability (and thus adoption) from the beginning, even if it is at the
expense of other desirable characteristics, such as speed. We aim for our system to a) be able
to support the whole language by design; b) be suitable for any \index{build environment}build
environment where OCaml programs can already be built; and c) be abstracted from
the compiler, and thus be robust to advances in the language and run-time
environment. Thus, instead of imagining the perfect debugger, writing a toy system, and worrying about how to extend it to a practical one later, we will make design decisions based on the practicalities, and work backward from our goal. Even if our system is initially a toy in the sense that it does not support the full language, it is not a toy in terms of its integration with the language and runtime, and so extending it to the full language should be technologically straightforward (though a sizeable piece of work). Can our system be used to debug any OCaml program where source is available, even if it uses external libraries? Can our system support development of a complex system such as the OCaml compiler itself? Most importantly, of course, do people actually choose to use it?
\label{tests}

\subsection*{The extremist approach}

We choose to build a debugger which puts this notion of accessibility first (it is the core of usability), and everything else second. We claim that, without such universal availability, our debugger would be added to the growing pile of debuggers for functional languages which lie unused. And so this extremism is, in fact, in the service of practicality.

We can contrast the accessibility of a low-level debugger, such as GDB, which allows any executable to be debugged to the much more limited accessibility of various high-level debuggers. We need to bridge this gap. So now we define precisely what trade-offs we are willing to accept.

\section{Scope}

It is important to demarcate our area of interest. Whilst we are adamant that our debugger design must be applicable in all reasonable circumstances in order to be usable and used by programmers, that does not mean that it will cover everything one may mean by the term debugging, used in its widest sense. For example, programs which statically analyse memory leaks or audit code against security threats such as timing attacks might be considered ways to identify (and therefore remove) bugs from a program, but they are not typically called debuggers. It is with the more everyday sense of the word debugger that we shall operate. Let us list some of our principal aims here. Our eventual debuggers should:

\begin{itemize}
\item be useable with any build system;
\item work with \index{mixing OCaml and C}mixed C/OCaml code;
\item be able to debug libraries, not just the programmer's own code;
\item be easy to keep in sync with the OCaml toolchain, so a new version can be released with each OCaml version;
\item require no patches to the target toolchain -- an independent development;
\item be suitable for debugging the development of the OCaml compiler itself, and similar complex code.
\end{itemize}

\noindent What restrictions will we insist upon? What latitude will we give ourselves? Just like \index{GDB}GDB, we will insist upon the executable being compiled with a special flag. In fact, the requirement is slightly  stronger than GDB, since GDB is of some use on an executable not so compiled. We shall not make any attempt to provide for the debugging of code from other languages (so, for example, C code linked into a primarily-OCaml executable will not be debuggable, only the OCaml portion). But such executables will run properly. This means that, for example, if one suspects a bug exhibited in an OCaml executable is really a bug in the \index{OCaml!compiler}OCaml compiler or \index{OCaml!runtime}runtime, debugging it may require GDB in addition to our debugger.

There will doubtless be small ways in which the interpreter differs, whilst still being equivalent given the OCaml semantics. For example, its exact limits for \index{stack overflow}stack overflows on non-tail-recursive code may be different, or the order of execution of threads may be different. This might mean that some bugs go away (or become apparent) when switching to the interpreted version of OCaml. This might damage reproducibility when debugging certain kinds of low-level problems. 

\label{decision3}\addition{In the previous chapter, we looked at debuggers with graphical interfaces and a little at those which used visualization tools such as charting to show, for example, the memory behaviour of programs. We will stick to text-based tools for our initial implementation for several reasons. First, it is possible to add GUI tools later, just as DDD builds upon GDB. Second, we know from our literature review that it has been widely observed that programmers prefer to think in the source language. Third, as we go on to discuss in section \ref{interplit}, the very idea of software visualization is considered dubious by some of its researchers.}

\correction{We believe that this is the sensible approach for a high-level debugger. A debugger which sits within these parameters ought to embody the spirit of accessibility and so fulfil our requirements.}

\section{Correctness and maintenance}
\label{correctness}

How will we know that the step-by-step interpreter is correct? \index{correctness}Correctness is the major concern in any implementation of a programming language, but we have a special extra concern. We must make sure that the interpreter matches the semantics of the OCaml compiler. It is no use trying to debug a program, only to find that the bug changes or disappears when using the interpreter. This concern exists also in the core OCaml distribution, unusually, since there are two compilers -- bytecode and native code.

Our interpreter will share the \index{compiler!front end}front end (parser, typechecker, etc.) with the OCaml bytecode and native code compilers, so we need not worry about correctness there -- its behaviour should match exactly their behaviour. In the actual evaluation, though, we must ensure the semantics of the language are obeyed to the same standard as in the compiled implementations. \index{formal proof}\index{proof}Formal proof will not work. There is no formal semantics of OCaml (save for a subset \cite{owens2008sound}) to prove adherence to and, in any event, without also formally proving the OCaml compiler, we could not show the two are semantically equivalent.

\correction{The OCaml distribution does come, however, with a large test suite, which we could use to test the interpreter.} \addition{We would choose a subset of its tests removing, for example, any which depend upon code generation,  and automatically test them against the expected results. Such a modification to the OCaml test suite, which has support for regression testing too, would be relatively simple and have the advantage of making a connection between the OCaml source tree and our interpreter. We will also test the interpreter manually against the corpus of simple programs in the author's textbook \cite{whitington2013ocaml}. In the future this could be automated. Another,  smaller corpus of programs is included in Appendix B -- this is the set we will use to test the speed of the interpreter, as described in section \ref{testspeed}.} 

How do we know that our debugger will continue to work when the OCaml compiler is updated? Programming languages change, and new language features are added. Of course, the best solution would be to eventually have our step-by-step interpreter included in the OCaml distribution itself. The reason for doing so would be social, not technical -- it would ensure that for each release the interpreter would be updated along with the compilers. It is not unusual for debuggers to be included in the core distribution of a language. Failing such an inclusion, the debugger will have to be updated for each major release of OCaml. But these changes are likely to be rather easier than one might expect, due to the sharing of the front end. And some, such as the changes to typechecking internals in the front end which frequently appear in OCaml compiler change logs, may require no changes to the interpreter.

The maintenance load is composed also of the complexity of the interpreter itself. There is a marked difference in complexity between writing a simple evaluator for an abstract syntax tree and its step-by-step counterpart. Recent work on an interpreter for a subset of OCaml \cite{cong2016implementing,furukawa2019} suggests a scheme whereby one may \textit{``implement a stepper concisely by writing an evaluator that is close to a standard big-step interpreter''}, using the same \index{continuation marks}continuation mark techniques as \index{Racket}Racket -- so there is reason to believe that this difficulty may also one day be overcome.

\section{Interpretation in the literature}
\label{interplit}

Given the long history of debuggers and debugging which we have already explored, and our stated intention to create a debugger which works by the direct interpretation of the program source, it is legitimate to ask: ``If it is such a good idea, why hasn't anyone done it before?'' As part of our work to address this, we take a quick tour of some of the few naively-interpreting programming language implementations, and the use of interpretation in debugging tools past and present.

Early computers were so slow (and machine time so precious) that little consideration seems to have been given to interpreted high-level languages -- the benefits of writing even a simplistic compiler were clear. The loss of debugging information which compilation embodies was an unfortunate side effect. Interpretation as part of a language was perhaps most embodied by Lisp, which provides an interesting case study in interactive debugging -- a retrospective by Sandewall \cite{sandewall1978programming} covers this in detail. So often, though, the many idiosyncrasies of Lisp set it apart from the mainstream, and so it can be difficult to apply the same techniques to more conventional languages.

There is some interesting work by Hoffman and O'Donnel on the automatic generation of interpreters for functional programs (or mathematical expressions) whose syntactic structures are tree-like \cite{hoffmann1979interpreter,hoffmann1982pattern}. However, we expect that modern automatic techniques for pattern matching \cite{Maranget} will allow these efficiencies to be automatically applied to our interpreter by the OCaml compiler itself.

Of direct interpreters in use today, there is little mainstream evidence. One interesting memory is that of a commercial product of the 1980s, the C interpreter Saber-C. The personal recollections \cite{ranum} of Marcus Ranum, a pioneer of network security, contain these rather tantalising passages:

\begin{quotation}
For me, using Saber-C was an eye-opener. It gave me a whole new approach to development, since I could use the interpreter to directly call functions from a command line, without having to write a test harness with a \texttt{main\!\! ()} routine and controlled inputs and outputs. \ldots without having to go through a compile/link/debug cycle, my code-creation sped up dramatically and I was catching bugs in `real-time' as I wrote each block of code. After a little while, I can safely say that the quality of my code skyrocketed. \ldots\ 
 I used to use Saber-C as my secret weapon to convince my friends I had sold my soul to the Devil: whenever they were dealing with a weird memory leak or a wild pointer that was making their programs crash with a corrupted stack or mangled free list. Usually Saber-C could pinpoint the problem in a single pass. \cite{ranum}
\end{quotation}

\noindent This software is still apparently available under the name CodeCenter. However, our  attempts to contact the company failed. The Software Preservation Group have made some resources available \cite{softpres} including the screenshot shown in figure 5.

\begin{figure}
\begin{center}
\includegraphics[width=0.6\textwidth]{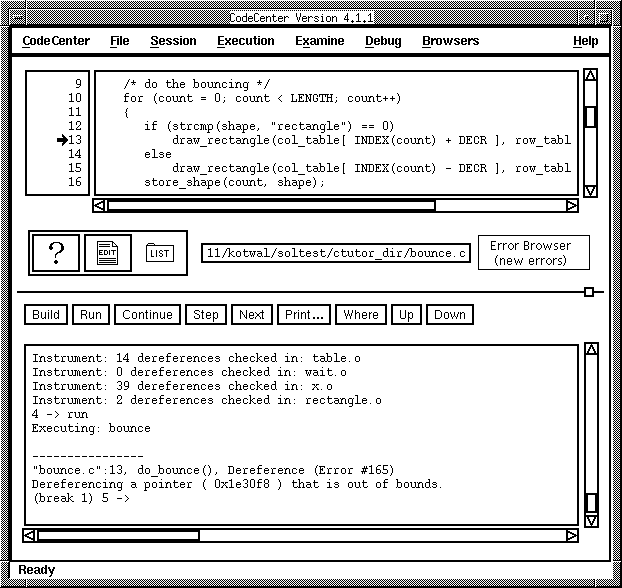}
\end{center}
{\centering\singlespacing\noindent\small\label{F5} Figure 5. The CodeCenter C Interpreter.\par}
\end{figure}

There are two interesting extant systems for \index{interpretation!of C/C++}interpretation of C/C++. A commercial project intended for beginning students, with an emphasis on engineering and numerical methods is SoftIntegrations's CH system \cite{ch, ChengBook}. It is provided as a sort of `platform', bundled with numerical and graph-plotting facilities. It does not support all modern C/C++, requires extra work to interface with any new library and is not derived from any other C compiler. The result of these things, taken together, is that it appears unsuitable for the working programmer.

Another educational system is Thetis \cite{freund1996thetis}, designed for a first-year course at Stanford. The authors say \textit{``On further examination, however, we made an important discovery: student frustration was less a function of the language than of the programming environment.''} and \textit{``Interactive debuggers typically require students to understand advanced concepts before they are ready to assimilate them''.} The Thetis system provides run-time error detection, a set of enhanced syntactic restrictions for beginners to prevent common confusion (e.g.\ {\small\texttt{if\! (i\! =\! 0)}} vs {\small\texttt{if\! (i\! ==\! 0)}}), and debugging and visualization tools. It supports, however, only the subset of the language required for the course.

The Cling interpreter \cite{vasilev2012cling} is a C++ interpreter, built on Clang and LLVM, providing a REPL for interactive C++ development.  Its aim is to provide a development environment for the data processing library ROOT used in the field of high-energy physics. The use of the modular structure of modern compiler components like Clang and \index{LLVM}LLVM (compared with Cling's predecessor Cint) allows Cling to keep up to date with modern language developments:

\begin{quotation}
One can see that using Clang and LLVM as libraries helps us to build a complete interactive C++ interpreter with far less effort than building it from scratch. \ldots\ This allows us to maintain only that part of the code and delegate the rest of the maintenance to the LLVM community. \cite{vasilev2012cling}
\end{quotation}

\noindent A similar REPL, for C\#, is available \cite{csharprepl}, based on the Mono .NET platform.

In chapter 6, we shall address the problem of \index{interpretation!speed of}speed of execution in interpreted languages by using \index{selective interpretation}selective interpretation. That is to say we shall have most of the program running natively, at full speed, while only the part which is being debugged will be interpreted. This has been written about before \cite{chase1987selective,gough1994interpretive} though, as we shall see, a number of factors combine to make our system somewhat simpler in its implementation.

In addition to this work in practical systems, there is also plenty of interest in the field of \index{software visualization}\index{visualization}software visualization, which seems to have its own, largely separate literature. Two useful surveys \cite{UrquizaFuentes,survey2009} give a general overview of recent developments in this area, the first specific to functional programming, the second with wider scope. A very broad introduction \cite{PetredeQuincey} gives background, and Sorva et al.\ give a comprehensive survey \cite{sorva2013review} of education systems for program visualization. We pick out a few recent systems for further discussion.

The WinHIPE system \cite{winhipe} is a recent incarnation of these ideas for the HOPE \cite{burstall1980hope} language. It uses a step-by-step evaluation system, and explicitly addresses the problems of scale by elision of information and a focusing mechanism. The emphasis, however, is on \index{visualization!graphical}graphical (tree-based) representations, an approach we shall not take, being of the belief that trees can often be, in fact, harder to read than good \index{prettyprinting}prettyprinted program representations. The Visual Miranda Machine \cite{visual-miranda} provides a trace of the evaluation of a lazy functional program, together with a commentary showing the reason for choosing each evaluation step. There is a discussion of granularity, taking the example of the ``list comprehension'' language feature. Touretsky  describes a Lisp-based system \cite{touretzky} that produces mainly textual traces, but with some use of graphical elements to indicate the different scoping mechanisms peculiar to Lisp. The presentation of ZStep95 \cite{zstep95} begins by noting that debugging is, essentially, a human interface problem. The authors concentrate on the concept of \textit{immediacy} (temporal, spatial, and so on), which they see as essential, and exhibit a stepping debugger for a functional language which can go back and forth through time. 

Another approach to this problem is as a special case of the more general concept of a \textit{calculator} \cite{Reeves95thecalculator,miracalc}, showing how it pertains to various logical systems with a mathematical basis, not just functional programs. Prospero \cite{taylor-thesis} is a more fully-developed system, again for a lazy language. It includes methods for filtering the evaluation trace to elide information and a careful discussion of usability issues. The MLExplain system \cite{mlexplain} is a recent interpreter for a subset of OCaml, based upon the JavaScript ``double debugger'' JSExplain \cite{jsexplain}.

These systems are mostly concerned with program visualization for teaching; we wish to bias ourselves towards the task of general debugging, hoping that some educational uses will be subsumed by it. The authors of DrScheme \cite{drscheme}, a precursor to Racket, urge caution here, choosing instead to build a `tower' of syntactically restrictive variants of Scheme specifically for educational purposes. We would prefer to avoid this in the name of universality.

It is worth pointing out that much research in software visualization concerns overtly graphical approaches. We take a simpler line, sticking to prettyprinting. We claim that the most important aspect of a successful visualization is elision -- reducing the information visible to just what is required so that large datasets may be understood easily -- whether interactively or not. Programmers are used to seeing their program as text, and visualizing its evaluation as, for example, a graphical tree structure, is less useful for debugging large programs (it can be useful, of course, for visualizing program source code structure as opposed to evaluation traces). A useful discussion \cite{patel1997comparison} of this topic of text-vs-graphics in visualization systems for programming education shows support for our choice, amid a mixed picture. Finally on this theme, a refreshing read, in which the author tears apart the failure of his own endeavours and that of his field, in order to do better in the future, is the paper ``The Paradox of Software Visualization'':

\begin{quotation}
Software visualization seems like such a logical and helpful concept with obvious benefits and advantages. But after decades of work, it has yet to be successful in any mainstream development environment. \cite{reiss2005paradox}
\end{quotation}

\noindent Some of Reiss' complaints echo those we found in our review of the debugging literature, about the ability to use these tools in any situation:

\begin{quotation}
Some tools require extensive configuration to get a program into an environment and get it understood by the environment. \ldots\ Some require that the user work with specific languages or subsets or convert portions of the system for compatibility. \cite{reiss2005paradox}
\end{quotation}

\noindent So we can feel reasonably confident in our conservative safety-first choice of text-based `visualization', at least for our first attempt on the problem of debugging.

\section{Summary}

We have introduced our central, rather unintuitive idea for debugging -- to interpret our programs naively -- based upon our previous review of the literature on debugging. We have given a few examples, and a literature review  of areas relevant to the chosen design, to add to our historical literature review in \cref{chap:related}. \label{summary3}\addition{We have decided on an appropriate motive, rationale and scope for our first attempt at implementation, but we shall only be able to evaluate its success or lack of success later in this document. There are, of course, other choices which could have been made, and ours is certainly a bold one. This boldness increases the risk in execution, of course, but makes the prize greater.}

Now it is time to build a prototype and find out if the idea can work.


\chapter{An interpreter for OCaml}
\label{chap:interpreter}

\begin{quotation}\textit{\large You campaign in poetry. You govern in prose. \textrm{\begin{flushright}--- \textup{Mario Cuomo}\end{flushright}}}\end{quotation}

\vspace{10pt}

\noindent This chapter presents the technical development and practical use of our prototype \index{OCamli@\textsf{OCamli}}\textsf{OCamli} interpreter for OCaml. It is the foundation upon which our eventual debugger will be built. Presently, it has a prototype command line interface which, on its own, is suitable for debugging little OCaml programs. Later, we shall introduce a system which is more widely applicable.

\addition{This interpreter was written in an experimental manner, not an overtly planned one. The aim is to build from tiny programs up to larger ones, using more and more parts of the OCaml language until we can load the Standard Library and run many programs. We have not attempted to implement OCaml's object-oriented subset of features, since they do not appear in the Standard Library. We have implemented only such parts of the module language as are strictly required to load the Standard Library. But we have implemented the vast majority of the core language. As we shall learn in the next chapter, this implementation was superseded, and so did not reach the stage where formal testing for correctness, as described in section \ref{correctness}, was appropriate.}

A primer on OCaml for the uninitiated can be found in appendix \ref{primer}.

\section{Architecture}

We should like to have a command {\small\texttt{ocamli}} so we may write:

\medskip
\begin{verbatim}
$ ocamli test.ml
\end{verbatim}
\medskip

\noindent The program whose source code is {\small\texttt{test.ml}} will be interpreted, functioning in the same way as it would if compiled and then run. We will allow the flag {\small\texttt{-show}} to show the final result of the evaluation of the program, and {\small\texttt{-show-all}} to show all the stages of computation:

\medskip
\begin{Verbatim}[commandchars=|\[\]]
$ ocamli -show test.ml
7

$ ocamli -show-all test.ml
1 + |underline[2 * 3]
|underline[1 + 6]
7
\end{Verbatim}
\medskip

\noindent Such a program will need to read the source code, convert it to a representation suitable for direct interpretation, interpret it in a fashion which allows for the printing of each individual step, and print those steps out in a readable way.

\section{A new representation for OCaml programs}

It is possible to produce a step-by-step interpreter for OCaml which operates directly upon the parse tree data type exposed by \index{compiler-libs@\textsf{compiler-libs}}\textsf{compiler-libs}, the OCaml toolchain's library form. However, the data type is far from ideal. It holds much information which is not needed for interpretation, complicating \index{pattern matching}pattern matching. At each step, we must then reconstruct such extra information to ensure that it is a valid parse tree again. A very  early version of our interpreter was constructed using this method. The intent was to avoid introducing a new data type (with maintenance issues), and to enable use of the existing OCaml prettyprinter. However, it quickly became apparent that the disadvantages outweighed the advantages. Consider, for example, the following code to add two integers, from a very early version of \textsf{OCamli}:

\medskip
\begin{verbatim}[commandchars=\\\{\}]
| Pexp_apply (expr, args) ->
    \textbf{if} List.for_all (\textbf{fun} (_, arg) -> is_value arg) args \textbf{then}\hfill\textrm{\textit{if all arguments are values}}
      \textbf{begin match} expr.pexp_desc \textbf{with}
      | Pexp_ident \{txt = Lident (("*" | "/" | "+" | "-") \textbf{as} op)\} -> \hfill\textrm{\textit{if an integer op}}
          \textbf{begin match} args \textbf{with}
            [(_, \{pexp_desc = Pexp_constant (Const_int a)\});
             (_, \{pexp_desc = Pexp_constant (Const_int b)\})] -> \hfill\textrm{\textit{extract integer values}}
              \textbf{let} result = calculate a b op \textbf{in}
                \{e \textbf{with} pexp_desc = Pexp_constant (Const_int result)\} \hfill\textrm{\textit{rebuild parse tree node}}
          | _ -> malformed __LOC__
          \textbf{end}
      \textbf{end}
    \textbf{else}
      \textrm{\textit{(cases where one or more arguments not yet values)}}
\end{verbatim}
\medskip

\noindent We have to check each item in the list of arguments is a value, match against strings representing operators, and deal with many nested records, taking them apart and building them back up once we have evaluated a single step. Instead, we should like to be able to just write:

\begin{verbatim}
| Op (op, Int x, Int y) -> Int (calculate op x y)
\end{verbatim}

\noindent This is the aim of our new representation for OCaml programs. We call it \index{TinyOCaml@\textsf{TinyOCaml}}\textsf{TinyOCaml}. We shall now exhibit the bulk of the main type, skipping about a third for brevity \addition{(for example, the module system)}, together with some of the types to which it refers. Unlike the OCaml parse tree data type, which is a complicated set of mutually-recursive data type definitions, we have only a few here. Let us begin with very simple enumerations for mathematical operators, comparison operators, and kinds of \index{for loop}\texttt{\textbf{for}} loop:

\begin{verbatim}[commandchars=\\\{\}]
\textbf{type} op = Add | Sub | Mul | Div

\textbf{type} cmp = LT | EQ | GT | EQLT | EQGT | NEQ

\textbf{type} forkind = UpTo | DownTo
\end{verbatim}

\noindent Now, a definition for patterns in OCaml, used with \texttt{\textbf{match}}:

\begin{verbatim}[commandchars=\\\{\}]
\textbf{type} pattern =
  PatAny\hfill\textit{\textrm{the pattern \_ which matches anything}}
| PatVar \textbf{of} string\hfill\textit{\textrm{atomic patterns}}
| PatInt \textbf{of} int
| PatBool \textbf{of} bool
| PatChar \textbf{of} char
| PatString \textbf{of} string
| PatUnit
| PatTuple \textbf{of} pattern list\hfill\textit{\textrm{compound patterns}}
| PatArray \textbf{of} pattern array
| PatNil\hfill\textit{\textrm{list patterns}}
| PatCons \textbf{of} pattern * pattern
\vdots
\end{verbatim}

\noindent Pattern match cases and bindings:

\begin{verbatim}[commandchars=\\\{\}]
\textbf{and} case = pattern * t option * t\hfill\textrm{\textit{pattern, guard, right-hand side}}

\textbf{and} binding = pattern * t
\end{verbatim}

\noindent We shall discuss environment items (which are used for actual evaluation) later, but we show them here for completeness:
\begin{verbatim}[commandchars=\\\{\}]
\textbf{and} envitem =
  EnvBinding \textbf{of} bool * binding list ref \hfill\textrm{\textit{value}}
| EnvType \textbf{of} (bool * Parsetree.type_declaration list)\hfill\textit{\textrm{type declaration}}

\textbf{and} env = envitem list
\end{verbatim}

\noindent And so we come to the main type:

\begin{Verbatim}[commandchars=\\\{\}]
\textbf{type} t =
  Unit                        \hfill\textrm{\textit{atomic types}}
| Int \textbf{of} int         
| Bool \textbf{of} bool       
| Float \textbf{of} float
| Char \textbf{of} char    
| String \textbf{of} string   
| Record \textbf{of} (string * t ref) list \hfill\textrm{\textit{record}}
| Tuple \textbf{of} t list    \hfill\textrm{\textit{tuple}}
| Cons \textbf{of} (t * t)    \hfill\textrm{\textit{list}}
| Nil                       
| Array \textbf{of} t array   \hfill\textrm{\textit{array}}
| Constr \textbf{of} int * string * t option \hfill\textrm{\textit{user-defined data type constructor}}
| Fun \textbf{of} (label * pattern * t * env)  \hfill\textrm{\textit{function}}
| Function \textbf{of} (case list * env) \hfill\textrm{\textit{function with pattern-match}}  
| Var \textbf{of} string               \hfill\textrm{\textit{variable}}
| Op \textbf{of} (op * t * t)          \hfill\textrm{\textit{binary operator}}
| And \textbf{of} (t * t)              \hfill\textrm{\textit{boolean operator}}
| Or \textbf{of} (t * t)               
| Cmp \textbf{of} (cmp * t * t)        \hfill\textrm{\textit{comparison operator}}
| If \textbf{of} (t * t * t option)    \hfill\textrm{\textit{conditional statement}}
| Let \textbf{of} (bool * binding list * t) \hfill\textrm{\textit{let binding}}
| LetDef \textbf{of} (bool * binding list) \hfill\textrm{\textit{let binding structure item}}
| TypeDef \textbf{of} (bool * Parsetree.type_declaration list) \hfill\textrm{\textit{user-defined type definition}}
| App \textbf{of} (t * t)              \hfill\textrm{\textit{function application}}
| Seq \textbf{of} (t * t)              \hfill\textrm{\textit{imperative \texttt{;} operator}}
| While \textbf{of} (t * t * t * t)    \hfill\textrm{\textit{while loop}}
| For \textbf{of} (string * t * forkind * t * t * t) \hfill\textrm{\textit{for loop}}
| Field \textbf{of} (t * string)       \hfill\textrm{\textit{access and set record field}}
| SetField \textbf{of} (t * string * t)
| Raise \textbf{of} (string * t option)\hfill\textrm{\textit{raise exception}}
| Match \textbf{of} (t * case list)    \hfill\textrm{\textit{pattern match}}
| TryWith \textbf{of} (t * case list)  \hfill\textrm{\textit{\texttt{\textbf{try}}\ldots\texttt{\textbf{with}}} \textit{block}}
| ExceptionDef \textbf{of} (string * Parsetree.constructor_arguments) \hfill\textrm{\textit{exception definition}}
| CallBuiltIn \textbf{of} \hfill\textrm{\textit{built-in primitive}}
   (typ option * string * t list * (env -> t list -> t))
| Struct \textbf{of} (bool * t list)   \hfill\textrm{\textit{module implementation}}
| Sig \textbf{of} t list               \hfill\textrm{\textit{module signature}}
\end{Verbatim}

\noindent Let us look again at an example program, and see its evaluation as it may be printed on screen by \textsf{OCamli}:

\medskip
\begin{Verbatim}[commandchars=|\[\]]
   1 + 2 > |underline[3 + 4]
=> |underline[1 + 2] > 7
=> |underline[3 > 7]
=> false
\end{Verbatim}
\medskip

\noindent Here is what is going on inside \textsf{OCamli} -- much simpler than directly manipulating the OCaml parse tree itself:

\medskip
\begin{Verbatim}
   Cmp (GT, Op (Add, Int 1, Int 2), Op (Add, Int 3, Int 4))
=> Cmp (GT, Op (Add, Int 1, Int 2), Int 7)
=> Cmp (GT, Int 3, Int 7)
=> Bool false 
\end{Verbatim}
\medskip

\noindent We shall now consider how to convert the OCaml parse tree into our new, easier to manipulate type for OCaml programs. Consider the following extract of the {\small\texttt{of\_real\_ocaml}} reader for converting an OCaml parse tree into its \textsf{TinyOCaml} representation:

\medskip
\begin{verbatim}[commandchars=\\\{\}]
| Pexp_construct ({txt = Lident "[]"}, _) -> Nil
| Pexp_construct ({txt = Lident "::"}, Some ({pexp_desc = Pexp_tuple [e; e\textquotesingle{}]})) ->
    Cons (of_real_ocaml env e, of_real_ocaml env e\textquotesingle{})
\end{verbatim}
\medskip

\noindent This deals with the standard OCaml list syntax. Similar code deals with each other part of the OCaml syntax. Closure conversion is done at the same time (this is the {\small\texttt{env}} argument above), since it is convenient and avoids another pass.

Converting the other way (from {\textsf{TinyOCaml}} to OCaml's parse tree type) can be useful too, for example if we wish to use the built-in OCaml prettyprinter:

\medskip
\begin{verbatim}
| Unit -> Pexp_construct ({txt = Longident.Lident "()"; loc = Location.none}, None)
| Int i -> Pexp_constant (Pconst_integer (string_of_int i, None)) 
| String s -> Pexp_constant (Pconst_string (s, None))
| Bool b ->
    Pexp_construct
      ({txt = Longident.Lident (string_of_bool b); loc = Location.none},
        None)
\end{verbatim}
\medskip

\noindent Note again the stark difference in verbosity between our type {\small\texttt{Tinyocaml.t}} (to the left of each arrow) and OCaml's parse tree type (to the right of each arrow).

\section{Evaluating expressions}\index{evaluation!of an expression}

For this proof of concept, a very simple step-by-step interpreter has been produced. It has no pretensions towards performance, either by preserving \index{space complexity}\index{time complexity}space and time efficiency vis-a-vis the same program compiled and executed, or with regard to constant overheads. Its job is to provide a minimal working example for experimentation. 

\subsection{Evaluation strategy}
\label{evaluationstrategy}

To evaluate a step of a program (that is, something of type {\small\texttt{Tinyocaml.t}}), we must first determine if the program is a \index{value}value. If it is, there is no evaluation to be done. If not, we find the reducible expression \index{reducible expression}(redex), following the OCaml order of evaluation. We perform one step of evaluation only. This new expression may now be returned for printing, and we continue with the next step.

Let us look at a simple example, comparing with a traditional interpretive evaluator (whose job is to evaluate down to a value in one continuous operation). We choose the \index{operator!short-circuiting}\index{short-circuit}short-circuiting \index{operator!boolean}\index{value!boolean}boolean conjunction operator {\small\texttt{\&\&}}. Here is a snippet from an all-at-once interpreter: 

\medskip
\begin{Verbatim}[commandchars=\\\{\}]
\textbf{let rec} eval = \textbf{function}
  | And (a, b) ->
      \textbf{match} eval a \textbf{with}
      | Bool false -> Bool false
      | Bool true -> eval b
\end{Verbatim}
\medskip

\noindent We evaluate the left-hand side {\small\texttt{a}} to a \textsf{TinyOCaml} representation of  a boolean (either {\small\texttt{Bool\negthinspace\ true}} or {\small\texttt{Bool\negthinspace\ false}}). If it is {\small\texttt{Bool\negthinspace\ false}}, this is the result. If it is {\small\texttt{Bool\negthinspace\ true}}, the code evaluates the right-hand side to a value, and returns it.
 Contrast with the following, which evaluates just a single step:

\medskip
\begin{Verbatim}[commandchars=\\\{\}]
\textbf{let rec} eval_step = \textbf{function}
  | And (Bool false, _) -> Bool false
  | And (Bool true, Bool b) -> Bool b
  | And (Bool true, b) -> eval_step b
  | And (a, b) -> And (eval_step a, b)
\end{Verbatim}
\medskip

\noindent The first line of the pattern match in the step-by-step example is used when the left-hand side has already been fully evaluated and is false: this is the short circuit. The second deals with a fully-evaluated true left-hand side, and a fully evaluated right-hand side. The third is the same as the second, but for a right-hand side not yet fully evaluated: we have found the step which requires evaluation. The fourth and last is for an unevaluated left-hand side: we evaluate the left-hand side one step and leave the right-hand side alone. By similar mechanisms it is possible to write a step-by-step evaluator for each other part of the language. We consider some of the more interesting ones now by way of further example.

\subsection{Imperative programs}\index{imperative program}\index{program!imperative}

Whilst OCaml is a functional language first, there is occasional use of imperative features, and we need to display them in a way which fits in. Consider the evaluation of the OCaml {\small\textbf{\texttt{for}}} construct. When compiled, the following piece of code will print {\small\texttt{12345}}:

\medskip\index{for loop}
\begin{Verbatim}[commandchars=\\\{\}]
\textbf{for} y = 0 + 1 \textbf{to} 6 - 1 \textbf{do} print_int y \textbf{done}
\end{Verbatim}
\medskip

\noindent But how do we show it? Like anything else in OCaml it is simply a kind of expression, so the natural visualization, omitting the internals of {\small\texttt{print\_int}}, is this:

\medskip
\begin{Verbatim}[commandchars=|\[\]]
$ ocamli -e 'for y = 0 + 1 to 6 - 1 do print_int y done' -fast-for -show-all
   |textbf[for] y = |underline[0 + 1] |textbf[to] 6 - 1 |textbf[do] print_int y |textbf[done]
=> |textbf[for] y = 1 |textbf[to] |underline[6 - 1] |textbf[do] print_int y |textbf[done]
=> |underline[|textbf[for] y = 1 |textbf[to] 5 |textbf[do] print_int y |textbf[done]]
1=> |underline[|textbf[for] y = 2 |textbf[to] 5 |textbf[do] print_int y |textbf[done]]
2=> |underline[|textbf[for] y = 3 |textbf[to] 5 |textbf[do] print_int y |textbf[done]]
3=> |underline[|textbf[for] y = 4 |textbf[to] 5 |textbf[do] print_int y |textbf[done]]
4=> |underline[|textbf[for] y = 5 |textbf[to] 5 |textbf[do] print_int y |textbf[done]]
5=> |underline[|textbf[for] y = 6 |textbf[to] 5 |textbf[do] print_int y |textbf[done]]
=> ()
\end{Verbatim}
\medskip

\noindent Helpfully, the semantics of OCaml are such that {\small\texttt{\textbf{for}\negthinspace\ y\negthinspace\ =\negthinspace\ 6\negthinspace\ \textbf{to}\negthinspace\ 5\negthinspace\ \textbf{do}\negthinspace\ \ldots\negthinspace\ \textbf{done}}} is legal and does not execute the body, so we have a proper terminating condition. How is this implemented? The {\small\texttt{For}} constructor of the {\small\texttt{Tinyocaml.t}} data type looks like this:

\medskip
\begin{Verbatim}[commandchars=\\\{\}]
For \textbf{of} string * t * forkind * t * t * t\hfill\textit{\textrm{see previous definition of \texttt{forkind}}}
\end{Verbatim}
\medskip

\noindent Our example would be represented like this:

\medskip
\begin{Verbatim}
  For ("y",
       Op (Add, Int 0, Int 1),
       UpTo,
       Op (Sub, Int 6, Int 1),
       App (Var "print_int", Var "y"),
       App (Var "print_int", Var "y"))
\end{Verbatim}  
\medskip

\noindent We need two copies of the body, so that one may be evaluated step-by-step, and then, when it has been reduced to a value, the spare copy can be moved into place, and we go round again. Here are all the cases needed for step-by-step evaluation of the {\small\textbf{\texttt{for}}} construct:

\medskip
\begin{Verbatim}[commandchars=\\\{\}]
| For (v, e, ud, e\textquotesingle{}, e\textquotesingle{}\textquotesingle{}, copy) \textbf{when} not (is_value e) ->        \hfill\textrm{\textit{evaluate from part}}
    For (v, eval env e, ud, e\textquotesingle{}, e\textquotesingle{}\textquotesingle{}, copy)
| For (v, e, ud, e\textquotesingle{}, e\textquotesingle{}\textquotesingle{}, copy) \textbf{when} not (is_value e\textquotesingle{}) ->       \hfill\textrm{\textit{evaluate to part}}
    For (v, e, ud, eval env e\textquotesingle{}, e\textquotesingle{}\textquotesingle{}, copy)
| For (_, Int x, UpTo, Int y, _, _) \textbf{when} x > y -> Unit          \hfill\textrm{\textit{end condition}}
| For (_, Int x, DownTo, Int y, _, _) \textbf{when} y > x -> Unit        \hfill\textrm{\textit{end condition}}
| For (v, Int x, ud, e\textquotesingle{}, e\textquotesingle{}\textquotesingle{}, copy) \textbf{when} is_value e\textquotesingle{}\textquotesingle{} ->\hfill\textrm{\textit{advance the \textbf{\texttt{for}} loop using the copy}}
    For (v, Int (x + 1), ud, e\textquotesingle{}, copy, copy)
| For (v, x, ud, e\textquotesingle{}, e\textquotesingle{}\textquotesingle{}, copy) ->                                       \hfill\textrm{\textit{evaluate the body}}
    For (v, x, ud, e\textquotesingle{}, eval (EnvBinding (false, ref [(PatVar v, x)])::env) e\textquotesingle{}\textquotesingle{}, copy)
\end{Verbatim}
\medskip

\noindent Note the final case, where the variable is bound for the next step of  evaluation. Now consider how to deal with another imperative construct: the reference. A \index{reference}reference in OCaml is a mutable cell containing a value. Here is a possible visualization of a simple imperative program using a reference:

\medskip
\begin{verbatim}[commandchars=\\\{\}]
    \textbf{let} x = \underline{ref 0} \textbf{in} x := !x + 1
=>  \textbf{let} x = \{contents = 0\} \textbf{in} x := \underline{!x} + 1
=>  \textbf{let} x = \{contents = 0\} \textbf{in} x := \underline{0 + 1}
=>  \textbf{let} x = \{contents = 0\} \textbf{in} \underline{x := 1}
=>  \underline{\textbf{let} x = \{contents = 1\} \textbf{in} ()}
=>  ()
\end{verbatim}
\medskip

\noindent Note that, even though the new value of the reference is lost in the final expression {\small\texttt{()}}, it is visible in the penultimate step, which is good enough. On paper, we would probably represent the reference cell in a graphical way rather than writing {\small\verb!{contents = ...}!}:

\bigskip
{\centering\noindent\includegraphics[width=0.5\textwidth]{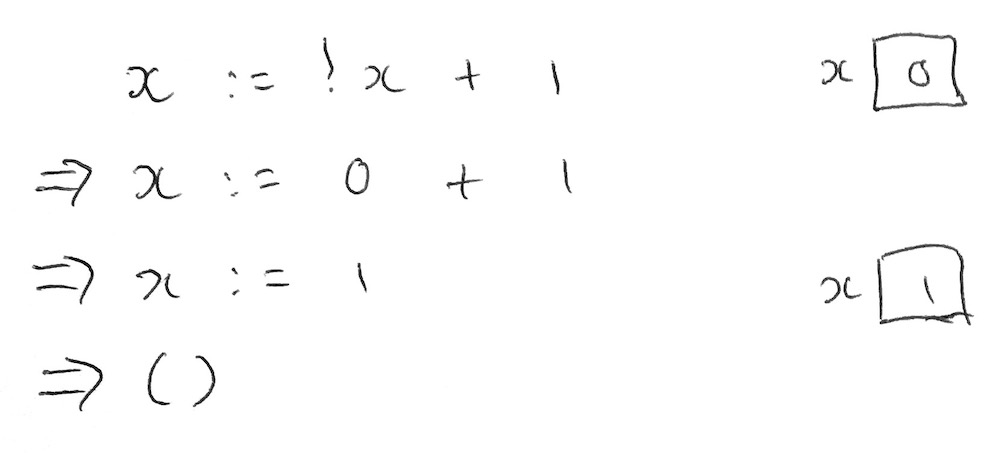}\par}
\medskip

\subsection{Currying}\index{currying}

When we teach functional programming we often say ``every function only has one argument'' but really, except in cases of \index{partial application}partial application, programmers think of curried functions as a single function of multiple arguments. And how the programmer thinks is how the debugger must behave. Consider the default evaluation of {\small\texttt{(\textbf{fun}\negthinspace\ x\negthinspace\ y\negthinspace\ ->\negthinspace\ x\negthinspace\ +\negthinspace\ y)\negthinspace\ 4\negthinspace\ 5}}:

\medskip
\begin{verbatim}[commandchars=\\\{\}]
$ ocamli -e \textquotesingle{}(fun x y -> x + y) 4 5\textquotesingle\ -show-all
    \underline{(\textbf{fun} x y -> x + y) 4} 5
=>  \underline{(\textbf{let} x = 4 \textbf{in fun} y -> x + y)} 5
=>  \underline{(\textbf{fun} y -> \textbf{let} x = 4 \textbf{in} x + y) 5}
=>  \textbf{let} y = 5 \textbf{in let} x = 4 \textbf{in} \underline{x + y}
=>  \textbf{let} y = 5 \textbf{in} \underline{4 + y}
=>  \underline{4 + 5}
=>  9
\end{verbatim}
\medskip

\noindent We always print {\small\texttt{\textbf{fun}\negthinspace\ x\negthinspace\ y\negthinspace\ ->}} instead of {\small\texttt{\textbf{fun}\negthinspace\ x\negthinspace\ ->\negthinspace\ \textbf{fun}\negthinspace\ y\negthinspace\ ->}}, since they are indistinguishable in the OCaml parse tree. There are a small number of such places where similar abstract syntax forms are not distinguished in the concrete syntax, and we would want eventually to modify the OCaml parser to retain information about the original form.

Returning to currying, the evaluation above is excessively verbose. When another option is added to the command line, the arguments will be applied at once:

\medskip
\begin{verbatim}[commandchars=\\\{\}]
$ ocamli -e \textquotesingle{}(fun x y -> x + y) 4 5\textquotesingle\ -show-all -fast-curry
    \underline{(\textbf{fun} x y -> x + y) 4 5}
=>  \textbf{let} x = 4 \textbf{in let} y = 5 \textbf{in} \underline{x + y}
=>  \textbf{let} y = 5 \textbf{in} \underline{4 + y}
=>  \underline{4 + 5}
=>  9
\end{verbatim}
\medskip

\noindent This involves a more complicated matching on the program to identify all the arguments which can be applied. In fact, combined with another option which pulls out let bindings to the side, we get an evaluation which is better still:

\medskip
\begin{verbatim}[commandchars=\\\{\}]
$ ocamli -e \textquotesingle{}(fun x y -> x + y) 4 5\textquotesingle\ -show-all -fast-curry -side-lets
    \underline{(\textbf{fun} x y -> x + y) 4 5}
x = 4 y = 5 =>  \underline{x + y}
      y = 5 =>  \underline{4 + y}
            =>  \underline{4 + 5}
            =>  9
\end{verbatim}
\medskip

\noindent If we could go further, and do away with the step-by-step \index{variable lookup}lookup of variables, we can imagine the optimal visualisation:

\medskip
\begin{verbatim}[commandchars=\\\{\}]
    \underline{(\textbf{fun} x y -> x + y) 4 5}
x = 4 y = 5 => \underline{x + y}
            => \underline{4 + 5}
            => 9
\end{verbatim}
\medskip

\noindent or even:

\medskip
\begin{verbatim}[commandchars=\\\{\}]
    \underline{(\textbf{fun} x y -> x + y) 4 5}
=>  \underline{4 + 5}
=>  9
\end{verbatim}
\medskip

\noindent This is perhaps what we might write if we were to do this on paper -- when we write such evaluations informally we naturally skip `obvious' steps. It is the same when doing mathematics. We can see that most of the job of improving upon the naive visualization consists of removing information, rather than adding it. Here is our default paper visualization when teaching beginning students:

\medskip
{\centering\noindent\includegraphics[width=0.5\textwidth]{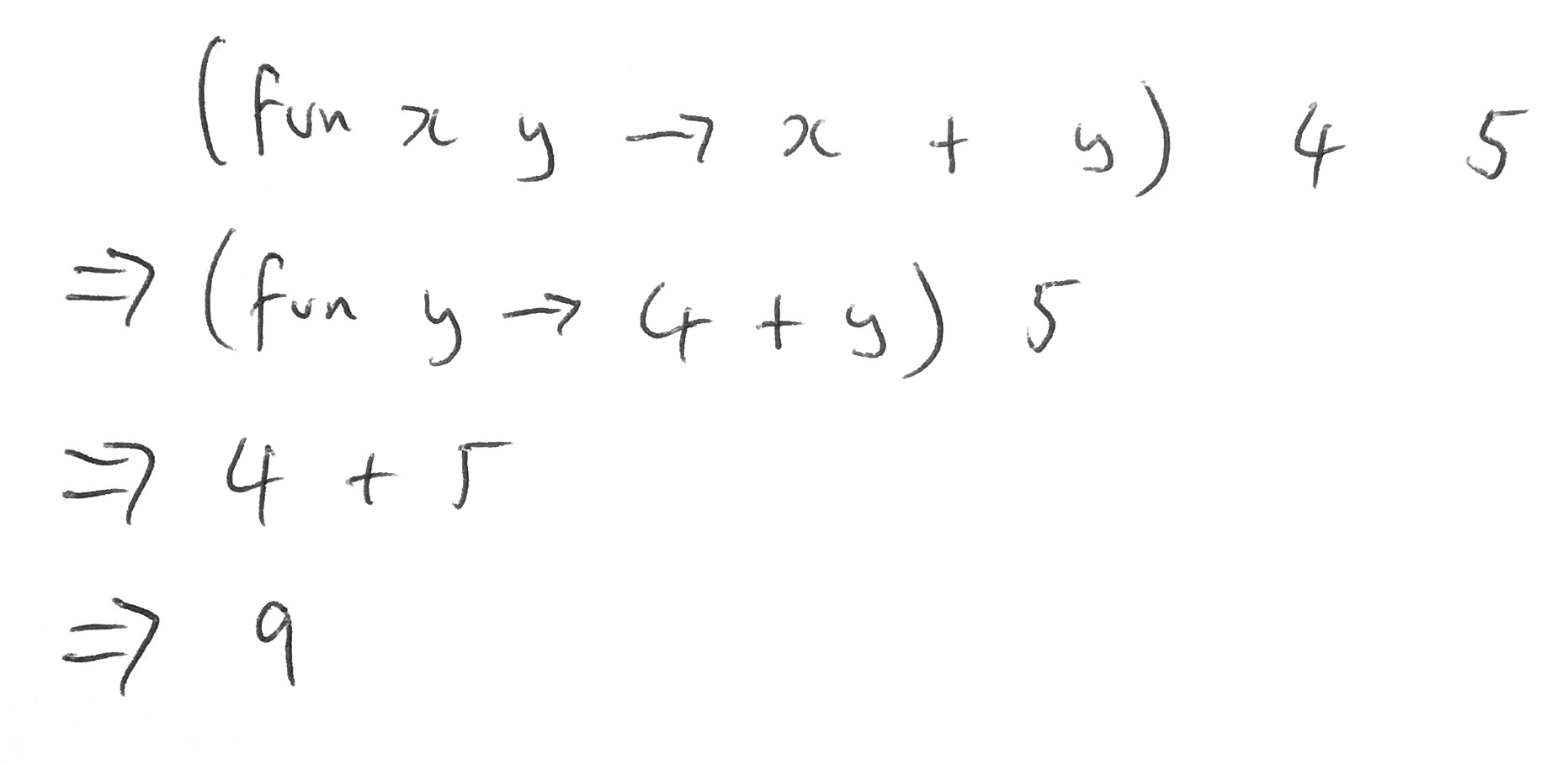}\par}

\noindent Again we see that our intuition in writing paper diagrams is to what corresponds to quite radical elision for the computer-generated diagram.
 
\subsection{Exceptions}\index{exception}

Exceptions are thought of as a sort of break in the evaluation of a program, causing evaluation to `jump' from one place to another. And so exceptions can cause a little trouble when it comes to the visualization of a program's execution, especially if some parts of the program are elided. To exacerbate the issue, OCaml programmers often use exceptions for control flow, rather than only for truly exceptional situations. This is possible because OCaml exceptions are very fast compared with most languages. When an exception is caught, one expression (the one which raised the exception) is replaced by another (the right-hand side of the catching case), and it can seem to appear from nowhere. For some background, let us look at the paper visualization of the execution of a program which uses exceptions for control flow. Here is the example program:  

\begin{verbatim}[commandchars=\\\{\}]
\textbf{type} \textquotesingle{}a tree =
  Lf
| Br \textbf{of} \textquotesingle{}a tree * \textquotesingle{}a * \textquotesingle{}a tree

\textbf{exception} E

\textbf{let rec} path = \textbf{function}
  Lf -> \textbf{raise} E
| Br (l, v, r) ->
    \textbf{try}
      \textbf{if} v = 7 \textbf{then} [] \textbf{else} 1 :: path l
    \textbf{with}
      E -> 2 :: path r
\end{verbatim}

\noindent Our example binary tree will be this:

\bigskip
{\centering \Tree [.3 [.5 2 7 ] [.7 \edge[draw=none]; {\phantom{7}} 7 ] ]\par}
\bigskip

\noindent This program was adapted from a Standard ML exam question set by Lawrence C. Paulson at Cambridge. The function \texttt{path} finds the first 7 it can in a binary tree, giving a path (1 = left, 2 = right) to its position. Figure 6 shows the pen and paper visualization of its operation on the given tree. Notice that we annotate both raising and catching of exceptions here. The visualization is very stripped-down, showing almost no code, just the accumulation of the result list. This is a good example of how radical such elisions often are when we do things on paper, with our human intuition about the specific case at hand. 

\begin{figure}
\begin{center}
\includegraphics[width=0.5\textwidth]{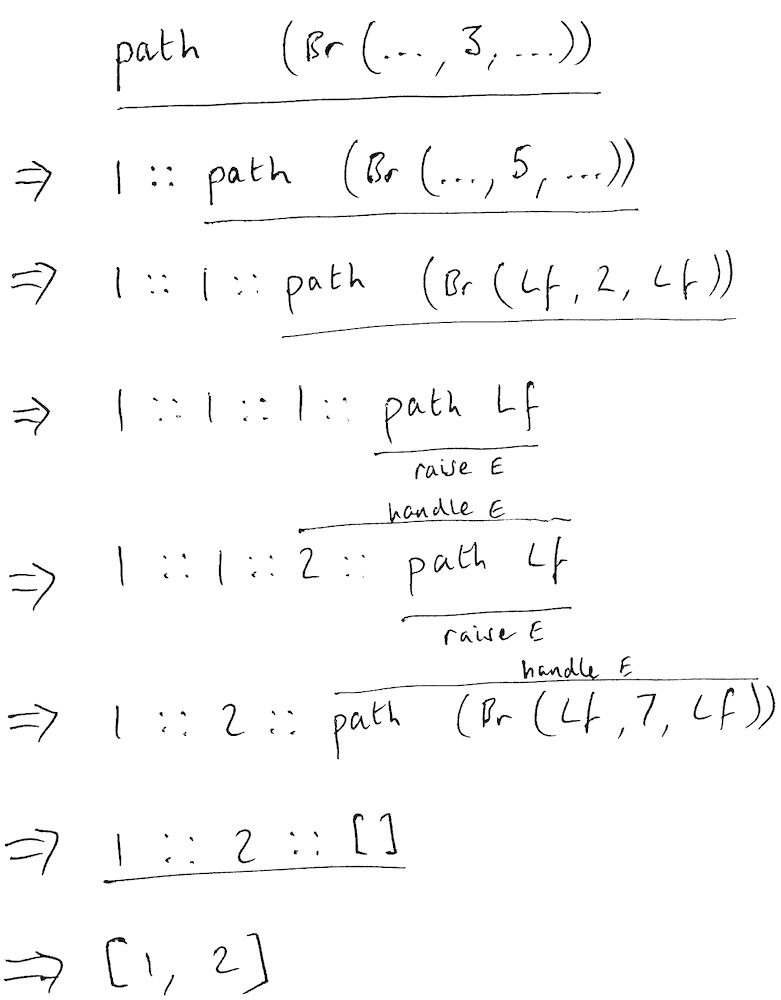}
\end{center}
{\centering\singlespacing\noindent\small\label{F6} Figure 6. Handwritten diagram for a function using exceptions for control flow.\par}
\end{figure}

As one would expect, exceptions (which interrupt the flow of evaluation) require a similar mechanism inside an interpreter. The complication of a step-by-step interpreter is that exceptions must be modelled in a step-by-step way too: we cannot let uncaught exceptions cascade all at once. The solution to this is to model exceptions in two ways: as a special {\small\texttt{Tinyocaml.t}} constructor {\small\texttt{Raise}} and using actual exceptions in the interpreter. Here is the exception definition we will use, which represents, for example, the result of evaluating {\small\texttt{\textbf{raise}\negthinspace\ (Failure\negthinspace\ "broken")}} as {\small\texttt{ExceptionRaised\negthinspace\ ("Failure",\negthinspace\ Some\negthinspace\ (String\negthinspace\ "broken"))}}:

\medskip
\begin{verbatim}[commandchars=\\\{\}]
\textbf{exception} ExceptionRaised \textbf{of} string * t option
\end{verbatim}
\medskip

\noindent Here is the {\small\texttt{Tinyocaml.t}} constructor used to represent exceptions which need to be raised:

\medskip
\begin{verbatim}[commandchars=\\\{\}]
| Raise \textbf{of} string * t option
\end{verbatim}
\medskip

\noindent See how it mirrors the exception definition above. Now, let us consider the case of dividing two numbers, where the second may be zero. Here is the code from the interpreter:

\medskip
\begin{verbatim}[commandchars=\\\{\}]
| Op (op, Int a, Int b) ->
    \textbf{begin try} Int (calc op a b) \textbf{with}
      Division_by_zero -> Raise ("Division_by_zero", None)
    \textbf{end}
\end{verbatim}
\medskip

\noindent We use OCaml exception handling to check for {\small\texttt{Division\_by\_zero}} in the {\small\texttt{calc}} function, and if we see it, we build the {\small\texttt{Raise}} constructor as the result of evaluating this expression one step. This freezes the exception. What happens when, in the next step of evaluation, this {\small\texttt{Raise}} is found?

\medskip
\begin{verbatim}[commandchars=\\\{\}]
| Raise (e, payload) -> \hfill\textrm{\textit{\texttt{payload} is the data carried with the exception}}
    \textbf{match} payload \textbf{with}
    | Some x \textbf{when} not (is_value x) ->
        Raise (e, Some (eval_step env x)) \hfill\textrm{\textit{if payload not a value, evaluate one step}}
    | _ ->
      \textbf{raise} (ExceptionRaised (e, payload)) \hfill\textrm{\textit{otherwise, the exception may be raised}}
\end{verbatim}
\medskip

\noindent We may need to evaluate the expression in the {\small\texttt{Raise}} one step if it is not a value (it might be {\small\texttt{\textbf{raise}\negthinspace\ (Fail\negthinspace\ (1\negthinspace\ +\negthinspace\ 2))}}, for example). Thus, the {\small\texttt{Raise}} may take several steps to be processed. If it is a value, though, we can raise the actual exception. This will be caught in the evaluator, and mirrors the effect of the exception occurring in a compiled program. Here is code for the {\small\texttt{\textbf{try}\negthinspace\ \ldots\negthinspace\ \textbf{with}}} construct:

\medskip
\begin{verbatim}[commandchars=\\\{\}]
| TryWith (e, cases) ->
    \textbf{if} is_value e \textbf{then} e \textbf{else} \hfill\textrm{\textit{if body a value, return}}
      \textbf{begin try} TryWith (eval_step env e, cases) \textbf{with} \hfill\textrm{\textit{evaluate body one step}}
        ExceptionRaised (x, payload) -> \hfill\textit{\textrm{if this step caused an exception}}
          \textbf{match} eval_match_exception env x payload cases \textbf{with} \hfill\textit{\textrm{see if it matches a case}}
          | FailedToMatch -> Raise (x, payload) \hfill\textit{\textrm{if not, recreate the raise node}}
          | Matched e\textquotesingle{} -> e\textquotesingle{} \hfill\textit{\textrm{otherwise, return the body of the matched case}}
      \textbf{end}
\end{verbatim}
\medskip

\noindent If the exception is not surrounded by a {\small\texttt{\textbf{try}\negthinspace\ \ldots\negthinspace\ \textbf{with}}}, it is not caught, and so is printed at the top-level and the interpreter exits:

\smallskip
\begin{verbatim}[commandchars=\\\{\}]
    1 + 1 / (\underline{1 - 1})
=>  1 + \underline{1 / 0}
=>  1 + \underline{\textbf{raise} Division_by_zero}
Exception: Division_by_zero.
\end{verbatim}
\smallskip

\noindent Let us add a {\small\texttt{\textbf{try}\negthinspace\ \ldots\negthinspace\ \textbf{with}}}:

\smallskip
\begin{verbatim}[commandchars=\\\{\}]
    \textbf{try} 1 + 1 / \underline{(1 - 1}) \textbf{with} Division_by_zero -> 2 + 2 
=>  \textbf{try} 1 + \underline{1 / 0} \textbf{with} Division_by_zero -> 2 + 2 
=>  \textbf{try} 1 + \underline{\textbf{raise} Division_by_zero} \textbf{with} Division_by_zero -> 2 + 2 
=>  \underline{2 + 2}
=>  4
\end{verbatim}
\smallskip

\noindent Now we can see the whole process. As an improvement, we might like to annotate the penultimate step to indicate which expression matched.

The problem of visualizing exceptions is discussed by Shah et al.\ in an interesting paper \cite{shah} showing several mechanisms for exceptions in the Java programming language.

\subsection{Opening modules}\index{module}\index{module!opening}\index{opening a module}

To use a function from another module, one must either type its full path e.g.\ {\small\texttt{Char.code}} (function {\small\texttt{code}} in module {\small\texttt{Char}}) or use the {\small\texttt{\textbf{open}}} keyword to bring the name into scope. OCaml has so called local opens -- writing {\small\texttt{Char.(code \textquotesingle x\textquotesingle\negthinspace\ +\negthinspace\ code\negthinspace\ \textquotesingle y\textquotesingle)}} is the same as writing {\small\texttt{Char.code\negthinspace\ \textquotesingle x\textquotesingle\negthinspace\ + Char.code\negthinspace\ \textquotesingle y\textquotesingle}}. These are relatively easy to handle. We introduce a new {\small\texttt{LocalOpen\negthinspace\ \textbf{of}\negthinspace\ string\negthinspace\ *\negthinspace\ t}} node in the \textsf{TinyOCaml} data type, and upon encountering it, open the module up and bring its names into the environment at top-level, for example adding the name {\small\texttt{hd}} bound to the definition of {\small\texttt{List.hd}}:

\medskip
\begin{verbatim}[commandchars=\\\{\}]
$ ocamli -e \textquotesingle{}List.(hd [1; 2; 3])\textquotesingle\ -show-all
    List.(\underline{hd [1; 2; 3]})
=>  List.(\underline{(\textbf{function} a::_ -> a ) [1; 2; 3]})
=>  List.(\textbf{let} a = 1 \textbf{in} \underline{\texttt{a}})
=>  List.(1)
\end{verbatim}
\medskip

\noindent We could remove the local open when there are no longer any uses of its symbols, and improve the underlining:

\medskip
\begin{verbatim}[commandchars=\\\{\}]
    List.(\underline{hd [1; 2; 3]})
=>  \underline{(\textbf{function} a::_ -> a ) [1; 2; 3]}
=>  let a = 1 in \underline{a}
=>  1
\end{verbatim}
\medskip

\noindent We cannot really use the same approach for the normal (non-local) {\small\textbf{\texttt{open}}} keyword, which brings all definitions of a module up to top-level. Here is an example, with four structure items (a structure item is OCaml parse tree parlance for a top-level definition in a source file):

\medskip
\begin{verbatim}[commandchars=\\\{\}]
\textbf{open} Sys

\textbf{let} () = Printf.printf "%s" argv.[0]
\end{verbatim}
\medskip

\noindent We are using the function {\small\texttt{Sys.argv}} by just typing {\small\texttt{argv}} after opening the {\small\texttt{Sys}} module with the {\small\texttt{\textbf{open}}} keyword. In our data type, we would have to represent this as a parenthesised structure {\small\texttt{Open(x, \ldots\ Open(y, \ldots))}}, which would be awkward to work with. So instead, we keep {\small\textbf{\texttt{open}}} as a structure item, dealing with it at every step. For example:

\medskip
\begin{verbatim}[commandchars=\\\{\}]
$ ocamli -e \textquotesingle{}open List let x = hd [1; 2; 3]\textquotesingle\ -show-all
    \textbf{open} List
  
    \textbf{let} x = \underline{hd} [1; 2; 3]

=>  \textbf{open} List
    
    \textbf{let} x = \underline{(\textbf{function} a::l -> a) [1; 2; 3]}

=>  \textbf{open} List
  
    \textbf{let} x = \underline{\textbf{let} a = 1 \textbf{in} a}

=>  \textbf{open} List

    \textbf{let} x = 1
\end{verbatim}
\medskip

\noindent Of course, there is no need to print it out at each step. We might prefer:

\medskip
\begin{verbatim}[commandchars=\\\{\}]
    \textbf{open} List
  
    \textbf{let} x = \underline{hd} [1; 2; 3]

=>  \textbf{let} x = \underline{(\textbf{function} a::l -> a) [1; 2; 3]}
=>  \textbf{let} x = \underline{\textbf{let} a = 1 \textbf{in} a}
=>  \textbf{let} x = 1
\end{verbatim}

\noindent This is shorter. However each step is no longer necessarily a valid program, since the {\small\texttt{\textbf{open}}} has been removed.

\subsection{Pattern matching}\index{pattern matching}

How can we visualize pattern matching, one of the most widely used, and praised, features of functional programming? Do we show the whole pattern, then jump to the right-hand side of the chosen match? Do we show how the match matches? Consider an example:

\medskip
\begin{verbatim}[commandchars=\\\{\}]
$ ocamli -e \textquotesingle{}match 1 + 2 with 4 -> 0 | 3 -> 1 + 2 | _ -> 1\textquotesingle{} -show-all
    \textbf{match} \underline{1 + 2} \textbf{with} 4 -> 0 | 3 -> 1 + 2 | _ -> 1 
=>  \underline{\textbf{match} 3 \textbf{with} 4 -> 0 | 3 -> 1 + 2 | _ -> 1} 
=>  \underline{\textbf{match} 3 \textbf{with} 3 -> 1 + 2 | _ -> 1} 
=>  \underline{1 + 2}
=>  3
\end{verbatim}
\medskip

\noindent In this method of visualisation, we simply show the whole match expression with all its cases, and each time a case does not match, we drop it from the front.  We can imagine wanting to skip this process, and show just the case that matched. Figure 7 shows two possible on-paper evaluations of this program snippet.
\begin{figure}
{\centering\noindent\includegraphics[width=0.8\textwidth]{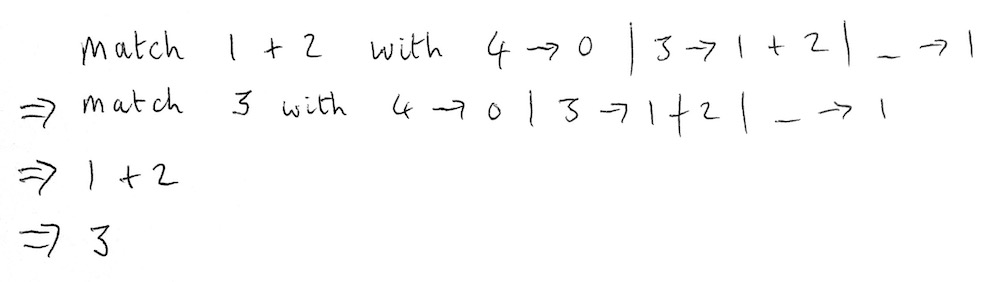}

\noindent\includegraphics[width=0.6\textwidth]{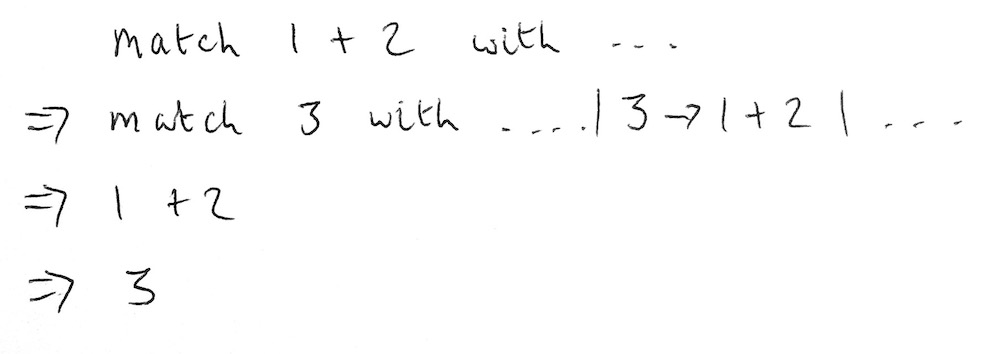}\par}

{\centering\singlespacing\noindent\small\label{F7} Figure 7. Two handwritten diagrams of a pattern-match\par}

\end{figure}
In the second case, the evaluation would have to be explained orally alongside the full snippet. This is often the case in teaching situations.

The evaluator, as one would expect, being implemented in the language it is interpreting, exhibits a certain self-similarity. For example, figure 8 shows part of the implementation of pattern-matching defined, of course, using pattern matching.

\begin{figure}
\begin{verbatim}[commandchars=\\\{\}]
\textbf{let rec} matches expr pattern rhs =
  \textbf{let} yes = Some rhs \textbf{in}
  \textbf{let} no = None \textbf{in}
    \textbf{match} expr, pattern \textbf{with}
      _, PatAny -> yes
    | x, PatConstraint (p, _) -> matches x p rhs
    | Unit, PatUnit -> yes
    | Bool b, PatBool b\textquotesingle{} \textbf{when} b = b\textquotesingle{} -> yes
    | Int i, PatInt i\textquotesingle{} \textbf{when} i = i\textquotesingle{} -> yes
    | Int32 i, PatInt32 i\textquotesingle{} \textbf{when} i = i\textquotesingle{} -> yes
    | Int64 i, PatInt64 i\textquotesingle{} \textbf{when} i = i\textquotesingle{} -> yes
    | NativeInt i, PatNativeInt i\textquotesingle{} \textbf{when} i = i\textquotesingle{} -> yes
    | String s, PatString s\textquotesingle{} \textbf{when} s = s\textquotesingle{} -> yes
    | Char c, PatChar c\textquotesingle{} \textbf{when} c = c\textquotesingle{} -> yes
    | Char x, PatCharRange (c, c\textquotesingle{}) \textbf{when} x >= c && x <= c\textquotesingle{} -> yes 
    | e, PatVar v -> Some (Let (false, [(PatVar v, e)], rhs))
    | Nil, PatNil -> yes
    | Cons (h, t), PatCons (ph, pt) ->
        \textbf{begin match} matches h ph rhs \textbf{with}
        | Some rhs\textquotesingle{} -> matches t pt rhs\textquotesingle{}
        | None -> no
        \textbf{end}
    | Tuple es, PatTuple ps ->
        match_many_binders es ps rhs
    | Record es, PatRecord (_, ps) ->
        match_many_binders (List.map (!) (List.map snd es)) (List.map snd ps) rhs
    | Array es, PatArray ps ->
        match_many_binders (Array.to_list es) (Array.to_list ps) rhs
    | e, PatAlias (a, p) ->
        matches e p (Let (false, [(PatVar a, e)], rhs))
    | e, PatOr (a, b) ->
        \textbf{begin match} matches e a rhs \textbf{with}
          Some _ -> yes
        | _ -> matches e b rhs
        \textbf{end}
    | Constr (_, y, None), PatConstr (x, None) \textbf{when} x = y -> yes
    | Constr (_, y, Some yp), PatConstr (x, Some xp)
        \textbf{when} x = y -> matches yp xp rhs
    | _ -> no

\textbf{and} match_many_binders es ps rhs =
  \textbf{match} es, ps \textbf{with}
    [], [] -> Some rhs
  | eh::et, ph::pt ->
      \textbf{begin match} matches eh ph rhs \textbf{with}
        None -> None
      | Some rhs\textquotesingle{} -> match_many_binders et pt rhs\textquotesingle{}
      \textbf{end}
  | _ -> None
\end{verbatim}

{\centering\singlespacing\noindent\small\label{F8} Figure 8. \index{pattern matching}Pattern matching by pattern matching\par}

\end{figure}

\subsection{Functor application}\index{functor}

We show this as an example of the more complicated work that \textsf{OCamli} must do at run-time. In this case, it is something which OCaml does at run-time too: applying a functor. Consider the following example, culled from the popular textbook ``Real World OCaml'' \cite{minsky2013real}:

\begin{verbatim}[commandchars=\\\{\}]
\textbf{module} type X_int = \textbf{sig val} x : int \textbf{end}\hfill\textit{\textrm{type}}

\textbf{module} Three : X_int = \textbf{struct let} x = 3 \textbf{end}\hfill\textit{\textrm{module}}

\textbf{let} y = Three.x                                  \hfill\textit{\textrm{test}}

\textbf{module} Increment (M : X_int) : X_int =           \hfill\textit{\textrm{define functor}}
  \textbf{struct}
    \textbf{let} x = M.x + 1
  \textbf{end}

\textbf{module} Four = Increment(Three)                   \hfill\textit{\textrm{apply functor}}

\textbf{let} z = Four.x                                   \hfill\textit{\textrm{test}}
\end{verbatim}

\noindent Functor definitions, like type definitions, must be kept in the environment, to be looked up when a functor is applied. So, each time \textsf{OCamli} encounters a functor definition, it adds it to the environment, and moves on, trying to find the next redex. The functor application itself is done by what is effectively textual substitution. Here is the final state:

\begin{verbatim}[commandchars=\\\{\}]
\textbf{module} Three : X_int = 
\textbf{struct} 
  \textbf{let} x = 3
\textbf{end}
    
\textbf{let} y = 3
    
\textbf{module} Increment = 
\textbf{functor} (M : X_int) : X_int ->
  \textbf{struct} 
    \textbf{let} x = M.x + 1
  \textbf{end}
    
\textbf{module} Four = 
\textbf{struct} 
  \textbf{let} x = 4
\textbf{end}
    
\textbf{let} z = 4
\end{verbatim}

\noindent The functor application has occurred, and \texttt{z} has been fully evaluated. We could, of course, have removed the functor definition now it is no longer needed.

\subsection{Summary}

We now have a function which, given a program, can evaluate it one step. By calling the function repeatedly, feeding its own output back in as the next input, the program can be evaluated completely, step by step. With an appropriate prettyprinter, each step may be printed out.

\section{Dealing with size by elision}\index{elision!of debugger output}
\label{prob-elision}

This section concerns the important task of making the output readable. We discuss searching, which also reduces the output, in section \ref{searching}. So, what remains? Three things: 1) the showing or eliding of whole steps  for things like simple arithmetic and variable lookup;  2) the hiding or showing of parts of the expression at each step; and 3) the default heuristics for eliding parts of individual expressions (for example, the internals of built-in functions).

This topic has attracted some attention in the literature. Foubister's Spatial and temporal filters \cite{foubister1995graphical}, and Penney's  `reactive' and `shadow' zooms and ``hierarchical closing'', a tree-based mechanism similar to folding in text editors \cite{penney2000augmenting}. There is, inevitably, overlap between searching and elision, especially in systems which operate on-line. For offline trace browsers, the division is more clear-cut: elision or filtering is suppressing trace output, searching is picking out pieces of the trace file to examine.

We have developed simple mechanisms for elision. There is much more to be done, but even the simplest elision mechanisms can be expected to suppress plenty of unneeded trace material, so it makes sense to conquer those first, and then consider what remains. For an example, we shall consider how to automatically abridge the following arithmetic evaluation, of a type which frequently occurs at the end of a non-tail-recursive function application:

\medskip
\begin{verbatim}[commandchars=\\\{\}]
   1 * (2 * (\underline{3 * 4}))
=> 1 * (\underline{2 * 12})
=> \underline{1 * 24}
=> 24 
\end{verbatim}
\medskip

\noindent We wish to remove the middle two steps, leaving just:

\medskip
\begin{verbatim}[commandchars=\\\{\}]
   \underline{1 * (2 * (3 * 4))}
=> 24
\end{verbatim}
\medskip

\noindent This can be done by a mechanism we call peeking.

\subsection*{Peeking}\index{peeking}
In order to decide whether to show the current state, it is sometimes important to know the next state, and to remember the previous state. But how can we know the next state without evaluating it? One way, of course, would be to evaluate the whole program and print out its steps of execution offline. But we may wish to stop evaluation based on what is about to happen, and we cannot do this with a real running program with side effects, since we cannot roll back a side effect such as a network communication with a third party.

The solution is to add to the step-by-step evaluator the notion of \textit{peeking}. In this mode, the evaluator identifies the reducible expression, but does not evaluate it. The calling function can then interrogate the interpreter to ask ``What kind of operation would have been performed?''. Presently, the answer is one of a short list, giving just enough information to provide for some elisions the \textsf{OCamli} prototype can perform:

\medskip
\begin{verbatim}[commandchars=\\\{\}]
\textbf{type} last_op =
    Arith                   \textrm{\textit{simple arithmetic}}
  | Boolean                 \textrm{\textit{\texttt{\&\&}, \texttt{||}}}
  | Comparison              \textrm{\textit{comparison operators}}
  | IfBool                  \textrm{\textit{\texttt{\textbf{if}\! true}, \texttt{\textbf{if}\! false}}}
  | InsideBuiltIn           \textrm{\textit{evaluation inside an external piece of code}}
  | VarLookup               \textrm{\textit{variable lookup}}
\end{verbatim}
\medskip

\noindent In our example, we print the step if and only if a) the next state is a value, or b) the current state is a value, or c) {\small\texttt{Arith}} is not present for the previous state, or d) {\small\texttt{Arith}} is not present for the next state. These four conditions, taken together, elide just enough steps of the arithmetic, but do not remove information we might want to see. Similar conditions have been devised for the other kinds of elision listed in the {\small\texttt{last\_op}} type. Of course, as we have mentioned, such elision may be interactive in many reasonable interfaces to such a system.

\subsection*{Eliding within a step}\index{elision!within a single step}

Consider the following example with multiple structure items (remember, a structure item in the parlance of the OCaml parse tree is a type definition or a top-level let binding):

\begin{verbatim}[commandchars=\\\{\}]
\textbf{let} x = 1 + 2

\textbf{let} y = x + x

\textbf{let} z = 1 + y
\end{verbatim}

\noindent First, of course, we begin by evaluating {\small\texttt{1 + 2}}, and proceed from there. However, a lot of screen space is used by printing out these five lines (three code, two blank) for each step, and it can be hard for the user to follow along. Should we remove a structure item when it is no longer needed, assuming that the user is interested only in the final result of {\small\texttt{z}}? This results in a shorter but arguably incomplete trace. Or, instead, only show the structure item which is currently being evaluated? Most likely, this would be a configurable option with a sensible default, which is probably to reduce the trace as much as possible.

\section{The Standard Library}\index{Standard Library}\index{OCaml!Standard Library}
\label{ppxautoearlier}
\label{prob-c-ocaml}

OCaml comes with a small but useful library of routines. These fall broadly into three categories:

\begin{enumerate}
\item Those which are simply there to provide a selection of common routines, useful for many programs, but which the user could write themselves -- entirely in OCaml -- if they wanted. For example {\small\texttt{List.map}}.
\item Those which are in the Standard Library because they are used in the implementation of the OCaml toolchain, but seemed to the authors to be generic enough as to be useful for the general programmer, such as the {\small\texttt{Arg}} module for handling command line arguments. When a programming language is in its infancy, the general programmer and the compiler author are one.
\item Those which must be in the Standard Library because they provide facilities which pure OCaml programs could not provide, or use an external symbol, or talk to the runtime. For example, the function {\small\texttt{output}} which writes to standard output, or the value {\small\texttt{Sys.word\_size}} which is the word size of the machine upon which the program is running.
\end{enumerate}

\noindent Categories 1 and 2 are easy to deal with -- we are just interpreting standard OCaml code, so it is as if the user had themselves supplied the code. The \textsf{OCamli} interpreter can simply load the source code for the OCaml Standard Library before loading the main program's source. The \textsf{OCamli} interpreter also knows how to load multiple modules as libraries using command line arguments. For example, the following command line loads modules {\small\texttt{A}} and {\small\texttt{B}}, performing any module initialisation code, then executes the code given in the {\small\texttt{-e}} argument in an environment in which such modules exist:

\medskip
\begin{verbatim}[commandchars=\\\{\}]
$ ocamli a.ml b.ml -e \textquotesingle{}let () = B.calc 10\textquotesingle 
\end{verbatim}
\medskip

\noindent It is the third category above which requires special treatment. Functions which are external to OCaml are introduced like this in a source file:

\medskip
\begin{verbatim}[commandchars=\\\{\}]
\textbf{external} word_size : unit -> int = "%word_size"
\end{verbatim}
\medskip

\noindent This name might be exported directly or might be used in the definition of a Standard Library function which is then exported. In the example above, it indicates that a function of type {\small\textsf{\textbf{unit} $\rightarrow$ \textbf{int}}} is expected to be available at link time under the symbol {\small\texttt{\%word\_size}} and that it is to be given the name {\small\texttt{word\_size}}. When we come across such an {\small\texttt{\textbf{external}}} declaration in a {\small\texttt{.ml}} source file (such as when loading the Standard Library), how should \textsf{OCamli} deal with it? What we do is to write (or generate) a binding for it. The \textsf{TinyOCaml} data type already exhibited contains the constructor {\small\texttt{CallBuiltIn}}:

\medskip
\begin{verbatim}[commandchars=\\\{\}]
CallBuiltIn \textbf{of} \ldots\ * (env -> t list -> t)
\end{verbatim}
\medskip

\noindent This inclusion of a native OCaml function into the \textsf{TinyOCaml} data type for programs is the mechanism by which the gap between the interpreted and native worlds is bridged. It represents an OCaml function which takes an environment and a list of \textsf{TinyOCaml} arguments, calls some external native function and returns a \textsf{TinyOCaml} result.

We can use this {\small\texttt{CallBuiltIn}} mechanism to build an interface to our function, and a way to look up such an interface by name so that, at run-time, it may be located and called by the interpreter:

\medskip
\begin{verbatim}[commandchars=\\\{\}]
\textbf{external} word_size : unit -> int = "%word_size"

\textbf{let} percent_word_size =
  \textbf{let} f =
    (\textbf{function} [Unit] ->
      \textbf{begin try} Int (word_size ()) \textbf{with} e -> exception_from_ocaml e \textbf{end}
     | _ -> failwith "%word_size")
  \textbf{in}
    ("%word_size",
     Fun (NoLabel, PatVar "*x", CallBuiltIn (None, "%word_size", [Var "*x"], f), []))
\end{verbatim}
\medskip

\noindent Notice the {\small\texttt{\textbf{external}}} declaration is retained. We then create an entry {\small\texttt{("\%word\_size",\negthinspace\ x)}} in the table of primitives  where {\small\texttt{x}} is a function containing a {\small\texttt{CallBuiltIn}}. This table will be used for lookup when an {\small\texttt{\textbf{external}}} declaration is found in a {\small\texttt{.ml}} source file being interpreted. In this case, it is a function of one argument {\small\texttt{*x}}. The asterisk is a crude mechanism to mark such functions so they are not printed, since they are not part of the original source code. This function can then be applied to an argument in the interpreted world. The argument will be assigned the name {\small\texttt{*x}} and used by the native function -- the result will be returned to the interpreted world. Now we need to look at the function {\small\texttt{f}} itself. It pattern-matches on the input argument list, requiring just one argument {\small\texttt{[Unit]}}. It tries to produce the output {\small\texttt{Int\negthinspace\negthinspace\ (word\_size\negthinspace\negthinspace\  ())}} by applying the native function {\small\texttt{word\_size}} as defined by the {\small\texttt{\textbf{external}}}. This is the result. Should an exception be raised during the execution of {\small\texttt{word\_size}} (either in OCaml code or C code) it comes into the OCaml \index{OCaml!runtime}runtime as an OCaml \index{exception}exception, and is then converted into a \textsf{TinyOCaml} representation of an exception by the function {\small\texttt{exception\_of\_ocaml}}. Curried functions may be defined using a helper for each arity. For example, for arity three:

\medskip
\begin{verbatim}[commandchars=\\\{\}]
\textbf{let} mk3 name f =
  (name,
   Fun (NoLabel, PatVar "*x",
     Fun (NoLabel, PatVar "*y",
       Fun (NoLabel, PatVar "*z",
            CallBuiltIn (None, name, [Var "*x"; Var "*y"; Var "*z"], f), []), []), []))
\end{verbatim}
\medskip

\noindent A function defined by this method may be partially applied as usual: only when all the arguments are actually applied in the interpreter will the native function {\small\texttt{f}} be run. To avoid writing all these bindings for the Standard Library by hand, a system has been developed which allows one to write, instead:

\begin{verbatim}[commandchars=\\\{\}]
[%%auto \textbf{external} string_of_float : float -> string = "%string_of_float"]
\end{verbatim}

\noindent The binding is then generated automatically. This system, which we describe in section \ref{ppxauto}, works for most of the Standard Library functions, and so reduces \textsf{OCamli}'s Standard Library file to a third of its previous size. Thus, we keep the part of \textsf{OCamli} which may need updating when OCaml is updated as small as possible.

\textsf{OCamli} ordinarily emulates the low-level \index{primitive operation}\index{OCaml!primitive}primitives used to implement some of OCaml's basic language features. Recall our references example:

\medskip
\begin{verbatim}[commandchars=\\\{\}]
    \textbf{let} x = \underline{ref 0} \textbf{in} x := !x + 1
=>  \textbf{let} x = \{contents = 0\} \textbf{in} x := \underline{!x} + 1
=>  \textbf{let} x = \{contents = 0\} \textbf{in} x := \underline{0 + 1}
=>  \textbf{let} x = \{contents = 0\} \textbf{in} \underline{x := 1}
=>  \underline{\textbf{let} x = \{contents = 1\} \textbf{in} ()}
=>  ()
\end{verbatim}
\medskip

\noindent The \texttt{ref} function is emulated, rather than taken from the Standard Library. If we were instead to show the low-level primitives involved in the use of references, we would see a much longer visualization:

\medskip
\begin{verbatim}[commandchars=\\\{\}]
    \textbf{let} x = \underline{ref 0} \textbf{in} x := (!x + 1)
=>  \textbf{let} x = \textbf{let} x = 0 \textbf{in} \underline{<<%makemutable>>} \textbf{in} x := (!x + 1)
=>  \textbf{let} x = \{contents = 0\} \textbf{in} \underline{x} := (!x + 1)
=>  \textbf{let} x = \{contents = 0\} \textbf{in} \underline{\{contents = 0\} := (!x + 1)}
=>  \textbf{let} x = \{contents = 0\} \textbf{in}
      \underline{(\textbf{let} x = \{contents = 0\} \textbf{in fun} y -> <<%setfield0>>)} (!x + 1)
=>  (\textbf{fun} y -> \textbf{let} x = \{contents = 0\} \textbf{in} <<%setfield0>>) (\underline{!\{contents = 0\}} + 1)
=>  (\textbf{fun} y -> \textbf{let} x = \{contents = 0\} \textbf{in} <<%setfield0>>)
    ((\textbf{let} x = \{contents = 0\} \textbf{in} \underline{<<%field0>>}) + 1)
=>  (\textbf{fun} y -> \textbf{let} x = \{contents = 0\} \textbf{in} <<%setfield0>>) \underline{(0 + 1)}
=>  \underline{(\textbf{fun} y -> \textbf{let} x = \{contents = 0\} \textbf{in} <<%setfield0>>)} 1
=>  \textbf{let} y = 1 \textbf{in let} x = \{contents = 0\} \textbf{in} \underline{<<%setfield0>>}
=>  ()
\end{verbatim}
\medskip

\noindent Most users will not want this longer output by default, but it is helpful when we wish to see, for example, exactly what I/O calls are triggered by Standard Library functions. This sort of elision is often called ``trusted code''. Trusted in our sense refers not to security, but that we trust (or assume) it does not contain the bug. Bugs in standard libraries are less common than in code we are in the process of writing and debugging, and compiler bugs even less common. Our first investigation should involve suspecting our code and trusting the third-party components it uses.

\section{An oddity: polymorphic comparison}\index{polymorphic comparison}

The OCaml runtime deals with \index{memory allocation}\index{heap!allocation}memory allocation and \index{garbage collection}garbage collection, and handling signals and threads and other low-level concerns. However, there are some surprising jobs which one might expect to be handled in the language, but in fact require the connivance of the runtime. One example is polymorphic comparison, the ability to compare for equality or ordering two items of like type, where the type is not known at compile time. For example, a sorting function capable of sorting items of any type might use the polymorphic comparison operator \texttt{<} to order them.

OCaml's polymorphic comparison operator is implemented in C, and simply traverses the heap representations of the two items it is given, checking that their structure and contents are equal or comparing them by order. It may raise an exception if a closure is found, since functions may not be compared for equality.

We must replicate this system, for the data type which represents OCaml data as part of OCaml programs in \textsf{OCamli}. For simple types, using the OCaml comparison operator on the data types representing the values in the interpreter works for many types, but we have not shown that this is always the case (and do not expect it to be so). Similar work is required to emulate OCaml's standard hash functions, which again require support from the runtime. We are providing, effectively, a partial  alternative runtime. One thing we do not need to provide, happily, is a garbage collector: the OCaml garbage collector will happily collect the garbage of the interpreted program in just the same fashion as it will collect the interpreter's garbage. However, the space-efficiency of the running program may be different, or its garbage may be longer-lived if, for example, the storage of environments in the interpreter is inefficient.

Later, we shall see how to avoid bringing this complication into our interpreter at all.

\section{Prettyprinting}\index{prettyprinting}

We have not yet discussed how to print out the individual steps of evaluation of a program. Each step of evaluation is a valid OCaml program itself. The printed representation, of course, may elide much of this for brevity, but the task is essentially the same: convert the \textsf{TinyOCaml} tree representing the current step of evaluation to a string. At first glance, we might decide to convert the \textsf{TinyOCaml} representation back into a valid OCaml parse tree, and print that using the prettyprinter provided by the OCaml toolchain. Many modern language toolchains include such functionality. The advantage of not having to write our own prettyprinter is compelling, but the straitjacket is ultimately too tight: we cannot control the line-breaking adequately, add bold or underlining, and so on. Thankfully, writing a prettyprinter for \textsf{TinyOCaml} is relatively easy, given the table of precedences and associativities from the OCaml language manual. We have a prototype prettyprinter which can underline the redex, use bold for keywords and highlight search results. This is enough for our experimentation purposes for now.

It is worth discussing one issue: to what extent should the prettyprinter ape the actual layout of the code in the source file? Programmers are very attached to the layout they choose for their code (although some systems have automatic formatters \cite{gofmt} \cite{ocamlformat}). It would reduce the impedance mismatch between source and debugger output if this formatting could be kept. On the other hand, printing each stage of the evaluation in the original format would take up many many lines, making it perhaps harder to follow, and as the expression evaluates, the original formatting would be somewhat lost anyway. So, for now, we make no attempt to follow source formatting.

The techniques the prettyprinter uses are decades old, the line breaking is aided by the OCaml Standard Library's comprehensive \textsf{Format} module, and the basics of printing are no real consequence to our core aims, so the implementation is not discussed in detail here.

\section{Searching}\index{searching!in debugger output}\index{debugger!searching output of a}
\label{searching}

We have discussed various mechanisms for making sure that \textsf{OCamli}'s output is reasonable in the default case, and that there are options for deciding what information to display. But we will want a proper searching mechanism too, especially for interactive scenarios. Of course, one way is to use standard command line tools like {\small\texttt{grep}}. How well would that work for a typical search on a typical program? We can foresee problems -- for example, patterns may need to match independent of parenthesisation. In essence, we are searching the text not the program's syntactic structure.

The problem of searching in program code, either in textual or AST form, is known in the literature. Paul and Prakash's SCRUPLE system \cite{paul1994framework} uses an extended form of the programming language's own grammar, an approach from which we shall draw inspiration. Devanbu's GENOA \cite{devanbu1999genoa} also reuses the language's parser in the context of source code analysis. Crew's ASTLOG \cite{crew1997astlog} has similar aims. The distinction between ``lexical matchers'' (such as regular expressions) and ``syntactic matchers'' (which know the syntactic structure of what they are searching) is explored in Griswold et al's TAWK system \cite{griswold1996fast}. Specifically searching through program traces (rather than program source code) is also known in the literature, for example Watson and Salzman's work on the offline processing of traces of the evaluation of lazy functional programs \cite{watson1997trace} which allows not just single-stepping but the ability to search for given variables, successful pattern matches or return values.

Taylor distinguishes \cite{taylor-thesis} between filtering (preventing information appearing at all) and searching (moving through an evaluation to discover things). What we shall be calling searching is roughly what he calls filtering. He provides temporal filters, allowing information to be displayed only when a starting condition is met, and to be suppressed when an ending condition is met. We shall incorporate a similar facility: it allows the removal of swathes of information from a trace at a single stroke -- the remaining trace can then be reduced using ordinary searching mechanisms.

Penney applies  the well-known concepts of \index{temporal locality-of-reference}\index{spatial locality-of-reference}temporal and spatial immediacy (sometimes called temporal and spatial locality-of-reference) to the interface of an interactive tracing debugger. He says \textit{``Temporal immediacy implies that a user should expect to find certain core facilities that support efficient reasoning by means of conceptual step sizes that bring the user quickly to the points of interest in the trace.''} and \textit{``Spatial immediacy concerns the manner of presentation. Items that are conceptually linked should clearly be seen to be linked, perhaps by placing them near one another.''} \cite{penney2000augmenting}. He links this to Eisenstadt's  diagnosis, which we have already discussed, that the hardest bugs to fix are those which \textit{``were made hard because of a large temporal or spatial distance between their root cause and observable effect''} \cite{eisenstadt1997my}.

We have already discussed the elision of information in our interpreter. Here we provide a searching mechanism which learns the basic lessons of earlier systems. It is no grand design, but for experimentation. Until we know the final form of the interface to our debugger, it is unwise to commit too deeply. If we are to provide simple tools of our own, what facilities might be useful? Here are the basic options provided in \textsf{OCamli}:

\begin{verbatim}[commandchars=\\\{\}]
-search                  \textrm{\textit{show only matching evaluation steps}}
-highlight               \textrm{\textit{highlight the matching part of each matched step}}
-no-parens               \textrm{\textit{ignore parentheses when matching}}
-regexp                  \textrm{\textit{search terms are regular expressions rather than the built-in system}}
-upto <n>                \textrm{\textit{show the three lines up to each result line}}
\end{verbatim}

\noindent For example, consider the evaluation:

\medskip
\begin{verbatim}[commandchars=\\\{\}]
$ ocamli -e \textquotesingle{}List.map (fun x -> x + 1) [1; 2; 3]\textquotesingle\ -search \textquotesingle{}4::\textquotesingle\ -remove-rec-all\\
=>  2::3::\textbf{let} l = [] \textbf{in let} f x = x + 1 \textbf{in} 4::map \underline{f} l
=>  2::3::\textbf{let} l = [] \textbf{in} 4::\underline{map (\textbf{fun} x -> x + 1)} l
=>  2::3::\textbf{let} l = [] \textbf{in}
      4::(\underline{\textbf{let} f x = x + 1 \textbf{in function} [] -> [] | a::l -> \textbf{let} r = f a \textbf{in} r::map f l}) l
=>  2::3::\textbf{let} l = [] \textbf{in} 
      4::(\underline{\textbf{function} [] -> [] | a::l -> \textbf{let} f x = x + 1 \textbf{in let} r = f a \textbf{in} r::map f l}) l
=>  2::3::\textbf{let} l = [] \textbf{in}
      4::(\textbf{function} [] -> [] | a::l -> \textbf{let} f x = x + 1 \textbf{in let} r = f a \textbf{in} r::map f l ) \underline{l}
=>  2::3::4::
  \ \ \ \ \underline{(\textbf{function} [] -> [] | a::l -> \textbf{let} f x = x + 1 in \textbf{let} r = f a \textbf{in} r::map f l ) []}
\end{verbatim}
\medskip

\noindent This shows only the evaluation steps containing the text ``{\small\texttt{4::}}'', that is the ones where the list has almost been processed. Our search syntax is tailored to the job of searching \textsf{OCamli}'s output. The \index{searching!with patterns}search pattern is parsed using OCaml's \index{lexing}lexer, and then we allow any amount of whitespace between tokens, skip parentheses (if {\small\texttt{-no-parens}} is set), and allow the underscore character {\small\texttt{\_}} to stand for any token. A \index{regular expression}regular expression is generated to represent this, and searching proceeds. For example, we can search for only those steps of evaluation which contain lists of length three with \texttt{1} as the first element:

\medskip
\begin{verbatim}[commandchars=\\\{\}]
$ ocamli -e \textquotesingle{}List.map (fun x -> x + 1) [1; 2; 3]\textquotesingle -search \textquotesingle{}[1; _; _]\textquotesingle\ -remove-rec-all 
    \underline{List.map (\textbf{fun} x -> x + 1)} [1; 2; 3]
=>  \underline{(\textbf{let} f x = x + 1 \textbf{in function} [] -> [] | a::l -> \textbf{let} r = f a \textbf{in} r::map f l)}
\ \ \ \ \ \ [1; 2; 3]
=>  \underline{(\textbf{function} [] -> [] | a::l -> \textbf{let} f x = x + 1 \textbf{in let} r = f a \textbf{in} r::map f l)}
\ \ \ \ \ \ [1; 2; 3]
=>  \underline{(\textbf{function} [] -> [] | a::l -> \textbf{let} f x = x + 1 \textbf{in let} r = f a \textbf{in} r::map f l)}
\underline{\ \ \ \ \ \ [1; 2; 3]}
=>  \underline{(\textbf{function} a::l -> \textbf{let} f x = x + 1 \textbf{in let} r = f a \textbf{in} r::map f l ) [1; 2; 3]}
\end{verbatim}
\medskip

\noindent The search results may be highlighted with {\small\texttt{-highlight}}:\index{highlighting a search}\index{searching!highlighting}

\medskip
\begin{verbatim}[commandchars=\\\{\}]
$ ocamli -e \textquotesingle{}List.map (fun x -> x + 1) [1; 2; 3]\textquotesingle -search \textquotesingle{}[1; _; _]\textquotesingle\ 

  -highlight -remove-rec-all
    \underline{List.map (\textbf{fun} x -> x + 1)} \colorbox{black}{\color{white}[1; 2; 3]}
=>  \underline{(\textbf{let} f x = x + 1 \textbf{in function} [] -> [] | a::l -> \textbf{let} r = f a \textbf{in} r::map f l)}
\ \ \ \ \ \ \colorbox{black}{\color{white}[1; 2; 3]}
=>  \underline{(\textbf{function} [] -> [] | a::l -> \textbf{let} f x = x + 1 \textbf{in let} r = f a \textbf{in} r::map f l)}
\ \ \ \ \ \ \colorbox{black}{\color{white}[1; 2; 3]}
=>  \underline{(\textbf{function} [] -> [] | a::l -> \textbf{let} f x = x + 1 \textbf{in let} r = f a \textbf{in} r::map f l)}
\underline{\ \ \ \ \ \ \colorbox{black}{\color{white}[1; 2; 3]}}
=>  \underline{(\textbf{function} a::l -> \textbf{let} f x = x + 1 \textbf{in let} r = f a \textbf{in} r::map f l ) \colorbox{black}{\color{white}[1; 2; 3]}}
\end{verbatim}
\medskip

\noindent There are also options to alter the type and number of results:

\medskip
\begin{verbatim}[commandchars=\\\{\}]
-invert-search            \textrm{\textit{invert the search, showing non-matching steps}}
-n                        \textrm{\textit{show only n results}}
-until                    \textrm{\textit{show only until this matches a printed step}}
-after                    \textrm{\textit{show only after this matches a printed step}}
-until-any                \textrm{\textit{show only until this matches any step}}
-after-any                \textrm{\textit{show only after this matches any step}}
-invert-after             \textrm{\textit{invert the after condition}}
-invert-until             \textrm{\textit{invert the until condition}}
-stop                     \textrm{\textit{stop computation after final search results}}
-repeat                   \textrm{\textit{allow the after \ldots until result to be repeated}}
\end{verbatim}
\medskip

\noindent These options allow the programmer to show output only after a search matches, and only until another search matches. For example:

\medskip
\begin{verbatim}[commandchars=\\\{\}]
$ ocamli -e \textquotesingle{}List.map (fun x -> x + 1) [1; 2; 3]\textquotesingle\ -after \textquotesingle{}3 + 1\textquotesingle\ -until \textquotesingle{}2::3::4\textquotesingle\ \\
  -remove-rec-all
=>  2::3::\textbf{let} l = [] \textbf{in let} f x = x + 1 \textbf{in let} r = \underline{3 + 1} \textbf{in} r::map f l
=>  2::3::\textbf{let} l = [] \textbf{in let} f x = x + 1 \textbf{in let} r = 4 \textbf{in} \underline{r}::map f l
=>  2::3::\textbf{let} l = [] \textbf{in let} f x = x + 1 \textbf{in} 4::map \underline{f} l
=>  2::3::\textbf{let} l = [] \textbf{in} 4::\underline{map (\textbf{fun} x -> x + 1)} l
=>  2::3::\textbf{let} l = [] \textbf{in}
      4::\underline{(\textbf{let} f x = x + 1 \textbf{in function} [] -> [] | a::l -> \textbf{let} r = f a \textbf{in} r::map f l)} l
=>  2::3::\textbf{let} l = [] \textbf{in}
      4::\underline{(\textbf{function} [] -> [] | a::l -> \textbf{let} f x = x + 1 \textbf{in let} r = f a \textbf{in} r::map f l)} l
=>  2::3::\textbf{let} l = [] \textbf{in}
      4::(\textbf{function} [] -> [] | a::l -> \textbf{let} f x = x + 1 \textbf{in let} r = f a \textbf{in} r::map f l ) \underline{l}
=>  2::3::4::
      \underline{(\textbf{function} [] -> [] | a::l -> \textbf{let} f x = x + 1 \textbf{in let} r = f a \textbf{in} r::map f l ) []}
\end{verbatim}
\medskip

\noindent These searching mechanisms were arrived at through conjecture about and exploration of the most likely useful tools. It remains to be seen what the best interface for our interpreter or debugger will be.

We have said that an essential element of the notion of accessibility is that our interpreter should be usable with any project: no matter the build system, no matter what external libraries it uses. We now have an OCaml interpreter, \textsf{OCamli}. Assuming that \textsf{OCamli} were to be extended to support the full language, such that it can run any pure OCaml code, is that good enough? No. We still cannot deal with OCaml code which calls into C, we still cannot deal with arbitrary build systems (what happens if a build system has a preprocessor for some of its code?), in short we cannot cope with any project which is much more than a directory of plain {\small\texttt{.ml}} source files. A programmer will, in such circumstances, not turn to \textsf{OCamli} for debugging anything other than toy programs.

We shall need to interface with code written in C through the OCaml Foreign Function Interface (FFI).

\section{From interpreted to native and back again}

Most language toolchains have a \textit{Foreign Function Interface} (FFI) \index{FFI}allowing interfacing with C  and thence to any other language. This involves  calling functions in C from our language, calling functions in our language from C, and doing any work required to convert data types. Sometimes C is the host language, with the other language embedded, sometimes the other language controls the compilation and links in the C parts. The result of the compilation might be an executable or a static or dynamic shared library.

Let us look in turn at three parts of the OCaml toolchain which we will need to understand building this new, more accessible interpreter: the FFI itself, \index{OCaml!memory model}the memory model, and OCaml's built-in preprocessing mechanism PPX\index{PPX}.

\subsection{The OCaml/C FFI}
\label{prob-values}

The OCaml/C FFI is somewhat complicated, and difficult to write correct bindings for, though there is now an automatic system \cite{yallop2017modular}. Happily, our job is only to use bindings which have already been written -- they are part of any C code linked into an OCaml program, including the C parts of the Standard Library.

Let us look at a very simple example. Here, in a {\small\texttt{.ml}} file, we use the {\small\texttt{\textbf{external}}} keyword to define a function {\small\texttt{f}} of type \textsf{\textbf{int}} $\rightarrow$ \textsf{\textbf{float}} which will not be defined in the same file, but is provided elsewhere and will exist at link-time:

\medskip
\begin{verbatim}[commandchars=\\\{\}]
\textbf{external} f : int -> float
\end{verbatim}
\medskip

\noindent The implementation in C is defined thus:

\medskip
\begin{verbatim}[commandchars=\\\{\}]
#include <mlvalues.h> \hfill\textrm{\textit{load OCaml's special macros}}

CAMlprim value f(value x) \hfill\textrm{\textit{the function takes and returns an OCaml value}}
\{
  CAMLparam(x); \hfill\textrm{\textit{OCaml macros to mark values}}
  CAMLvalue(result);
  result = Val_float(ceil(foi Int_val(x))); \hfill\textrm{\textit{OCaml value to integer, process, back to OCaml as float}}
  CAMLreturn(result); \hfill\textrm{\textit{another macro to return the result}}
\}
\end{verbatim}
\medskip

\noindent The {\small\texttt{value}} type is used in C to describe OCaml values as they are stored in the OCaml heap. Our function in C does not really take an integer and return a floating-point value -- it takes a {\small\texttt{value}} which represents an OCaml \textsf{\textbf{int}} and returns a {\small\texttt{value}} which represents an OCaml \textsf{\textbf{float}}. The {\small\texttt{CAMLprim}}, {\small\texttt{CAMLparam}}, {\small\texttt{CAMLvalue}} and {\small\texttt{CAMLreturn}} macros take care of the other requirements of the FFI, such as maintaining the fidelity of the heap (and thus allowing garbage collection, which involves walking the heap).

What of the other direction -- how can we call back from C into OCaml? This is done by registering a function in the OCaml part at run-time using the Standard Library function {\small\texttt{Callback.register}}:

\medskip
\begin{verbatim}[commandchars=\\\{\}]
\textbf{let} g x = floor x \hspace{60mm}\textit{g}\negthinspace\ \textit{\textrm{has type \textbf{\textsf{float}} \(\rightarrow\) \textbf{\textsf{int}}}}

\textbf{let} () = Callback.register "g" g
\end{verbatim}
\medskip

\noindent It will be the programmer's responsibility to make sure the types match and the whole program remains type-safe. Now, we look up the function from the C side, at run-time also, and call the function:

\medskip
\begin{verbatim}
double c_g (int x)
{
  double d = Double_val(caml_callback(*caml_named_value("g"), Val_int(x)));
  return d; 
}
\end{verbatim} 
\medskip

\noindent Again, we need to ensure the rules of the FFI (especially with regard to the {\small\texttt{value}} type) are respected. The OCaml/C FFI is, of course, more complicated than these simple examples show.

\subsection{Modelling OCaml heap values}\label{heap}
We shall have values in our interpreter (of type {\small\texttt{Tinyocaml.t}}) such as {\small\texttt{Int\negthinspace\ 5}} or {\small\texttt{Cons(Int\negthinspace\ 5,\negthinspace\ Cons(Int\negthinspace\ 6, Nil))}}. When we call into a C function we shall need to make sure these values are in the OCaml heap, and represented in the correct fashion. Conversely, when reading the results of such a call, we will have \index{OCaml!heap}OCaml heap values which we need to read back into the interpreted world.

OCaml, at run-time, has an untyped but tagged representation of data, providing just enough information to allow the \index{garbage collection}\index{OCaml!garbage collector}garbage collector to traverse the heap, finding what is no longer used, and moving memory around from the minor heap to the major heap and to compact the heap. To give some examples:

\begin{itemize}
\item An integer is represented unboxed, but shifted left one bit, with a tag bit of 1 in the lowest bit. This distinguishes such immediate integers from boxed representations, since there can never be a 1 in the lowest bit of a memory address.
\item The booleans {\small\texttt{false}} and {\small\texttt{true}} are represented in the same way (like the integers 0 and 1 would be) so they are indistinguishable from integers. But, again, the garbage collector and runtime do not need to know the type.
\item The tuple (1, 3, (4, 5)) is represented by a heap block with tag 0, followed by two immediate integers representing 1 and 3, and a pointer to another block, tagged 0, with immediate integers 4 and 5.
\item Floating-point numbers are stored boxed, but there is a special case for arrays of floating-point numbers, which are stored in one block without further boxing for each array element.
\item Strings are stored as a valid C string in a block.
\end{itemize}

\noindent The following data type can be used to represent such OCaml heap values:

\begin{verbatim}[commandchars=\\\{\}]
\textbf{type} untyped_ocaml_value =
  UInt \textbf{of} int
| UBlock \textbf{of} int * untyped_ocaml_value array
| UString \textbf{of} string
| UFloat \textbf{of} float
| UFloatArray \textbf{of} float array
\end{verbatim}

\noindent User-defined data types are also represented in the same way on the OCaml heap. Consider, for example, the type:

\begin{verbatim}[commandchars=\\\{\}]
\textbf{type} colour =
  Red
| Green
| Blue
| RGB \textbf{of} int * int * int
| Transparent
\end{verbatim}

\noindent Nullary constructors are represented as integers 0,1\ldots and non-nullary ones as blocks tagged 0,1\ldots\ So, in our example, we have:

\begin{verbatim}[commandchars=\\\{\}]
\textbf{type} colour =
  Red                             \hspace{10mm}\textit{\textrm{integer 0}}
| Green                           \hspace{10mm}\textit{\textrm{integer 1}}
| Blue                            \hspace{10mm}\textit{\textrm{integer 2}}
| RGB \textbf{of} int * int * int          \hspace{10mm}\textit{\textrm{block with tag 0}}
| Transparent                     \hspace{10mm}\textit{\textrm{integer 3}}
\end{verbatim}

\noindent We shall retain these tag numbers when reading the program into a \textsf{TinyOCaml} one. We require a function to convert any \textsf{TinyOCaml} value to an OCaml heap value:

\begin{verbatim}[commandchars=\\\{\}]
\textbf{external} to_ocaml_value : t -> \textquotesingle{}a = "to_ocaml_value"
\end{verbatim}

\noindent
 The implementation must be in C, since we cannot construct the heap value using OCaml operations, but only using the C macros provided by the OCaml FFI. This is why the output type of {\small\texttt{to\_ocaml\_value}} is the polymorphic {\small\texttt{\textquotesingle a}}.

The type definitions of any user-defined data types are required for this function to work, so that it knows the tag numbers for each constructor. We read these upon initial conversion from the OCaml parse tree to the \textsf{TinyOCaml} representation and store the tag with each data constructor. Other than this complication, this function, at least for immutable values, is straightforward.

The inverse operation has two parts: first we want to read the heap value into the OCaml data type we created for untyped heap values, then we need to convert that to a \textsf{TinyOCaml} representation. The first part is simple, and again must be written in C:

\begin{verbatim}[commandchars=\\\{\}]
\textbf{external} untyped_of_ocaml_value :
  \textquotesingle{}a -> untyped_ocaml_value = "untyped_of_ocaml_value"
\end{verbatim}

\noindent We need to convert, say, the heap value {\small\texttt{UInt\negthinspace\ 0}} to the interpreter value {\small\texttt{Int\negthinspace\ 0}} if we expect an integer, but to the interpreter value {\small\texttt{Bool\negthinspace\ false}} if we expect a boolean. Thus, we need the expected type available if we are to do the conversion. So we provide a function which takes the type and the value of type {\small\texttt{Tinyocaml.untyped\raisebox{0.5mm}{\_}ocaml\raisebox{0.5mm}{\_}value}} and yields a \textsf{Tinyocaml} result.

\subsection{An introduction to OCaml PPX: \textsf{ppx\raisebox{0.8mm}{\_}auto}}\index{PPX}
\label{ppxauto}
We said in \cref{ppxautoearlier} that, to generate a binding to a C function, we can now write:

\medskip
\begin{verbatim}[commandchars=\\\{\}]
[%%auto \textbf{external} word_size : unit -> int = "%word_size"]
\end{verbatim}
\medskip

\noindent This will be converted automatically by our preprocessor \textsf{ppx\raisebox{0.5mm}{\_}auto} into:

\medskip
\begin{verbatim}[commandchars=\\\{\}]
\textbf{external} word_size : unit -> int = "%word_size"

\textbf{let} percent_word_size =
  \textbf{let} f =
    (\textbf{function} [Unit] ->
      \textbf{begin try} Int (word_size ()) \textbf{with} e -> exception_from_ocaml e \textbf{end}
     | _ -> failwith "%word_size")
  \textbf{in}
    ("%word_size",
     Fun (NoLabel, PatVar "*x", CallBuiltIn (None, "%word_size", [Var "*x"], f), []))
\end{verbatim}
\medskip

\noindent Such a system minimises the size of the part of \textsf{OCamli} which is special, that is to say the part which must be updated carefully each time the OCaml version number is increased.

The conversion from the first to the second piece of code above is done by the use of what is called a PPX (PreProcessor eXtension) -- this is a preprocessing mechanism built directly into the OCaml toolchain. It operates not source-text-to-source-text but AST-to-AST. This occurs after parsing but prior to typechecking. Thus, only syntactically valid programs can arrive at and leave the preprocessor, and the result will be typechecked as usual. The PPX invocation is denoted by a so-called extension node {\small\texttt{[\%\%<name>\negthinspace\ <contents>]}} where the contents in our case is the {\small\texttt{\textbf{external}}} declaration. To compile this, we just add {\small\texttt{-ppx <ppx\negthinspace\ name>}} to the compiler command line, or use whatever facility our favourite build system provides. The AST will be passed through the PPX processor, which recognises the name {\small\texttt{auto}} and processes only that part of the \index{Abstract Syntax Tree}AST, returning the rest unchanged. Thus, one can have several PPX processors operating in turn on a given file. The only requirement is that, at the end of all of them, there should be no {\small\texttt{[\%\% ]}} extension nodes left.

Let us look at the example above in detail. What information does \textsf{ppx\raisebox{0.5mm}{\_}auto} have? It knows the name of the function provided ({\small\texttt{word\_size}}), the name of the symbol which is expected to be available at link-time ({\small\texttt{\%word\_size}}), and the types of the input to the function (\textsf{\textbf{unit}}) and the output from the function (\textbf{\textsf{\textbf{int}}}). Our PPX processor builds a copy of the external declaration -- it is still needed -- and a {\small\textsf{\textbf{string} $\times$ Tinyocaml.t}} pair consisting of the symbol name and a generated {\small\texttt{CallBuiltIn}} to actually call the {\small\texttt{\textbf{external}}}. This {\small\texttt{CallBuiltIn}} knows to take the argument {\small\texttt{Unit}}, call the function, check for an exception (and process it if need be) and construct the {\small\texttt{Int}} return value.

So now we can generate bindings to C functions for our interpreter automatically, so long as a binding for native OCaml exists.

\section{Summary}

\addition{We have exhibited a prototype interpreter for OCaml, showing that it is plausible. The interpreter can load the OCaml Standard Library (with the exception of the complicated {\small\texttt{Printf}} and {\small\texttt{Scanf}} modules), and we have manually verified that it can execute all the programs in the textbook \textit{OCaml from the Very Beginning} \cite{whitington2013ocaml}. Further suggestions on possible future testing mechanisms are given in section \ref{correctness}.} 

\label{summary4}\addition{This work has succeeded in important ways: first, of course, it shows that such an interpreter can be reasonably written using the API provided by the OCaml compiler's public components. Second, we can see that the work required to support many of OCaml's complex language concepts in interpretation is much less difficult that compiling them. This leads us to glimpse a future, working tool for the first time: we have something with which to experiment in practice, rather than merely think about in theory.}

\correction{However, as a debugger for the working programmer (as opposed to the student) it is less compelling, failing to meet many of our concerns about accessibility. In addition, its interface was designed ad hoc as part of the prototyping process, rather than in advance as a cohesive whole.} \addition{We cannot, therefore, consider it as anything other than a partial success.}

Nonetheless, building such an experimental interpreter is an essential part of the process, allowing us to discover the boulders which are in our way, before moving on to design a better interface for our debugger. \addition{The state of the implementation of this work is given in section \ref{stateimpl}}. \addition{As we shall see in the next two chapters, lessons learned from this work lead to a better implementation.}

\chapter{An improved interpreter}

\label{chap:efficient}

\begin{quotation}\textit{\large The management question, therefore, is not whether to build a pilot system and throw it away. You will do that. \ldots\ Hence plan to throw one away; you will, anyhow.\textrm{\begin{flushright}--- \textup{Fred Brooks} The Mythical Man Month\end{flushright}}}\end{quotation}

\vspace{10pt}

\noindent As we have built \textsf{OCamli} we have glossed over a number of concerns about \index{time complexity}\index{space complexity}time and space usage, both in terms of \index{complexity}\index{algorithmic complexity}algorithmic complexity, and just raw \index{interpretation!speed of}speed. We have also learned of several significant technical problems with the previous approach, which a new design may fix. We concentrated on building an interpreter which could cope with most of the language, and allow real examples to be tried, loading the Standard Library and other modules. This was the right approach at the time, but now we must put it to one side, and begin again.  This time, we shall implement only a small subset of OCaml, but make sure that every other practical aspect is catered to; that is to say, this time, we shall be deep and narrow, rather than broad and shallow. This is not a repudiation of the previous approach, but a complement to it. Together, we think they lay the foundation for an eventual, full implementation of our debugging concept -- an implementation which, sadly, we leave for future work.

In this chapter we enumerate the problems of both efficiency and technical design which were encountered during the work described in the previous chapter. We shall discuss each in some detail, and develop a design for a more practical interpreter, \textsf{OCamli2}\index{OCamli2@\textsf{OCamli2}}.

\section{Problems}

Let us now review the process of interpretation as it stands in \textsf{OCamli}, identify the most serious technical and efficiency concerns, and then address them in turn. Some we have workable solutions for, some are discussed more briefly and left for future work. We do not wish to fall into the well-known trap of premature optimisation, but some inefficiencies really are too glaring to ignore: without fixing them, we could not have a usable tool, and without a usable tool, our ability to evaluate our work will be hampered. Here is the present process:

\begin{itemize}
\item A program is read and converted into an internal representation, \index{TinyOCaml@\textsf{TinyOCaml}}\textsf{TinyOCaml}. This happens only once per program.
\item The program is evaluated step by step. For each step, we must first identify the \index{reducible expression}reducible expression (redex). This involves traversing the tree representing the \textsf{TinyOCaml} program. This tree can be deep, especially in the case of code which is not tail-recursive. In fact, it is easy to see how such a process can add a factor of $O(n)$ to the complexity of running some programs. It is presently not possible to find the redex in any other way than starting at the top of the program tree.
\item As the redex is located, moving down the tree representing the program, an environment of bindings is built up. This environment is then used when evaluating the expression a single step. It is thrown away before evaluating the next step, even if the next redex is likely to have an identical \index{environment}environment.
\item The evaluation is advanced one step. The new program tree has now been built. Some of these operations will have no efficiency concerns -- adding two numbers is $O(1)$ just like in a compiled program, but some may be much more complex than the equivalent compiled form, for example functor application.

\item Now that the next step is ready for printing, we decide whether to print it. These considerations are not typically computationally complex. If we decide to print it, however, we must clean it up by, for example, removing unused let bindings. This involves multiple passes over the tree, which is expensive.
\item We now \index{prettyprinting}print the step itself. This is complicated compared with the interpretation of the actual step of computation (which might be as simple as adding two numbers). However, it is less complicated than finding the reducible expression and building up the environment. In addition, the work is largely unavoidable, and prettyprinting techniques are well understood.
\item Some programs require the use of the \index{FFI}Foreign Function Interface involving the copying of arbitrary amounts of data between the \index{OCaml!heap}\index{heap}OCaml heap and the \textsf{TinyOCaml} program representation. We should like to avoid this, for simplicity, for correctness,  as well as for performance.
\end{itemize}

\noindent\label{decision5-1}\addition{Why it is important to fix these problems? They prevent our having a system which is (a) tractable in implementation, due to complexity of the previous approach; (b) usable by programmers in terms of the functionality it can support (for example, high quality elision); and (c) usable by programmers due to being fast enough. The alternative, to plough on with the original \textsf{OCamli}, looks doomed to slow suffocation.}

Let us now look at some of these problems in more detail, and how \textsf{OCamli2} approaches them.

\section{A better data structure for programs}

In writing our first interpreter, we chose a data structure for representing programs and their data which, as it turned out, was somewhat flawed. Here are some difficulties with the \textsf{TinyOCaml} data structure used in \textsf{OCamli}:

\begin{itemize}
\item Being based upon the untyped OCaml parse tree, not the typed OCaml tree, types for each subexpression are not always  available, complicating C/OCaml interaction and the provision of external features such as the Standard Library. Types are also required for printing, so any lack of availability of type information is a concern. This was probably the key mistake of our first implementation. [see section \ref{prob-c-ocaml}]
\item Values (for example data) in the type were not represented in the same way as a compiled OCaml program would represent them, making easy interaction with the OCaml runtime and external functions accessed through the FFI difficult (and, we suspect, probably not even correct in all circumstances). [see section \ref{prob-values}]
\item Pattern matching was difficult when one or more let bindings were interspersed between structures one wished to match. [see section \ref{prob-patterns}]
\item Elision of information for printing was a largely ad hoc affair, with little direct support in the data structure. [see section \ref{prob-elision}]
\end{itemize}

\noindent Some of these problems could only have been discovered through the process of writing \textsf{OCamli} and some were simply errors which might have been foreseen. In any event, we now exhibit a new data type for \textsf{OCamli2} which remedies them.

The first change is to represent values as integers or pointers into the heap, exactly as a compiled OCaml program would. That is to say instead of {\small\texttt{Int\negthinspace\ 5}} or {\small\texttt{Cons(Int\negthinspace\ 1,\negthinspace\ Nil)}} in \textsf{OCamli}, we just have the {\small\texttt{Value}} constructor to represent all values. This means that interaction between C and interpreted OCaml and between interpreted OCaml and compiled OCaml becomes much simpler. Data need no longer be copied. The two modes of execution (native and interpreted)  simply live in the same executable, using the same heap. We shall see later how the existence of types in our new design allows such heap values to be printed out even though the OCaml heap does not contain types itself.

The second change is to decorate each node of the program tree with a record containing ancillary information: most importantly for our new design, the type of that node from the typed tree. We shall also see how it allows an elegant solution to the problem of pattern matching over let bindings.

Here is the main part of the new type for representing OCaml programs. Notice that a pair of mutually recursive data types is used to have a record decorate each node of the tree.

\begin{verbatim}[commandchars=\\\{\}]
\textbf{type} t\textquotesingle{} =
  Value \textbf{of} Obj.t\hfill\textrm{\textit{values in OCaml native representation}}
| Function \textbf{of} case list * env\hfill\textrm{\textit{functions}}
| Apply \textbf{of} t * t list\hfill\textrm{\textit{function application}}
| Var \textbf{of} string\hfill\textrm{\textit{variables}}
| Cons \textbf{of} t * t\hfill\textrm{\textit{cons part of list literal which is not yet a value e.g.\ \texttt{[1\! +\! 2;\! 3]}}}
| Append \textbf{of} t * t\hfill\textrm{\textit{append lists}}
| IntOp \textbf{of} op * t * t\hfill\textrm{\textit{integer operations}}
| FOp \textbf{of} op * t * t\hfill\textrm{\textit{floating-point operations}}
| Compare \textbf{of} cmpop * t * t\hfill\textrm{\textit{polymorphic comparison}}
| BoolOp \textbf{of} boolop * t * t\hfill\textrm{\textit{boolean operations}}
| Let \textbf{of} bool * binding * t\hfill\textrm{\textit{let bindings}}
| Struct \textbf{of} t list\hfill\textrm{\textit{structures}}
| LetDef \textbf{of} bool * binding\hfill\textrm{\textit{top-level let bindings}}

\textbf{and} t =
  \{typ : Types.type_expr;\hfill\textrm{\textit{type from OCaml typed tree}}
   e : t\textquotesingle{};\hfill\textrm{\textit{expression}}
   lets : env;\hfill\textrm{\textit{let bindings with value right-hand-sides}}
   peek : peekinfo option;\hfill\textrm{\textit{info for peeking support}}
   printas : string option\}\hfill\textrm{\textit{substitute printed representation}}
\end{verbatim}

\noindent Consider the following program:

\begin{verbatim}[commandchars=\\\{\}]
\textbf{let} x = 3 \textbf{in}
  \textbf{let} y = 1 + 2 \textbf{in}
    x + y
\end{verbatim}

\noindent The representation of this program in our new type might be:

\begin{verbatim}[commandchars=\\\{\}]
\{typ = \textrm{\textit{<OCaml compiler type representation of \textsf{int}}>};
 e = Let (false,
       ("y", IntOp (Add,
                    \{typ = \textrm{\textit{<OCaml compiler type representation of \textsf{int}}>};
                      e = Value \textrm{\textit{<OCaml heap representation of 1>}};
                      lets = []; peek = None; printas = None\},
                    \{typ = \textrm{\textit{<OCaml compiler type representation of \textsf{int}}>};
                      e = Value \textrm{\textit{<OCaml heap representation of 2>}};
                      lets = []; peek = None; printas = None\}));
        \{typ = \textrm{\textit{<OCaml compiler type representation of \textsf{int}}>};
         e = IntOp (Add,
                    \{typ = \textrm{\textit{<OCaml compiler type representation of \textsf{int}}>};
                      e = Var "x"; lets = []; peek = None; printas = None\},
                    \{typ = \textrm{\textit{<OCaml compiler type representation of \textsf{int}}>};
                      e = Var "y"; lets = []; peek = None; printas = None\});
         lets = [];
         peek = None;
         printas = None
        \});
 lets = [("x", Value \textrm{\textit{<OCaml heap representation of 3>}})];
 peek = None;
 printas = None\}
\end{verbatim}

\noindent\label{decision5-2}\correction{We have not been explicit thus far about why this new data structure fixes the problems we have identified: in the next few sections we shall explain the reasons that this particular structure improves evaluation, printing, and interaction with compiled code.}

\section{Values in the new representation}

\index{value!representation}\index{heap!values}How do we convert a program represented by the type-checked tree provided by the OCaml compiler into our representation? It is not quite so straightforward as before. We must identify items which ``should be values'' in our system and convert them to the OCaml compiled memory representation to build a {\small\texttt{Value}}. For example, a constant integer represented in the typed tree must be converted to a real integer in memory. Similarly, a cascade of {\small\texttt{Cons}} parse tree nodes where each element of the list should be a value, must itself be converted into a heap representation of such a list, along with its elements. Tediously, we must of course also have constructors in our type to represent the relatively rare case where a list literal should not be a value, for example {\small\texttt{[f\! a;\! f\! b;\! f\! c]}}. \correction{When a non-value array such as {\small\texttt{[|1;\! 2\! +\! 3|]}} becomes the value {\small\texttt{[|1;\! 5|]}} during evaluation, it is added to the OCaml heap, and converted to a {\small\texttt{Value}}}. \addition{Consider the following program:}

\begin{verbatim}[commandchars=\\\{\}]
\addition{([1; 2; 3], List.map (fun x -> x + 1) [1; 2 + 3])}
\end{verbatim}

\noindent \addition{It is a tuple of two items: the first is the list {\small\texttt{[1; 2; 3]}}, which is considered to be value in \textsf{OCamli2}. Heap constructs representing this list will be built upon loading the program, and it will be represented in \textsf{OCamli2}'s program representation as a pointer into the heap, i.e {\small\texttt{Value}} in the type given above. The second part of the tuple is plainly not a value, being an unevaluated function application. The list {\small\texttt{[1; 2 + 3]}} is not a  value, either. And so, it must be represented in \textsf{OCamli2} as {\small\texttt{Cons(Value ... IntOp(Add, Value ..., Value ...))}} where the three values represent 1, 2, and 3 respectively in OCaml's native representation. Now we may proceed to evaluate one step:}

\begin{verbatim}[commandchars=\\\{\}]
\addition{([1; 2; 3], List.map (fun x -> x + 1) [1; 5])}
\end{verbatim}

\noindent\addition{The list is now {\small\texttt{[1; 5]}}. This is a value, and is immediately converted to a {\small\texttt{Value}} on the heap before the end of this step of interpretation. Thus the interpreter, at the next step, may assume any such value-normalisation has been done, and the program is in this normal form, just as it may do on the first step, such normalisation having been done upon loading the program. A few steps of interpretation later, we have reduced the program fully to a value:}

\begin{verbatim}[commandchars=\\\{\}]
\addition{([1; 2; 3], [2; 6])}
\end{verbatim}

\noindent\addition{Now, the entire tuple becomes a heap object represented by a {\small\texttt{Value}} constructor. Since the program is now a value, no more evaluation is possible, and interpretation ends.}

Reading the type given above, and considering our definition of what should constitute values, one might wonder why the {\small\texttt{Function}} constructor exists at all: are not functions also values? We shall certainly have to have a way of building native OCaml functions for higher order use -- for example, to evaluate {\small\texttt{List.map\! (\textbf{fun}\! x\! ->\! x\! +\! 1)\! [1;\! 2;\! 3]}} when {\small\texttt{List.map}} is a native function, we shall have to produce a native version of the function {\small\texttt{\textbf{fun}\! x\! ->\! x\! +\! 1}} for {\small\texttt{List.map}} to use. The reason is that functions in OCaml are used for pattern-matching, and we need to see those pattern matches when single-stepping interpreted code. So, in the present design, functions are only  converted to values when being applied to a native function such as {\small\texttt{List.map}}. Otherwise, they are considered values only when required: for example, when deciding whether to end execution with the expression {\small\texttt{[(\textbf{fun}\! x\! ->\! x\! +\! 1);\! (\textbf{fun}\! x\! ->\! x\! +\! 2)]}}, it is considered a value; when deciding whether to convert it to an OCaml heap object upon reading the program, it is not.

Our previous solution for dealing with code written in C and accessed through the C/OCaml \index{FFI}FFI is flawed. The copying back and forth of heap data between the interpreted and compiled worlds is cumbersome and inefficient. It may not even be possible for it to work in all cases, or for it to alter program behaviour. Consider the case where a memory location is written to both from the OCaml program and the C program (e.g.\ a shared mutable array)\addition{, or where the program is operating in parallel execution.} In the new scheme, all values are opaque under the {\small\texttt{Value}} constructor, and only unevaluated parts of the program are represented in the data type proper. This ensures that no conversion is needed when transferring values to and from the OCaml heap and the interpreted representation -- the same values are simply shared. \label{decision5-3}\addition{We are, in effect, following the OCaml memory model wherever possible, and the model of OCaml's longstanding FFI. That is to say the model OCaml uses itself to interface with its runtime, with external code, and with dynamic loading. This is the correct level of abstraction to work at, with regard to not only correctness but future compatibility: it ties us into a union with the compiler reducing or removing the need to update this part of the codebase with each successive OCaml version.}

We have removed the possibility of a fragile interpreter-native interface. We know that such a hybrid system is workable since the OCamlJIT project \cite{meurer2010ocamljit}, which provides a JIT system for compiling parts of OCaml bytecode programs into executable code at run-time, is a working system with similar requirements.

\section{Printing}

\label{decision5-4}\correction{The reader might wonder why we need \index{types at run-time}types at run-time at all. After all, compiled OCaml programs have no types available. We need them because we are printing the steps of execution, and so we need to print values, and to print them we need their types \addition{-- there is simply not enough information in the OCaml heap structure to reconstruct the type, it being designed for traversal by the runtime and in particular the garbage collector only}. In the original \textsf{OCamli}, we represented data in programs in our own way, where the types were mostly clear: it is obvious how to print {\small\texttt{Cons\! (Int\! 5,\! Nil)}} as {\small\texttt{[5]}}, for example. In \textsf{OCamli2}, the data is simply {\small\texttt{Value\! x}} where {\small\texttt{x}} is an immediate integer or heap pointer in the standard OCaml manner. We cannot print it without knowledge of the type.} \correction{This is where the type annotation attached to each node comes in. We already know how to build a structured value from a heap value given its type, as we showed in section \ref{heap}.}

\correction{Our starting point, we remember, is the typed tree produced by the OCaml compiler after parsing and typechecking. It might, then, be tempting to assume that every node has a defined type which can be used, as we have noted, for printing. However, polymorphism means this is not necessarily true: $\alpha\ \textsf{\textbf{list}}$ is no good for printing a list of items inside the evaluation of a polymorphic function unless we know what $\alpha$ is. We know only at run-time, when the function is called. Again, an evaluator which did not have to print the steps would not have to care what the type of the list elements is -- that is the very definition of what it means for such a function to be \index{polymorphic function}\index{function!polymorphic}polymorphic. If we fail to get this right, we may end up with having to print {\small\texttt{<poly>}} or {\small\texttt{\_}} in lieu of such values, something we earlier noticed as a fault of many other debugging approaches for functional programming.}

\correction{When we look up a variable denoting a polymorphic function, we pull its type apart, and substitute in the known types of the arguments. If the polymorphic function is used again, at a different call site, different types will be substituted in.} \addition{Consider the following simple program making use of polymorphism:}

\medskip
\begin{verbatim}[commandchars=\\\{\}]
\addition{\textbf{let} hd = \textbf{function} h::_ -> h}

\addition{\textbf{let} a = hd [1; 2; 3]}

\addition{\textbf{let} b = hd [[1]; [2]; [3]]}
\end{verbatim}
\medskip 

\noindent\addition{The {\small\texttt{hd}} function has type $\alpha\ \textsf{\textbf{list}} \rightarrow \alpha$ and is use here twice, once with an input of type \textsf{\textbf{int list}} and once with an input of type \textsf{\textbf{int list list}}. That is to say, with \textsf{\textbf{int}} and \textsf{\textbf{int list}} as concrete values for $\alpha$.}

\correction{For a user-defined type, printing a value of that type may need not just the type itself, but the type definition, for example to find constructor names. Such information must be loaded from compiler artefacts. It would be more pleasant if the OCaml typed tree could include such information directly.}

\correction{The other printing innovation in \textsf{OCamli2} is the use of the decorating record in our new data structure to store an annotation giving an alternative printing of the node in question. The evaluator may set the {\small\texttt{printas}} entry to a string which is printed instead of attempting to print the expression or value itself. This simple little hack should subsume a number of the complicated elision methods described in \cref{chap:interpreter}, leading to shorter and better traces with little alteration or complication of the evaluator itself.}

The evaluator sets up {\small\texttt{printas}} information in two ways. First, when a variable lookup results in a function (for example when a function is being passed to a higher-order function) the {\small\texttt{printas}} is set to simply print the name of the variable. This substitution is silent. We thus avoid printing the whole function body inline in subsequent steps of evaluation, a significant source of visual noise in the original \textsf{OCamli}, even on small programs. Second, on \index{partial evaluation}partial evaluation of a function passed to another. For example in the program {\small\texttt{List.map\! ((\! +\! )\! 2)\! [1;\! 2;\! 3]}} applying the argument {\small\texttt{2}} to the addition operator is a step of evaluation, but its result is best printed as {\small\texttt{((\! +\! )\! 2)}} since that is what the reader will understand.

\correction{Presently, the evaluator never overwrites an existing {\small\texttt{printas}} instruction. As with several aspects of the experimental \textsf{OCamli2} it is not yet clear what the optimal scheme will be.
For this and other reasons, example programs produced by \textsf{OCamli2} often have traces less than half the length of their \textsf{OCamli} versions, and these traces are much more readable. This is despite much less time  having been spent on the implementation of \textsf{OCamli2} -- it is simply the result of learning the lessons of earlier design mistakes. \addition{An example is given at the end of this chapter.}}

\section{Making native functions from interpreted ones}

When the program {\small\texttt{\textbf{let}\! y\! =\! 1\! \textbf{in}\! List.map\! (\textbf{fun} x\! ->\! x\! +\! y)\! [1;\! 2;\! 3]}} is encountered we have need to pass the function {\small\texttt{\textbf{fun}\! x\! ->\! x\! +\! y}} to the (native)  {\small\texttt{List.map}} function. This is a problem because {\small\texttt{\textbf{fun}\! x\! ->\! x\! +\! y}} is code we wish to interpret. What is required is to convert this function to a natively-callable version which {\small\texttt{List.map}} can use but which, internally, will call the interpretive evaluator so the steps of evaluation can be shown, returning the resultant value back to \addition{the native} {\small\texttt{List.map}}. Notice that such a function must also have access to the environment from the interpreter, to access the value of {\small\texttt{y}}.

Writing this {\small\texttt{make\_native}} function is actually relatively simple, though we must be careful to give the correct result if the application of it is partial: the result will be another (interpreted) function which itself must be made native by a recursive call to {\small\texttt{make\_native}}. \addition{Here it is:}

\medskip
\begin{verbatim}[commandchars=\\\{\}]
\addition{\textbf{let rec} make_native impl_lets funexpr = \hfill\textit{\textrm{make native function given implicit lets and definition}}}
\addition{  \textbf{match} funexpr \textbf{with}}
\addition{  | \{e = Function ([PatVar vname, None, rhs], fenv);\hfill\textit{\textrm{match expression and its type}}}
\addition{     typ = \{desc = Tarrow (_, alpha, beta, _)\}\} ->}
\addition{      \textbf{let} fn = \hfill\textit{\textrm{what will become the natively-callable function}}}
\addition{        \textbf{let} expr = \hfill\textit{\textrm{build the function application}}}
\addition{          \{rhs \textbf{with} lets = impl_lets; e = Apply (funexpr, [{rhs with e = Var vname}])\}}
\addition{        \textbf{in}}
\addition{          \textbf{fun} x ->}
\addition{            \textbf{let} newlet = (false, ref [(vname, \{expr \textbf{with} e = Value x; typ = beta\})]) \textbf{in}}
\addition{              \textbf{match} ((eval_full (newlet::fenv)) expr).e \textbf{with}}
\addition{                Value ret -> ret \hfill\textrm{\textit{full application, return the result}}}
\addition{              | Function (f', fenv') -> \hfill\textrm{\textit{this was a partial application, keep binding for next time}}}
\addition{                  make_native}
\addition{                    (newlet::impl_lets)}
\addition{                    \{expr \textbf{with} typ = beta; e = (Function (f', fenv'))\}}
\addition{              | _ -> failwith "didn't make a value"}
\addition{      \textbf{in}}
\addition{        (Obj.magic fn : Obj.t) \hfill\textrm{\textit{build the heap object}}}
\addition{  | _ -> \textbf{failwith} "make_native: not a function"}
\end{verbatim}
\medskip

\section{The Standard Library}\index{OCaml!Standard Library}\index{Standard Library}

Our new representation does away with almost all the complications of interacting with native functions. Recall that we had to write the following boilerplate to bridge between the interpreted and native worlds for a single function:

\begin{verbatim}[commandchars=\\\{\}]
\textbf{external} word_size : unit -> int = "%word_size"

\textbf{let} percent_word_size =
  \textbf{let} f =
    (\textbf{function} [Unit] ->
      \textbf{begin try} Int (word_size ()) \textbf{with} e -> exception_from_ocaml e \textbf{end}
     | _ -> failwith "%word_size")
  \textbf{in}
    ("%word_size",
     Fun (NoLabel, PatVar "*x", CallBuiltIn (None, "%word_size", [Var "*x"], f), []))
\end{verbatim}

\noindent We wrote a \index{PPX}PPX extension to automate this in most circumstances, writing instead:

\begin{verbatim}[commandchars=\\\{\}]
[%%auto \textbf{external} word_size : unit -> int = "%word_size"]
\end{verbatim}

\noindent We had to give the whole {\small\texttt{\textbf{external}}} declaration, including specifying the type. Such a type may not always be available (if we are accessing pre-compiled code, for example). In \textsf{OCamli2}, we may write simply:

\begin{verbatim}[commandchars=\\\{\}]
addfun "List.map" List.map
\end{verbatim}

\noindent The {\small\texttt{addfun}} function is pure OCaml, and no PPX is needed. This works for any OCaml function of any type and, of course, for external functions too. \label{decision5-6}\addition{The advantages for this design decision are first that it makes the code base much more maintainable due to the reduction in new work required when the OCaml version changes; and second, that it reduces traces in the ordinary case. Small traces which can then be iteratively explored by the user are, as we shall see in the next chapter, a good interface model.}

\section{Finding the redex and building the environment}

Consider, by way of example, the following:

\begin{verbatim}[commandchars=\\\{\}]
\textbf{let} x = 1 \textbf{in}
  \textbf{let} y = 2 + x \textbf{in}
    x + y
\end{verbatim}

\noindent The next step of evaluation is this:

\begin{verbatim}[commandchars=\\\{\}]
\textbf{let} x = 1 \textbf{in}
  \textbf{let} y = 2 + 1 \textbf{in}
    x + y
\end{verbatim}

\noindent To find the \index{reducible expression}reducible expression, as \textsf{OCamli} is presently conceived, we begin at the top of the program tree. First, we encounter the binding {\small\texttt{\textbf{let}\! x\! =\! 1}} whose right-hand side is already a value. Thus, we need a function to test if an expression is a value. Of course, this function might have to recurse a great deal to find out if a big structured piece of data really is a value. It being so, we have the environment $\{x = 1\}$ and move inside the let binding, but retaining it so we can build the expression back up when we have evaluated one step. The next binding {\small\texttt{\textbf{let}\! y\! =\! 2\! +\! x}} has a non-value right-hand side, so the redex is in its right-hand side, and we go inside the addition operation, to find that its right-hand side is the non-value {\small\texttt{x}}. We look this up in the environment, and our step is done. All the information we learned about the environment and the position of the redex is lost: we must begin at the top of the expression for the next step.

Can we do better? Since the right-hand side of the addition is now a value, we know exactly what the next reducible expression is -- it is the addition {\small\texttt{2 + 1}} itself. Such situations, in which the redex is in a known place, are relatively frequent. We could return the next step, together with a continuation allowing easy calculation of the step after that. If the next redex cannot be determined easily, we default back to beginning at the top of the expression. It would be interesting to see if this mechanism can be achieved without undue complication to the code, and to measure its effect -- that is to say, statistically, how many reducible expression sequences can be dealt with in this way, and how much do they speed things up?

At this point the reader might wonder whether our methods are just too juvenile, given, of course, the well-known literature on \index{abstract machine}abstract machines for the evaluation of the \index{lambda calculus}lambda calculus such as the \index{SECD machine}SECD machine \cite{danvy2004rational,ager2003functional} and its many successors and variants: they make sure that the next \index{reducible expression}redex is always known. Why do we not just modify such an abstract machine to hold enough extra information such that expressions may always be reconstructed step-by-step? In \cref{chap:misc} we will show one way this might be done (by modifying an OCaml-style bytecode), but we believe that it is important to get our interpreter and debugger design right, and working correctly, before jumping to such a scheme. Once we make the choice to use an abstract machine, it may be that various things we should like to do become technically difficult; sticking with inefficient, simple interpretation is best for now, however theoretically suboptimal it seems. We have already discussed, in \cref{correctness}, Cong, Asai and Furkawa's mechanism \cite{cong2016implementing,furukawa2019} for writing a step-by-step interpreter in terms of a big-step one using a variant of Racket's continuation marks. This may provide a less disruptive way to make such an improvement.

\section{Let bindings}\index{let binding}
\label{prob-patterns}

\label{decision5-8}\correction{One of the difficulties of the implementation of the original \textsf{OCamli} is that matching on the program data structure to find the \index{reducible expression}redex is impeded by there being a non-zero number of let-bindings between layers of the data structure. So, we should like to change the data structure to make this easier.} For example, when evaluating the following program, there are lets surrounding the function we wish to apply:

\begin{verbatim}[commandchars=\\\{\}]
(\textbf{let} x = 1 \textbf{in} \textbf{let} y = 2 \textbf{in} (\textbf{fun} p -> p + x + y)) 3
\end{verbatim}

\noindent The first problem is that this expression is not of the form {\small\texttt{App\! (<fun>,\! x)}} but {\small\texttt{App\! (Let\! (Let\! (\_,\! \_),\! <fun>),\! x)}} so is hard to pattern-match. We do not want to do lexical substitution, because there may be much bigger values than 1 and 2 and this would make the visualization awkward.

The connection to our work on the efficiency of the step-by-step interpreter is that, when a program is being executed, names disappear, and the let bindings become orphaned, the name they define being unused in the program. So the second problem is that such names must be removed efficiently when printing the program. To search through the whole program each step to remove unused lets (as \textsf{OCamli} does) is plainly a source of inefficiency. Can we devise a mechanism which prevents this?

The first problem is solved by introducing a new type which wraps each expression (and all its subexpressions) in a record containing what we call \index{let binding!implicit}``implicit lets''. This is a list of bindings where the right-hand sides are values. These are ones whose right-hand sides require no further evaluation. When the program is read (from a real OCaml one), let bindings which are already values are put in the implicit lets. When the program is being evaluated, a let binding's right-hand side will be evaluated step by step. When its right-hand side is a value, it ceases to be a {\small\texttt{Let}} but is placed in the implicit lets surrounding its expression. Thus, all lets with fully-evaluated right-hand sides are always in the list of implicit lets, and all ones with unevaluated right-hand sides are explicit in the data type. Now, when we want to, say, find the redex by matching, there are no longer interleaved {\small\texttt{Lets}} in the way -- the implicit ones cannot contain the redex since they are fully evaluated by definition. And we will never need to evaluate underneath an explicit {\small\texttt{Let}}, since the redex will always be in the right-hand side of the outermost such {\small\texttt{Let}}.

During evaluation, the implicit lets are appended to the environment whenever they are encountered. Similarly, during prettyprinting both the implicit lets and those still in the program text can be printed.

The second problem, as we mentioned, is how to remove an implicit let which is no longer used in the enclosed expression, so it is not printed. The current mechanism (in \textsf{OCamli}) traverses the tree, removing any lets whose names appear in the enclosed program text -- save for shadowing. For example, consider the following partly-evaluated program:

\begin{verbatim}[commandchars=\\\{\}]
\textbf{let} x = 5 \textbf{in}
  \textbf{let} y = 6 \textbf{in}
    5 + y
\end{verbatim}

\noindent The name {\small\texttt{x}} no longer appears, but the let binding is not removed. In particular, the \index{prettyprinting}prettyprinter will not discover that the name is unused until the let binding has been printed. So, removal must be a separate process before the expression is printed. This is not just an issue of printed output either -- a let binding which is not removed constitutes a space leak. A more efficient solution might be to keep a mutable flag with each implicit let, and have the printer set the flag for each name encountered when printing. Then, implicit lets with flags not so set are removed from the expression, and evaluation continues. This removes the unused let binding at the next stage.

\section{Speed}\index{speed of interpretation}\index{interpretation!speed of}
\label{testspeed}

We know that interpretation is slow, but how slow? And is it the computation or the printing out of steps which is slow? What is the cost of doing the interpretation step by step even if we only print out the pertinent steps? What are the costs of choices in data structure representation? 

\label{decision5-9}\addition{We might pause to remind ourselves why speed matters. It is for two reasons. First, in some cases a slow debugging tool might make debugging a particular program (or a program for a particular input) simply intractable. Second, in the case that it is tractable, it makes debugging simply more pleasant. As we saw in our literature review in chapter 2, human factors like this can be critical in the uptake of a debugging tool.}

\addition{First, we look at some early performance results from the beginning of the project. Then, we shall benchmark \textsf{OCamli} against OCaml using a series of test programs, with regard to time and space usage.} \correction{The following table shows some time benchmark results for a tiny program (calculating the addition of two numbers) run using the OCaml compilers, \textsf{OCamli}, and various early experimental precursors to \textsf{OCamli}, each more naive than \textsf{OCamli}. The times given are relative to the time for the OCaml native code compiler. The discussion follows the table.}

\bigskip
\noindent\begin{tabular}{@{}lll}\toprule
Letter & Time & Description\\\midrule
A & 1 & OCaml native code compiler\\
B & 7 & OCaml bytecode code compiler\\
C & 1960 & \textsf{OCamli} interpreter\\
D & 2055 & \correction{Substitution, not step-by-step}, \textsf{TinyOCaml} tree\\
E & 5754 & Substitution, step-by-step, \textsf{TinyOCaml} tree\\
F & 36,627 & Substitution, step-by-step, OCaml parse tree\\
G & 27,264,567 & Substitution, step-by-step, OCaml parse tree, printing to {\small\texttt{/dev/null}}\\\bottomrule
\end{tabular}
\bigskip

\noindent Let us take these in turn. Letter A is the OCaml native code compiler, whose evaluation time is defined as unity. Letter B is the OCaml bytecode compiler, which is seven times slower on our benchmark. Then, at Letter C, our \textsf{OCamli} interpreter which is about 2000 times slower than native code. This is without prettyprinting the steps of execution -- just interpreting the code silently, step by step. The latter four are timings taken from very early precursors to \textsf{OCamli}, in the initial experimentation stages of this project. \correction{The first, letter D, is not a step-by-step interpreter but reduces the expression to a value all at once}. It uses the same \textsf{TinyOCaml} representation of programs which eventually became the basis of \textsf{OCamli} and a very naive model of execution which textually substitutes instead of looking things up in an environment of bindings. Compare this with letter E, which is the same but operates step by step. The slowdown is about three times. Letter F is the same, but is of an even earlier vintage when we were still using the usual OCaml parse tree as our main data structure. The cost of this is about six times. Finally, we add prettyprinting of each step. The cost is about 750 times. \correction{Overall, Letter G is more than 27,000,000 times slower than Letter A, demonstrating the cumulative effect of all the inefficiencies in A\ldots G.} \addition{Printing itself, unsurprisingly, slows the running time by about 750 times.}

\addition{\subsection*{Time and space usage of some test programs}}

\noindent\addition{A small corpus of test programs has been prepared; they are listed in Appendix B. Due to the great disparity in time between native code compiler \textsf{OCaml} and \textsf{OCamli}, testing for time involves:}

\bigskip
\begin{enumerate}
\item \addition{Picking input sizes (values of n below) which generate a run time not too long (so we do not wait for hours) and not too short (so the time measured can be considered accurate).}
\item \addition{Testing each program three times and taking the mean.}
\item \addition{Calculating the startup time for OCaml and \textsf{Ocamli} programs, and removing it from the measured time.}
\item \addition{Dividing through to calculate a ratio showing how much slower each \textsf{OCamli} program is that its OCaml equivalent.}
\end{enumerate}
\bigskip

\noindent \addition{Here are the results:}

\vspace{10mm}

\addition{
\noindent\begin{tabular}{@{}llllll}\toprule
Program & OCaml n & OCaml t & \textsf{OCamli} n & \textsf{OCamli} t & Ratio\\\midrule
\texttt{reference\_swap.ml} & 1,000,000,000 & 0.045 & 1000 & 0.986 & 21911111\\
\texttt{table.ml} & 25,000,000 & 3.603 & 10000 & 22.471 & 15591 \\
\texttt{factorial.ml} & 5,000,000 & 0.006 & 4000 & 84.468 & 17597500 \\
\texttt{factorialacc.ml} & 5,000,000,000 & 7.793 & 4000 & 3.628 & 581932 \\
\texttt{helloworld.ml} & 50,000,000 & 0.743 & 5000 & 4.22 & 56796 \\
\texttt{exception.ml} & 5,000,000 & 0.124 & 50000 & 0.653 & 526 \\
\texttt{tree.ml} & 10000 & 0.104 & 1 & 4.171 & 382596 \\\bottomrule
\end{tabular}}

\vspace{10mm}

\noindent\addition{The first observation is, of course, that the ratios are enormous. In particular, for the {\small\texttt{reference\_swap.ml}} and recursive {\small\texttt{factorial.ml}} programs. In the case of the program {\small\texttt{reference\_swap.ml}} this is due to the simple fact that the execution of the compiled version reduces to very simple memory accesses, where the \textsf{OCamli} interpreted execution is creating and modifying compound data structures (those representing the references). It is worth noting also the contrast between the recursive and iterative versions of the factorial calculation: the recursive one generates a large intermediate expression which not only affects memory usage (as it would in the compiled version) but also affects speed since \textsf{OCamli}'s step-by-step nature means finding the redex becomes quadratic.}

\addition{Moving on to space usage, then, may give us further insight. In this test, we again have to remove a baseline, the space usage of the {\small\texttt{donothing.ml}} program in both \textsf{OCaml} and \textsf{OCamli}. This time, however, we can use equal values of n for both \textsf{OCaml} and \textsf{OCamli}. There is a complication: the measurement is maximum memory usage, which of course depends on how the garbage collector runs. In the future we might like to find a way of measuring the number of bytes allocated by the program, rather than just its highest memory usage. Here are the results, all space numbers in bytes:}

\vspace{10mm}

\addition{\noindent\begin{tabular}{@{}llll}\toprule
Program & n & OCaml excess & OCamli excess\\\midrule
\texttt{reference\_swap.ml} & 1000 & 0 & 10,600,448\\
\texttt{table.ml} & 10000 & 212,992 & 337,518,592\\
\texttt{factorial.ml} & 4000 & 45,056 & 920,727,552\\
\texttt{factorialacc.ml} & 4000 & 0 & 13,910,016\\
\texttt{helloworld.ml} & 5000 & 65,536 & 50,286,592\\
\texttt{exception.ml} & 50000 & 0 & 8,986,624\\
\texttt{tree.ml} & 1 & 20,480 & 54,091,776 \\\bottomrule
\end{tabular}}

\vspace{10mm}

\noindent\addition{We can see that there are several programs which do not appear to allocate any memory in the compiled case (in fact, this just means that they did not need to allocate a page, they may allocate a token amount of memory). In each of these three cases ({\small\texttt{reference\_swap.ml}}, {\small\texttt{factorialacc.ml}}, and {\small\texttt{exception.ml}}), \textsf{OCamli} needs significant extra memory. The worst offenders, however, are the recursive factorial program, and the times-table printer. In the case of the times-table printer this is to do with interpreting the insides of the Standard Library {\small\texttt{print\_string}} function. As we have mentioned, a future \textsf{OCamli2} need not do this: it will be able to use the function directly. We have illustrated the evaluation of {\small\texttt{print\_string}} in Figure 9. In the case of the recursive factorial, the space used is for the expression itself, and for its repeated modification by the step-by-step interpreter.}

\begin{sidewaysfigure}
\begin{landscape}
\scalebox{0.65}{
\begin{minipage}{3\textwidth}
{\ttfamily
\addition{~~~print\_string "Hello, World$\textbackslash\negthinspace\negthinspace\negthinspace$ n"}

\addition{=>  \textbf{let} string\_length a = <<\%string\_length x>> \textbf{in let} unsafe\_output\_string a b c d = <<caml\_ml\_output x>> \textbf{in}}

\addition{~~~~~\textbf{let} output\_string oc s = unsafe\_output\_string oc s 0 (string\_length s) \textbf{in let} s = "Hello, World!"$\textbackslash\negthinspace\negthinspace\negthinspace$ n" \textbf{in} output\_string <out\_channel> s}

\addition{=>  \textbf{let} string\_length a = <<\%string\_length x>> \textbf{in let} unsafe\_output\_string a b c d = <<caml\_ml\_output x>> \textbf{in}}

\addition{~~~~~\textbf{let} output\_string oc s = unsafe\_output\_string oc s 0 (string\_length s) \textbf{in let} s = "Hello, World!"$\textbackslash\negthinspace\negthinspace\negthinspace$ n" \textbf{in} output\_string <out\_channel> s}

\addition{=>  \textbf{let} s = "Hello, World!$\textbackslash\negthinspace\negthinspace\negthinspace$ n" \textbf{in} (\textbf{let} string\_length a = <<\%string\_length x>> \textbf{in}}

\addition{~~~~~\textbf{let} unsafe\_output\_string a b c d = <<caml\_ml\_output x>> \textbf{in let} oc = <out\_channel> \textbf{in fun} s -> unsafe\_output\_string oc s 0 (string\_length s)) s}

\addition{=>  \textbf{let} s = "Hello, World!$\textbackslash\negthinspace\negthinspace\negthinspace$ n" \textbf{in} (\textbf{let} string\_length a = <<\%string\_length x>> \textbf{in}}

\addition{~~~~~\textbf{let} unsafe\_output\_string a b c d = <<caml\_ml\_output x>> \textbf{in fun} s -> \textbf{let} oc = <out\_channel> \textbf{in} unsafe\_output\_string oc s 0 (string\_length s)) s}

\addition{=>  \textbf{let} s = "Hello, World!$\textbackslash\negthinspace\negthinspace\negthinspace$ n" \textbf{in} (\textbf{let} string\_length a = <<\%string\_length x>> \textbf{in}}

\addition{~~~~~\textbf{fun} s -> \textbf{let} unsafe\_output\_string a b c d = <<caml\_ml\_output x>> \textbf{in let} oc = <out\_channel> \textbf{in} unsafe\_output\_string oc s 0 (string\_length s)) s}

\addition{=>  \textbf{let} s = "Hello, World!$\textbackslash\negthinspace\negthinspace\negthinspace$ n" \textbf{in} (\textbf{fun} s -> \textbf{let} string\_length a = <<\%string\_length x>> \textbf{in}} 

\addition{~~~~~\textbf{let} unsafe\_output\_string a b c d = <<caml\_ml\_output x>> \textbf{in let} oc = <out\_channel> \textbf{in} unsafe\_output\_string oc s 0 (string\_length s)) s}

\addition{=>  (\textbf{fun} s -> \textbf{let} string\_length a = <<\%string\_length x>> \textbf{in} \textbf{let} unsafe\_output\_string a b c d = <<caml\_ml\_output x>> \textbf{in}}

\addition{~~~~~\textbf{let} oc = <out\_channel> \textbf{in} unsafe\_output\_string oc s 0 (string\_length s)) "Hello, World!$\textbackslash\negthinspace\negthinspace\negthinspace$ n"}

\addition{=>  \textbf{let} s = "Hello, World!$\textbackslash\negthinspace\negthinspace\negthinspace$ n" \textbf{in let} string\_length a = <<\%string\_length x>> \textbf{in}}

\addition{~~~~~\textbf{let} unsafe\_output\_string a b c d = <<caml\_ml\_output x>> \textbf{in let} oc = <out\_channel> \textbf{in} unsafe\_output\_string oc s 0 (string\_length s)}

\addition{=>  \textbf{let} s = "Hello, World!$\textbackslash\negthinspace\negthinspace\negthinspace$ n" \textbf{in let} string\_length a = <<\%string\_length x>> \textbf{in}}

\addition{~~~~~\textbf{let} unsafe\_output\_string a b c d = <<caml\_ml\_output x>> \textbf{in} unsafe\_output\_string <out\_channel> s 0 (string\_length s)}

\addition{=>  \textbf{let} s = "Hello, World!$\textbackslash\negthinspace\negthinspace\negthinspace$ n" \textbf{in let} string\_length a = <<\%string\_length x>> \textbf{in} (\textbf{let} a = <out\_channel> \textbf{in fun} b c d -> <<caml\_ml\_output x>>) s 0 (string\_length s)}

\addition{=>  \textbf{let} s = "Hello, World!$\textbackslash\negthinspace\negthinspace\negthinspace$ n" \textbf{in let} string\_length a = <<\%string\_length x>> \textbf{in}}

\addition{~~~~~(\textbf{fun} b -> \textbf{let} a = <out\_channel> \textbf{in fun} c d -> <<caml\_ml\_output x>>) "Hello, World!$\textbackslash\negthinspace\negthinspace\negthinspace$ n" 0 (string\_length s)}

\addition{=>  \textbf{let} s = "Hello, World!$\textbackslash\negthinspace\negthinspace\negthinspace$ n" \textbf{in let} string\_length a = <<\%string\_length x>> \textbf{in}}

\addition{~~~~~(\textbf{let} b = "Hello, World!$\textbackslash\negthinspace\negthinspace\negthinspace$ n" \textbf{in let} a = <out\_channel> \textbf{in fun} c d -> <<caml\_ml\_output x>>) 0 (string\_length s)}

\addition{=>  \textbf{let} s = "Hello, World!$\textbackslash\negthinspace\negthinspace\negthinspace$ n" \textbf{in} \textbf{let} string\_length a = <<\%string\_length x>> \textbf{in}}

\addition{~~~~~(\textbf{let} c = 0 \textbf{in let} b = "Hello, World!$\textbackslash\negthinspace\negthinspace\negthinspace$ n" \textbf{in let} a = <out\_channel> \textbf{in fun} d -> <<caml\_ml\_output x>>) (string\_length s)}

\addition{=>  \textbf{let} string\_length a = <<\%string\_length x>> \textbf{in}}

\addition{~~~~~(\textbf{fun} d -> \textbf{let} c = 0 \textbf{in let} b = "Hello, World!$\textbackslash\negthinspace\negthinspace\negthinspace$ n" \textbf{in let} a = <out\_channel> \textbf{in} <<caml\_ml\_output x>>) (string\_length "Hello, World!$\textbackslash\negthinspace\negthinspace\negthinspace$ n")}

\addition{=>  (\textbf{fun} d -> \textbf{let} c = 0 \textbf{in let} b = "Hello, World!$\textbackslash\negthinspace\negthinspace\negthinspace$ n" \textbf{in let} a = <out\_channel>\textbf{in} <<caml\_ml\_output x>>) (\textbf{let} a = "Hello, World!$\textbackslash\negthinspace\negthinspace\negthinspace$ n" \textbf{in} <<\%string\_length x>>)}

\addition{=>  (\textbf{fun} d -> \textbf{let} c = 0 \textbf{in let} b = "Hello, World!$\textbackslash\negthinspace\negthinspace\negthinspace$ n" \textbf{in let} a = <out\_channel> \textbf{in} <<caml\_ml\_output x>>) 14}

\addition{=>  \textbf{let} d = 14 \textbf{in let} c = 0 \textbf{in let} b = "Hello, World!$\textbackslash\negthinspace\negthinspace\negthinspace$ n" \textbf{in let} a = <out\_channel> \textbf{in} <<caml\_ml\_output x>>}

\addition{Hello, World!}

\addition{=>  ()}\par}
\bigskip
\bigskip
\bigskip

\end{minipage}}
\bigskip

\label{sideways}
\singlespacing\noindent \addition{\label{F9}Figure 9. A simple program to print to the screen, which results in a lengthy trace, due to \textsf{OCamli}'s habit of interpreting inside Standard Library functions. Such intricacy is almost never needed by the user.}\end{landscape}\end{sidewaysfigure}

\onehalfspacing

\noindent  \addition{\subsection*{Growth in time and space}}

\noindent\addition{We have considered the time and space requirements of programs interpreted with \textsf{OCamli} and \textsf{OCaml}. However, this misses out an important part of the story: the question whether the performance of a given program with regard to time and space scales in the same way when interpreted with \textsf{OCamli} rather than compiled with \textsf{OCaml}. Take, for example, our recursive and iterative {\small\texttt{factorial.ml}} and {\small\texttt{factorialacc.ml}} programs. When compiled with \textsf{OCaml}, they are both linear in time. The recursive {\small\texttt{factorial.ml}} is linear also in space, but {\small\texttt{factorialacc.ml}} is constant in space. In Figure 10, the top and middle graphs show the time growth of the two programs in \textsf{OCamli}: the time behaviour of {\small\texttt{factorial.ml}} becomes polynomial, due to the repeated traversing of an increasingly large expression to find the redex at each step and other similar processing. In the case of {\small\texttt{factorialacc.ml}} the linear behaviour is however retained. The bottom graph shows a comparison of the space usage of {\small\texttt{factorial.ml}} and {\small\texttt{factorialacc.ml}} in \textsf{OCamli}. The recursive case is clearly polynomial, compared with the linear behaviour of the same program compiled with \textsf{OCaml}. It remains to be seen how tractable the improvement of these space and time behaviours will be as we improve our implementation.}

\begin{minipage}{\textwidth}
\begin{center}
\includegraphics[width=10cm]{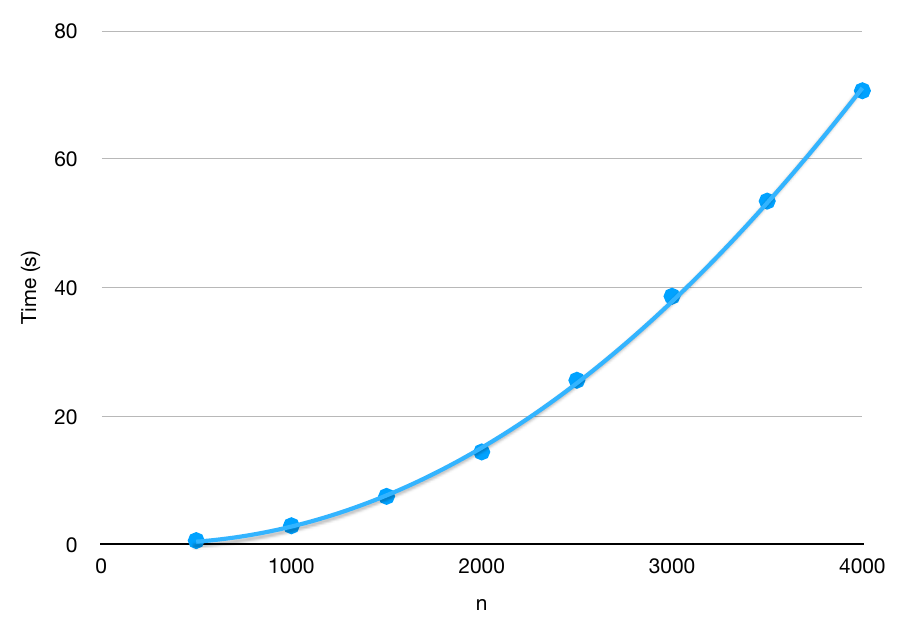}

\includegraphics[width=10cm]{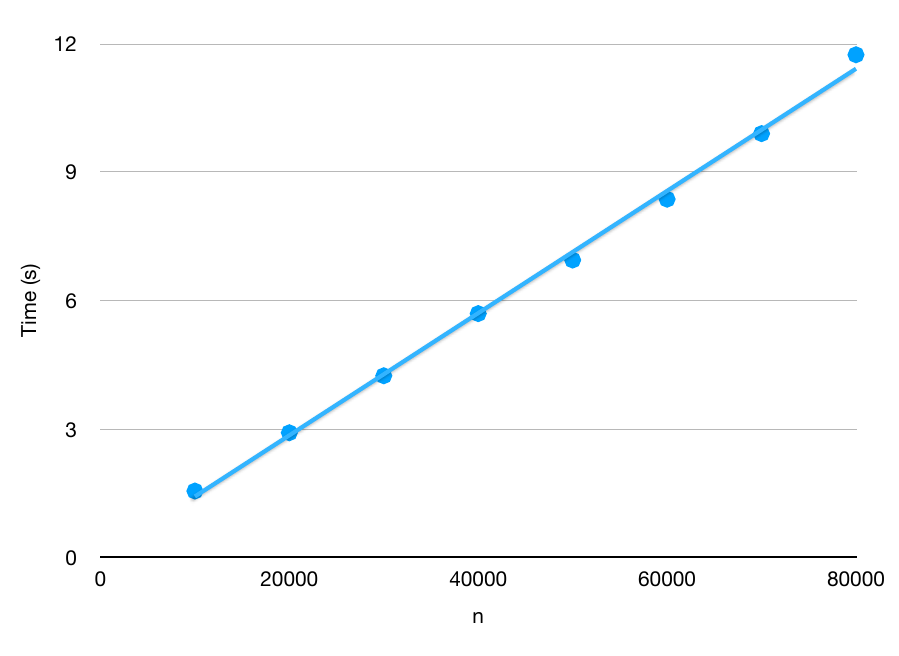}

\includegraphics[width=10cm]{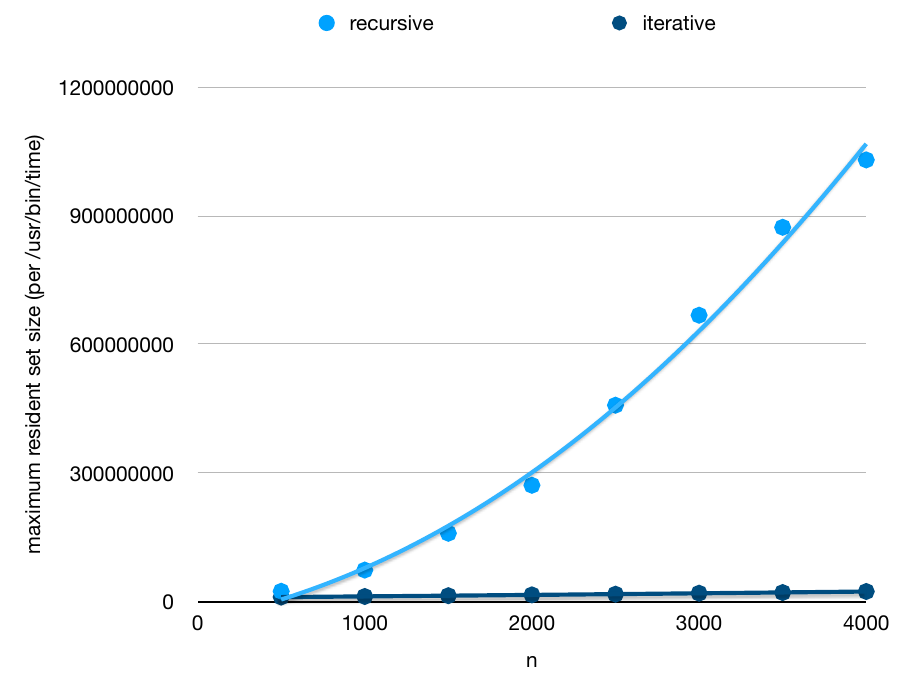}
\end{center}

\label{F10}\addition{Figure 10. Time and space growth. Top: the growth of time usage in {\small\texttt{factorial.ml}}; middle: the growth of time usage in {\small\texttt{factorialacc.ml}}; bottom: space usage for the recursive {\small\texttt{factorial.ml}} and iterative {\small\texttt{factorialacc.ml}} compared.}
\clearpage

\end{minipage}

\addition{We have mentioned that the cost of printing can be significant. There are two reasons: the job of creating the string to be printed for each step and preparing for such by eliding information; and the fact that the printed representation of the step can scale in size by a factor of $O(n)$ turning, for example, an algorithm of $O(n)$ into one of $O(n^2)$ when its steps are printed. Concretely, when testing the {\small\texttt{factorial.ml}} and {\small\texttt{factorialacc.ml}} programs, printing slows the former by about three times, and the latter by less than two times. As the interpreter becomes more efficient, we might expect this multiple to increase, since the printing will consume more of the overall time budget.}\par

\bigskip

\subsection*{Discussion}

These sorts of magnitudes are similar to those encountered elsewhere -- in reviewing the \index{C interpreter}Ch system \cite{ch}, which is a mature C interpreter (and which does not print out the steps of execution) Huber says \textit{``In running several simple benchmark programs, code executes at about 1/1,000 the speed of compiled C code.''} \cite{huber}. So this is perhaps the best we can hope for with pure interpretation. In fact, we might expect an OCaml interpreter to be slower due to the higher-level nature of the language.

We can consider the efficiency of a given step-by-step, prettyprinting interpreter in two fairly standard ways. Ideally, we should like an interpreter whose time and space behaviour is, modulo a constant factor, the same as that of the program compiled by the OCaml bytecode or native code compiler and executed in the usual fashion, that is when it prints none of the steps. A corollary of the space requirement is that such an `efficient' interpreter must have the same general behaviour with regard to tail recursion (though it may cause a stack overflow earlier or later than a compiled version of the same program). In reality, some functionality of step-by-step evaluation (or, indeed the idea of step-by-step evaluation itself) may preclude this goal, and we may have to settle for some small increase in algorithmic complexity in some situations. Of course, when we start prettyprinting most or all the steps of the evaluation, time-efficiency becomes moot -- the act of printing the line itself probably multiplies the complexity of the program in question by at least $O(n)$. And second, of course, we should simply like the interpreter to be as fast as possible in absolute terms.

\addition{There is an OCaml benchmarking suite in development \cite{sandmark}, which we would hope to modify to support \textsf{OCamli}, leading to a lower friction way of keeping time and space benchmarks a part of our development cycle.}

\section{Example}

Many of the improvements of \textsf{OCamli2} over \textsf{OCamli} discussed in this chapter are about making the implementation simpler and more robust, not about visualization itself. However, some of these technical decisions (such as not interpreting the Standard Library) have a bearing on the visualization itself. Consider the following output of the original \textsf{OCamli} on the program {\small\texttt{List.map\! (\textbf{function}\! x\! ->\! x\! +\! 1)\! [1; 2]}}:

\begin{verbatim}[commandchars=\\\{\}]
$ ./ocamli -e 'List.map (fun x -> x + 1) [1; 2]' -show-all -remove-rec-all
    \underline{List.map (\textbf{fun} x -> x + 1)} [1; 2]
=>  \underline{(\textbf{let} f x = x + 1 \textbf{in function} [] -> [] | a::l -> \textbf{let} r = f a \textbf{in} r::map f l)} [1; 2]
=>  \underline{(\textbf{function} [] -> [] | a::l -> \textbf{let} f x = x + 1 \textbf{in let} r = f a \textbf{in} r::map f l)} [1; 2]
=>  \underline{(\textbf{function} a::l -> \textbf{let} f x = x + 1 \textbf{in} \textbf{let} r = f a \textbf{in} r::map f l) [1; 2]}
=>  \textbf{let} l = [2] \textbf{in} \textbf{let} a = 1 \textbf{in} \textbf{let} f x = x + 1 \textbf{in} \textbf{let} r = f \underline{a} \textbf{in} r::map f l
=>  \textbf{let} l = [2] \textbf{in} \textbf{let} f x = x + 1 \textbf{in} \textbf{let} r = \underline{f 1} \textbf{in} r::map f l
=>  \textbf{let} l = [2] \textbf{in} \textbf{let} f x = x + 1 \textbf{in} \textbf{let} r = \textbf{let} x = 1 \textbf{in} \underline{x + 1} \textbf{in} r::map f l
=>  \textbf{let} l = [2] \textbf{in} \textbf{let} f x = x + 1 \textbf{in} \textbf{let} r = \underline{1 + 1} \textbf{in} r::map f l
=>  \textbf{let} l = [2] \textbf{in} \textbf{let} f x = x + 1 \textbf{in} \textbf{let} r = 2 \textbf{in} \underline{r}::map f l
=>  \textbf{let} l = [2] \textbf{in} \textbf{let} f x = x + 1 \textbf{in} 2::map \underline{f} l
=>  \textbf{let} l = [2] \textbf{in} 2::\underline{map (\textbf{fun} x -> x + 1)} l
=>  \textbf{let} l = [2] \textbf{in}
      2::\underline{(\textbf{let} f x = x + 1 \textbf{in} \textbf{function} [] -> [] | a::l -> \textbf{let} r = f a \textbf{in} r::map f l)} l
=>  \textbf{let} l = [2] \textbf{in}
      2::\underline{(\textbf{function} [] -> [] | a::l -> \textbf{let} f x = x + 1 \textbf{in} \textbf{let} r = f a \textbf{in} r::map f l)} l
=>  2::\underline{(\textbf{function} [] -> [] | a::l -> \textbf{let} f x = x + 1 \textbf{in} \textbf{let} r = f a \textbf{in} r::map f l) [2]}
=>  2::\underline{(\textbf{function} a::l -> \textbf{let} f x = x + 1 \textbf{in} \textbf{let} r = f a \textbf{in} r::map f l) [2]}
=>  2::\textbf{let} l = [] \textbf{in} \textbf{let} a = 2 \textbf{in} \textbf{let} f x = x + 1 \textbf{in} \textbf{let} r = f \underline{a} \textbf{in} r::map f l
=>  2::\textbf{let} l = [] \textbf{in} \textbf{let} f x = x + 1 \textbf{in} \textbf{let} r = \underline{f 2} \textbf{in} r::map f l
=>  2::\textbf{let} l = [] \textbf{in} \textbf{let} f x = x + 1 \textbf{in} \textbf{let} r = \textbf{let} x = 2 \textbf{in} \underline{x + 1} \textbf{in} r::map f l
=>  2::\textbf{let} l = [] \textbf{in} \textbf{let} f x = x + 1 \textbf{in} \textbf{let} r = \underline{2 + 1} \textbf{in} r::map f l
=>  2::\textbf{let} l = [] \textbf{in} \textbf{let} f x = x + 1 \textbf{in} \textbf{let} r = 3 \textbf{in} \underline{r}::map f l
=>  2::\textbf{let} l = [] \textbf{in} \textbf{let} f x = x + 1 \textbf{in} 3::map \underline{f} l
=>  2::\textbf{let} l = [] \textbf{in} 3::\underline{map (\textbf{fun} x -> x + 1)} l
=>  2::\textbf{let} l = [] \textbf{in}
      3::\underline{(\textbf{let} f x = x + 1 \textbf{in} \textbf{function} [] -> [] | a::l -> \textbf{let} r = f a \textbf{in} r::map f l)} l
=>  2::\textbf{let} l = [] \textbf{in}
      3::\underline{(\textbf{function} [] -> [] | a::l -> \textbf{let} f x = x + 1 \textbf{in} \textbf{let} r = f a \textbf{in} r::map f l)} l
=>  2::3::\underline{(\textbf{function} [] -> [] | a::l -> \textbf{let} f x = x + 1 \textbf{in} \textbf{let} r = f a \textbf{in} r::map f l) []}
=>  [2; 3]
\end{verbatim}

\noindent Without the extra command line option {\small\texttt{-remove-rec-all}} it would be longer still. Here, in contrast, is the default output from \textsf{OCamli2} on the same program.

\begin{verbatim}[commandchars=\\\{\}]
$./ocamli2 -e 'List.map (fun x -> x + 1) [1; 2]'
   \underline{List.map (\textbf{function} x -> x + 1) [1; 2]}
\{entering List.map\}
\{entering \textbf{function} x -> x + 1\} 
=> (\textbf{function} x -> x + 1) \underline{x}
=> \underline{(\textbf{function} x -> x + 1) 1}
=> \textbf{let} x = 1 \textbf{in} \underline{x} + 1
=> \underline{1 + 1}
=> 2
\{leaving \textbf{function} x -> x + 1\}
\{entering \textbf{function} x -> x + 1\}
=> (\textbf{function} x -> x + 1) \underline{x}
=> \underline{(\textbf{function} x -> x + 1) 2}
=> \textbf{let} x = 2 \textbf{in} \underline{x} + 1
=> \underline{2 + 1}
=> 3
\{leaving \textbf{function} x -> x + 1\}
\{leaving List.map\}
=> [2; 3]
\end{verbatim}

\noindent Much of the reduction comes from not tracing the insides of the Standard Library function {\small\texttt{List.map}}, just its callbacks to the interpreted function {\small\texttt{\textbf{function}\! x\! ->\! x\! +\! 1}}. Until we have a fuller implementation of our new, more efficient interpreter, we will not know where the balance lies between detail and conciseness in debugger output -- this will require returning to our work on visualization itself.

\section{Summary}

Reflecting on our original implementation of \textsf{OCamli}, we have addressed a number of issues which were put aside during the rush to rapid prototyping. This further exploration gives us some confidence that a ``real world'' implementation of \textsf{OCamli2} is tractable. \addition{The state of the implementation of this work is given in section \ref{stateimpl}}.

In \cref{chap:misc}, we pick over another kind of discard. But first, we shall discuss a practical interface for our debugger.

\chapter{An interface for debugging}
\label{chap:interface}

\begin{quotation}\textit{\large The customer is never wrong.\textrm{\begin{flushright}--- \textup{C\'esar Ritz}\end{flushright}}}\end{quotation}

\vspace{10pt}

\noindent (Or, \textit{``if people don't use your debugger, it's not their fault''}).  In \cref{chap:interpreter} we described the OCaml PPX rewriting mechanism and used it to simplify the \textsf{OCamli} Standard Library. In this chapter, we suggest a possible interface for interpretive debugging based upon the same kind of PPX annotations.
Then we show a failed attempt to build such a system for the original \textsf{OCamli}. It was the out-of-hand complexity leading to this failure which alerted us to the need to redesign from the ground up. We shall see how the new design of \textsf{OCamli2} allows for a much simpler implementation of such a system, leading us to believe that our approach is now technically feasible.

\label{decision6}\addition{The interface described in this chapter was not chosen as part of a formal process, comparing and ranking possible designs. It was simply an idea which occurred one day. And, as we shall see, it seems to be a very natural and almost beautiful concept. In the future, we would wish to consider other designs for and compare and contrast them. Let us look at it now.}

\section{Choosing what to interpret}

We have already seen how annotations can be used to invoke \index{PPX}PPX functionality in OCaml programs. What if we simply put an {\small\texttt{[@interpret]}} \index{PPX!annotation}annotation on any part of the code we wish to have interpreted step by step, and write a PPX extension to facilitate this? All other (unannotated) code in the source file (and indeed the rest of the program) would be natively compiled as usual, \label{annotationspeed}\addition{running at full speed despite the annotations}.

The use of {\small\texttt{[@interpret]}} annotations to control which parts of the code are executed natively and which parts are interpreted (and so have their steps of evaluation displayed on the screen) is motivated by our observation that a tool like \textsf{OCamli} on its own would not fulfil our usability or accessibility needs, in particular  our requirement that we must get it `inside' the build process. The improvement in speed, by interpreting only what we need to debug, is a side effect -- but a pleasant one. Is another such side effect of this mechanism a natural and pliable human interface for debugging? If it is, we may be within sight of achieving our original aim of a usable debugger. Here is the debugger interface we envisage:

\begin{enumerate}
  \item Notice that a misbehaviour is occurring.
  \item Knowing or speculating upon the location of the root cause, insert one or more appropriate {\small\texttt{[@interpret]}} annotations in the code.
  \item Recompile and run the program. The evaluation of the parts chosen will be shown.
  \item If the source or nature of the bug is now clear:
   \begin{enumerate}
     \item Change the source to fix the bug.
     \item Build and run again and inspect the output to be sure it is fixed.
     \item Remove the {\small\texttt{[@interpret]}} annotation(s).
   \end{enumerate}
  \item If the source or nature of the bug is not yet clear, due to a wrong or insufficient choice of {\small\texttt{[@interpret]}} annotations, return to step 2.
\end{enumerate}

\noindent Let us explore the design of such debugging annotations. It will not be until we have implemented some of these (or until they have been used in anger) that we will be able to know if they are the best choices. But consider the following possibilities:

\begin{description}

\item [{\small\texttt{\textmd{[@interpret]}}}] The piece of code annotated is interpreted, but functions it calls into are not. Consider the following buggy function on lists:

\begin{Verbatim}[commandchars=\\\{\}]
\textbf{let rec} pairs f a l =
  \textbf{match} l \textbf{with}
    [] -> rev a
  | [_] -> []
  | h::h'::t -> pairs f (f h h'::a) t

\textbf{let} x = pairs ( + ) [] [1; 2; 3; 4]
\end{Verbatim}

\noindent It is supposed to take, for example, {\small\texttt{[1;\! 2;\! 3;\! 4]}} to {\small\texttt{[1\! +\! 2;\! 2\! +\! 3;\! 3 +\! 4]}} if the function {\small\texttt{f}} is addition. The argument {\small\texttt{a}} is an accumulator to make the function tail-recursive. There are two bugs in this program. First, the final case should read {\small\texttt{pairs\! f\! (f\! h\! h\textquotesingle{}::a)\! (h\textquotesingle{}::t)}}. Second, for the case of the single-item list, the result should be {\small\texttt{rev\! a}}, as for the empty list.  We can add an {\small\texttt{[@interpret]}} annotation to the outer invocation of {\small\texttt{pairs}}:

\begin{Verbatim}[commandchars=\\\{\}]
\textbf{let rec} pairs f a l =
  \textbf{match} l \textbf{with}
    [] -> rev a
  | [_] -> []
  | h::h'::t -> pairs f (f h h'::a) t

\textbf{let} x = pairs ( + ) [] [1; 2; 3; 4] [@interpret]
\end{Verbatim}

\noindent Now, upon compiling the program the interpreter is embedded, and calls to {\small\texttt{pairs}} (but not the insides of {\small\texttt{pairs}}) are shown on screen. We think this is the sensible default, both for elision of information and elision of computation. The output upon running the program would be the following:

\begin{Verbatim}[commandchars=\\\{\}]
   pairs ( + ) [] [1; 2; 3; 4]
=> pairs ( + ) [3] [3; 4]
=> pairs ( + ) [7; 3] []
=> [3; 7]
\end{Verbatim}

    \noindent The first three lines are generated from the pairs function call itself, the last line from the returned value.

Note that the default elision also does not show as much detail as the output of \textsf{OCamli} we showed in \cref{chap:interpreter}. We could annotate the body of the function {\small\texttt{pairs}} too, in order to show the step-by-step execution of the recursive parts. For now, we will just add a simpler {\small\texttt{[@showmatch]}} annotation to show which case matches each time, without interpretation:

\begin{Verbatim}[commandchars=\\\{\}]
\textbf{let rec} pairs f a l =
  \textbf{match} l [@showmatch] \textbf{with}
    [] -> rev a
  | [_] -> []
  | h::h'::t -> pairs f (f h h'::a) t

\textbf{let} x = pairs ( + ) [] [1; 2; 3; 4] [@interpret]
\end{Verbatim}

\noindent This is a more pleasing output:

\begin{Verbatim}[commandchars=\\\{\}]
   pairs ( + ) [] [1; 2; 3; 4]
\{matches h::h'::t\}
=> pairs ( + ) [3] [3; 4]
\{matches h::h'::t\}
=> pairs ( + ) [7; 3] []
\{matches []\}
=> [3; 7]
\end{Verbatim}

\noindent The bug is plain to see -- the list is being reduced in size by two each time not one -- so we correct it, replacing {\small\texttt{t}} with {\small\texttt{(h\textquotesingle{}::t)}} in the final case of the pattern match:

\begin{Verbatim}[commandchars=\\\{\}]
\textbf{let rec} pairs f a l =
  \textbf{match} l [@showmatch] \textbf{with}
    [] -> rev a
  | [_] -> []
  | h::h'::t -> pairs f (f h h\textquotesingle{}::a) (h\textquotesingle{}::t)

\textbf{let} x = pairs ( + ) [] [1; 2; 3; 4] [@interpret] 
\end{Verbatim}

\noindent We compile the code again, with the annotations in the same places, and try again:

\begin{Verbatim}[commandchars=\\\{\}]
   pairs ( + ) [] [1; 2; 3; 4]
\{matches h::h\textquotesingle{}::t\}
=> pairs ( + ) [3] [2; 3; 4]
\{matches h::h\textquotesingle{}::t\}
=> pairs ( + ) [5; 3] [3; 4]
\{matches h::h\textquotesingle{}::t\}
=> pairs ( + ) [7; 5; 3] [4]
\{matches [_]\}
=> []
\end{Verbatim}

\noindent Still there is a bug. Since the accumulator looks correct during evaluation, only for the output to disappear at the last moment, we deduce it must be the match case \texttt{[\_]} which is wrong, and we correct it:

\begin{Verbatim}[commandchars=\\\{\}]
\textbf{let rec} pairs f a l =
  \textbf{match} l [@showmatch] \textbf{with}
    [] | [_] -> rev a
  | h::h'::t -> pairs f (f h h'::a) (h'::t)

\textbf{let} x = pairs ( + ) [] [1; 2; 3; 4] [@interpret]
\end{Verbatim}

\noindent Here is the final, correct output:

\begin{Verbatim}[commandchars=\\\{\}]
   pairs ( + ) [] [1; 2; 3; 4]
\{matches h::h\textquotesingle{}::t\}
=> pairs ( + ) [3] [2; 3; 4]
\{matches h::h\textquotesingle{}::t\}
=> pairs ( + ) [5; 3] [3; 4]
\{matches h::h\textquotesingle{}::t\}
=> pairs ( + ) [7; 5; 3] [4]
\{matches [_]\}
=> [3; 5; 7]
\end{Verbatim}

\noindent Now, we may remove our annotations:

\begin{Verbatim}[commandchars=\\\{\}]
\textbf{let rec} pairs f a l =
  \textbf{match} l \textbf{with}
    [] | [_] -> rev a
  | h::h'::t -> pairs f (f h h\textquotesingle{}::a) (h\textquotesingle{}::t)

\textbf{let} x = pairs ( + ) [] [1; 2; 3; 4]
\end{Verbatim}

\noindent Our debugging is complete. As a matter of style, we notice now, of course, the two match cases having been coalesced, their order may be reversed to reduce their number:

\begin{Verbatim}[commandchars=\\\{\}]
\textbf{let rec} pairs f a l =
  \textbf{match} l \textbf{with}
    h::h\textquotesingle{}::t -> pairs f (f h h\textquotesingle{}::a) (h\textquotesingle{}::t)
  | _ -> rev a

\textbf{let} x = pairs ( + ) [] [1; 2; 3; 4]
\end{Verbatim}

\noindent Now, the method can be summarised as ``If there are two numbers left to process them, do so and remember the result, otherwise we can do no more, and we return the result.''

\item [{\small\texttt{\textmd{[@interpret-deep <level>]}}}] The piece of code annotated is interpreted, and so is every function it calls, if in the same module. Recall our example program:

\begin{Verbatim}[commandchars=\\\{\}]
\textbf{let rec} pairs f a l =
  \textbf{match} l [@showmatch] \textbf{with}
    h::h\textquotesingle{}::t -> pairs f (f h h\textquotesingle{}::a) (h\textquotesingle{}::t)
  | _ -> rev a

\textbf{let} x = pairs ( + ) [] [1; 2; 3; 4] [@interpret-deep 1]
\end{Verbatim}

\noindent When using {\small\texttt{[@interpret-deep <level>]}}, calls to {\small\texttt{rev}} would also be treated as if they had an {\small\texttt{[@interpret]}} annotation attached to them, if the level is 1, and calls from {\small\texttt{rev}} to if the level is 2 and so on. For example, take a buggy definition of {\small\texttt{rev}}, using {\small\texttt{[@interpret-deep 1]}}, as follows:

\begin{Verbatim}[commandchars=\\\{\}]
\textbf{let rec} rev_inner a l =
  \textbf{match} l \textbf{with}
    [] -> []
  | h::t -> rev_inner (h::a) l

\textbf{let} rev = rev_inner []
\end{Verbatim}

\noindent We might see:

\begin{Verbatim}[commandchars=\\\{\}]
   pairs ( + ) [] [1; 2; 3; 4]
\{matches h::h'::t\}
=> pairs ( + ) [3] [2; 3; 4]
\{matches h::h'::t\}
=> pairs ( + ) [5; 3] [3; 4]
\{matches h::h'::t\}
=> pairs ( + ) [7; 5; 3] [4]
\{matches [_]\}
=> rev [7; 5; 3]
\{return from rev\}
=> []
\{return from pairs\}
=> []
\end{Verbatim}

\noindent This is not useful if the bug is in fact in {\small\texttt{rev}}, since {\small\texttt{rev}} immediately calls {\small\texttt{rev\_inner}} which is one level deeper, so we increase the level by one and write {\small\texttt{[@interpret-deep 2]}}. We will then see this:

\begin{Verbatim}[commandchars=\\\{\}]
   pairs ( + ) [] [1; 2; 3; 4]
\{matches h::h\textquotesingle{}::t\}
=> pairs ( + ) [3] [2; 3; 4]
\{matches h::h\textquotesingle{}::t\}
=> pairs ( + ) [5; 3] [3; 4]
\{matches h::h\textquotesingle{}::t\}
=> pairs ( + ) [7; 5; 3] [4]
\{matches [_]\}
=> rev [7; 5; 3]
=> rev_inner [] [7; 5; 3]
\{matches h::t\}
=> rev_inner [7] [5; 3]
=> rev_inner [5; 7] [3]
=> rev_inner [3; 5; 7] []
\{return from rev_inner\}
=> []
\{return from rev\}
=> []
\{return from pairs\}
=> []
\end{Verbatim}

\noindent Now we can see the source of the bug. This sort of interactive deepening of the search space for a bug allows us to begin with small manageable traces, and explore lower-level code only when required. It is especially useful in the case of well-used libraries, which are unlikely to be the source of a bug. If the trace becomes overwhelming, the annotation may be moved from the {\small\texttt{pairs}} function to the {\small\texttt{rev}} function, now that the programmer knows that it is {\small\texttt{rev}} which is at fault. We correct the code, check by running it again to yield:

\begin{Verbatim}[commandchars=\\\{\}]
   pairs ( + ) [] [1; 2; 3; 4]
\{matches h::h\textquotesingle{}::t\}
=> pairs ( + ) [3] [2; 3; 4]
\{matches h::h\textquotesingle{}::t\}
=> pairs ( + ) [5; 3] [3; 4]
\{matches h::h\textquotesingle{}::t\}
=> pairs ( + ) [7; 5; 3] [4]
\{matches [_]\}
=> rev [7; 5; 3]
=> rev_inner [] [7; 5; 3]
\{matches h::t\}
=> rev_inner [7] [5; 3]
=> rev_inner [5; 7] [3]
=> rev_inner [3; 5; 7] []
\{return from rev_inner\}
=> [3; 5; 7]
\{return from rev\}
=> [3; 5; 7]
\{return from pairs\}
=> [3; 5; 7]
\end{Verbatim}

\noindent The astute reader will notice that in this example, much of the effect is similar to simply inserting a print statement at each recursive call, and we are not really using the ability to show the fine-grained step-by-step execution of an expression. Nonetheless, the {\small\texttt{[@interpret]}} annotation does provide this functionality, as and when required.

\item [{\small\texttt{\textmd{[@interpret-logto <filename>]}}}]\index{logging} The output is not written to standard output or standard error, but appended to a file. This can be used to separate the output of several annotations, or several runs of the same program, or as a crude logging mechanism. For example, if we have a bug which seems to be caused by a {\small\texttt{Not\_found}} exception, but the bug is not reproducible, occurring only on certain runs (for example a threaded program where the threads may run in a different order each time), we may leave a {\small\texttt{[@interpret-logto]}} annotation in place, and check the log later to find what data led to the exception being raised. Since each separate run appends rather than replaces the data, we can collect even from multiple runs of a command line tool. The logs need not be as enormous as one might think, either: the annotation mechanism reduces the trace, and a streaming compression algorithm may be used.

\item [{\small\texttt{\textmd{[@interpret-env <variable>]}}}] Interpret only if an \index{environment variable}environment variable is set, otherwise run natively. This would allow code to remain unaltered after debugging, leaving the annotation in place in case the bug is not really fixed. This allows the shipping of test executables which operate as quickly as one might expect, but which can, upon instruction, be used to generate debug information.

One can imagine, in fact, scenarios in which it might be sensible to leave multiple such annotations in code, each having a unique identifier, to be triggered by environment variables or command line flags. An executable with \textsf{ppx\raisebox{0.8mm}{\_}interpret} embedded in it is about 4Mb larger than usual, so this sort of usage would not be suitable in all environments.

\item [{\small\texttt{\textmd{[@interpret-sub]}}}] Pause the program at the given point, printing the current expression and allowing the programmer to substitute their own. This can be used when we think that an intermediate result in a large program may be wrong, but do not yet know how to fix the code. We wish to stop the program, edit the data structure or program, and restart execution. Should the execution continue and produce correct output, we have evidence that the intermediate result was in fact wrong, and that we have narrowed down the source of the bug.   There might, of course, be technical limitations on this substitution, for example a limitation on which symbols might be included.

\item [{\small\texttt{\textmd{[@show-only]}}}] Simply show a given expression, but run it natively. This has two purposes. First, for data structures, it is a way of adding a \index{generic value printer}generic printer to OCaml (like Java's {\small\texttt{toString}} or Haskell's {\small\texttt{show}}). Second, it allows an extreme form of elision -- we do not show any of the evaluation of the expression, just the expression itself. Now, in a recursive function, this is not so different a result as for an imperative one, but it is likely to be a good first step in debugging. For example, if we think some data is wrong somewhere, and wish to see it printed out.

\item [{\small\texttt{\textmd{[@interpret-matching <search term>]}}}] Give a \index{searching}search term (just like in \textsf{OCamli}) and show only those lines of the evaluation when interpreting. This allows output to be reduced from within the program, rather than having to do it by piping through another program, or in batch mode on the log file afterwards. This helps to ensure accessibility, because we may not always be able to rely upon the standard Unix tools such as {\small\texttt{grep}} being available -- for example in embedded environments. In any event, our own search syntax is specialised to the task.

\item [{\small\texttt{\textmd{[@interpret-n <n>]}}} / {\small\texttt{\textmd{[@exit-after <n>]}}}] Show only {\small\texttt{<n>}} times through this code point. After that, be silent whilst continuing to run, or exit. The purpose is to produce a given, small, amount of debug output. Then, the output may be inspected manually with ease. In the case of a single debug annotation, this may not add much -- we could just produce more output and inspect the beginning of it. However, when there are several annotations, it is important to make sure that a single frequently-reached annotation does not overwhelm the output. Only with experience will we be able to design this sort of functionality in detail.

\item [{\small\texttt{\textmd{[@interpret-until <pattern>]}}} / {\small\texttt{\textmd{[@exit-when <pattern>]}}}] As above, but decide when to finish interpreting or when to exit based not upon a number but upon matching the given pattern. We can use this when we already know which value indicates a bug and are not interested in seeing output after the bug has already occurred. We wish to see only what led up to it. 

\item [{\small\texttt{\textmd{[@interpret-interactive <n>]}}}] Upon reaching this program point, dump into an interface which acts as an interactive debugger, setting and clearing \index{breakpoint}breakpoints and so on. So, we might see, in our {\small\texttt{pairs}} example:

\begin{Verbatim}[commandchars=\\\{\}]
   pairs ( + ) [] [1; 2; 3; 4]
?next
\{matches h::h\textquotesingle{}::t\}
=> pairs ( + ) [3] [2; 3; 4]
?next 5
\{matches h::h\textquotesingle{}::t\}
=> pairs ( + ) [5; 3] [3; 4]
\{matches h::h\textquotesingle{}::t\}
=> pairs ( + ) [7; 5; 3] [4]
\{matches [_]\}
=> rev [7; 5; 3]
=> rev_inner [] [7; 5; 3]
\{matches h::t\}
=> rev_inner [7] [5; 3]
?next
=> rev_inner [5; 7] [3]
?next
=> rev_inner [3; 5; 7] []
\{return from rev_inner\}
=> []
?exit
\end{Verbatim}

\noindent Here we step through the code interactively, then exit the process upon finding the bug. All sorts of standard debugging tools could be included here, such as breakpoints. This is an example of embedding a whole interface for debugging inside the \textsf{ppx\raisebox{0.8mm}{\_}interpret} mechanism. Note that the whole debugger is part of the executable, just like the interpreter -- so still no external tools are required, and the debugger remains accessible.

\item [{\small\texttt{\textmd{[@expected <pattern>]}}}] We expect the result of evaluating this expression to match the given pattern. If it does not, the suspect evaluation may be printed out. These \index{expect test}``expect tests'' are a common method of software testing. We could also, instead of a pattern, use a predicate to be run on the value to see if it matches. We might write the annotation {\small\texttt{[@expected\! [3;\! 5;\! 7]]}} into our {\small\texttt{pairs}} example, and should a bug be introduced in a future version of the code and running with debugging turned on (perhaps using the environment variable method already described), the mismatch would be reported.

\item [{\small\texttt{\textmd{[@stoprepeat]}}}] Stop after a duplicate expression is encountered. For detecting bugs caused by non-termination. Such bugs can cause huge amounts of output before execution can be interrupted, making it hard to see the steps leading to the \index{evaluation!non-termination of}\index{non-termination}non-termination. For example, we might see:

\begin{verbatim}
f [1; 2; 3]
f [1; 2]
f [1]
f [1]
f [1]
ppx_interpret: output ended on [@stoprepeat]
\end{verbatim}

\end{description}

\noindent This concludes our tour of possible annotation types. We can see that many well-known mechanisms of debugging, such as breakpointing, find a new home here. The approach is in general a low-impact one: the programmer need use only the parts of the debugging system they wish to, or which suit their mental model or debugging style. We hope that this makes the debugger more likely to be used by more people. It is possible to imagine other interface models, of course.  But the annotation-based one we have alighted upon seems promising, so we persevere with it for now.

\section{An early failure}

In this section, we discuss an attempted implementation of our scheme, using the original \textsf{OCamli} architecture from \cref{chap:interpreter}. This sheds further light upon some technical difficulties which made the transition to the typed \textsf{OCamli2} necessary. We shall go on to describe the successful implementation in due course.

Recall that we wish to debug a program, and think that the problem is in a certain module, or that the problem would be best identified by examining the innards of a certain module. Instead of inserting print statements, we could write the PPX annotation {\small\texttt{[@interpret]}} to show the steps of evaluation of all the code in a module as if it had been interpreted by \textsf{OCamli}:

\medskip
\begin{verbatim}[commandchars=\\\{\}]
[@interpret]

\textbf{let} f x = A.double x

\textbf{let} g y = A.double (f y) + 2

\textbf{let} h = g 1
\end{verbatim}
\medskip

\noindent Call this module {\small\texttt{B}}. It uses {\small\texttt{A.double}}. Suppose also that there is another module, {\small\texttt{C}}, which uses {\small\texttt{B.f}}. Then the \index{module!dependencies}module dependencies are {\small\texttt{C}} $\rightarrow$ {\small\texttt{B}} $\rightarrow$ {\small\texttt{A}} where $\rightarrow$ means ``depends upon''. We wish to be able to compile this program using whatever commands and build system we are used to, and for the result to be an ordinary executable. Then, when we run the executable, only the code in module {\small\texttt{B}} is interpreted and displayed on the screen in the manner of \textsf{OCamli}. The code in modules {\small\texttt{A}} and {\small\texttt{C}} is run natively and silently. 

In fact, it would be better if we could interpret only exactly the expression which we are interested in. PPX annotations such as {\small\texttt{[@interpret]}} may be placed in any position which corresponds to a node of the parse tree (to which they then become attached). This is what we described in the previous section. So we might write:

\medskip
\begin{verbatim}[commandchars=\\\{\}]
\textbf{let} f x = A.double x

\textbf{let} g y = A.double (f y) [@interpret] + 2

\textbf{let} h = g 1
\end{verbatim}
\medskip

\noindent Now, only {\small\texttt{A.double\! (f\! y)}} will be interpreted, the rest of module  {\small\texttt{B}} being natively executed, together with modules {\small\texttt{A}} and {\small\texttt{C}} as before. We have a very lightweight but powerful version of ``adding print statements'', with the added ability to show steps of execution. Of course, the annotations could be silently added by some other interface such as an IDE, rather than by the user manually typing and then removing {\small\texttt{[@interpret]}} annotations.

How is this achieved technically, making use of the tools we have already developed in \textsf{OCamli}  for interfacing between the interpreted and native worlds? We have four tasks:

\begin{itemize}
\item Calling into native code from interpreted code, which happens if the part of the program annotated with {\small\texttt{[@interpret]}} calls into native code -- in this instance the second call to {\small\texttt{A.double}} in our latter example.
\item Calling into interpreted code from native code -- for example if module {\small\texttt{C}} calls the interpreted function {\small\texttt{B.g}}.
\item Calling into native code from interpreted code in the same module. For example, when {\small\texttt{B.g}} uses the value of {\small\texttt{f\negthinspace\ y}} in the second example.
\item Calling into interpreted code from native code in the same module, for example when {\small\texttt{h}} calls {\small\texttt{g}} in the second example.
\end{itemize}

\setcounter{figure}{1}
\begin{figure}
\begin{verbatim}[commandchars=\\\{\}]
\textbf{let} f x =
  \textbf{let module} A =                                \hfill\textrm{\textit{shim for calling into module \texttt{A}}}
    \textbf{struct}
      \textbf{let} double env =
        \textbf{function}
        | x::[] ->
            \textbf{let} heap_x = Tinyexternal.to_ocaml_value x \textbf{in} \hfill\textrm{\textit{shim for }\texttt{\textit{A.double}}}
            \textbf{let} result = A.double heap_x \textbf{in}
            Tinyexternal.of_ocaml_value env result "int"
        | _ -> failwith "A.double: arity" 
    \textbf{end in}
    \textbf{let} \textbf{open} Tinyocaml \textbf{in}
      \textbf{let} tiny_x = Tinyexternal.of_ocaml_value [] x "int" \textbf{in} \hfill\textrm{\textit{read argument from native world}}
      \textbf{let} (_,program) =
        Tinyocamlrw.of_string "let rec f x = A.double x in f x" \textbf{in} \hfill\textrm{\textit{interpreted version of \texttt{f}}}
      \textbf{let} env =
        [EnvBinding (false, (ref [((PatVar "x"), tiny_x)]))] @ \hfill\textrm{\textit{environment for interpreter}}
          ([EnvBinding
              (false,
                (ref [((PatVar "A.double"), (mk "A.double" A.double))]))]
             @ [])
         \textbf{in}
      \textbf{let} tiny_result =
        Eval.eval_until_value true false (env @ (!Eval.lib)) program \hfill\textrm{\textit{interpret the code}}
      \textbf{in} 
        (Tinyexternal.to_ocaml_value tiny_result : int) \hfill\textrm{\textit{result back to native world}}
\end{verbatim}
{\centering\singlespacing\noindent\small\label{F11} Figure 11. A shim for the function \texttt{f}\par}

\label{fig2}
\end{figure}

\noindent The job of our PPX extension \textsf{ppx\raisebox{0.5mm}{\_}interpret} must be to take a parse tree representing an {\small\texttt{.ml}} file with one or more {\small\texttt{[@interpret]}} annotations, and transform it into one with exactly the same interface, which interprets code marked with {\small\texttt{[@interpret]}} and runs all other code natively. The shim produced by our system for the function {\small\texttt{f}} in our example is displayed in figure 9. The attempted implementation of this early version of \textsf{ppx\raisebox{0.5mm}{\_}interpret}, however, only served to highlight further the faults of the untyped nature of \textsf{OCamli}, discussed in \cref{chap:efficient}. The interface between the native and interpreted parts of the running program is fragile, and the difficulty of the implementation, were it to be completed, would be overwhelming. Large parts of the OCaml front end would have to be re-implemented to make sure type definitions could be found, for example.

Now let us describe a feasible implementation of this annotation-based debugging interface, using our new interpreter \textsf{OCamli2}.

\section{Typed \textsf{ppx\raisebox{1mm}{\_}interpret} with \textsf{OCamli2}}

\index{typed PPX}\index{PPX!typed}We have written a patch to the OCaml compiler to provide a typed analog to PPX. This allows typed-tree-to-typed-tree rewriting in addition to the standard PPX mechanism of parse-tree-to-parse-tree rewriting. We did not want to have to patch the compiler, but this patch is very small and unobtrusive, and need not change between OCaml versions. It may also prove worthwhile to other developers -- there are many uses for a typed PPX. Here is the OCaml compiler architecture with PPX and typed PPX mechanisms in place:

\vphantom{fpoo}\vspace{0mm}

\bigskip
\noindent\includegraphics[width=\textwidth]{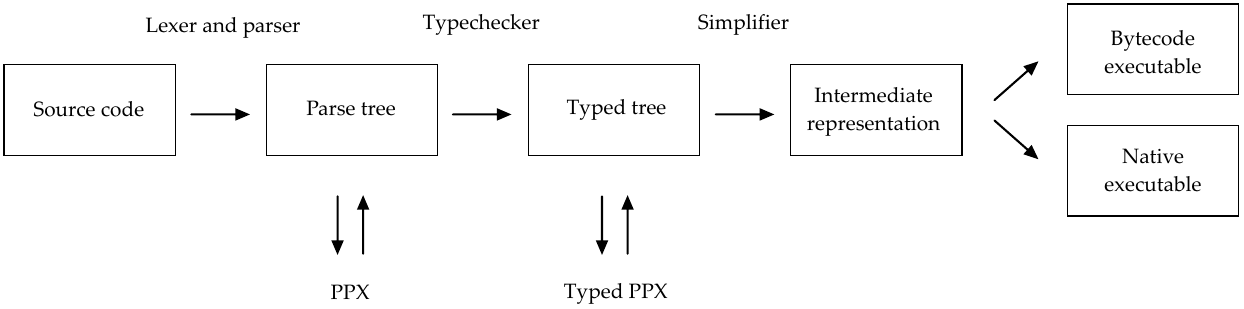}\medskip
\bigskip

\vspace{0mm}

\noindent Consider the following program, with an {\small\texttt{[@interpret]}} annotation:

\medskip
\begin{verbatim}[commandchars=\\\{\}]
\textbf{let} global x = x * 2

\textbf{let} x () =
  \textbf{let} y = 3 - 4 \textbf{in}
    (Random.int 50 + y + 5 + global 6 [@interpret]) + 7

\textbf{let} _ =
  Printf.printf "Result is %i\textbackslash\negthinspace\negthinspace\negthinspace n" (x ())
\end{verbatim}
\medskip

\noindent The program will be compiled as normal, and run natively. However, the section  marked by {\small\texttt{[@interpret]}} will be interpreted step by step. In this case, it is {\small\texttt{Random.int 50 + y + 5 + global 6}}. Note that adding the {\small\texttt{[@interpret]}} annotation does not imply that the innards of {\small\texttt{global}} will be interpreted too. So we have a native function {\small\texttt{x}}, which natively calculates the value of {\small\texttt{y}}, and then has an interpreted section which needs that value of {\small\texttt{y}} and also uses the native \index{Standard Library}\index{OCaml!Standard Library}Standard Library function {\small\texttt{Random.int}}. The result of {\small\texttt{x}} must be returned natively. We have both native code and interpreted code, and the interpreted code needs to be able to call native functions and use the results of natively calculated values.

It is the job of our new \textsf{OCamli2}-based, typed version of \textsf{ppx\raisebox{0.8mm}{\_}interpret} to arrange all this. Altering the \index{typed tree}typed tree is a delicate affair, since there are so many invariants required for a valid tree, so we arrange for most of the work to be done by a plain PPX  prior to typechecking. The typed PPX which operates after \index{typechecking}typechecking is limited to transporting the type of expressions from compile-time to run-time. Here is the output of the plain PPX on our example, before the typed PPX acts:

\medskip
\begin{verbatim}[commandchars=\\\{\}]
\textbf{let} env = ref Lib.stdlib\hfill\textrm{\textit{initial environment}}
\textbf{let} () = Tppxsupport.init ()\hfill\textrm{\textit{set up support code}}
\textbf{let} eval_full = Tppxsupport.eval_full env\hfill\textrm{\textit{the interpreter}}

\textbf{let} global x = x * 2

\textbf{let} () = Tppxsupport.addenv env "global" global ""\hfill\textrm{\textit{make \texttt{global} available to the interpreter}}

\textbf{let} x () =
  \textbf{let} y = 3 - 4 \textbf{in}
    \textbf{let} () = Tppxsupport.addenv env "y" y "" \textbf{in}\hfill\textrm{\textit{make value of \texttt{y} available to the interpreter}}
      (Random.int 50 + y + 5 + global 6 [@interpret]) + 7

\textbf{let} () = Tppxsupport.addenv env "x" x ""\hfill\textrm{\textit{make function \texttt{x} available to the interpreter}}

\textbf{let} _ =
  Printf.printf "Result is %i\textbackslash\negthinspace\negthinspace\negthinspace n" (x ())
\end{verbatim}
\medskip

\noindent Two things have happened. First, a preamble has been added, for the use of the typed PPX, to keep it as simple as possible. Second, both local and global \index{let binding}let bindings have been followed by calls to {\small\texttt{Tppxsupport.addenv}}, whose job is to put native values into an environment for the interpreted code to refer to. The empty string {\small\texttt{""}} will be replaced by the subsequent typed PPX with a marshalled representation of the type of the value. Marshalling is used to minimise interference in the typed tree. We need not build new nodes for the typed tree, just change one constant string for another.

Once typechecking has occurred, the typed PPX part of \textsf{ppx\raisebox{0.8mm}{\_}interpret} runs. It has two jobs: to build the interpreted sections, and to fill in the types of the environment items. Here is the output:

\medskip
\begin{verbatim}[commandchars=\\\{\}]
\textbf{let} env = ref Lib.stdlib\hfill\textrm{\textit{initial environment}}
\textbf{let} () = Tppxsupport.init ()\hfill\textrm{\textit{set up support code}}
\textbf{let} eval_full = Tppxsupport.eval_full env\hfill\textrm{\textit{the interpreter}}

\textbf{let} global x = x * 2

\textbf{let} () = Tppxsupport.addenv env "global" global "\textrm{\textit{<marshalled type of function global>}}"

\textbf{let} x () =
  \textbf{let} y = 3 - 4 \textbf{in}
    \textbf{let} () = Tppxsupport.addenv env "y" y "\textrm{\textit{<marshalled type of value y>}}" \textbf{in}
      eval_full env "\textrm{\textit{<marshalled \textsf{TinyOCaml} representation of \texttt{Random.int\! 50\! +\! y\! +\! 5\! +\! global\! 6}>}}"
      + 7

\textbf{let} () = Tppxsupport.addenv env "x" x "\textrm{\textit{<marshalled type of function x>}}"

\textbf{let} _ =
  Printf.printf "Result is %i\textbackslash\negthinspace\negthinspace\negthinspace n" (x ())
\end{verbatim}
\medskip

\noindent Now, at run-time, not only are the types of the interpreted parts available (as they are in command-line \textsf{OCamli2}) but the types of environment elements from the natively-executed parts of the enclosing program are available too: the step-by-step interpreter has everything it needs to print out the steps of evaluation. Here is the output for a run of our example program:

\medskip
\begin{verbatim}[commandchars=\\\{\}]
$ ./example
   \underline{Random.int 50} + y + 5 + global 6
=> 44 + \underline{y} + 5 + global 6
=> \underline{44 + -1} + 5 + global 6
=> \underline{43 + 5} + global 6
=> 48 + \underline{global 6}
=> \underline{48 + 12}
=> 60
Result is 67\end{verbatim}
\medskip

\noindent The file {\small\texttt{example}} is a native code executable, containing within it the main program, executed natively, the phrase {\small\texttt{Random.int\! 50\! +\! y\! +\! 5\! +\! global\! 6}} interpreted step by step, and the interpreter itself. The native code portion and the interpreted portion share the same \index{OCaml!runtime}OCaml runtime, and the same \index{heap}heap.

One other aspect of the implementation of this annotation-based interface to debugging is worth discussion. Can we now dispense with our special \index{Standard Library}\index{OCaml!Standard Library}Standard Library? We should not need to load the Standard Library and interpret its module initialisation, since the interpreter is in the same process as the OCaml runtime in which the native code is running and in which the Standard Library has already been loaded and its initialisation run. So, we should be able to reduce the footprint of our changes greatly, simply by interfacing with the Standard Library as if it were any other piece of code. This means, of course, that we can no longer show the steps of interpretation inside the Standard Library (since it was not compiled with {\small\texttt{[@interpret]}}).  This should not be too much of a problem, since it is unlikely that a programmer would like to inspect the inside of the Standard Library. In any case, we should be able to provide an alternative Standard Library package compiled with interpretation turned on, if required.

\section{Use in the REPL}\index{REPL}

Another pleasant side effect of the way in which we have been able to use the PPX system to embed our step-by-step interpreter into bytecode and native code programs is that, of course, it could be extended to work with the normal OCaml REPL (which functions by compiling phrases to bytecode and then executing them). And so, to the REPL's playground is added a new toy. We may put {\small\texttt{[@interpret]}} annotations to see the steps of execution.

\section{Summary}

\correction{We have found a pleasant interface to our step-by-step interpreter which might make debugging a better experience, meeting many of the challenges of accessibility with which we have been most concerned. Along the way, we have confirmed that our new \textsf{OCamli2} seems to fix the technical problems of our original \textsf{OCamli} implementation with regard to the interpreted/native interface. We have sketched an implementation to show that the idea is technically feasible (\addition{the state of the implementation of this work is given in section \ref{stateimpl}}). 
\label{summary6}\addition{Once we have a working implementation of both \textsf{OCamli} and this PPX-based interface, the debugger will be usable, and so may be evaluated more fully. Only when the debugger is in use, as we have repeatedly said, can we know for certain that it is useful. But we have been careful to design it to be so.}}

\chapter{Roads not taken}
\label{chap:misc}

\begin{quotation}\textit{\large One man's rubbish may be another's treasure.\textrm{\begin{flushright}--- \textup{Proverb}\end{flushright}}}\end{quotation}

\vspace{10pt}

\noindent Along the way, we turn down some blind alleys. One of the pleasant peculiarities of the thesis as a form, when compared with conference or journal papers, is the space to include what was observed in those alleys, before we were forced by failure or the constraints of time to return to the main path.

In this chapter we describe, in turn, an interesting way to improve error messages in an interpreter by allowing it to attempt to evaluate programs which are ill-typed, a mechanism for including arbitrary interpreted code in ordinary OCaml programs to be run at compile-time, and a way of visualising the execution of OCaml bytecode programs without needing interpretation at all.

\section{Turning off the typechecker}\index{typechecking}\index{typechecking!turning off}

As an aid to debugging \textsf{OCamli} itself, the command line option {\small\texttt{-no-typecheck}} was introduced. This option, when enabled, parses and interprets the code without running the OCaml typechecker. The interpreter, which normally relies on programs having been typechecked for correct operation, will then run an incorrect program as far as it can, concluding either with an incorrect result, or an exception raised from the interpreter.

Might this option have legitimate uses outside of debugging the interpreter itself? A program which will not compile is a kind of bug, in a sense. Especially so when other languages with weaker type systems would have compiled the same source code to a buggy executable without complaint. The question "Why doesn't my code compile?" is a variant on "Why does my code fail at run-time?" because they are both, really, "Why doesn't my code work?".

The usual mantra when debugging is that the failure should be as close to the source of the problem as possible, helping us to pinpoint the bug. However, we think that, in the case of recursive functions, it might be helpful sometimes to find errors slightly later -- when a badly typed value is received by code which does not expect it, rather than when it might be created (as a type inference engine would find the error). Turning the typechecker off allows such ill formed values to be created and propagated.

Presently, some ill-typed programs fail because no case is there to handle them in the interpreter. For example, take the expression {\small\texttt{1 + false}}, and try to evaluate it in the normal OCaml REPL:

\medskip
\begin{verbatim}[commandchars=\\\{\}]
# 1 + \underline{false};;
Error: This expression has type bool but an expression was expected of type int
\end{verbatim}
\medskip

\noindent In \textsf{OCamli}, the addition operator looks for two integers, or an unevaluated left or right side. For now, there is no case to handle the type mismatch, and so we see an  unhelpful error message:

\medskip
\begin{verbatim}[commandchars=\\\{\}]
$ ocamli -e '1 + false' -no-typecheck -show-all
    \underline{1 + false}
Error in Eval.next Failure("already a value or unimplemented: false")
\end{verbatim}
\medskip

\noindent The actual text of this error message is of no use except to the interpreter-writer. It indicates that, not being able to find two integers to add, the interpreter tried to evaluate the right-hand-side. Since the evaluator is never run on values, and {\small\texttt{false}} is a value, this error results. Let us look at another example. A classic beginner's mistake in functional languages is to confuse list consing and list concatenation:

\begin{verbatim}[commandchars=\\\{\}]
# [1] :: [\underline{2}] @ [3];;
Error: This expression has type int but an expression was expected of type int list
\end{verbatim}

\noindent In this instance, \textsf{OCamli} with {\small\texttt{-no-typecheck}} runs the program with no problem at all:

\begin{verbatim}[commandchars=\\\{\}]
$ ocamli -e '[1] :: [2] @ [3]' -no-typecheck -show-all
    \underline{[[1]; 2] @ [3]}
=>  [[1]; 2; 3]
\end{verbatim}

\noindent This is, of course, because both {\small\texttt{::}} and {\small\texttt{@}} are \index{polymorphism}polymorphic. So this is an example where the interpreter cannot currently spot the type error at run-time.

We should like to alter \textsf{OCamli} to properly report \index{type error}type errors, such as adding an integer to a boolean, where the error message is currently unhelpful. In addition, we should like to add errors for things like the cons/concatenate confusion above. Do we detect this kind on value creation or value consumption? Presently, the code for arithmetic operators in our interpreter looks like this:

\medskip
\begin{verbatim}[commandchars=\\\{\}]
| Op (op, Int a, Int b) -> \hfill\textit{\textrm{we have two integers}}
    \textbf{begin try} Int (calc op a b) \textbf{with}
      Division_by_zero -> Raise ("Division_by_zero", None)
    \textbf{end}
| Op (op, Int a, b) -> Op (op, Int a, eval peek env b)\hfill\textit{\textrm{evaluate the right-hand side}}
| Op (op, a, b) -> Op (op, eval peek env a, b)\hfill\textit{\textrm{evaluate the left-hand side}}
\end{verbatim}
\medskip

\noindent The final cases are too generic in an \index{program!ill-typed}\index{ill-typed program}ill-typed program. They are those  cases where the left or right-hand side are not integers, and not compound expressions, but non-integer values. If {\small\texttt{a}} and {\small\texttt{b}} are both values, but not both integers, we can give the appropriate message:

\medskip
\begin{verbatim}[commandchars=\\\{\}]
| Op (op, Int a, Int b) -> \hfill\textrm{\textit{we have two integers}}
    \textbf{begin try} Int (calc op a b) \textbf{with}
      Division_by_zero -> Raise ("Division_by_zero", None)
    \textbf{end}
| Op (op, a, b) \textbf{when} is_value a && is_value b ->\hfill\textrm{\textit{values not both integers}}
    \textbf{raise}
      (RuntimeTypeError
         (Printf.sprintf
            "operation %s can operate only on integers"
            (string_of_op op)))
| Op (op, a, b) \textbf{when} is_value a -> Op (op, Int a, eval peek env b)\hfill\textrm{\textit{any value on left}}
| Op (op, a, b) -> Op (op, eval peek env a, b)
\end{verbatim}
\medskip

\noindent The cons/append example is harder. It is difficult to see how polymorphic operations can be typechecked at run-time. Do we need a type inference engine? Or just a ``same-type'' check to make sure that the new item being consed on to the front has the same type as the elements already in the list? As another example, how might we deal with generic comparison? We cannot actually check the types if we are using the real generic comparison operator:

\medskip
\begin{verbatim}[commandchars=\\\{\}]
$ ocamli -e '1 < 2 < 3' -no-typecheck -show-all
    \underline{1 < 2} < 3
=>  \underline{true < 3}
=>  false
\end{verbatim}
\medskip

\noindent We can indeed develop a ``same-type'' approximation, applying to just values (excepting functions). Due to OCaml's order of evaluation, we can only find the run-time type error when the expression has been reduced to a value. This is a shame, because it means we see it quite late in the process. But, as we have discussed, this can be useful to the beginner too. So, for example, we now get a useful error:

\medskip
\begin{verbatim}[commandchars=\\\{\}]
$ ocamli -e '(fun x y -> x :: y) 2 ['a']' -no-typecheck -show-all
    \underline{(\textbf{fun} x y -> x::y) 2} ['a']
=>  \underline{(\textbf{let} x = 2 \textbf{in fun} y -> x::y)} ['a']
=>  \underline{(\textbf{fun} y -> \textbf{let} x = 2 \textbf{in} x::y) ['a']}
=>  \textbf{let} y = ['a'] \textbf{in let} x = 2 \textbf{in} \underline{x}::y
=>  \textbf{let} y = ['a'] \textbf{in} 2::\underline{y}
Run time type error:
  Cannot cons onto this list: differing element types
\end{verbatim}
\medskip

\noindent Now, for our generic comparison example on what is, to a beginner, an innocuous piece of code -- because it is correct mathematics:

\medskip
\begin{verbatim}[commandchars=\\\{\}]
$ ocamli -e '1 < 2 < 3' -no-typecheck -show-all
    \underline{1 < 2} < 3
=>  \underline{true < 3}
Run time type error:
  Comparison between values of differing types
\end{verbatim}
\medskip

\noindent This is better for the beginner than the static type error:

\medskip
\begin{verbatim}[commandchars=\\\{\}]
$ ocamli -e '1 < 2 < 3' -show-all
File "", line 1, characters 8-9:
Error: This expression has type int but an expression was expected of type bool
\end{verbatim}
\medskip

\noindent Consider another example. Here, the programmer has left out {\small\texttt{f}} in the recursive call to {\small\texttt{map}}\medskip
\begin{verbatim}[commandchars=\\\{\}]
\textbf{let} g x = x + 1

\textbf{let rec} map f = \textbf{function}
  | [] -> []
  | h::t -> f h :: map t

\textbf{let} l = map g [1; 2; 3]
\end{verbatim}
\medskip

\noindent This leads to a type error in the OCaml REPL:

\medskip
\begin{verbatim}[commandchars=\\\{\}]
# let rec map f = function [] -> [] | h::t > f h :: map \underline{t};;
Error: This expression has type 'a list
       but an expression was expected of type 'a -> 'b
\end{verbatim}
\medskip

\noindent What we get in our modified \textsf{OCamli} with run-time type detection  is:

\medskip
\begin{verbatim}[commandchars=\\\{\}]
=>  \textbf{let} g x = x + 1
    
    \textbf{let rec} map f = \textbf{function} [] -> [] | h::t -> f h :: map t 
    
    \textbf{let} l = \textbf{let rec} map f = \textbf{function} [] -> [] | h::t -> f h :: map t \textbf{in} 2 :: map [2; 3]
Run time type error:
  Attempt to cons onto non-list
\end{verbatim}
\medskip

\noindent Whilst this is true ({\small\texttt{map [2; 3]}} is a function, and so a value), it is not quite the intuitive error message we would like. A much wider study of this phenomenon vis-a-vis common beginner errors would be needed to see if this approach of turning the typechecker off has wide applicability.

Another approach  to improving OCaml error messages is to modify the \index{typechecking!error reporting}\index{type error}typechecker itself \cite{chargueraud2015improving}. Trying to improve error messages in type inference has a long record of research -- two recent examples are the localization approach \cite{Pavlinovic} and the type-debugger \cite{Lerner2006}.

\addition{Our work, in any event, is not useable under \textsf{OCamli2}, where full typechecking is required -- recall, from chapter 5, that types must be available at every node so that heap values can be converted to strings to be printed by the interpreter during step-by-step debugging. Hence its presence in this chapter. 
\label{decision7-1}}

\section{Compile-time interpretation}

Now that we have an interpreter for OCaml, what can we use it for, other than our intended purpose of debugging? Of course, \textsf{OCamli} can be built as a library, and so can be linked into other programs. For example, linking the OCaml REPL with \textsf{OCamli} allows one to evaluate arbitrary OCaml source at run-time:

\medskip
\begin{verbatim}[commandchars=\\\{\}]
# let s = Runeval.eval_string "List.split [(1, 2); (3, 4)]";;
val s : string = "([1; 3], [2; 4])"
\end{verbatim}
\medskip

\noindent Unfortunately, the result is a string. However, we have functions to deal with that already, from our work interfacing with C. We can convert this string to a real OCaml value, though the user must provide the type in the REPL, and it must be correct:

\medskip
\begin{verbatim}[commandchars=\\\{\}]
# let x : int list * int list =
    Tinyexternal.to_ocaml_value (snd (Tinyocamlrw.of_string s));;
val x : int list * int list = ([1; 3], [2; 4])
\end{verbatim}
\medskip

\noindent Now, we can build OCaml expressions as strings or ASTs at run-time and typecheck and evaluate them. This is fun, but a better \index{metaprogramming}metaprogamming solution for OCaml already exists in the form of MetaOCaml \cite{metaocaml} which provides guarantees of well-typedness -- that is to say, a well-typed MetaOCaml program can generate only well-typed OCaml programs at run-time.

A more practical application of this kind of ad hoc code generation takes the form of another \index{PPX}PPX extension, to allow arbitrary code to be executed at compile-time and the resultant value inserted in source code of the program under compilation. With the \textsf{ppx\raisebox{0.8mm}{\_}eval} extension we can write the following:

\medskip
\begin{verbatim}[commandchars=\\\{\}]
\textbf{let} compiler\_command = [\%compiletime "Sys.argv.(0)"]
\end{verbatim}
\medskip

\noindent This code, in a normal compiled OCaml program with \textsf{ppx\raisebox{0.8mm}{\_}eval} might generate this, the string representing the compiler command itself:

\medskip
\begin{verbatim}[commandchars=\\\{\}]
\textbf{let} compiler\_command = "ocamlopt"
\end{verbatim}
\medskip

\noindent This can do jobs normally performed by \index{preprocessing}preprocessing tools, but in a way which respects the OCaml grammar. Another example might be to include in the file the date of its compilation. In many codebases (including the OCaml compiler itself), there are one or more places where a preprocessor such as {\small\texttt{sed}} is used, and each time the behaviour of such a tool on a given source file must be checked, and re-checked when the file is significantly altered. Depending upon an external command-line tool like {\small\texttt{sed}}  for preprocessing is fraught, since it can behave differently on each operating system. By bringing the functionality into OCaml, we reduce such concerns.

\label{decision7-2}\addition{This is interesting work, which merits further study, but is not directly relevant to our core aim of debugging OCaml programs, so it was not taken further. It is merely a consequence of our having written an interpreter, not a prerequisite for our debugging work.}

\section{Debuggable bytecode by decompilation}\index{bytecode}\index{decompilation}

Instead of writing an interpreter for OCaml, could we instead simply modify OCaml's own bytecode interpreter to print the current step of evaluation at any given moment, thus producing a trace of the program's execution without affecting execution speed unduly?

OCaml's bytecode system is based on the ZINC machine \cite{leroy1990zinc}. We shall consider the simplified version designed by the same author for pedagogical purposes \cite{leroynotes}, look at the compilation scheme it uses, and see if we can derive a decompilation scheme.

\subsection{Programs}

Our programs are defined using a tiny subset of OCaml. Variable accesses have been converted to deBruijn indices when the program was converted from the OCaml parse tree. Here is the type for programs:

\begin{verbatim}[commandchars=\\\{\}]
\textbf{type} op = Add | Sub | Mul | Div

\textbf{type} prog =
  Int \textbf{of} int
| Bool \textbf{of} bool
| Var \textbf{of} int
| Eq \textbf{of} prog * prog
| Op \textbf{of} prog * op * prog
| Apply \textbf{of} prog * prog
| Lambda \textbf{of} prog
| Let \textbf{of} prog * prog
| If \textbf{of} prog * prog * prog\end{verbatim}

\noindent For example, the OCaml program

\medskip
\begin{verbatim}[commandchars=\\\{\}]
\textbf{let} x = 5 \textbf{in} \textbf{if} x = 4 \textbf{then} 1 \textbf{else} (\textbf{fun} x -> x + 1) 2
\end{verbatim}
\medskip

\noindent may be represented as:

\medskip
\begin{verbatim}[commandchars=\\\{\}]
Let (Int 5,
     If (Eq (Var 1, Int 4),
         Int 1,
         Apply (Lambda (Op (Var 1, Add, Int 1), Int 2))))
\end{verbatim}

\subsection{Compilation scheme}

The abstract machine instructions follow \cite{leroynotes} but we add booleans, IF and the equality test EQ:

\bigskip
\noindent
EMPTY\hfill\textit{program ends}\\
INT(integer)\hfill\textit{integer}\\
BOOL(boolean)\hfill\textit{boolean}\\
OP(op)\hfill\textit{arithmetic operators}\\
EQ\hfill\textit{test for equality}\\
ACCESS(integer)\hfill\textit{fetch value of name}\\
CLOSURE(instructions)\hfill\textit{closures}\\
LET\hfill\textit{let bindings}\\
ENDLET\\
APPLY\hfill\textit{function application}\\
RETURN\hfill\textit{last instruction in closure}\\
IF\hfill\textit{if\ldots\ then\ldots\ else \ldots\ }
\bigskip

\noindent Here is the compilation scheme $\mathcal{C}$, again extended from Leroy:
\begin{align*}
\mathcal{C}(\texttt{Int}(i)) &= \text{INT}(i)\\
\mathcal{C}(\texttt{Bool}(b)) &= \text{BOOL}(b)\\
\mathcal{C}(\texttt{Op}(a, \oplus, b)) &= \mathcal{C}(a); \mathcal{C}(b); \text{OP}(\oplus)\\
\mathcal{C}(\texttt{Eq}(a, b)) &= \mathcal{C}(a); \mathcal{C}(b); \text{EQ}\\
\mathcal{C}(\texttt{Var}(n)) &= \text{ACCESS}(n)\\
\mathcal{C}(\texttt{Lambda}(a)) &= \text{CLOSURE}(\mathcal{C}(a); \text{RETURN})\\
\mathcal{C}(\texttt{Let}(a, b)) &= \mathcal{C}(a); \text{LET}; \mathcal{C}(b); \text{ENDLET}\\
\mathcal{C}(\texttt{Apply}(a, b)) &= \mathcal{C}(a); \mathcal{C}(b); \text{APPLY}\\
\mathcal{C}(\texttt{If}(a, b, c)) &= \mathcal{C}(\texttt{Lambda}(b)); \mathcal{C}(\texttt{Lambda}(c)); \mathcal{C}(a); \text{IF}
\end{align*}

\noindent So our example

\medskip
\begin{verbatim}[commandchars=\\\{\}]
\textbf{let} x = 5 \textbf{in} \textbf{if} x = 4 \textbf{then} 1 \textbf{else} (\textbf{fun} x -> x + 1) 2
\end{verbatim}
\medskip

\noindent compiles to:

\bigskip
\noindent INT 5\\
LET\\
CLOSURE\\
\phantom{\ \ }\ INT 1\\
\phantom{\ \ }\ RETURN\\
CLOSURE\\
\phantom{\ \ } CLOSURE\\
\phantom{\ \ \ \ } ACCESS 1\\
\phantom{\ \ \ \ } INT 1\\
\phantom{\ \ \ \ } OP +\\
\phantom{\ \ \ \ } RETURN\\
\phantom{\ \ } INT 2\\
\phantom{\ \ } APPLY\\
\phantom{\ \ } RETURN\\
ACCESS 1\\
INT 4\\
EQ\\
BRANCH\\
ENDLET\\
EMPTY
\bigskip

\noindent Notice the EMPTY inserted at the end. We have used indentation here to make it easier to see the structure of the CLOSUREs but it is, in reality, a simple list of instructions with no structure. The compilation process is illustrated in figure 10.

\subsection{Evaluation scheme}

Here is the evaluation scheme $\mathcal{E}$, again extended from \cite{leroynotes} with  our new instructions.

\bigskip

\begin{tabular}{l|l|l||l|l|l}
\multicolumn{3}{c}{Machine state before}&\multicolumn{3}{c}{Machine state after}                       \\
Code                    & Env   & Stack                          & Code   & Env    & Stack             \\
$\text{INT}(i);c$       & $e$   & $s$                            & $c$    & $e$    & $i.s$             \\
$\text{BOOL}(b);c$      & $e$   & $s$                            & $c$    & $e$    & $b.s$             \\
$\text{OP}(\oplus);c$   & $e$   & $i.i'.s$                       & $c$    & $e$    & $\oplus(i, i').s$ \\
$\text{EQ};c$           & $e$   & $i.i'.s$                       & $c$    & $e$    & $(i = i').s$      \\
$\text{ACCESS}(n);c$    & $e$   & $s$                            & $c$    & $e$    & $e(n).s$          \\
$\text{CLOSURE}(c');c$  & $e$   & $s$                            & $c$    & $e$    & $c'[e].s$         \\
$\text{LET};c$          & $e$   & $v.s$                          & $c$    & $v.e$  & $s$               \\
$\text{ENDLET};c$       & $v.e$ & $s$                            & $c$    & $e$    & $s$               \\
$\text{APPLY};c$        & $e$   & $v.c'[e'].s$                   & $c'$   & $v.e'$ & $c.e.s$           \\
$\text{RETURN};c$       & $e$   & $v.c'.e'.s$                    & $c'$   & $e'$   & $v.s$             \\
$\text{IF};c$           & $e$   & $\textsf{T}.c'[e'].c''[e''].s$ & $c'$   & $e'$   & $c[e].s$          \\
$\text{IF};c$           & $e$   & $\textsf{F}.c'[e'].c''[e''].s$ & $c''$  & $e''$  & $c[e].s$          \\
\end{tabular}

\bigskip

\noindent There are three components: a code pointer $c$ (instructions yet to be reached), an environment $e$, and a stack $s$ (intermediate results and pending function calls). The notation $c[e]$ is the closure of code $c$ with environment $e$.

The final result is at the top of the stack when the code is EMPTY. The evaluation process is illustrated in figure 11.

\subsection{Decompilation scheme}\index{decompilation}

In order to print the evaluation step by step, we need to be able to decompile:

\begin{itemize}
\item Any program which has been compiled by the compilation scheme above.
\item Certain incomplete evaluations under the evaluation scheme above. That is to say, given $(c, s)$ we can decompile a program which represents the evaluation at that stage. We need not be able to decompile arbitrary $(c, e, s)$ triples.
\end{itemize}

\noindent  We add names to {\small\texttt{Var}}, {\small\texttt{Lambda}} and {\small\texttt{Let}}, since we shall need to recover variable names during decompilation:

\begin{verbatim}[commandchars=\\\{\}]
\textbf{type} prog =
  Int \textbf{of} int
| Bool \textbf{of} bool
| Var \textbf{of} name * int
| Eq \textbf{of} prog * prog
| Op \textbf{of} prog * op * prog
| Apply \textbf{of} prog * prog
| Lambda \textbf{of} name * prog
| Let \textbf{of} name * prog * prog
| If \textbf{of} prog * prog * prog\end{verbatim}

\noindent So we must add names to the ACCESS, CLOSURE and LET instructions. These are not required for evaluation, of course, but only for decompilation:

\bigskip
\noindent
EMPTY\hfill\textit{program ends}\\
INT(integer)\hfill\textit{integer}\\
BOOL(boolean)\hfill\textit{boolean}\\
OP(op)\hfill\textit{arithmetic operators}\\
EQ\hfill\textit{test for equality}\\
ACCESS(name, integer)\hfill\textit{fetch value of name}\\
CLOSURE(name, instructions)\hfill\textit{closures}\\
LET(name)\hfill\textit{let bindings}\\
ENDLET\\
APPLY\hfill\textit{function application}\\
RETURN\hfill\textit{last instruction in closure}\\
IF\hfill\textit{if\ldots\ then\ldots\ else \ldots\ }
\bigskip

\noindent Decompilation is performed by going through the instructions in order, holding a stack a little like the evaluation stack, but which may also contain decompiled program fragments -- the empty stack is written $\{\}$. When we have gone through all the instructions, the final program is at the top of the stack. We do not need the environment, since we are not running the code, just decompiling it. Here is the decompilation scheme $\mathcal{D}$:
\allowdisplaybreaks
\begin{align*}
\mathcal{D}(\text{EMPTY}, v.s) &= v\\
\mathcal{D}(\text{INT}(i); c, s) &= \mathcal{D}(c, \texttt{Int}(i).s)\\
\mathcal{D}(\text{BOOL}(i); c, s) &= \mathcal{D}(c, \texttt{Bool}(b).s)\\
\mathcal{D}(\text{OP}(\oplus); c, i.i'.s) &= \mathcal{D}(c, \texttt{Op}(i, \oplus, i').s) \\
\mathcal{D}(\text{EQ}; c, i.i'.s) &= \mathcal{D}(c, \texttt{Eq}(i, i').s)\\
\mathcal{D}(\text{ACCESS}(n, l); c, s) &= \mathcal{D}(c, \texttt{VarAccess}(n, l).s)\\
\mathcal{D}(\text{CLOSURE}(n, c'); c, s) &= \mathcal{D}(c, c'[n, \{\}].s)\\
\mathcal{D}(\text{LET}(n); c, v.s) &= \texttt{Let}(n, v, \mathcal{D}(c, s))\\
\mathcal{D}(\text{ENDLET}; c, s) &= \mathcal{D}(c, s)\\
\mathcal{D}(\text{APPLY}; c, v.c'[n, e'].s) &= \texttt{Apply}(\texttt{Lambda(n}, \mathcal{D}(c', \{\}\texttt{))}, v)\\
\mathcal{D}(\text{RETURN}; c, v.c'.e'.s) &= \mathcal{D}(c', v.s)\\
\mathcal{D}(\text{RETURN}; c, s) &= \mathcal{D}(c, s)\\
\mathcal{D}(\text{IF}; c, e.c'[e'].c''[e''].s) &= \mathcal{D}(c, \texttt{If}(e, \mathcal{D}(c', s), \mathcal{D}(c'', s)).s)
\end{align*}

\noindent This decompiler works for:

\begin{itemize}
\item Any program-stack pair (P, \{\}) where P was compiled by $\mathcal{C}$ above.
\item A program-stack pair (P, S) which is an intermediate state of the evaluation procedure $\mathcal{E}$ (minus the environment) where P begins with an instruction such as OP or APPLY.
\end{itemize}

\noindent Our example program decompiles properly from bytecode, both as a whole, and when part-evaluated. Two such examples are shown in figures 12 and 13.

\subsection{Prototype}

Once we have such a decompilation regime, we can build a step-by-step interpreter very easily. We compile the program to bytecode, then evaluate it one bytecode instruction at a time, decompiling after each step, and displaying the resulting program. There would be many repeated lines, for example when an instruction simply puts something onto the stack. We need a notion of what constitutes an `interesting' instruction execution. Presently we consider a step `interesting' if it follows immediately the execution of an ACCESS, BRANCH, OP, or EQ instruction. For our program, this gives the following.

\medskip
\begin{verbatim}[commandchars=\\\{\}]
$ ./bytecode tests/example.ml
let x = 5 in if x = 4 then 1 else (fun x -> x + 1) 2
if 5 = 4 then 1 else (fun x -> x + 1) 2
if false then 1 else (fun x -> x + 1) 2
(fun x -> x + 1) 2
2 + 1
3
\end{verbatim}
\medskip

\noindent It is somewhat unintuitive that such a simple scheme should work so well -- eliding more than our own interpreter by default and mimicking so well what we might write on paper, given that it operates at such a low level. However, since the process of compilation is intended to reduce a program to an efficient form where the fewest or fastest instructions are generated and each instruction or little sequence of instructions does something to make definite progress in a computation, it is perhaps not so surprising.

\label{decision7-3}\correction{This work has produced some intriguing results -- its properties we have just described, and its undoubted speed, but we have put it aside. Our new mechanism for selective interpretation, described in the previous chapter, probably makes the speed increase unimportant in most cases, and plain interpretation continues to have the compelling advantage of full information at all times. It will be interesting, though, to compare the results of interpreting (with regard to quality of output) sample programs such as those in Appendix B, between \textsf{OCamli}, \textsf{OCamli2} and the bytecode decompiler described, once the projects are all in a state in which such a comparison is possible.}

\section{Summary}

\label{summary7}\addition{In this chapter we have exhibited three pieces of work which, whilst we consider them to be interesting and worth describing for an audience, were either dead ends due to design decisions taken elsewhere, or which simply became surplus to requirement for the same reason.}

In \cref{chap:approach}, we have justified our work. In chapters \ref{chap:interpreter}, \ref{chap:efficient}, \ref{chap:interface} and \ref{chap:misc} we have described it. Now it is time to step back, and evaluate it in as dispassionate and impartial way as is possible.

\begin{sidewaysfigure}
\begin{landscape}
{\small

\noindent$\mathcal{C}$(\texttt{Let(Int 5, If(Eq(Var 1, Int 4), Int 1, Apply(Lambda(Op(Var 1, Add, Int 1), Int 2))))})

\smallskip
\hspace{10mm} Rule $\mathcal{C}$-\texttt{Let}
\smallskip

\noindent $\mathcal{C}$(\texttt{Int 5}); LET; $\mathcal{C}$(\texttt{If(Eq(Var 1, Int 4), Int 1, Apply(Lambda(Op(Var 1, Add, Int 1), Int 2)))}); ENDLET

\smallskip
\hspace{10mm} Rule $\mathcal{C}$-\texttt{Int}
\smallskip

\noindent INT 5; LET; $\mathcal{C}$(\texttt{If(Eq(Var 1, Int 4), Int 1, Apply(Lambda(Op(Var 1, Add, Int 1), Int 2)))}); ENDLET

\smallskip
\hspace{10mm} Rule $\mathcal{C}$-\texttt{If}
\smallskip

\noindent INT 5; LET; $\mathcal{C}$(\texttt{Lambda (Int 1)}); $\mathcal{C}$(\texttt{Lambda(Apply(Lambda(Op(Var 1, Add, Int 1), Int 2)))}); $\mathcal{C}$(\texttt{Eq(Var 1, Int 4)}); IF; ENDLET

\smallskip
\hspace{10mm} Rule $\mathcal{C}$-\texttt{Eq} then Rule $\mathcal{C}$-\texttt{Eq} then Rule $\mathcal{C}$-\texttt{Eq}
\smallskip

\noindent INT 5; LET; $\mathcal{C}$(\texttt{Lambda (Int 1)}); $\mathcal{C}$(\texttt{Lambda(Apply(Lambda(Op(Var 1, Add, Int 1), Int 2)))}); ACCESS 1; INT 4; EQ; IF; ENDLET

\smallskip
\hspace{10mm} Rule $\mathcal{C}$-\texttt{Lambda} then Rule $\mathcal{C}$-\texttt{Int}
\smallskip

\noindent INT 5; LET; CLOSURE [INT 1; RETURN]; $\mathcal{C}$(\texttt{Lambda(Apply(Lambda(Op(Var 1, Add, Int 1), Int 2)))}); ACCESS 1; INT 4; EQ; IF; ENDLET

\smallskip
\hspace{10mm} Rule $\mathcal{C}$-\texttt{Lambda}
\smallskip

\noindent INT 5; LET; CLOSURE [INT 1; RETURN]; CLOSURE [$\mathcal{C}$(\texttt{Apply(Lambda(Op(Var 1, Add, Int 1), Int 2))}); RETURN]; ACCESS 1; INT 4; EQ; IF; ENDLET

\smallskip
\hspace{10mm} Rule $\mathcal{C}$-\texttt{Apply}
\smallskip

\noindent INT 5; LET; CLOSURE [INT 1; RETURN]; CLOSURE [$\mathcal{C}$(\texttt{Lambda(Op(Var 1, Add, Int 1))}; $\mathcal{C}$(\texttt{Int 2}); APPLY; RETURN]; ACCESS 1; INT 4; EQ; IF; ENDLET

\smallskip
\hspace{10mm} Rule $\mathcal{C}$-\texttt{Int} then Rule $\mathcal{C}$-\texttt{Lambda} then Rule $\mathcal{C}$-\texttt{Op} then Rule $\mathcal{C}$-\texttt{Var} then Rule $\mathcal{C}$-\texttt{Int}
\smallskip

\noindent INT 5; LET; CLOSURE [INT 1; RETURN]; CLOSURE [CLOSURE [ACCESS 1; INT 1; OP +; RETURN]; INT 2; APPLY; RETURN]; ACCESS 1; INT 4; EQ; IF; ENDLET

}
{\centering\singlespacing\noindent\small\label{F12} Figure 12. Compilation of the program {\small\texttt{\textbf{let}\! x\! =\! 5\! \textbf{in}\! \textbf{if}\! x\! =\! 4\! \textbf{then}\! 1\! \textbf{else}\! (\textbf{fun}\! x\! ->\! x\! +\! 1)\! 2}} with $\mathcal{C}$.\par}
\label{compilation}
\end{landscape}
\end{sidewaysfigure}

\begin{sidewaysfigure}
\begin{landscape}

{\small
\begin{tabular}{l||p{9.5cm}|l|p{9.5cm}}
& \multicolumn{3}{c}{Machine state after}                       \\\\
Instruction & Code      & Env   & Stack\\
   -        & INT 5; LET; CLOSURE [INT 1; RETURN]; CLOSURE [CLOSURE [ACCESS 1; INT 1; OP +; RETURN]; INT 2; APPLY; RETURN]; ACCESS 1; INT 4; EQ; IF; ENDLET & \{\}  & \{\}\\

INT & LET; CLOSURE [INT 1; RETURN]; CLOSURE [CLOSURE [ACCESS 1; INT 1; OP +; RETURN]; INT 2; APPLY; RETURN]; ACCESS 1; INT 4; EQ; IF; ENDLET& \{\}   & \{5\}\\

LET & CLOSURE [INT 1; RETURN]; CLOSURE [CLOSURE [ACCESS 1; INT 1; OP +; RETURN]; INT 2; APPLY; RETURN]; ACCESS 1; INT 4; EQ; IF; ENDLET & \{5\}   & \{\}\\

CLOSURE & CLOSURE [CLOSURE [ACCESS 1; INT 1; OP +; RETURN]; INT 2; APPLY; RETURN]; ACCESS 1; INT 4; EQ; IF; ENDLET & \{5\}   & \{[INT 1; RETURN]\{5\}\}\\

CLOSURE & ACCESS 1; INT 4; EQ; IF; ENDLET & \{5\} & \{[CLOSURE [ACCESS 1; INT 1; OP +; RETURN]\{5\}; INT 2; APPLY; RETURN]; [INT 1; RETURN]\{5\}\}\\

ACCESS & INT 4; EQ; IF; ENDLET & \{5\}   & \{5; [CLOSURE [ACCESS 1; INT 1; OP +; RETURN]\{5\}; INT 2; APPLY; RETURN]; [INT 1; RETURN]\{5\}\}\\

INT & EQ; IF; ENDLET & \{5\}   & \{4; 5; [CLOSURE [ACCESS 1; INT 1; OP +; RETURN]\{5\}; INT 2; APPLY; RETURN]; [INT 1; RETURN]\{5\}\}\\

EQ & IF; ENDLET & \{5\}   & \{false; [CLOSURE [ACCESS 1; INT 1; OP +; RETURN]\{5\}; INT 2; APPLY; RETURN]; [INT 1; RETURN]\{5\}\}\\

IF & CLOSURE [ACCESS 1; INT 1; OP +; RETURN]; INT 2; APPLY; RETURN & \{5\} & \{[ENDLET]; \{5\}\}\\

CLOSURE & INT 2; APPLY; RETURN & \{5\} & \{[ACCESS 1; INT 1; OP +; RETURN]\{5\}; [ENDLET]; \{5\}\}\\

INT & APPLY; RETURN & \{5\} & \{2; [ACCESS 1; INT 1; OP +; RETURN]\{5\}; [ENDLET]; \{5\}\}\\

APPLY & ACCESS 1; INT 1; OP +; RETURN & \{2; 5\} & \{[RETURN]; \{5\}; [ENDLET]; \{5\}\}\\

ACCESS & INT 1; OP +; RETURN & \{2; 5\} & \{2; [RETURN]; \{5\}; [ENDLET]; \{5\}\}\\

INT & OP +; RETURN & \{2; 5\} & \{1; 2; [RETURN]; \{5\}; [ENDLET]; \{5\}\}\\

OP & RETURN & \{2; 5\} & \{3; [RETURN]; \{5\}; [ENDLET]; \{5\}\}\\

RETURN & EMPTY & \{2; 5\} & \{3; [RETURN]; \{5\}; [ENDLET]; \{5\}\}\\

RETURN & EMPTY & \{5\} & \{3; [ENDLET]; \{5\}\}\\

ENDLET & EMPTY & \{\} & \{3\}\\
EMPTY
\end{tabular}
}
\bigskip

{\singlespacing\noindent\small\label{F13}Figure 13. Evaluation of {\small\texttt{\textbf{let}\! x\! =\! 5\! \textbf{in}\! \textbf{if}\! x\! =\! 4\! \textbf{then}\! 1\! \textbf{else}\! (\textbf{fun}\! x\! ->\! x\! +\! 1)\! 2}} under $\mathcal{E}$. Stacks and environments are written \{items\}, and a closure on the stack is written [instructions]\{environment\}. Environments may be put on the stack.\par}


\label{evaluation}
\end{landscape}
\end{sidewaysfigure}

\begin{sidewaysfigure}
\begin{landscape}

{\scriptsize

\noindent $\mathcal{D}$(INT 5; LET x; CLOSURE [INT 1; RETURN]; CLOSURE [CLOSURE [ACCESS (x, 1); INT 1; OP +; RETURN]; INT 2; APPLY; RETURN]; ACCESS (x, 1); INT 4; EQ; IF; ENDLET, \{\})

\smallskip
\hspace{10mm} \noindent Rule $\mathcal{D}$-CONST
\smallskip

\noindent $\mathcal{D}$(LET x; CLOSURE [INT 1; RETURN]; CLOSURE [CLOSURE [ACCESS (x, 1); INT 1; OP +; RETURN]; INT 2; APPLY; RETURN]; ACCESS (x, 1); INT 4; EQ; IF; ENDLET, \{\texttt{Int 5}\})

\smallskip
\hspace{10mm} \noindent Rule $\mathcal{D}$-LET
\smallskip

\noindent \texttt{Let(x, }\texttt{Int 5}, $\mathcal{D}$(CLOSURE [INT 1; RETURN]; CLOSURE [CLOSURE [ACCESS (x, 1); INT 1; OP +; RETURN]; INT 2; APPLY; RETURN]; ACCESS (x, 1); INT 4; EQ; IF; ENDLET, \{\})\texttt{)}

\smallskip
\hspace{10mm} \noindent Rule $\mathcal{D}$-CLOSURE
\smallskip

\noindent \texttt{Let(x, }\texttt{Int 5}, $\mathcal{D}$(CLOSURE [CLOSURE [ACCESS (x, 1); INT 1; OP +; RETURN]; INT 2; APPLY; RETURN]; ACCESS (x, 1); INT 4; EQ; IF; ENDLET, \{CLOSURE [INT 1; RETURN]\})\texttt{)}

\smallskip
\hspace{10mm} \noindent Rule $\mathcal{D}$-CLOSURE
\smallskip

\noindent \texttt{Let(x, }\texttt{Int 5}, $\mathcal{D}$(ACCESS (x, 1); INT 4; EQ; IF; ENDLET, \{ CLOSURE [ACCESS (x, 1); INT 1; OP +; RETURN; INT 2; APPLY; RETURN]; CLOSURE [INT 1; RETURN]\})\texttt{)}

\smallskip
\hspace{10mm} \noindent Rule $\mathcal{D}$-ACCESS
\smallskip

\noindent \texttt{Let(x, }\texttt{Int 5}, $\mathcal{D}$(INT 4; EQ; IF; ENDLET, \{\texttt{VarAccess(x, 1)}; CLOSURE [ACCESS (x, 1); INT 1; OP +; RETURN; INT 2; APPLY; RETURN]; CLOSURE [INT 1; RETURN]\})\texttt{)}

\smallskip
\hspace{10mm} \noindent Rule $\mathcal{D}$-INT
\smallskip

\noindent \texttt{Let(x, }\texttt{Int 5}, $\mathcal{D}$(EQ; IF; ENDLET, \{\texttt{Int 4}; \texttt{VarAccess(x, 1)}; CLOSURE [ACCESS (x, 1); INT 1; OP +; RETURN]; INT 2; APPLY; RETURN; CLOSURE [INT 1; RETURN]\})\texttt{)}

\smallskip
\hspace{10mm} \noindent Rule $\mathcal{D}$-EQ
\smallskip

\noindent \texttt{Let(x, }\texttt{Int 5}, $\mathcal{D}$(IF; ENDLET, \{\texttt{Eq(VarAccess (x, 1), Int 4)};  CLOSURE [ACCESS (x, 1); INT 1; OP +; RETURN]; INT 2; APPLY; RETURN; CLOSURE [INT 1; RETURN]\})\texttt{)}

\smallskip
\hspace{10mm} \noindent Rule $\mathcal{D}$-IF
\smallskip

\noindent \texttt{Let(x, }\texttt{Int 5}, $\mathcal{D}$(ENDLET, \{\texttt{If(Eq(VarAccess (x, 1), Int 4)}, $\mathcal{D}$(CLOSURE [ACCESS (x, 1); INT 1; OP +; RETURN]; INT 2; APPLY; RETURN, \{\}), $\mathcal{D}$(CLOSURE [INT 1; RETURN], \{\})\texttt{)}

\smallskip
\hspace{10mm} \noindent Rule $\mathcal{D}$-ENDLET
\smallskip

\noindent \texttt{Let(x, Int 5, If(Eq(VarAccess (x, 1), Int 4)}, $\mathcal{D}$(CLOSURE [ACCESS (x, 1); INT 1; OP +; RETURN]; INT 2; APPLY; RETURN, \{\}), $\mathcal{D}$(CLOSURE [INT 1; RETURN], \{\})\texttt{)}

\smallskip
\hspace{10mm} \noindent Rule $\mathcal{D}$-CLOSURE then $\mathcal{D}$-INT then $\mathcal{D}$-RETURN
\smallskip

\noindent \texttt{Let(x, Int 5, If(Eq(VarAccess (x, 1), Int 4)}, $\mathcal{D}$(CLOSURE [ACCESS (x, 1); INT 1; OP +; RETURN]; INT 2; APPLY; RETURN, \{\}), \texttt{Int 1})\texttt{)}

\smallskip
\hspace{10mm} \noindent Rule $\mathcal{D}$-CLOSURE
\smallskip

\noindent \texttt{Let(x, Int 5, If(Eq(VarAccess (x, 1), Int 4)}, $\mathcal{D}$(INT 2; APPLY; RETURN, \{CLOSURE [ACCESS (x, 1); INT 1; OP +; RETURN]\}), \texttt{Int 1})\texttt{)}

\smallskip
\hspace{10mm} \noindent Rule $\mathcal{D}$-INT
\smallskip

\noindent \texttt{Let(x, Int 5, If(Eq(VarAccess (x, 1), Int 4)}, $\mathcal{D}$(APPLY; RETURN, \{\texttt{Int 2}; CLOSURE [ACCESS (x, 1); INT 1; OP +; RETURN]\}), \texttt{Int 1})\texttt{)}

\smallskip
\hspace{10mm} \noindent Rule $\mathcal{D}$-APPLY
\smallskip

\noindent \texttt{Let(x, Int 5, If(Eq(VarAccess (x, 1), Int 4)}, \ \texttt{Apply(}$\mathcal{D}$(ACCESS (x, 1); INT 1; OP +; RETURN, \{\}), \texttt{Int 2)}, \texttt{Int 1})\texttt{)}

\smallskip
\hspace{10mm} \noindent Rule $\mathcal{D}$-ACCESS then $\mathcal{D}$-INT then $\mathcal{D}$-OP then $\mathcal{D}$-RETURN
\smallskip

\noindent \texttt{Let(x, Int 5, If(Eq(VarAccess (x, 1), Int 4)}, \ \texttt{Apply(Lambda(Op(Var 1, Add, Int 1)), \texttt{Int 2)}, \texttt{Int 1})\texttt{)}}

}
\end{landscape}
{\singlespacing\noindent\small\label{F14}Figure 14. Decompilation under $\mathcal{D}$ of the program {\small\texttt{\textbf{let}\! x\! =\! 5\! \textbf{in}\! \textbf{if}\! x\! =\! 4\! \textbf{then}\! 1\! \textbf{else}\! (\textbf{fun}\! x\! ->\! x\! +\! 1)\! 2}} compiled under $\mathcal{C}$, with an empty stack to begin, since the program is unexecuted.\par}

\label{decomp1}
\end{sidewaysfigure}

\begin{sidewaysfigure}
\begin{landscape}
{\small

\noindent $\mathcal{D}$(IF; ENDLET, {false; [CLOSURE [ACCESS(x, 1); INT 1; OP +; RETURN]; INT 2; APPLY; RETURN]; [INT 1; RETURN]})

\smallskip
\hspace{10mm} \noindent Rule $\mathcal{D}$-IF
\smallskip

\noindent $\mathcal{D}$(ENDLET, \texttt{If (false,} $\mathcal{D}$([INT 1; RETURN], \{\})) $\mathcal{D}$([CLOSURE [ACCESS(x, 1); INT 1; OP +: RETURN]; INT 2; APPLY; RETURN], \{\}), \{\})

\smallskip
\hspace{10mm} \noindent Rule $\mathcal{D}$-ENDLET
\smallskip

\noindent \texttt{If (Bool false,} $\mathcal{D}$([INT 1; RETURN], \{\}), $\mathcal{D}$([CLOSURE [ACCESS(x, 1); INT 1; OP +: RETURN]; INT 2; APPLY; RETURN]\{\}), \texttt{)}

\smallskip
\hspace{10mm} \noindent Rule $\mathcal{D}$-INT then $\mathcal{D}$-RETURN
\smallskip

\noindent \texttt{If (Bool false,} \texttt{Int 1}, $\mathcal{D}$([CLOSURE [ACCESS(x, 1); INT 1; OP +: RETURN]; INT 2; APPLY; RETURN]\{\}),\texttt{)}

\smallskip
\hspace{10mm} \noindent Rule $\mathcal{D}$-CLOSURE
\smallskip

\noindent \texttt{If (Bool false,} \texttt{Int 1}, $\mathcal{D}$([INT 2; APPLY; RETURN], \{[ACCESS(x, 1); INT 1; OP +: RETURN]\})\texttt{)}

\smallskip
\hspace{10mm} \noindent Rule $\mathcal{D}$-INT
\smallskip

\noindent \texttt{If (Bool false,} \texttt{Int 1}, $\mathcal{D}$([APPLY; RETURN], \{\texttt{Int 2}; [ACCESS(x, 1); INT 1; OP +: RETURN]\})\texttt{)}

\smallskip
\hspace{10mm} \noindent Rule $\mathcal{D}$-APPLY
\smallskip

\noindent \texttt{If (Bool false}, \texttt{Int 1}, \texttt{Apply (Lambda(x, }$\mathcal{D}$(ACCESS(x, 1); INT 1; OP +: RETURN, \{\}), \texttt{Int 2))}\texttt{)}

\smallskip
\hspace{10mm} \noindent Rule $\mathcal{D}$-ACCESS
\smallskip

\noindent \texttt{If (Bool false,} \texttt{Int 1}, \texttt{Apply (Lambda(x, }$\mathcal{D}$(INT 1; OP +; RETURN, \{\texttt{VarAccess(x, 1)}\}), \texttt{Int 2))}\texttt{)}

\smallskip
\hspace{10mm} \noindent Rule $\mathcal{D}$-INT
\smallskip

\noindent \texttt{If (Bool false,} \texttt{Int 1}, \texttt{Apply (Lambda(x, }$\mathcal{D}$(OP +; RETURN, \{\texttt{Int 1; VarAccess(x, 1)}\}), \texttt{Int 2))}\texttt{)}

\smallskip
\hspace{10mm} \noindent Rule $\mathcal{D}$-OP
\smallskip

\noindent \texttt{If (Bool false,} \texttt{Int 1}, \texttt{Apply (Lambda(x, }$\mathcal{D}$(RETURN, \{\texttt{Op(VarAccess(x, 1), Add, Int 1)}\}), \texttt{Int 2))}\texttt{)}

\smallskip
\hspace{10mm} Rule $\mathcal{D}$-RETURN
\smallskip

\noindent \texttt{If (Bool false,} \texttt{Int 1}, \texttt{Apply (Lambda(x, Op(VarAccess(x, 1), Add, Int 1)}, \texttt{Int 2))}\texttt{)}

\bigskip

\noindent This is the program {\small\texttt{\textbf{if}\! false\! \textbf{then}\! 1\!  \textbf{else}\! (\textbf{fun}\! x\! ->\! x\! +\! 1)\! 2}}, as required.

}
\end{landscape}
{\centering\singlespacing\noindent\small\label{F15} Figure 15. Decompilation under $\mathcal{D}$ of the program {\small\texttt{\textbf{let}\! x\! =\! 5\! \textbf{in}\! \textbf{if}\! x\! =\! 4\! \textbf{then}\! 1\! \textbf{else}\! (\textbf{fun}\! x\! ->\! x\! +\! 1)\! 2}} compiled under $\mathcal{C}$, and partly evaluated with $\mathcal{D}$.\par}
\label{decomp2}
\end{sidewaysfigure}

\chapter{Evaluation}
\label{chap:evaluation}
\begin{quotation}\textit{\large In Africa a thing is true at first light and a lie by noon and you have no more respect for it than for the lovely, perfect weed-fringed lake you see across the sun-baked salt plain. You have walked across the plain in the morning and you know that no such lake is there. But now it is there absolutely true, beautiful and believable.\textrm{\begin{flushright}--- \textup{Hemingway}, True at First Light\end{flushright}}}\end{quotation}

\vspace{10pt}

\noindent We have looked again at the problem of debugging, identified what we believe to be a step forward which may lead to more people using debugging tools, and built a proof-of-concept debugger for the functional language OCaml. As we look back upon the prototype we have built and plan how to turn it into a finished and usable tool, can we evaluate what has been done to inform our future work? By what criteria might we measure success?

The primary measure, of course, is whether the tool, once finished, is widely used. There is little we can do about this type of evaluation now, but it must remain our most important concern. Quantitatively, we can measure two things: how many people use the tool in preference to another, and how many in preference to nothing at all or more often than they used their previous tool. Does it replace or merely complement other tools? 

For now, we can evaluate only what has been done, and the design of the whole tool as it is envisaged. We shall do this by qualitatively measuring our progress against our stated aims. First, we shall give a broad, narrative discussion of the results of our work, to provide a general assessment. Then, we shall use more structured approaches -- comparing our progress with the research questions, and against various criteria for successful debuggers given in the literature. Then, we will discuss what methods of evaluation might be possible once the work is more advanced. Finally, we give details of the implementation status of our work.

\subsection*{\addition{Summary of Work}}
\addition{In chapter 1, we describe twin motivations for the work: experience in teaching and in real-world programming tasks. We decide upon research questions, and a thesis. Chapter 2 contains two literature reviews: of historical literature in debugging and of debugging today, both in functional and imperative settings. In chapter 3 we decide upon an approach to the work, taking lessons from our literature review. We review literature in the related field of software visualization. Chapter 4 describes our first, experimental implementation of a step-by-step interpreter and explores some of its difficulties and deficiencies. In chapter 5, we re-assess the interpreter from the previous chapter and produce a new design with numerous improvements, such that it constitutes a practical system. Chapter 6 describes an annotation-based design for an interface to our interpreter, forming the heart of our eventual approach to debugging. In chapter 7, we talk about pieces of work which did not make the final cut, and why they did not. In this chapter, chapter 8, we evaluate the work and update the research questions. Finally, in chapter 9, we give concluding remarks.}

\addition{A statement of our contributions to research is given in section 1.5.}

\section{Narrative discussion}

Our plan was to look at the literature and practice of debugging, trying to discern those qualities which separate a debugger which is useful and used from one which lies unused. This was motivated by observing the widespread feeling amongst programmers that using a debugger would be more common if only it were more easily applicable to their problems. That is to say, the feeling that there is nothing fundamentally impossible about producing a widely used debugger. We looked at this in the context of \index{functional programming!vs  imperative programming}functional programming especially, working on the assumption that functional programming is different enough from imperative programming that there are likely to be significant differences in the debugging process.

We have identified what we believe to be the key requirements for a usable debugger -- that it should be available all the time, whenever the programmer needs it, and that it should be sufficiently flexible so as to be unobtrusive when not in use. There are, of course, many other requirements for a good debugger. But we have claimed that without this accessibility requirement being fulfilled, the rest is in vain. 

We chose a radical approach to these key requirements: to build an interpreter for our chosen language, OCaml. The supposed advantages were that this would result in accessibility by default, that there would be no information loss (since there is no compilation process), and that the obvious downside of interpretation -- slowness -- could be mitigated. We produced two systems, \textsf{OCamli} and \textsf{OCamli2/ppx\raisebox{0.5mm}{\_}interpret}. How did we fare? Let us take them in turn.

Our first program, \textsf{OCamli} (see \cref{chap:interpreter}) was written, first, to answer the question ``What would an interpreter for OCaml look like?'', and second to begin to explore the design of a visualizer -- and thereby a debugger -- for OCaml programs. However, it fails on several counts to pass our tests of what a good debugger would look like -- and not only because it does not yet support the whole language. Let us suppose that \textsf{OCamli} were to be finished to support the whole of OCaml. What would it still lack, with regard to our principles and, in particular, the tests we set out in \cref{tests}? \textsf{OCamli}, as presently constructed, fails the most important of our  tests. It cannot be used as an alternative to the compiler except for the simplest of projects, and cannot be used with mixed C/OCaml code. These are design flaws, and impact upon usability -- in particular the notion of accessibility we have been concerned with. It is clear that the other test we set, about being easy to keep in sync with the \index{toolchain}toolchain, has not been fully achieved at this stage. Using the library form of the OCaml compiler helps us a lot, but there would still be significant work to make the software function with each new compiler version.

Let us compare the \textsf{OCamli} model with the typical debugging methods we have talked about, both in imperative and functional languages: \index{breakpoint}breakpointing, \index{debugging!by print statement}\index{print statement}inserting print statements, using a \index{REPL}REPL, and \index{tracing}tracing. It is clear that breakpointing, either in its traditional form, or by search mechanisms, can be replicated. \textsf{OCamli} effectively introduces universal printing of values (which OCaml does not have by default), a clear boon for debugging. However, the \textsf{OCamli} command line interface is clearly inferior to the REPL (save for the ability to show the steps of evaluation). An interactive REPL-like version of \textsf{OCamli} should be explored as an alternative debugging interface. We consider the online behaviour of \textsf{OCamli}-style interpretation as likely more useful than offline tracing (e.g.\ Haskell's Hood \cite{Gill00debugginghaskell}), due to the ability of the user to refine and experiment interactively.

\textsf{OCamli}, then, is an interesting exposition of our central idea of debugging by interpretation, but flawed with regard to usability, both in its choice of interface, and by dint of its failure to address fully the notion of \index{accessibility}accessibility.

\correction{Let us now consider our second program \textsf{OCamli2/ppx\raisebox{0.5mm}{\_}interpret} (see chapters \ref{chap:efficient} and \ref{chap:interface}) in the same way. This, we believe, will meet our principles of usability when it is complete. It should make a lightweight debugging mechanism which may be invoked at will no matter the environment, build system, or other circumstances. It meets the tests set out in section \ref{tests}, due to a happy coincidence in the way that the \index{PPX}PPX mechanism works. It can be used with any build system, works with mixed C/OCaml code, is relatively simple to keep in sync with compiler toolchain releases, requires only a tiny patch to OCaml itself, and is flexible enough that it should be suitable for debugging large projects, even the OCaml toolchain itself. Just like \textsf{OCamli}, \textsf{ppx\raisebox{0.7mm}{\_}interpret} provides the opportunity for universal printing, exploration of the steps of evaluation, and opportunities for interaction. We have identified a number of encouraging avenues for interface design of \textsf{ppx\raisebox{0.7mm}{\_}interpret} functionality (see chapter 6), and we would expect some or many of those options to provide a pleasurable debugging experience. So, \textsf{ppx\raisebox{0.7mm}{\_}interpret} is promising, and a definite improvement to \textsf{OCamli} with regard to our aims and the tests which we have set ourselves.}

\section{With regard to teaching}

The impetus for this research was the author's experience of \index{functional programming!teaching}\index{teaching functional programming}teaching functional programming to beginners, some of whom had never written a computer programs before. The step-by-step diagrams which our interpreter produces are frequently drawn out by hand in such lessons because it demonstrates the semantic operation of the program text so well. However, once we decided our focus must be debugging for the working programmer, we hoped merely that the teaching uses would be subsumed by the work on debugging -- that a good debugger for the working programmer could be a good debugger for the student. Has that happened? The prototype \textsf{OCamli} has been written to cover enough of the language to execute almost all the programs in the author's own introductory OCaml textbook ``OCaml from the Very Beginning'' \cite{whitington2013ocaml}, and so we can try it on those programs. Whilst it is useful for many of them, the evaluator's ability at elision is plainly not good enough to replace pen and paper examples properly yet. This is not surprising, though, since we curtailed our work on visualization somewhat to attack the central problem of debugging more thoroughly.

However, we believe the problems are solvable, and therefore, that our contention that our design of debugger (both in the form of \textsf{OCamli} and \textsf{ppx\raisebox{0.8mm}{\_}interpret}) will eventually be useful to the beginner as well as the experienced programmer will turn out to be substantially true. 

\section{Against the research questions and claims}

\correction{In \cref{chap:introduction}, we wrote down two separate lists. The first list consisted of a number of research questions, decided right at the beginning of the project, and modified since only by a little deletion for time pressure; the text of the remaining entries has not been changed. The second was a list of `claims', which was written after significant prototyping of the interpreter, but still only part way through the research presented here. It remains unmodified from that time. We summarised the claims as our thesis: \textit{an interpretive debugger for a functional language is technically practicable and might be expected to make debugger use more widespread, routine and productive.} What we shall do now is to look back on those two lists, and discuss to what extent the questions were answered or claims met as the research stands now, and as it may stand in the future. \addition{In addition, we shall decide upon new or modified research questions and claims which we can use to take the work forward, and set them in context.}}

\subsection{The research questions}\index{research questions}

\setlength{\fboxsep}{8pt}

Taking the research questions in turn:

\begin{quotation}
\noindent\textbf{It appears that debuggers are not as widely used as one might expect, despite being common for decades. Why?} This appears to be doubly true for functional programming languages. How does debugging practice vary among languages (compiled vs interpreted, stateful vs stateless)? What can we learn from debugging theory and practice since the dawn of the computer age? \textup{\addition{[original research question]}}
\end{quotation}

\noindent We gave an account of the surprising lack of widespread use of debuggers in our literature review in \cref{chap:related}. We found debugger use did, indeed, seem to be even less widespread in the functional programming community. Many functional programmers never use a debugger, even where the toolchain provides it. We also discovered research over several decades discussing the problem, including much recent work in the \index{Haskell}Haskell community, which seemed to be very well thought out with regard to the kinds of usability concerns which turn out to be central, we believe, to the lack of use of debuggers.

The difficulty here, of course, is that we might ask ``If these problems are so longstanding, why would you believe you can solve them?''. The answer, first, is that we do not seek some miraculous, instant solution to the whole field of debugging, but merely a concrete step forward, built on proper foundations. Second, whilst we can never be sure of the efficacy of our approach until the tool is finished and widely used, we can -- thanks to the breadth of our literature review -- be sure that our thoughts on the nature and character and deficiencies of debugging are not unusual, but shared by many through the years. So, we can be confident we are attacking the right problem, and have some confidence we are attacking it in the right way. Of course, there is a small chance that the dream of the usable, applicable, universal debugger which we described is simply a mirage. But we believe that, given the advances in so much of our field, that much more exploration of debugging is required before giving too much credence to that distressing conclusion. \addition{Now that we are moving on to the next stage of this project, to finish the implementation and release a debugger which can be used and tested, we replace this research question with a more practical one, which may be more tangibly evaluated in the future. This replaces our wide questions about the history of debugging with somewhat more narrow ones about the utility of our creation:}

\begin{quotation}
\noindent\addition{\textbf{Is our debugger suitable for widespread use?} Is it used by beginners when they come into contact with the language? Is it used by ordinary working programmers? What proportion of such users? Do they consider it better than existing solutions for debugging OCaml? Than simple solutions like print statements? Do they use it only occasionally, treating debugging as a separate activity, or seamlessly as part of everyday programming activities? \textup{[new research question]}}
\end{quotation}

\noindent\addition{These are the key questions when we come to evaluate, in the round, our final implementation, once it is released and in use. Here is the next original research question:}

\begin{quotation}\noindent\textbf{Can we find a good way to visualize functional program execution?} Is the automatic production of such diagrams always going to be inferior to drawing them on paper? How can we deal with scale? How can we show exceptions? What about imperative and mutable features? What are the practicalities of directly interpreting an Abstract Syntax Tree? Can the direct interpretation of the AST of a program ever or always have the same time or space complexity as running the compiled program? \textup{\addition{[original research question]}}
\end{quotation}

\noindent We had, right at the beginning of this piece of research, intended it to be all about the \index{visualization}visualization of functional program execution, with a view to debugging. It became clear fairly quickly, though, that the challenge of debugging itself should be central. Thus, we have not done as much work on visualization itself as we might otherwise have. Nonetheless, the \textsf{OCamli} prototype addresses a number of questions of scale, and of the visualization of various tricky aspects like mutable state. For a final implementation of our ideas, especially after practical use, it would be appropriate to revisit this work.

We have explored the efficient implementation of a step-by-step interpreter but stopped well short of proving mathematically that such an interpreter can or cannot preserve the \index{time complexity}\index{space complexity}time and space complexity of the program it is interpreting. When such an interpreter is used for debugging, we believe that reducing the part of the program which need be interpreted (i.e.\ using \textsf{ppx\raisebox{0.8mm}{\_}interpret}) reduces the immediate concern with speed, by side-stepping it. But a robust final implementation would have to address the issue of efficiency with more formality. \addition{What is left unanswered of this research question and what must be altered? We have not adequately addressed in this thesis the time and space complexity. We have discussed it in general terms, but achieved nothing formal. So this must remain. The rest of the research question has been altered to shift the emphasis to the annotation-based debugging interface described in chapter 6.}

\begin{quotation}\noindent\addition{\textbf{What are the visualization characteristics of the annotation-based debugger in practice?} Is the automatic production of such diagrams always going to be inferior to drawing them on paper? How can we deal with scale in the annotation scenario, including a proper sense of granularity?  Can the direct interpretation of the AST of a program ever or always have the same time or space complexity as running the compiled program? In any case, does the selective nature of the annotation-based system obviate or reduce the need for the time and space complexity to be as good as compiled code? \textup{\addition{[new research question]}}}
\end{quotation}

\noindent\addition{These questions are somewhat related to the usability ones in the rewritten first research question above, but the emphasis on the characteristics of the visualization itself is, we think, worth keeping separate.}

\begin{quotation}
\noindent\textbf{Is such an interpreter useful for debugging?} Taking into account of the current practice of debugging, is an interpretive debugger better than what is already available? If so, why, and in what ways? \textup{\addition{[original research question]}}
\end{quotation}

\noindent We believe we made a good case that this is (or will be) true in chapter \ref{chap:interface}. The primary advantages are the accessibility of the debugger -- that it is available whenever required, does not need complex setup and is applicable to all programs no matter how compiled and that, in the form of \textsf{ppx\raisebox{0.8mm}{\_}interpret}, it mimics the common ``insert print statements'' method of debugging. In addition, we have shown how \textsf{ppx\raisebox{0.8mm}{\_}interpret} can encapsulate many techniques from traditional debuggers  such as breakpoints and selective printing and interactive debugging. That such flexibility is available in what is, in some ways, quite a novel implementation of a debugger is, we believe, a sign of success. \addition{This research question is subsumed into our first rewritten question above, so is deleted. Here is the next original research question:}

\begin{quotation}\noindent\textbf{Is there an alternative abstract machine which might allow for this kind of visualized debugging?} That is to say, are we condemned to interpret the AST in the simplest way if we want to be able to properly visualize the evaluation in a human-readable manner? We know that this may have greater complexity of running time than well-known abstract machines. Can we design a bytecode that retains the ability to produce source code for the running computation upon demand, but which is much faster than brute interpretation and maybe even close to that of a normal bytecode system? \textup{\addition{[original research question]}}
\end{quotation}

\noindent We briefly addressed this in \cref{chap:misc} by experimenting with a way to modify OCaml's bytecode interpreter to print steps of execution and found the results interesting but probably unworkable. We discussed abstract machine options in \cref{chap:efficient}. The question needs to be looked at again, especially if we cannot make the debugger efficient enough as a simple interpreter in practice. So we may leave this research question unaltered. \addition{Here is the last original research question:}

\begin{quotation}\noindent\textbf{Could we, instead, build an interpreter which can work alongside the native code execution of a program, interpreting only when required?} We could compile a program in a slightly different manner. It would run as usual, but when it comes to a part we wish to debug, it would begin interpretation. After this part, it would return to native code, as if nothing had happened. \textup{\addition{[original research question]}}
\end{quotation}

\noindent Indeed we can. And it turned out easier than expected, due to the recently-introduced PPX  mechanism in OCaml. We believe it will be the final interface to our debugger. \addition{We update the research question to remove the part already completed, and to include more formal criteria for success of the next stage of our work:}

\begin{quotation}\noindent\addition{\textbf{Could we, instead, build an interpreter which can work alongside the native code execution of a program, interpreting only when required?} Can we show formally (or convincingly) that it is robust with respect to the OCaml runtime, multithreaded environments, and the use of external linked code? Can we build a system which allows the interpreted section of code to be selected at runtime rather than compile time? \textup{\addition{[new research question]}}}\end{quotation}

\noindent\addition{Here are the new research questions in one place:}

\begin{quotation}
\noindent\addition{\textbf{Is our debugger suitable for widespread use?} Is it used by beginners when they come into contact with the language? Is it used by ordinary working programmers? What proportion of such users? Do they consider it better than existing solutions for debugging OCaml? Than simple solutions like print statements? Do they use it only occasionally, treating debugging as a separate activity, or seamlessly as part of everyday programming activities?}
\end{quotation}

\begin{quotation}\noindent\addition{\textbf{What are the visualization characteristics of the annotation-based debugger in practice?} Is the automatic production of such diagrams always going to be inferior to drawing them on paper? How can we deal with scale in the annotation scenario, including a proper sense of granularity?  Can the direct interpretation of the AST of a program ever or always have the same time or space complexity as running the compiled program? In any case, does the selective nature of the annotation-based system obviate or reduce the need for the time and space complexity to be as good as compiled code?}
\end{quotation}

\begin{quotation}\noindent\addition{\textbf{Is there an alternative abstract machine which might allow for this kind of visualized debugging?} That is to say, are we condemned to interpret the AST in the simplest way if we want to be able to properly visualize the evaluation in a human-readable manner? We know that this may have greater complexity of running time than well-known abstract machines. Can we design a bytecode that retains the ability to produce source code for the running computation upon demand, but which is much faster than brute interpretation and maybe even close to that of a normal bytecode system?}
\end{quotation}

\begin{quotation}\noindent\addition{\textbf{Could we, instead, build an interpreter which can work alongside the native code execution of a program, interpreting only when required?} Can we show formally (or convincingly) that it is robust with respect to the OCaml runtime, multithreaded environments, and the use of external linked code? Can we build a system which allows the interpreted section of code to be selected at runtime rather than compile time?}\end{quotation}

\noindent Having discussed the research questions, we move on to our more explicit claims, made after initial exploratory research had been performed.

\subsection{The claims}

\noindent We cannot say we have proven these claims, because many of them are social not scientific, only that we have provided some evidence for them.

\begin{quotation}
\noindent That not many programmers use debuggers, even though they exist.\end{quotation}

\noindent The literature review of \cref{chap:related} makes it clear that this is a perennial problem.

\begin{quotation}
\noindent That (almost) everyone could benefit from a debugger.\end{quotation}

\noindent It is harder to make the case for this in definite terms, but it seems to be widely accepted as an assumption by authors we excerpt in the literature review.

\begin{quotation}
\noindent
That the reasons for this disparity are frequently incidental rather than intrinsic and include:

\begin{itemize}
\item The inability of debuggers working on compiled programs to properly reflect their workings at a source code level.
\item The requirement to learn a new tool at exactly the moment one is trying to fix a bug.\end{itemize}\end{quotation}

\noindent This is largely backed up by the literature review and the results of our programmer survey and others. If we were to rewrite this statement now, with the benefit of hindsight from our widespread reading, we might formulate it differently, emphasising different priorities. And, we cannot say with certainty that those reasons listed are exhaustive. Each time a new debugger is launched on the unsuspecting programming public and they toss it to one side, we learn a new lesson.

\begin{quotation}
\noindent That many of these barriers melt away in the presence of an interpreter ranking equally to a native compiler, with the same language and toolchain support.\end{quotation}

\noindent Perhaps ``melt away'' is too strong a phrase. But we have managed to produce a design in \textsf{ppx\raisebox{0.8mm}{\_}interpret} which could indeed rank equally to a native compiler, and we expect this to solve (or almost solve) the problems given in the two bullet points above.

\begin{quotation}
\noindent That the huge disadvantage of slowness which comes with this approach:

\begin{itemize}
\item Can be ameliorated more that one might expect.
\item In any case, is not a show-stopper for most uses, since the external steps of debugging such as case reduction are still in place.
\item May be obviated by finding a way to produce mixed native/interpreted programs so that, in any case, the need for interpretation is much reduced.
\end{itemize}
\end{quotation}

\noindent As we have already mentioned, it may still be the case that we need a more radical approach to the problem of speed, such as a specialised \index{abstract machine}abstract machine which can display steps of evaluation, rather than a simple interpreter. But, we think that a combination of our mixed native/interpretative model amounts to much the same thing: if the part of the program we interpret is too slow, it is probably producing too much output to read sensibly, so we expect to narrow the bug down further anyway. Again, a final determination will require widespread use.

\begin{quotation}
\noindent That such an interpretive approach is particularly suited to functional programs due to the mental model of calculation.\end{quotation}

\noindent\correction{This is a little harder to be certain of: we stopped our visualization research short when we realised that the problem of debugging itself was worth tackling. But personal experience of teaching students using such step-by-step evaluation models is that they are the one of the most useful pedagogical techniques, at least for beginners. And as we have said, and as is supported by the literature review of chapter 2, there is much in common between the beginning programmer and the experienced programmer when tackling a bug.}

And so, what of our thesis? \textit{An interpretive debugger for a functional language is technically practicable and might be expected to make debugger use more widespread, routine and productive.} Technically practicable? Yes. Expected to make debugger use more widespread? At least for OCaml, we think so. Routine and productive? Perhaps, but a final determination will have to wait until the system is in use. Or not in use, as the case may be.

\addition{What criteria could be used in the future, when the project is further forward? Here are  new, updated claims, based upon an understanding and evaluation of the work so far. We would like to use to move this research forward to the next stage, towards a real implementation. We claim:}

\begin{itemize}
\item \addition{that the new OCaml interpreter, \textsf{OCamli2}, when finished:}

\begin{itemize}
      \item \addition{can be shown informally to be correct.}
      \item \addition{can support the whole language.}
      \item \addition{is performant, to the extent which can be expected.}
\end{itemize}

\item \addition{that the integration of our interpreter into the toolchain:}

\begin{itemize}
      \item \addition{is technically possible.}
      \item \addition{is robust against changes in each version of OCaml, requiring only minimal alterations.}
\end{itemize}

\item \addition{that the annotation-based interface we have described in chapter 6 provides a natural, malleable interface to debugging functional programs which:}

\begin{itemize}
    \item \addition{is likely to be widely used, compared with existing OCaml debugging tools.}
    \item \addition{suitable for both beginners and experts.}
    \item \addition{subsumes existing techniques such as the insertion of print statements.}
    \item \addition{scales to large programs, including the OCaml compiler itself.}
\end{itemize}

\end{itemize}

\noindent\addition{Evaluation of these claims can proceed fully only when we have a working, used implementation.}

\section{Against the literature}

In \cref{chap:related} we gave an in-depth review of the debugging literature of the last seventy years. It is instructive in our evaluation to go back to this literature. We shall re-examine the section we entitled \textit{``What makes a good debugger''} to see if we can compare our solution against the rules set by the researchers we drew inspiration from. Of course, we must be careful in this exercise, and make sure it is only one method of evaluation, as Sattherthwaite reminds us:

\begin{quotation}
\textit{Since debugging, as usually understood, is more a practical than a theoretical problem, proposed solutions must be evaluated within a framework of practical constraints.
\cite{satterthwaite1972debugging}}
\end{quotation}

\noindent Let us begin with one of the earliest pieces of source material in our literature review, Ira Deihm's contribution to the 1952 ACM national meeting ``Computer aids to code checking'' \cite{diehm1952computer}. One of our contentions has been that debugging is an abiding and persistent problem, so it makes sense to go back this far.

\begin{quotation}
What one tries to achieve in designing such auxiliary routines is to program the machine to select the pertinent information rather than to read out large quantities of data which must be searched through by the programmer.
\cite{diehm1952computer}
\end{quotation}

\noindent We have already discussed the tasks of reducing the trace output of \textsf{OCamli} by adjusting its default rules, by giving it extra command line flags, and by options which\index{searching} search through the trace explicitly. However, it is far from clear that this will be anywhere near enough to reduce the output of real-world programs in a way which will make debugging as painless as it could be on its own. One of the side-effects of our attempts to deal with the speed problems of interpretation is that \textsf{ppx\raisebox{0.8mm}{\_}interpret} is also a de facto mechanism for trace reduction -- what is not executed by the interpreter is not traced. This, together with \textsf{ppx\raisebox{0.8mm}{\_}interpret}'s effective emulation of print-statement-based debugging will, we believe, be a potent combination.

\begin{quotation}
The degree of difficulty the programmer experiences in isolating a bug once he has noticed an error depends on the nature of the bug and the ease with which he can obtain additional information about intermediate states in the computation.
\cite{Gaines:1969:DCP:905460} 
\end{quotation}

\begin{quotation}
\textit{In a debugging program it is of prime importance that the program be simple, flexible, and highly efficient to use.
\cite{brady1968writing}}
\end{quotation}

\noindent The whole purpose of our system is to give what Gaines describes as ``information about intermediate states in the computation''. As functional programmers, of course, we do not like calling them states -- but even functional programmers think \textit{as if} steps in the evaluation of an expression are states when debugging. If we could solve every bug in a functional program simply by staring at the source code, programmers would need only to locate the buggy function. But in reality, this approach frequently fails and we still need the debugger -- to step through the function. The \textsf{ppx\raisebox{0.8mm}{\_}interpret} system allows the programmer to move, change, add, or delete annotations to find further information, and fine-tune the information presented. We are cautiously optimistic that this approach will produce a pliable debugging tool which will fit the debugging workflow well.

Eisenstadt gives some more detailed qualifications a good debugging system must have, in terms of availability to the programmer, a sort of lightness of touch or unobtrusiveness:

\begin{quotation}
{\noindent\vspace{-7mm}\itshape\begin{itemize}
\item Allow full functionality at all times. Debugging environments that prevent access to certain facilities make matters worse.
\item Viewers should be provided for ``key players'' (any evaluable expression) rather than just for `variables'.
\item Provide a variety of navigation tools at different levels of granularity.\\ \cite{eisenstadt1997my}\end{itemize}}
\end{quotation}

\noindent And Evans and Darley, from thirty years earlier:

\begin{quotation}
\ldots a very selective and close control over the execution of portions of one's program and for the examination of intermediate results, together with the possibility of making on-the-spot changes based on them, as desired.
\cite{evans1966line}
\end{quotation}

\noindent Our \textsf{ppx\raisebox{0.8mm}{\_}interpret} system will, when it is finished, score on all three of Eisenstadt's criteria. On the first, because it is well-integrated into the toolchain, and can be used in any environment where the compiler is available. On the second and third, our system fulfils these by design, since the source code annotation scheme is the very essence of granularity -- the parse tree of a program is a granular structure, and annotations may be attached to any node.

\begin{quotation}
\ldots to facilitate maintenance, the same program was to be useable in both batch and interactive modes. Second, to facilitate distribution, the system had to be useable without any modification to the operating system.
\cite{grishman1970debugging}
\end{quotation}

\noindent Our debugger is, indeed, independent of the operating system, build system, and all other specifics. If a programmer can install the OCaml toolchain on a system (OCaml's only prerequisites are a C compiler and an implementation of {\small\texttt{make}}), then they can use the debugger. It demands nothing else. It will be suitable for use with the bytecode or native code compilers, the REPL, or anywhere else OCaml code can be used.

Looking back at these quotations from the literature review, conducted earlier in our research, gives us some reassurance that our system is in line with the aims we set out when beginning the implementation. Research projects involve, of course, experimentation, dead ends, and so on, but that our implementation echoes the literature review shows that we have not veered too far off course.

\section{Future evaluation}

As we have established, the only real test of our approach is whether OCaml programmers use our debugger in preference to others, or in preference to debugging methods involving no debugger at all. However, evaluation is not just useful as a binary test, but as a qualitative process too, shedding a sidelight on the work's qualities, and one can never know what insights will be gained without conducting such an evaluation. In addition, as we have mentioned before, this ultimate test can only apply when the tool is completely finished -- earlier evaluations help to guide us towards our goal.

\subsection*{Experimental evaluation}

It is worth taking stock of the ways in which other researchers have sought to evaluate debuggers -- both in the context of their own work, and in surveys of other debugging tools or mechanisms. We shall look at five papers, already cited in our literature review, and which contain such evaluations, and discuss to what extent these methods may be applied to our own work.

The paper ``An Analysis of Patterns of Debugging Among Novice Computer Science Students'' \cite{ahmadzadeh2005analysis} is an examination of students' mistakes in debugging their own and others' code. A Java compiler is instrumented to report each syntactic or semantic compiler error, together with the source code and time stamp. In the first phase, students are answering set questions by writing (and therefore debugging) their own programs. In the second, more tightly controlled,  phase, a single Java program containing both compiler errors and errors of logic only was presented to the students, and they were given a fixed time to correct it. The results are analysed quantitatively  by categorisation of the errors and qualitatively to categorise the misconceptions leading to each bug. The primary finding of the second phase of study is, as the authors say, rather surprising: \textit{``\ldots\ the majority (66\%) of the competent debuggers, that is those who were able to correct all three logical bugs, are also competent programmers. In contrast, only 39\% of the competent programmers were also competent debuggers.''}

Second, we consider ``Debugging: An Analysis of Bug-Location Strategies''  \cite{katz1987debugging}. This is another study of students (academics having ready access to students seems to make these studies much more common than industrial ones). In this case, the students are not beginners. Another  modified compiler, this time for Lisp, is used to feed information on student debugging behaviour back to the researchers. The bugs and debugging strategies are categorised to find the sources of errors and to discover how students debug. The authors then compare students debugging their own programs to the same students debugging programs written by others.

Another paper pertaining to debugging as a part of teaching is ``Debugging: The Good, the Bad, and the Quirky -- a Qualitative Analysis of Novices' Strategies'' \cite{Murphy:2008:DGB:1352322.1352191}, which takes a looser approach to setting the experimental parameters. Students were allowed to use any method of debugging, and any online resources they felt they required. The authors point out that, although teachers' interactions with students during lab sessions give plenty of anecdotal evidence about the kind of problems the students face, the actual debugging process is rarely observed. Students were given a program containing logic errors, and the qualitative evaluation took the form of semi-structured interviews with the students afterward, and a one-page survey about the debugging processes used. The data analysis consists of categorisation of the debugging strategies used by the students. The authors conclude that, while students used a variety of debugging strategies with some success, there was a lack of methodical, systematic thinking in the application of the strategy.

The authors of MiraCalc, a step-by-step interpreter for the lazy language Miranda \cite{turner1985miranda} write about an interesting debugging experiment in ``A Symbolic Calculator for Non-Strict Functional Programs'' \cite{miracalc}. The hypothesis is that such step-by-step interpreters ``can have a positive effect on learning formal subjects''. Data on the perceptions of students during their functional programming courses took the form of ``questionnaires, video-taped interviews, diaries and e-mail''. This was combined with data on their exam results, and the two were analysed. This data is partly qualitative and partly quantitative. The data collected in the first stages was used to refine the questionnaire to ``quantify the opinions expressed at interview''. This is an example of a relatively large multi-year study, whose methodology might be adapted to our situation.

Finally, an older but very extensive study of debugging in an industrial setting is ``Expertise in Debugging Computer Programs'' \cite{vessey1985expertise}. Subjects were divided based on ability (expert/novice), and then their debugging strategies evaluated by listening to them speak aloud into a tape recorder as they worked. The program in question was in FORTRAN, with a simple logic error introduced. These tape recordings were then transcribed and analysed. The strategies were compared to the expert/novice categorisation to look for a correlation. The paper is interesting because of its careful approach. For example, subjects debugged example programs at length first, to get used to describing their actions into the tape recorder. However, the authors point out an important limitation of their approach is that the method they used to categorise programmers into novice and expert is not tested independently of the data. Such concerns are common in these studies, together with the perennial issue of small sample sizes.

It is not yet clear, at this point in development, exactly what type of experiment might be suitable for our system, but we can take inspiration from those we have just described. We hope to test our debugger with some formality, in order to compare our system to others, or to see which kinds of bugs, or which kinds of programmers, it is suitable for.

\subsection*{A teaching test}

The author's introductory OCaml textbook \cite{whitington2013ocaml} is used by a number of universities as a recommended text for first-year undergraduates. It contains some diagrams of step-by-step evaluations, but not nearly as many as we might like, for space reasons. The book has numerous examples and exercises with answers. Our interpreter \textsf{OCamli} can already produce step-by-step evaluations for almost all the examples and exercises, but they are too verbose, especially at the default settings.

It would be interesting, when our debugger is complete, to produce accompanying material in the form of a portable web-based interactive workbook after the fashion of the popular iPython \cite{ipython} but with step-by-step evaluations. This should be technically feasible, compiling OCaml programs to JavaScript, for example with js\_of\_ocaml \cite{vouillon2014bytecode}. We could then see how much this added functionality helps learners of OCaml using the textbook, based on self-reporting of both the amount of time spent using the extra resources and how helpful they were found to be, or by reporting from lecturers and supervisors of the help (or lack of help) such an approach gave in teaching and learning.

\subsection*{Against our survey of functional debuggers}

It is not really sensible to compare our incomplete system existing to functional debuggers now, because the software is not finished, but we will be able to do so in the future. Each researcher or programmer who created those systems thought, of course, that their design would be widely used, and yet many are not. Our claims about our design being widely used may still, of course, turn out just as hollow. And so whilst we could compare our debugger's features and functionality against those others,  this would not really be an evaluation as such. We must wait until our debugger is complete. Once it is, we can evaluate against our literature review of existing debuggers for functional languages in the following ways:

\begin{description}
\item[Classification]\index{debugger!classification of a}For what users is each debugger intended? For teaching or the working programmer? How is it invoked? What is the interface? Is it batch or interactive?

\item[Accessibility]\index{accessibility} What demands does it make on the build environment or compilation settings? What platforms does it work on? How must it be updated when the compiler toolchain is updated? Is it the default debugger shipped with a compiler toolchain?

\item[Use] What is known about how many people use it and how much? Have there been formal studies? Why was it written, according to its authors? What were their criticisms of existing solutions?

\end{description}

\noindent In addition, of course, we could perform the same sort of evaluation with regard to debuggers for other idioms -- for imperative and object-oriented languages, for example.

\section{State of the implementations}
\label{stateimpl}

\correction{We are not able to claim that either of our principal efforts are complete, implement the whole OCaml language, or that they are anything other than flawed prototypes. Our first such prototype \textsf{OCamli}, described in \cref{chap:interpreter}, was written to provide enough coverage of the language to allow experimentation with a wide variety of programs (including loading the OCaml Standard Library). It served as a testbed for implementing visualization elision mechanisms, and more traditional debugging tools such as stepping. It is now mothballed in favour of the second implementation for the reasons we described in \cref{chap:efficient,chap:interface}. \addition{This second implementation is only at the proof-of-concept stage. The \textsf{OCamli2} implementation from  \cref{chap:efficient} can only handle small, simple programs. The annotation-based interface described in \cref{chap:interface} is a design only, with but a tiny working example. The work described in chapter 7 is, as we said at the time, abortive.}}

The idea is to finish the second system in two stages: first by making it narrow but deep -- that is to say covering a small part of the language, but fully integrating the mixed compilation/interpretation model described in \cref{chap:interface}. Then, once the technical implementation is secure, we can expand it to the full language. Finally, we can return to the question of better visualization of the output, in particular with regard to formatting and elision. Taken together, we believe that real-world usage, such as debugging the OCaml compiler itself, is not imminent, but we have shown it to be technically feasible.

We believe that, though we have exhibited only early implementations and prototypes thus far, enough insight has been gained that we may be sure of a radically different (and, we hope, successful) OCaml debugger which has a good chance at widespread adoption.

The version of our original \textsf{OCamli} prototype, described in \cite{eptcs} is available on GitHub \cite{ocamli}.

\section{Summary}

We have attempted to evaluate our work whilst it is still in progress. To the extent that it is possible to say, we are confident that \textsf{OCamli} alone is not sufficient to genuinely advance the state of the art in debugging, interesting though it is. We are optimistic that \textsf{ppx\raisebox{0.5mm}{\_}interpret}, when fully implemented, will represent such an advancement.

\chapter{Conclusion}
\label{chap:conclusion}

\begin{quotation}\textit{\large If you want a happy ending, that depends, of course, on where you\\ stop your story. \textrm{\begin{flushright}--- \textup{Orson Welles}\end{flushright}}}\end{quotation}

\vspace{10pt}

\noindent We have looked again at the history and present practice of debugging and tried to identify the essential characteristics which separate debuggers which are likely to be used from those which are likely not to be. Working from these principles, we have described a design for, and early prototype of, a new debugger for the functional language OCaml based on the concept of direct interpretation, and a mechanism for it to be embedded into the build process in such a way that it is always available. We believe it to be promising, but it is too early to say if it really represents a significant step forward -- the problem of debugging is too old and too intransigent to allow us too much confidence.

We have begun the process of evaluating the design insofar as it can be evaluated in the absence of a full, widely-used implementation. As we have discussed, the ultimate test is, of course, whether anyone uses it. So, simply put, our most important item of future work is to provide a complete implementation as a concrete way of supporting (or undermining) our thesis.

Some technical mechanisms we have used to create this debugger are rather specific to OCaml -- to what extent might the insights gained be useful in other languages? It would be interesting to see if this mixture of interpretation and compilation can be applied elsewhere.

We have made initial efforts to preserve the time and space complexity of programs under interpretation, even when the intermediate steps are not shown -- but no theoretical basis or proper proof has been provided. Is it possible to interpret programs in such a way that we can give a guarantee about the time and space complexity?

There are questions of interface too. In our prototype, the user has to choose which part of the program is to be interpreted by annotating the source file, either manually or automatically through an IDE. Could a system be devised where all the parts of the program are available both in compiled and interpreted form at all times, and interpretation can be switched on and off during the debugging session?  Can we automatically insert all breakpoint annotations and choose between them at run-time not compile-time? Zhang et al. have devised a system \cite{zhang2013automated} for automatic breakpoint generation which may be applicable, with some modifications, to the functional realm.

The kind of diagrams our interpreter draws are also useful for teaching, that is to say for testing little programs a student is writing rather than debugging large codebases. It would be interesting to look at how exactly learning to program and debugging are intertwined or equivalent tasks, and see if our step-by-step interpreter, in either or both of its guises, helps beginning students. Besides teaching and debugging, having an interpreter readily available for a language, especially one which ranks equally with the compiler and can be mixed with it at will, may have more uses which we have yet to discover.

\appendix
\chapter{An OCaml primer}
{\setlength\columnsep{15pt}
\begin{multicols}{2}
\label{primer}
\noindent OCaml is a statically-typed strict functional language, with optional imperative features.

\section*{Simple data types}

We have the integers {\small\texttt{min\_int} \ldots\ \texttt{-3}\ \texttt{-2}\ \texttt{-1}\ \texttt{0}\ \texttt{1}\ \texttt{2}\ \texttt{3}\ \ldots\ \texttt{max\_int}} of type \textbf{\textsf{int}}. The booleans are {\small\texttt{true}} and {\small\texttt{false}} of type \textbf{\textsf{bool}}. \noindent Characters have type \textsf{\textbf{char}} and are written like {\small\texttt{\upquote{X}}} and {\small\texttt{\upquote{!}}}.
\vspace{2mm}

\noindent The mathematical operators {\small\texttt{+} \texttt{-} \texttt{*} \texttt{/} \texttt{mod}} take integers and give another. Note that they do not work on floating-point numbers: we have other operators for those.
\begin{eqnarray*}
 & & {\small\texttt{6 * 2}} \\
 \Longrightarrow & & {\small\texttt{12}}
\end{eqnarray*}
\noindent The comparison operators \ \!\! {\small\texttt{=\ \!\!} \texttt{<\ \!\!} \texttt{<=\ \!\!} \texttt{>\ \!\!} \texttt{>=\ \!\!} \texttt{<>}} \ \!\! which compare two values of like type (except functions) and evaluate to either {\small\texttt{true}} or {\small\texttt{false}}.
\begin{eqnarray*}
 & & {\small\texttt{1 + 2 + 3 = 1 * 2 * 3}} \\
 \Longrightarrow & & {\small\texttt{true}}
\end{eqnarray*}

\noindent The conditional expression is {\small\textbf{\texttt{if}}} \textit{expression1} {\small\textbf{\texttt{then}}} \textit{expression2} {\small\textbf{\texttt{else}}} \textit{expression3}, where \textit{expresssion1} has type \textsf{\textbf{bool}} and \textit{expression2} and \textit{expression3} have the same type as one another.
\begin{eqnarray*}
 & & {\small\texttt{\pif 4\! *\! 3\! >\! 2\! *\! 2\! \pthen 1\! \pelse 0}} \\
 \Longrightarrow & & {\small\texttt{1}}
\end{eqnarray*}

\noindent The boolean operators {\small\texttt{\&\&}} (logical AND) and {\small\texttt{||}} (logical OR) allow us to build compound boolean expressions. They are short-circuiting (evaluating their right-hand side only when needed).
\begin{eqnarray*}
 & & {\small\texttt{1 = 2 || 2 = 2}} \\
 \Longrightarrow & & {\small\texttt{true}}
\end{eqnarray*}

\noindent Tuples combine a fixed number of elements {\small\texttt{(a,\! b)},\!    \ \texttt{(a,\! b,\! c)}} etc. with types \textsf{\textbf{$\alpha$ $\times$ $\beta$}},\ \  \textsf{\textbf{$\alpha$ $\times$ $\beta$ $\times$ $\gamma$}} etc. For example, {\small\texttt{(1, \!\textquotesingle \!1\textquotesingle)}} is a tuple of type \textbf{\textsf{int $\times$ char}}. On the screen, OCaml writes {\small\texttt{\textquotesingle a}} for $\alpha$, {\small\texttt{\textquotesingle b}} for $\beta$ etc.

\vspace{2mm}

\noindent Strings are sequences of characters written between double quotes and are of type \textbf{\textsf{string}}. For example, \texttt{"one"} has type \textbf{\textsf{string}}.

\section*{Names and functions}

We can assign a name to the result of evaluating an expression using the {\small\textbf{\texttt{let}}}\ \textit{name} \texttt{=}\ \textit{expression} construct.

\vspace{2mm}
{\small\texttt{\plet x = 5 > 2}}\hfill\textit{x is {\small\texttt{true}}}
\vspace{2mm}

\noindent We can build compound expressions using {\small\textbf{\texttt{let}}}\ \textit{name1} {\small\texttt{=}} \textit{expression1} {\small\textbf{\texttt{in}}}\ {\small\textbf{\texttt{let}}}\ \textit{name2} {\small\texttt{=}}\ \textit{expression2} {\small\textbf{\texttt{in}}}\ \ldots

\vspace{2mm}
$\texttt{\small\plet x = 4 \pin \plet y = 5 \pin x + y}$
\vspace{2mm}

\noindent Functions can be introduced by {\small\textbf{\texttt{let}}}\ \textit{name} \textit{argument1} \textit{argument2} \ldots\ {\small\texttt{=}}\ \textit{expression}. These have type $\alpha \rightarrow \beta$, $\alpha \rightarrow \beta \rightarrow \gamma$ etc. for some types $\alpha$, $\beta$, $\gamma$ etc. For example, {\small\texttt{\plet \!f \!a \!b \!= \!a > \!b}} is a function of type \textsf{\textbf{$\alpha$ $\to$ $\alpha$ $\to$ bool}}.

\vspace{2mm}

\noindent Recursive functions are introduced in the same way, but using {\small\textbf{\texttt{let\!\! rec}}} instead of {\small\textbf{\texttt{let}}}. For example, here is a function \texttt{g} which calculates the smallest power of two greater than or equal to a given positive integer, using the recursive function \texttt{f}:

\vspace{2mm}
\noindent${\small\texttt{\pletrec f x y =}}$\\
${\small\texttt{~~\pif y < x \pthen f x (2 * y) \pelse y}}$\\
\\
${\small\texttt{\plet g z = f z 1}}$
\vspace{2mm}

\noindent Mutually recursive functions are introduced by writing {\small\textbf{\texttt{let\!\! rec \textmd{f x =}\ \ldots\ \ \!\!and \textmd{g y =} \ldots\ and \ldots}}}

\vspace{2mm}

\noindent Anonymous (un-named) functions can be defined like this: {\small\textbf{\texttt{fun}}} \textit{name} \texttt{->} \textit{expression}.
\begin{eqnarray*}
 & & {\small\texttt{(\pfun x -> x * 2) 4}} \\
 \Longrightarrow & & {\small\texttt{8}}
\end{eqnarray*}
\noindent We can make operators into functions using parentheses, for example  {\small\texttt{(\!~<\!~)}} and {\small\texttt{(\!~+~\!)}}.

\vspace{2mm}
$\texttt{\small( + ) 1 2 $\Longrightarrow$ 3}$
\vspace{2mm}

\section*{Pattern matching}

The expression {\small\textbf{\texttt{match}}} \textit{expression1} {\small\textbf{\texttt{with}}} \textit{pattern1} {\small\texttt{|}} \ldots\ {\small\texttt{->}} \textit{expression2} {\small\texttt{|}} \textit{pattern2} {\small\texttt{|}} \ldots\ {\small\texttt{->}} \textit{expression3} \texttt{|}\ \ldots\ matches an expression against a number of patterns in turn, choosing the result expression whose pattern first matches. The expressions \textit{expression2}, \textit{expression3} etc.\ must have the same type as one another, and this is the type of the whole \,{\small\textbf{\texttt{match}  \ldots\ \texttt{with}}}\, expression. The special pattern {\small\texttt{\_}} matches anything.

\vspace{2mm}
\noindent${\small\texttt{\pmatch x \pwith}}$\\
${\small\texttt{~~0 -> 1}}$\\
${\small\texttt{| 1 | 2 -> 3}}$\\
${\small\texttt{| \_ -> 4}}$
\vspace{2mm}

\noindent We may match two or more things at once, using commas to separate as in {\small\texttt{\textbf{match}}\, \texttt{a,\!\! b}\, \textbf{\texttt{with}} \texttt{0,\!\! 0\! ->\! }\! \textit{\normalsize expression1} \texttt{|\ \!x,\!\! y\! ->}\ \ \textit{\normalsize expression2} \texttt{|}} \ldots

\vspace{2mm}
\noindent$\texttt{\small\pmatch x, y, z \pwith}$\\
$\texttt{\small~~0, 0, 0 -> true}$\\
$\texttt{\small | \_, \_, \_ -> false}$

\section*{Lists}
Lists are ordered collections of zero or more elements of like type. They are written between square brackets, with elements separated by semicolons e.g.\ {\small\texttt{[1;\! 2;\! 3;\! 4;\! 5]}}. If a list is non-empty, it has a head, which is its first element, and a tail, which is the list composed of the rest of the elements.

\vspace{2mm}

\noindent The {\small\texttt{::}}\ `cons' operator adds an element to the front of a list. The {\small\texttt{@}} `append' operator concatenates two lists together.

\vspace{2mm}
$\small\texttt{1 ::\ [2; 3] \normalsize $\Longrightarrow\ $\small [1; 2; 3]}$\\
\indent$\small\texttt{[1; 2] @ [3] \normalsize $\Longrightarrow\ $\small [1; 2; 3]}$
\vspace{2mm}

\noindent Lists and the \texttt{::}\ `cons' symbol may be used for pattern matching to distinguish lists of length zero, one, etc. and with particular contents. For example, we can calculate the length of a list:

\vspace{2mm}
\noindent$\small\texttt{\pletrec length l =}$\\
$\small\texttt{~~\pmatch l \pwith}$\\
$\small\texttt{~~~~[] -> 0}$\\
$\small\texttt{~~| \_::t -> 1 + length t}$

\section*{Exceptions}

Exceptions are defined with {\small\textbf{\texttt{exception}}} \textit{name}. They can carry extra information by adding {\small\textbf{\texttt{of}}} \textit{type}. Exceptions are raised with {\small\textbf{\texttt{raise}}}, and handled with {\small\textbf{\texttt{try}}} \ldots\ {\small\textbf{\texttt{with}}}\ \ldots
\vspace{2mm}

\noindent$\texttt{\small\pexception Problem \pof int}$\\
\\
$\texttt{\small\plet f x y =}$\\
$\texttt{\small~~\pif y = 0}$\\
$\texttt{\small~~~~\pthen \praise (Problem x)}$\\
$\texttt{\small~~~~\pelse x / y}$\\
\\
$\texttt{\small\plet g x y =}$\\
$\texttt{\small~~\ptry f x y \pwith Problem p -> p}$

\section*{Partial application}

Functions may be partially applied by giving fewer than the full number of arguments. 

\vspace{2mm}
\noindent$\texttt{\small\plet add x y = x + y}$\\
\\
\noindent$\texttt{\small List.map (add 3) [1; 2; 3]}$\\
$\Longrightarrow$ {\small\texttt{[4; 5; 6]}}\\
\\
\noindent$\small\texttt{List.map (( + ) 3) [1; 2; 3]}$\\
$\Longrightarrow$ {\small\texttt{[4; 5; 6]}}

\section*{New data types}

New types are introduced with \textbf{\small\texttt{type}} \textit{name} {\small\texttt{=}} \textit{constructor1} \textbf{\small\texttt{of}} \textit{type1} {\small\texttt{|}} \textit{constructor2} \textbf{\small\texttt{of}} \textit{type2} {\small\texttt{|}} \ldots\ We may pattern-match on them just as on the built-in types.

\vspace{2mm}
\noindent$\texttt{\small\ptype colour =}$\\
$\texttt{\small~~Red | Blue | Green | Grey \pof int}$\\
\\
$\texttt{\small [Red; Blue; Grey 16]}$\hfill\textit{type \textsf{colour \textbf{list}}}\\
\\
$\texttt{\small\ptype\ \textquotesingle a tree =}$\hfill\textit{a polymorphic type}\\
$\texttt{\small~~ Lf}$\\
$\texttt{\small\ | Br \pof\ \textquotesingle a tree * \textquotesingle a * \textquotesingle a tree}$\\
\\
For example, {\small\texttt{Br\! (Lf,\! \textquotesingle \!X\textquotesingle, \!Br \!(Lf, \!\textquotesingle \!Y\textquotesingle, \!Lf))}} has type \textsf{\textbf{char} tree}. A useful built-in data type is the \textsf{\textbf{option}} type, defined as {\small\texttt{\ptype \!\textquotesingle a \!option \!= \!None \!| \!Some \!\pof \textquotesingle a}}. A type can be polymorphic in more than one type parameter, for example {\small\texttt{(\textquotesingle a, \!\textquotesingle \!b) Hashtbl.t}}, as in the Standard Library.

\section*{Basic input / output}

The value\! {\small\texttt{()}}\! has type \textbf{\textsf{unit}}. Input channels of type \textsf{\textbf{in\raisebox{2pt}{\_}channel}} and output channels of type \textsf{\textbf{out\raisebox{2pt}{\_}channel}} are available. Built-in functions such as {\small\texttt{open\_in}, \texttt{close\_in}, \texttt{open\_out}, \texttt{close\_out}, \texttt{input\_char}, \texttt{output\_char}} etc. exist for reading from and writing to them respectively.

\section*{Mutable state}

References of type \textsf{\textbf{$\alpha$ ref}} are mutable cells containing values. They are built using  {\small\texttt{ref}}, their contents is accessed using {\small\texttt{!}}\! and they are updated by using the {\small\texttt{:=}}\, operator.

\begin{verbatim}
# let p = ref 0;;
val p : int ref = {contents = 0}
# p := 5;;
- : unit = ()
# !p;;
- : int = 5
\end{verbatim}

\noindent Arrays of type \textsf{\textbf{$\alpha$ array}} are written like {\small\texttt{[|1;~\!2;~\!3|]}}. An array is created with the built-in function {\small\texttt{Array.make}}. Elements are accessed with {\small\texttt{a.}\!(}\textit{subscript}\texttt{\!)}, and updated with \texttt{{\small a.\!(}}\textit{subscript}{\small\texttt{\!)\!\! <-}} \textit{expression}.

\vspace{2mm}
\noindent$\small\texttt{\plet swap a x y =}$\\
$\small\texttt{~~\plet t = a.(x) \pin}$\\
$\small\texttt{~~~~a.(x) <- a.(y); a.(y) <- t}$ 
\vspace{2mm}

\noindent An action may be performed many times based on a boolean condition with the \textbf{\small\texttt{while}} \textit{boolean expression} \texttt{\small\textbf{do}}  \textit{expression} \texttt{\small\textbf{done}} construct.

\vspace{2mm}
\noindent$\small\texttt{\pwhile\ !x < y \pdo x := !x * 2 \pdone}$
\vspace{2mm}

\noindent Performing an action a fixed number of times with a varying parameter is achieved using the \textbf{\small\texttt{for}} \textit{name} \texttt{\small=} \textit{start} \texttt{\textbf{\small to}}  \textit{end} \texttt{\small\textbf{do}} \textit{expression} \texttt{\small\textbf{done}} construct.

\vspace{2mm}
\noindent$\small\texttt{\pfor x = 1 \pto 10 \pdo print\_int x \pdone}$

\section*{Floating-point numbers}

Floating-point numbers are written {\small\texttt{min\_float}} \ldots\ {\small\texttt{max\_float}} and have type \textsf{\textbf{float}}. \noindent Floating-point operators \texttt{\small+.} \texttt{\small*.} \texttt{\small-.} \texttt{\small/.} \texttt{\small**} and built-in functions \texttt{\small sqrt} \texttt{\small log} etc. are available.

\vspace{2mm}
\noindent$\small\texttt{2.\ ** 0.2 $\Longrightarrow$ 1.1486983549970351}$

\section*{The OCaml Standard Library}

Functions from the OCaml Standard Library are used with the form \textit{Module}\hspace{-1pt}\texttt{.}\textit{function}. For example, \texttt{\small List.map}, \texttt{\small String.length}, \texttt{\small Array.copy} etc. 

\section*{Simple modules}

Modules are written in \texttt{\small .ml} files. Corresponding interfaces live in \texttt{\small .mli} files. For example, the \texttt{\small .ml} file with contents \texttt{\small\plet \!f \!x \!= \!x \!+ \!1} might have the interface \texttt{\small\pval \!f \!:\ \!int \!-> \!int}

\section*{Compiling programs}

In addition to the \texttt{ocaml} REPL, OCaml has two compilers, \texttt{\small ocamlc} for bytecode, and \texttt{\small ocamlopt} for native code. For example:\\

\noindent\texttt{\small ocamlc\! -o\! x\! x.ml} builds \texttt{\small x} (or \texttt{\small x.exe}) from \texttt{\small x.ml} with the bytecode compiler.\\

\noindent\texttt{\small ocamlopt\! -o\! x\! x.ml} builds \texttt{\small x} (or \texttt{\small x.exe}) from \texttt{\small x.ml} with the native code compiler.
\end{multicols}}

\addition{
\chapter{Test programs}}

\addition{\section*{\texttt{donothing.ml}}}

\noindent\addition{This program does nothing, consisting only of a single value which is immediately returned and ignored. Its purpose is to provide a baseline for the time and space usage of each compiler or interpreter.} 

\bigskip
\begin{verbatim}[commandchars=\\\{\}]
\addition{1}
\end{verbatim}

\addition{\section*{\texttt{factorial.ml}}}

\noindent\addition{Calculates a factorial. The \texttt{\textit{<count>}} in this and other examples is the input number n in our timing tables. It is used to ensure an appropriate running time for accurate measurement.} 

\bigskip
\begin{verbatim}[commandchars=\\\{\}]
\addition{\textbf{let rec} factorial n =}
\addition{  \textbf{if} n = 1 \textbf{then} 1 \textbf{else} n * factorial (n - 1)}
\addition{\textbf{in}}
\addition{  factorial \textit{<count>}}
\end{verbatim}

\addition{\section*{\texttt{factorialacc.ml}}}

\noindent\addition{An iterative, or accumlative version of the factorial function. This requires only constant stack space with an ordinary compiler.} 

\bigskip
\begin{verbatim}[commandchars=\\\{\}]
\addition{\textbf{let rec} factorial a n =}
  \addition{\textbf{if} n = 1 \textbf{then} a \textbf{else} factorial (a * n) (n - 1)}
\addition{\textbf{in}}
  \addition{factorial 1 \textit{<count>}}
\end{verbatim}

\addition{\section*{\texttt{helloworld.ml}}}

\noindent\addition{Prints a string to the screen a number of times.} 

\bigskip
\begin{verbatim}[commandchars=\\\{\}]
\addition{\textbf{for} x = 0 \textbf{to} \textit{<count>} \textbf{do} print\_string "Hello, World!\textbackslash\negthinspace\negthinspace\negthinspace n" \textbf{done}}
\end{verbatim}

\addition{\section*{\texttt{reference\_swap.ml}}}

\noindent\addition{Creates and swaps the values of two references (mutable cells).} 

\bigskip
\begin{verbatim}[commandchars=\\\{\}]
\addition{\textbf{let} swap () =}
\addition{  \textbf{let} x = ref 0 \textbf{in}}
\addition{    \textbf{let} y = ref 1 \textbf{in}}
\addition{      \textbf{let} t = !x \textbf{in}}
\addition{        x := !y;}
\addition{        y := t}

\addition{\textbf{let} \_ = \textbf{for} x = 0 \textbf{to} \textit{<count>} \textbf{do} swap () \textbf{done}}
\end{verbatim}

\addition{\section*{\texttt{exception.ml}}}

\noindent\addition{Raises and catches an exception.} 

\bigskip
\begin{verbatim}[commandchars=\\\{\}]
\addition{\textbf{for} x = 1 \textbf{to} \textit{<count>} \textbf{do}}
\addition{  \textbf{begin try raise} Exit \textbf{with} Exit -> 4 \textbf{end}}
\addition{\textbf{done}}
\end{verbatim}

\addition{\section*{\texttt{table.ml}}}

\noindent\addition{Prints a times-table. We multiply the counts to reach our input size n.}

\bigskip
\begin{verbatim}[commandchars=\\\{\}]
\addition{\textbf{for} x = 1 \textbf{to} \textit{<count>} \textbf{do}}
\addition{  \textbf{for} y = 1 \textbf{to} \textit{<count>} \textbf{do}}
\addition{    print_int (x * y);}
\addition{    print_string "\textbackslash\negthinspace\negthinspace\negthinspace t"}
\addition{  \textbf{done};}
\addition{  print_string "\textbackslash\negthinspace\negthinspace\negthinspace n"}
\addition{\textbf{done}}
\end{verbatim}

\addition{\section*{\texttt{tree.ml}}}

\noindent\addition{Defines a new data type for binary trees, and functions to insert a single item and multiple items, and exercises them. The order of the data is so as to lead to a relatively balanced tree: the pathological case for the algorithm is thus avoided.} 

\bigskip
\begin{verbatim}[commandchars=\\\{\}]
\addition{\textbf{type} 'a tree = Lf | Br \textbf{of} 'a tree * 'a * 'a tree}

\addition{\textbf{let rec} insert t i =}
\addition{  \textbf{match} t \textbf{with}}
\addition{    Lf -> Br (Lf, i, Lf)}
\addition{  | Br (l, i', r) ->}
\addition{      \textbf{if} i < i' \textbf{then} Br (insert l i, i', r) \textbf{else}}
\addition{      \textbf{if} i > i' \textbf{then} Br (l, i', insert r i) \textbf{else}}
\addition{        Br (l, i, r)}
         
\addition{\textbf{let rec} insert_many t vs =}
\addition{  \textbf{match} vs \textbf{with} [] -> t}
\addition{  | x::xs -> insert_many (insert t x) xs}

\addition{\textbf{let} _ =}
\addition{  \textbf{for} x = 1 \textbf{to} \textit{<count>} \textbf{do}}
\addition{    \textbf{let} t =}
\addition{      insert_many Lf}
\addition{        [29; 34; 71; 15; 100; 46; 66; 70; 92; 20; 37; 29;}
\addition{         26; 84; 77; 100; 3; 63; 73; 52; 36; 99; 30; 46;}
\addition{         13; 67; 79; 85; 6; 31; 73; 27; 94; 92; 63; 93;}
\addition{         49; 6; 39; 3; 10; 32; 26; 83; 97; 44; 90; 65; 55;}
\addition{         36; 90; 48; 38; 96; 46; 38; 70; 81; 63; 10; 67;}
\addition{         82; 81; 6; 74; 41; 69; 57; 10; 31; 28; 87; 77; 92;}
\addition{         90; 35; 12; 8; 37; 43; 68; 58; 74; 49; 52; 61;}
\addition{         100; 63; 72; 65; 55; 56; 31; 35; 86; 93; 82; 50;}
\addition{         39; 22]}
\addition{    \textbf{in}}
\addition{      ()}
\addition{    \textbf{done}}
\end{verbatim}

\newpage
\bibliographystyle{apalike}
\normalem
\renewcommand\eminnershape{\itshape\bfseries}
\addcontentsline{toc}{chapter}{Bibliography}
\bibliography{thesis}

\newgeometry{left=3.5cm, right=2.5cm, top=2.5cm, bottom=2.5cm}\thispagestyle{empty}

\phantomsection
\cleardoublepage
\addcontentsline{toc}{chapter}{\indexname}
\printindex

\end{document}